%% file: diss.tex
\documentclass[12pt,a4paper]{book}
\pdfoutput=1
\usepackage{a4wide}
\usepackage{fancyhdr}
\usepackage{graphicx}
\usepackage{booktabs}
\usepackage{cite}
\usepackage{mcite}
\usepackage{amsmath, amsfonts, amssymb}
\usepackage{dsfont}
\usepackage[vcentermath]{youngtab}
\usepackage{slashed}
\usepackage{xcolor}
\allowdisplaybreaks[1]

\usepackage{titlesec}
\titleformat{\part}[display]{\bf\LARGE\filcenter}{\titlerule[1.5pt] \vspace{1pc} \Huge\MakeUppercase{\partname} \thepart}{1pc}{\titlerule[1.5pt] \vspace{1pc} \Huge}
\titleformat{\chapter}[display]{\bf\Large\filcenter}{\titlerule[1pt] \vspace{1pc} \LARGE\MakeUppercase{\chaptertitlename} \thechapter}{1pc}{\titlerule[1pt] \vspace{1pc} \LARGE}

\usepackage[small,bf]{caption}
\def\ci{\mathrm{i}}
\def\e{\mathrm{e}}
\def\dd{\mathrm{d}}

\newcommand{\mone}{\mbox{-1}}
\newcommand{\mtwo}{\mbox{-2}}

\renewcommand{\chaptermark}[1]%
         {\markboth{\thechapter.\ #1}{}}
\renewcommand{\sectionmark}[1]%
         {\markright{\thesection\ #1}}
\newcommand{\LMUTitle}[9]{
  \thispagestyle{empty}
  \vspace*{\stretch{1}}
  {\parindent0cm
   \rule{\linewidth}{.7ex}}
  \begin{flushright}

    \vspace*{\stretch{1}}
    \sffamily\bfseries\Huge
    #1\\
    \vspace*{\stretch{1}}
    \sffamily\bfseries\large
    #2
    \vspace*{\stretch{1}}
  \end{flushright}
  \rule{\linewidth}{.7ex}
  \vspace*{\stretch{5}}
  \begin{center}
    \includegraphics[width=2in]{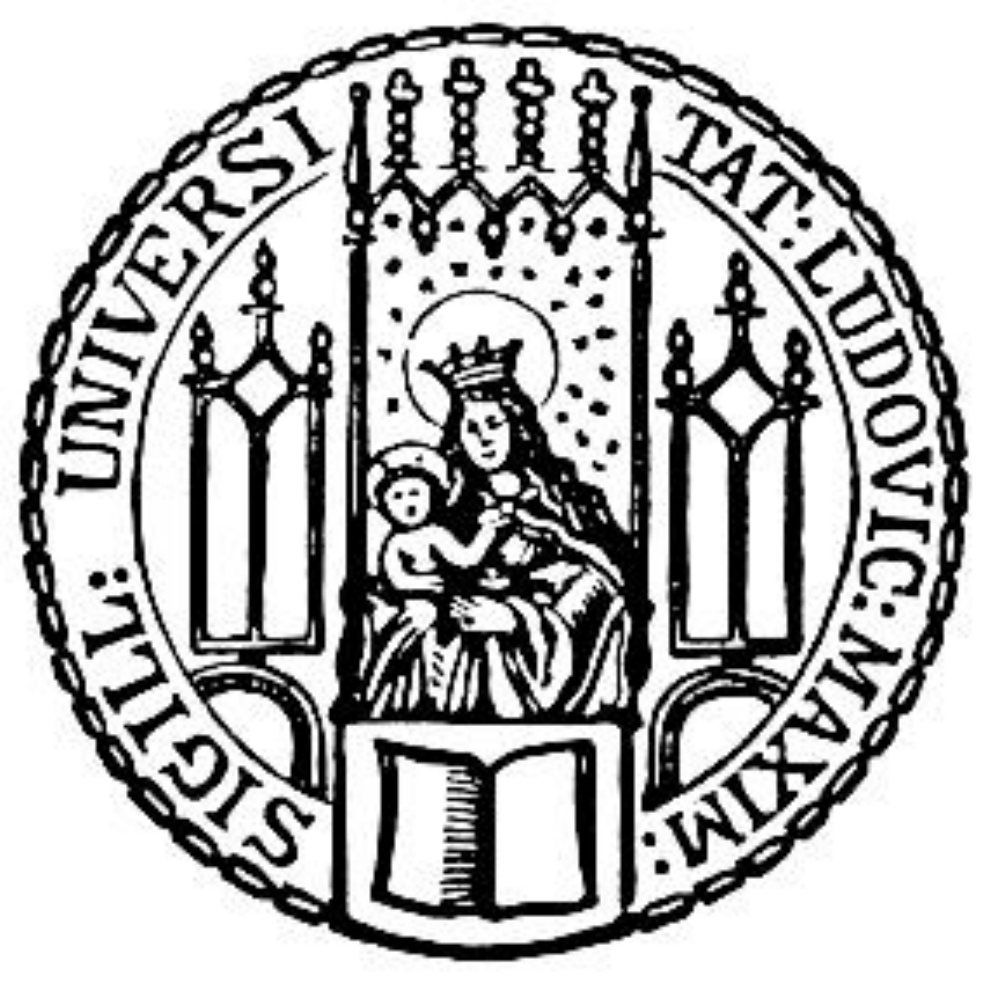} \hspace{2cm}
    \includegraphics[width=2in]{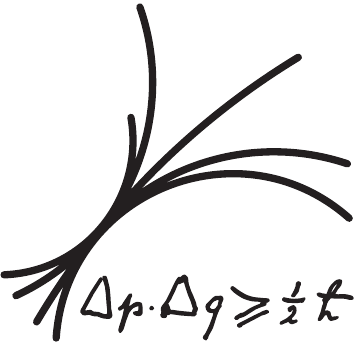}
  \end{center}
  \vspace*{\stretch{1}}
  \begin{center}\sffamily\LARGE{#5}\end{center}
  \newpage
  \thispagestyle{empty}

  \cleardoublepage
  \thispagestyle{empty}

  \vspace*{\stretch{1}}
  {\parindent0cm
  \rule{\linewidth}{.7ex}}
  \begin{flushright}
    \vspace*{\stretch{1}}
    \sffamily\bfseries\Huge
    #1\\
    \vspace*{\stretch{1}}
    \sffamily\bfseries\large
    #2
    \vspace*{\stretch{1}}
  \end{flushright}
  \rule{\linewidth}{.7ex}

  \vspace*{\stretch{3}}
  \begin{center}
    \Large Dissertation\\
    \Large an der #4\\
    \Large der Ludwig--Maximilians--Universit\"at\\
    \Large M\"unchen\\
    \vspace*{\stretch{1}}
    \Large vorgelegt von\\
    \Large #2\\
    \Large aus #3\\
    \vspace*{\stretch{2}}
    \Large M\"unchen, den #6
  \end{center}

  \newpage
  \thispagestyle{empty}

  \vspace*{\stretch{1}}

\begin{large}
\noindent This thesis is based on the author's work partly published in \cite{Antusch:2008tf, Antusch:2009gu, Antusch:2009hq, Antusch:2010es} conducted from December 2007 until May 2010 at the Max--Planck--Institut f\"ur Physik (Werner--Heisenberg--Institut), M\"unchen, under the supervision of Dr.\ Stefan Antusch.
\end{large}

  \vspace{36pt}

  \begin{flushleft}
    \large Erstgutachter:  #7 \\[1mm]
    \large Zweitgutachter: #8 \\[1mm]
    \large Tag der m\"undlichen Pr\"ufung: #9\\
  \end{flushleft}

  \cleardoublepage
}

\begin{document}

   \frontmatter

   \LMUTitle
      {New Aspects of Flavour Model Building in Supersymmetric Grand Unification}               
      {Martin Spinrath}                       
      {M\"onchengladbach}                             
      {Fakult\"at f\"ur Physik}                         
      {M\"unchen 2010}                          
      {19. Mai 2010}                            
      {PD Dr. Georg Raffelt}                          
      {Prof. Dr. Gerhard Buchalla}                         
      {23. Juli 2010}                         

  \tableofcontents
   \markboth{Table of Contents}{Table of Contents}

  \listoffigures
   \markboth{List of Figures}{List of Figures}

  \listoftables
   \markboth{List of Tables}{List of Tables}
   \cleardoublepage

   \include{zusammenfassung}
  \markboth{Zusammenfassung}{Zusammenfassung}

   \include{abstract}
  \markboth{Abstract}{Abstract}

  \mainmatter\setcounter{page}{1}
 \part{Introduction}
  \include{kap_01_Introduction}

 \part{Theoretical Framework}
  \include{kap_02_SM}
  \include{kap_03_SUSY}

  \include{kap_04_GUTs}

 \part[New GUT Predictions for Yukawa Coupling Ratios]{New GUT Predictions for\\ Yukawa Coupling Ratios}
  \include{kap_05_NewGUTRelations}

 \part{The Case of Medium and Large $\boldsymbol{\tan \beta}$}
  \include{kap_06_SUSYThresholdCorrections}

  \include{kap_07_FirstGlance}

  \include{kap_08_Pheno}

  \include{kap_09_GUTImplications}

 \part{The Case of Small $\boldsymbol{\tan \beta}$}
  \include{kap_10_Textures}

  \include{kap_11_Model}

 \part{Summary and Conclusions}
  \include{kap_12_SummaryConclusions}

 \part{Appendix}
  \appendix
  \include{app_01_Conventions}

  \include{app_02_SUNReps}
  \include{app_03_Plots}

  \include{app_04_Discrete}

  \backmatter
  \addcontentsline{toc}{chapter}{Bibliography}
  \bibliographystyle{JHEPmod}
  \bibliography{diss}
  \markboth{}{}

  \include{danksagung}

\end{document}

%% file: zusammenfassung.tex
\addcontentsline{toc}{chapter}{\protect Zusammenfassung}
\chapter*{Zusammenfassung}

Wir leiten Vorhersagen f\"ur Verh\"altnisse von Yukawa-Kopplungen in Gro\ss{}en Ver\-ein\-heit\-lichten Theorien her, die von Operatoren mit Massendimension vier und f\"unf erzeugt werden. Diese Relationen sind eine charakteristische Eigenschaft vereinheitlichter Flavourmodelle und k\"onnen die gro\ss{}e Anzahl freier Parameter im Flavoursektor des Standard Modells reduzieren.

Die Yukawa-Kopplungen der down-artigen Quarks und geladenen Leptonen erhalten in supersymmetrischen Modellen $\tan \beta$-verst\"arkte Schwellenkorrekturen, die gro\ss{} sein k\"onnen, wenn $\tan \beta$ gro\ss{} ist. In diesem Fall ist ihre sorgf\"altige Einbeziehung in die Renormierungsgruppenentwicklung obligatorisch. Wir analysieren diese Korrekturen und geben einfache analytische Ausdr\"ucke und numerische Absch\"atzungen f\"ur sie an.

Die Schwellenkorrekturen h\"angen empfindlich von den soften Supersymmetrie brechenden Parametern ab. Insbesondere bestimmen sie das globale Vorzeichen der Korrekturen und damit, ob die betroffenen Yukawa-Kopplungen verst\"arkt oder unterdr\"uckt werden. In der minimalen supersymmetrischen Erweiterung des Standard Modells f\"uhrt Supersymmetrie Brechung viele freie Parameter ein, \"uber die wir in unserem ersten, vereinfachten Ansatz einige plausible Annahmen machen. In einem zweiten, verfeinerten Ansatz verwenden wir stattdessen drei verbreitete Brechungsschemata, in denen alle soften Brechungsparameter an der elektroschwachen Skala aus einer Handvoll Parameter berechnet werden. In diesem Ansatz wenden wir verschiedene ph\"anomenologische Einschr\"ankungen auf die supersymmetrischen Parameter an und finden auf diese Weise neue zul\"assige Relationen f\"ur die Yukawa-Kopplungen, zum Beispiel $y_\mu/y_s = 9/2$ bzw.\ $6$ oder $y_\tau/y_b = 3/2$ in $SU(5)$.

Weiterhin untersuchen wir eine spezielle Klasse von Texturen von Quark-Massen\-ma\-tri\-zen f\"ur kleine $\tan \beta$, in denen $\theta_{13}^u = \theta_{13}^d = 0$. Wir leiten Summenregeln f\"ur die Quark-Mischungs\-para\-meter her und finden eine einfache Relation zwischen den beiden Phasen $\delta_{12}^u$ und $\delta_{12}^d$  und dem rechten Winkel $\alpha$ im Unit\"aritatsdreieck, die eine einfache Phasenstruktur f\"ur die Quark-Massenmatrizen suggeriert. Hierbei ist ein Matrixelement rein imagin\"ar und die restlichen rein reell.

Um die vorhergehenden \"Uberlegungen abzurunden, geben wir zwei explizite Flavourmodelle im $SU(5)$ Kontext an, eines f\"ur gro\ss{}e und eines f\"ur kleine $\tan \beta$, welche die zuvor erw\"ahnten Verh\"altnisse der Yukawa-Kopplungen implementieren. Die Modelle haben interessante ph\"anomenologische Konsequenzen wie zum Beispiel quasi-entartete Neutrinomassen im Fall kleiner $\tan \beta$.

%% file: abstract.tex
\addcontentsline{toc}{chapter}{\protect Abstract}
\chapter*{Abstract}

We derive predictions for Yukawa coupling ratios within Grand Unified Theories generated from operators with mass dimension four and five. These relations are a characteristic property of unified flavour models and can reduce the large number of free parameters related to the flavour sector of the Standard Model.

The Yukawa couplings of the down-type quarks and charged leptons are affected within supersymmetric models by $\tan \beta$-enhanced threshold corrections which can be sizeable if $\tan \beta$ is large. In this case their careful inclusion in the renormalisation group evolution is mandatory. We analyse these corrections and give simple analytic expressions and numerical estimates for them.

The threshold corrections sensitively depend on the soft supersymmetry breaking parameters. Especially, they determine the overall sign of the corrections and therefore if the affected Yukawa couplings are enhanced or suppressed. In the minimal supersymmetric extension of the Standard Model many free parameters are introduced by supersymmetry breaking about which we make some plausible assumptions in our first simplified approach. In a second, more sophisticated approach we use three common breaking schemes 
in which all the soft breaking parameters at the electroweak scale can be calculated from only a handful of parameters. Within the second approach, we apply various phenomenological constraints on the supersymmetric parameters and find in this way new viable Yukawa coupling relations, for example $y_\mu/y_s = 9/2$ or $6$ or $y_\tau/y_b = 3/2$ in $SU(5)$.

Furthermore, we study a special class of quark mass matrix textures for small $\tan \beta$ where $\theta_{13}^u = \theta_{13}^d = 0$. We derive sum rules for the quark mixing parameters and find a simple relation between the two phases $\delta_{12}^u$ and $\delta_{12}^d$ and the right unitarity triangle angle $\alpha$ which suggests a simple phase structure for the quark mass matrices where one matrix element is purely imaginary and the remaining ones are purely real.

To complement the aforementioned considerations, we give two explicit flavour models in a $SU(5)$ context, one for large and one for small $\tan \beta$ which implement the Yukawa coupling relations mentioned before. The models have interesting phenomenological consequences like, for example, quasi-degenerate neutrino masses in the case of small $\tan \beta$.

%% file: kap_01_Introduction.tex
\chapter{Introduction} \label{Ch:Introduction}

The Standard Model of elementary particle physics (SM) \cite{Glashow:1961tr, *Weinberg:1967tq, *Salam:1968rm, *Glashow:1970gm, Fritzsch:1973pi, Gross:1973ju, *Gross:1974cs, *Politzer:1973fx} has been tested to high accuracy and successfully describes the electroweak and strong interactions of all observed particles \cite{Amsler:2008zzb}. Despite these great achievements there remain unresolved issues such as the hierarchy problem, the non-unification of gauge couplings or the unknown source of dark matter in the universe, which point to new physics beyond the SM.

One possible solution to these problems is the idea of low energy supersymmetry (SUSY) \cite{Wess:1973kz, *Volkov:1973ix, Wess:1974tw}. In principle, it can solve all aforementioned issues. It stabilises the electroweak scale \cite{Witten:1981nf, *Kaul:1981hi}, changes the renormalisation group (RG) evolution of the gauge couplings in such a way that the gauge couplings almost exactly unify at a high energy scale \cite{Ibanez:1981yh, *Dimopoulos:1981yj} and the lightest SUSY particle is a viable candidate for dark matter as long as it does not carry colour or electromagnetic charge \cite{Ellis:1983ew, Hinshaw:2008kr}. Hence a supersymmetric extension of the SM is quite appealing, but it has yet to be proven that nature is indeed supersymmetric.

The main focus of this thesis, however, lies on a different intriguing puzzle of the SM, the origin and nature of quark and lepton masses and mixings. Within the SM, quark and lepton masses and mixings arise from Yukawa couplings which are essentially free and undetermined due to basis ambiguities. In this context, it is particularly challenging to explain the strong hierarchy among the masses of the three families of quarks and charged leptons as well as the strong suppression of the neutrino masses and the fact that quark mixings are small whereas there is large mixing between mass and flavour eigenstates in the lepton sector.

One hint towards a solution to this puzzle is the idea of  unification of the fundamental forces of the SM. In addition to providing a unified origin of the gauge interactions, Grand Unified Theories (GUTs) based, e.g.\ on the gauge symmetry groups $SU(5)$ \cite{Georgi:1974sy} or $SO(10)$ \cite{Georgi:1974my, *Fritzsch:1974nn} also unify quarks and leptons of the SM in representations of the unified gauge groups. This property makes them attractive frameworks to address the flavour puzzle since therein the Yukawa couplings within a particular family can be related. Indeed, an interesting observation in this context is that in supersymmetric theories with large  $\tan \beta$, the Yukawa couplings of the up-type quark, down-type quark and charged lepton of each generation are of similar order of magnitude at the GUT scale which is a prediction in a wide class of GUTs.

In (SUSY) GUTs the Yukawa couplings for different types of fermions of one generation can be generated from common operators involving the GUT representations. After GUT symmetry breaking the resulting Yukawa couplings typically have similar values. Depending on the specific operator, the group theoretical Clebsch--Gordan (CG) factors from GUT symmetry breaking can even lead to predictions for the ratios between the Yukawa couplings, see, e.g.\  \cite{Georgi:1979df, Antusch:2009gu}. Such relations, after evolving them from the GUT scale to low energies via their renormalisation group equations (RGEs) and including threshold effects \cite{Hall:1993gn, Blazek:1995nv, Carena:1994bv, *Hempfling:1993kv, Antusch:2008tf}, can be compared to experimental results for the quark masses and provide crucial tests of unified models of fermion masses and mixings in a top-down approach.

Another attractive feature of left-right symmetric GUTs is the appearance of right-handed neutrinos in their particle spectra, which become massive after spontaneous symmetry breaking to the SM and thereby lead to the small observed neutrino masses via the seesaw mechanism \cite{Minkowski:1977sc, *GellMann:1980vs, *Yanagida:1979as, *Glashow:1979nm, *Mohapatra:1980yp, *Schechter:1981cv}. In order to make the running gauge couplings meet at the so-called GUT scale $M_{\mathrm{GUT}} \approx 2 \times 10^{16}$ GeV, the idea of Grand Unification is typically combined with low-energy SUSY.

From the bottom-up perspective it is desirable to know the approximate GUT scale values of the quark and lepton masses and mixing parameters in order to construct successful GUT models of flavour. The experimental data on the masses of the strange quark and the muon, for example, extrapolated to the GUT scale by means of renormalisation group (RG) running within the SM, give rise to the so-called Georgi--Jarlskog (GJ) relations \cite{Georgi:1979df} $y_\mu/y_s = 3$ and $y_e/y_d = 1/3$ at the GUT scale. This can be realised from a CG factor after GUT symmetry breaking. The GJ relations have become a popular building block in many classes of unified flavour models.

In SUSY GUTs another intriguing possibility emerges, which is the unification of all third family Yukawa couplings, i.e.\ of $y_t$, $y_b$, $y_\tau$ and furthermore, in the context of the seesaw mechanism, of $y_\nu$ at the GUT scale.
It is well known that a careful inclusion of SUSY threshold corrections is required \cite{Hall:1993gn, Blazek:1995nv, Carena:1994bv, *Hempfling:1993kv, Antusch:2008tf}, in order to investigate whether this relation can be realised in a given model of low-energy SUSY.
These threshold effects are particularly relevant in the case of large $\tan \beta$ where $y_t = y_b = y_\tau$ seems achievable. Despite the possible importance of SUSY threshold effects, these effects are often ignored in studies which extrapolate the running fermion masses to the GUT scale, see, e.g.\  \cite{Fusaoka:1998vc,Xing:2007fb}.  In this work we include the SUSY threshold corrections and derive alternative Yukawa coupling relations besides the GJ relation and (partial) third family Yukawa coupling unification compatible with wide ranges of the SUSY parameter space in three common SUSY breaking schemes.

Besides SUSY GUTs there have to be more additional ingredients for a fundamental theory of nature. The mass hierarchy between different families is not yet explained and SUSY GUTs do not shed any light on this question either. Indeed, in the SM or GUTs, with or without SUSY, a specific structure of the Yukawa matrices has no intrinsic meaning due to basis ambiguities in flavour space. For example, one can always work in a basis in which, say, the up quark mass matrix is taken to be diagonal with the quark mixing arising entirely from the down quark mass matrix, or vice versa, and analogously in the lepton sector. This is symptomatic of the fact that neither the SM nor GUTs are candidates for a complete theory of flavour.

The situation changes somewhat once these theories are extended to include a family symmetry spontaneously broken by extra Higgs fields called flavon fields. These family symmetries can allow Yukawa couplings, in particular for the first and second generation, to be generated only via higher-dimensional effective operators leading to a certain suppression compared to apparently natural values of order one. Besides that,
there is an ad hoc approach for the structure of Yukawa coupling matrices pioneered some time ago by Fritzsch \cite{Fritzsch:1979zq,Fritzsch:1999ee} and currently represented by myriads of proposed effective Yukawa textures, see, e.g.\  \cite{Fritzsch:1979zq, Fritzsch:1999ee, Leontaris:2009pi, *Dev:2009he, *Adhikary:2009kz, *Goswami:2009bd, *Goswami:2008uv, *Choubey:2008tb, *Branco:2007nb, *Alhendi:2007iu, *Kaneko:2007ea, *Branco:2006wv, *Lam:2006wm, *Kaneko:2006wi, *Fuki:2006xw, *Haba:2005ds, *Kim:2004ki, *Jack:2003pb, *Jack:2003qg, *Caravaglios:2002br, *Everett:2000up, *Berezhiani:2000cg, *Kuo:1999dt, *Falcone:1998us, Roberts:2001zy, *Ramond:1993kv, Chiu:2000gw,Fritzsch:1999rb}. The starting assumption there is that (in some basis) the Yukawa matrices exhibit certain nice features such as symmetries or zeros in specific elements which have become known as \emph{texture zeros} and which we adopt later on to derive relations between quark masses and mixing parameters \cite{Antusch:2009hq}.

Furthermore, besides the discovery of neutrino masses and mixing, which may even be termed a \emph{neutrino revolution}, there are many questions about neutrinos which remain unanswered. Perhaps the most pressing of them is the origin, nature and magnitude of neutrino masses, since neutrino oscillations only provide information about the squared mass differences between neutrino species. These are independent of the absolute neutrino mass scale or the nature of the neutrino mass, i.e.\ Dirac or Majorana. In the absence of any confirmed experimental signal from either beta decay end-point experiments or neutrinoless double beta decay experiments, the most stringent limits on the absolute neutrino mass scale come, indirectly, from cosmology, where one typically obtains a limit on the absolute neutrino mass scale expressed in terms of the lightest neutrino mass as $m_{\mathrm{lightest}} \lesssim 0.2$ eV \cite{Jarosik:2010iu}. Thus, there remains the interesting possibility that neutrinos are quasi-degenerate, which one may roughly define as $m_{\mathrm{lightest}} > 0.05$ eV, where the lower limit is approximately set equal to the square root of the atmospheric neutrino mass squared difference.

In addition to the question of neutrino masses and their hierarchy, also the question of the neutrino mixing pattern arises. The phenomenologically viable idea of approximate tri-bimaximal (TB) lepton mixing \cite{Harrison:2002er, *Harrison:2003aw} strongly suggests that some kind of non-Abelian discrete family symmetry is at work. The observed symmetry may arise either directly or indirectly from a range of discrete symmetry groups \cite{King:2009ap, Altarelli:2010gt}.

If we assume a GUT-type structure relating quarks and leptons at a certain high energy scale, the mixings in the lepton and the quark sector are related to each other. For example, the SUSY GUT model based on $SO(10)$ with family symmetry $PSL(2,7)$ proposed in \cite{King:2009tj} is based on the type II seesaw mechanism, leads to TB mixing and allows quasi-degenerate neutrinos. On the other hand, the SUSY $A_4$ model in \cite{Altarelli:2005yx} based on the type I seesaw mechanism also leads to TB mixing and allows for quasi-degenerate neutrinos.

More generally, there is a huge literature on family symmetry models based on $A_4$ \cite{Ma:2001dn, Ma:2004zv, *Altarelli:2005yp, *Ma:2005qf, *Ma:2005mw, *Ma:2005sha, *Chen:2005jm, *Altarelli:2006kg, *Ma:2006vq, *Ma:2006wm,  *Ma:2006sk, *Adhikary:2006wi, *He:2006et, *King:2006np, *Ma:2007ku,  *Feruglio:2007uu,  *Chen:2009um, Altarelli:2008bg, Burrows:2009pi, Ciafaloni:2009ub, *Ciafaloni:2009qs}
or other symmetries \cite{Barbieri:1999km, *Xing:2002sw, *King:2003rf, *deMedeirosVarzielas:2005qg, *Kang:2005bg, *Luo:2005fc, *Xing:2005ur, *Xing:2006xa, *Haba:2006dz, *He:2006qd, *Hirsch:2006je, *King:2006me, *Mohapatra:2006pu, *Mohapatra:2006se, *Singh:2006dr, *deMedeirosVarzielas:2006fc, *Ma:2006ip, *Aranda:2007dp, *Luhn:2007sy, *Ma:2007wu, *Chan:2007ng,  *Chen:2007afa, *Chen:2007gp, *Chen:2008eq, *Plentinger:2008nv, *Csaki:2008qq, *Lin:2008aj, *Lam:2009hn, *Lin:2009ic, *King:2009mk,  *Hagedorn:2010th, *Cooper:2010ik, deMedeirosVarzielas:2005ax, Antusch:2004xd, King:2005bj} (with or without Grand Unification), some of which can accommodate TB mixing with quasi-degenerate neutrinos. However, to our knowledge, in all of the above examples the prediction of $m_{ee}$ as a function of the lightest neutrino mass is subject to phase uncertainties. This is resolved in our  flavour models \cite{Antusch:2010xx, Antusch:2010es}.

The thesis is organised as follows: In Part II we introduce the theoretical framework relevant for this thesis. We give a brief overview of the SM in Ch.~\ref{Ch:SM} with an emphasis on the flavour sector and we also discuss some of the open problems of the SM relevant for this work. After a short motivation we review some basic formal aspects of SUSY, introduce the Minimal Supersymmetric extension of the SM (MSSM) and give a short introduction to SUSY breaking  in Ch.~\ref{Ch:SUSY}. Afterwards, in Ch.~\ref{Ch:GUTs}, we explain basic features of GUTs based on the $SU(5)$ model.

In Part III We extend the approach by Georgi and Jarlskog \cite{Georgi:1979df} to higher dimensional operators and propose new candidate relations for Yukawa couplings at the GUT scale. To be more concrete, we discuss dimension five operators which can generate effective Yukawa couplings in $SU(5)$ and Pati--Salam (PS) \cite{Pati:1973uk} theories. These operators give alternative Yukawa coupling ratios besides the ones already known from dimension four operators.

GUT scale Yukawa couplings in the MSSM for medium and large $\tan \beta$ are discussed in Part IV. The inclusion of SUSY threshold corrections, which are discussed in Ch.~\ref{Ch:SUSYThresholdCorrections}, is of special importance in this $\tan \beta$ regime. In Ch.~\ref{Ch:AFirstGlance} we implement the SUSY threshold corrections in the RG evolution  of the Yukawa couplings in a somewhat simplified approach and cast a first glance at GUT scale Yukawa couplings. This provides the motivation for the more sophisticated approach in Ch.~\ref{Ch:Pheno} where we calculate full spectra in order to phenomenologically constrain viable GUT scale Yukawa ratios. This part concludes with a discussion of possible implications for alternative GUT scale Yukawa coupling ratios in Ch.~\ref{Ch:GUTImplications} where we also present a concrete application in form of a flavour model.

Part V focuses on the discussion of small $\tan \beta$. We start this discussion with sum rules for the quark mixing angles and phases and their relation to the unitarity triangle for a special class of quark mass textures in Ch.~\ref{Ch:Textures}. After that we present a predictive flavour model for small $\tan \beta$ in Ch.~\ref{Ch:Model}.

We summarise and conclude in Part VI.

The appendix can be found in Part VII where we fix notations and conventions in App.~\ref{App:Conventions} and discuss representations of $SU(N)$ in App.~\ref{App:SUNReps}. In App.~\ref{App:Plots} we give detailed plots of our phenomenological scan described in Ch.~\ref{Ch:Pheno}. The appendix is concluded with a brief discussion of the discrete symmetry groups $\mathbb{Z}_n$ and $A_4$ in App.~\ref{App:Discrete} which are used in our flavour models.

%% file: kap_02_SM.tex
\chapter{The Standard Model of Particle Physics} \label{Ch:SM}

The Standard Model of particle physics (SM) \cite{Glashow:1961tr, *Weinberg:1967tq, *Salam:1968rm, *Glashow:1970gm, Fritzsch:1973pi, Gross:1973ju, *Gross:1974cs, *Politzer:1973fx} is one of the most precisely tested physical theories \cite{Amsler:2008zzb}. It successfully describes all known elementary particles and their interactions to high accuracy except for gravity. In the following a short overview of the basic ingredients of the SM which are relevant for this work are given.

\section{Gauge Interactions and Field Content}

The SM is a renormalisable quantum-field theory whose gauge interactions are described by the SM gauge group $G_{\mathrm{SM}}$, which is a direct product of three groups $G_{\mathrm{SM}} = SU(3)_C \times SU(2)_L \times U(1)_Y$ and the space-time symmetry of the Poincar\'{e} group.

The strong interactions of quantum chromodynamics (QCD) are symmetric under $SU(3)_C$ transformations while the electroweak (EW) interactions obey the $SU(2)_L \times U(1)_Y$ symmetry. The generators of the non-Abelian groups $SU(3)_C$ and $SU(2)_L$ can be written as Hermitian matrices that fulfil the commutation relations of the corresponding Lie algebra. Since $U(1)_Y$ is Abelian and has only one generator the corresponding commutation relation is trivial. If we denote the generators of $SU(3)_C$ by $T^a$, $a=1, \ldots, 8$, the generators of $SU(2)_L$ with $I^i$, $i=1,\ldots,3$ and the generator of $U(1)_Y$ with $Y$, then the commutation relations are
\begin{equation}
 [ T^a, T^b ] = \ci f^{abc} T^c \;, \quad [ I^i, I^j ] = \ci \epsilon^{ijk} I^k \;, \quad \left[ Y, Y \right] = 0 \;,
\end{equation}
where the real totally antisymmetric tensors $f^{abc}$ and $\epsilon^{ijk}$  are the structure constants of the corresponding Lie algebras.

\begin{table}
\centering
\begin{tabular}{ccc}
\toprule
 Name & Particles & Quantum Numbers \\ \midrule 
Quarks  & $\begin{pmatrix} u \\ d \end{pmatrix}_L$, $\begin{pmatrix} c \\ s \end{pmatrix}_L$, $\begin{pmatrix} t \\ b \end{pmatrix}_L$ & $\left( \mathbf{3}, \mathbf{2}, +\frac{1}{3} \right)$  \\[1pc]
& $u^\dagger_R$, $c^\dagger_R$, $t^\dagger_R$  & $\left( \mathbf{\overline{3}}, \mathbf{1}, -\frac{4}{3} \right)$  \\[0.4pc]
& $d^\dagger_R$, $s^\dagger_R$, $b^\dagger_R$  & $\left( \mathbf{\overline{3}}, \mathbf{1}, +\frac{2}{3} \right)$  \\
\midrule 
Leptons & $\begin{pmatrix} \nu_e \\ e \end{pmatrix}_L$, $\begin{pmatrix} \nu_\mu \\ \mu \end{pmatrix}_L$, $\begin{pmatrix} \nu_\tau \\ \tau \end{pmatrix}_L$ & $\left( \mathbf{1}, \mathbf{2}, -1 \right)$  \\[1pc]
& $e^\dagger_R$, $\mu^\dagger_R$, $\tau^\dagger_R$  & $\left( \mathbf{1}, \mathbf{1}, +2 \right)$  \\
\midrule
Higgs Field & $H$  & $\left( \mathbf{1}, \mathbf{2}, +1 \right)$  \\
\midrule 
Gauge Bosons & $g$ & $\left(\mathbf{8},\mathbf{1},0\right)$ \\
& $W$ & $\left(\mathbf{1},\mathbf{3},0\right)$ \\
& $B$ & $\left(\mathbf{1},\mathbf{1},0\right)$ \\
\bottomrule
\end{tabular}
\caption[SM Field Content]{Field content of the SM. The representations and charges are given in the order $(SU(3)_C, SU(2)_L, U(1)_Y)$. While the quarks and leptons are fermions with spin $1/2$, the Higgs and the gauge fields are bosons with spin $0$ respectively spin $1$. \label{Tab:SMfields}}
\end{table}

Every particle of the SM can be written as a tensor with certain symmetry properties under $SU(3)_C$ and $SU(2)_L$ transformations. These properties define the representation of the field under the non-Abelian gauge groups. In App.~\ref{App:SUNReps} we give more details on these kind of representations. Here we classify the fields only according to the dimension of their representation of $SU(3)_C$ and $SU(2)_L$ and their hypercharge $q_Y$ which gives the transformation properties under the $U(1)_Y$ group. For example the left-handed doublet quarks are $SU(3)_C$ triplets $\mathbf{3}$ and $SU(2)_L$ doublets $\mathbf{2}$ with hypercharge $q_Y = 1/3$. The complete field content of the SM is listed in Tab. \ref{Tab:SMfields}. There we also see that quarks and leptons appear in each case in three copies, the so-called families or generations, which have the same quantum numbers and as we see later differ only in their Yukawa couplings.  The operator of electromagnetic charge is given by the Gell-Man--Nishijima relation
\begin{equation}
 Q = I^3 + \frac{Y}{2} \;.
\end{equation}

In a gauge theory with an unbroken symmetry the gauge bosons have to be strictly massless since a mass term for them is forbidden by the underlying symmetry. However, this turns out not to be the case for the EW symmetry. For example the $W^{\pm}$ bosons have a mass of  80.4~GeV \cite{Amsler:2008zzb, Aaltonen:2007ps}. In the SM we assume that the EW gauge symmetry $SU(2)_L \times U(1)_Y$ is \emph{spontaneously} broken by a non-vanishing vacuum expectation value (vev) of a scalar field, i.e.\ the so-called Higgs field.

We assume that the Higgs potential is of the form
\begin{equation}
 V(H) = \mu^2 H^\dagger H + \lambda (H^\dagger H)^2 \;,
\end{equation}
which is renormalisable and invariant under $SU(2)_L \times U(1)_Y$ transformations. For $\mu^2 < 0$ the minimum of the Higgs potential is not at the origin of the field space but shifted to a finite value. Since we observe the $U(1)_{\mathrm{em}}$ symmetry of electromagnetism in nature we have to demand that this symmetry remains unbroken. This can be achieved by giving the electrically neutral component of the Higgs doublet a non-vanishing vev $v$
\begin{equation}
 Q \, \langle 0 \vert H \vert 0 \rangle  = \left( I^3 + \frac{Y}{2} \right) \begin{pmatrix} 0 \\ v/\sqrt{2} \end{pmatrix} = 0 \;.
\end{equation}
By replacing the original Higgs field in the Lagrangian with a Higgs field expanded around this minimum the original invariance of the Lagrangian under $SU(2)_L \times U(1)_Y$ disappears. The ground state of the system does not respect the EW symmetry. Only the (smaller) symmetry of electromagnetism $U(1)_{\mathrm{em}}$ remains.

After electroweak symmetry breaking (EWSB) the original $SU(2)_L \times U(1)_Y$ quantum numbers have no meaning anymore. Under the  $U(1)_{\mathrm{em}}$ subgroup one of the components of the $SU(2)_L$ gauge boson triplet and the $B$ boson have the same quantum numbers and therefore can mix. Conveniently this is written as
\begin{equation}
 \begin{pmatrix} Z \\ \gamma \end{pmatrix} = \begin{pmatrix} \cos \theta_W & - \sin \theta_W \\ \sin \theta_W & \cos \theta_W \end{pmatrix} \begin{pmatrix} W^3 \\ B \end{pmatrix} \;, \quad W^{\pm} = \frac{1}{\sqrt{2}} (W^1 \mp \ci \, W^2) \;,
\end{equation}
where on the left-hand sides of the equations the mass (and electric charge) eigenstates are given and on the right-hand side the states in the unbroken EW theory. The weak mixing angle is defined as $\cos \theta_W = g_2/\sqrt{g_2^2 + g'^2}$ with $g_2$ being the coupling constant of the $SU(2)_L$ gauge theory and $g'$ being the coupling constant of the $U(1)_Y$ gauge theory. The charged $W^\pm$ bosons acquire the mass $M_W = (v/2) g_2$ and the $Z$ boson acquires the mass $M_Z = (v/2) \sqrt{g_2^2 + g'^2}$ while the photon $\gamma$ remains massless as it should be. The weak mixing angle is also related to the gauge boson masses via $\cos \theta_W = M_W/M_Z$.

\section{The Flavour Sector of the Standard Model}

The origin of the masses of the gauge bosons was discussed in the last section. Now we turn to the fermion masses. There are two bases for the fermions. On the one hand the mass eigenstates and on the other hand the gauge eigenstates which define the \emph{flavour} of a given particle. In the following we discuss the difference between these two bases. There are six quark flavours: up, down, charm, strange, top and bottom and six lepton flavours: electron, muon, tau and the corresponding three neutrino flavours. Flavour physics is the physics of the Yukawa matrices and subsequently the physics of transitions between different flavours. In the SM there is no mixing in the lepton sector although it is known that neutrinos mix. However, this is not included in the SM and in this section we therefore only discuss how quark mixing is introduced in the SM.

In the unbroken EW phase fermions as well as gauge bosons are strictly massless since a direct mass term for them is forbidden by gauge symmetry. For the SM fields in the EW broken phase only Dirac mass terms are allowed because of the $U(1)_{\mathrm{em}}$ gauge symmetry. But Dirac mass terms couple right-handed to left-handed fields which is forbidden by $SU(2)_L$ gauge invariance. It is possible to build gauge invariant combinations from the Higgs field, one $SU(2)_L$ doublet and one $SU(2)_L$ singlet field. We define the following vectors in flavour space
\begin{equation}
 Q' \equiv \begin{pmatrix} \begin{pmatrix} u'_L \\ d'_L \end{pmatrix} , & \begin{pmatrix} c'_L \\ s'_L \end{pmatrix} , & \begin{pmatrix} t'_L \\ b'_L \end{pmatrix} \end{pmatrix}^T \;, \quad L' \equiv \begin{pmatrix} \begin{pmatrix} \nu'_{e L} \\ e'_L \end{pmatrix}, & \begin{pmatrix} \nu'_{\mu L} \\ \mu'_L \end{pmatrix}, & \begin{pmatrix} \nu'_{\tau L} \\ \tau'_L \end{pmatrix} \end{pmatrix}^T \;,
\end{equation}
\begin{align}
 D'_L &\equiv \begin{pmatrix} d'_L \\ s'_L \\ b'_L \end{pmatrix} \;,\quad U'_L \equiv \begin{pmatrix} u'_L \\ c'_L \\ t'_L \end{pmatrix} \;, \quad E'_L \equiv \begin{pmatrix} e'_L \\ \mu'_L \\ \tau'_L \end{pmatrix} \;,\quad  N'_L \equiv \begin{pmatrix} \nu'_{e L} \\ \nu'_{\mu L} \\ \nu'_{\tau L} \end{pmatrix} \;, \\
 D'_R &\equiv \begin{pmatrix} d'_R \\ s'_R \\ c'_R \end{pmatrix} \;,\quad U'_R \equiv \begin{pmatrix} u'_R \\ c'_R \\ t'_R \end{pmatrix} \;, \quad E'_R \equiv \begin{pmatrix} e'_R \\ \mu'_R \\ \tau'_R \end{pmatrix}  \;,
\end{align}
where a subscript $L$ at $D'$, $U'$, $E'$ and $N'$ denotes the $SU(2)_L$ doublet fields and a subscript $R$ denotes the $SU(2)_L$ singlet fields. We have written all fields with a prime to label them as gauge eigenstates. Since in the SM no right-handed neutrinos are included because they are sterile under $G_{\mathrm{SM}}$ we have only the vector $N'_L$ and no $N'_R$ vector. Due to the absence of right-handed neutrinos in the SM no renormalisable coupling of the neutrinos to the Higgs field can be written down in the Lagrangian and neutrinos are strictly massless. Nevertheless, in experiment it was found out, that at least two neutrinos have to be massive, for reviews see, e.g.\  \cite{Altarelli:2010gt, Camilleri:2008zz, *Dore:2008dp, King:2007nw, *King:2003jb, *Mohapatra:2005wg, *Mohapatra:2006gs, *Albright:2009cn}.

Since the three generations of matter have the same quantum numbers under the SM gauge group the generations can mix. It turns out that the couplings of matter to gauge fields and the couplings to the Higgs field are twisted against each other. The kinetic terms are always diagonal in flavour space. Even after a rotation the new kinetic terms are still diagonal as long as the rotation is unitary which is the case in the SM. To discuss fermion mixings in the SM we need the gauge interaction terms
\begin{equation}
 \mathcal{L}_{\mathrm{int}} = - e J^\mu_{\mathrm{em}} A_\mu - \frac{e}{\sin \theta_W \cos \theta_W} J_{\mathrm{NC}}^\mu Z_\mu - \frac{e}{\sqrt{2} \sin \theta_W} (J_{\mathrm{CC}}^\mu W_\mu^+  + \mathrm{h. c.}) \;,
\end{equation}
where we have written down the interactions in the EW broken phase with the electromagnetic current
\begin{equation}
\begin{split}
 J^\mu_{\mathrm{em}} &= Q_u \left( \bar{U}'_L \gamma^\mu U'_L + \bar{U}'_R \gamma^\mu U'_R \right) + Q_d \left( \bar{D}'_L \gamma^\mu D'_L + \bar{D}'_R \gamma^\mu D'_R \right) \\ & + Q_e \left( \bar{E}'_L \gamma^\mu E'_L + \bar{E}'_R \gamma^\mu E'_R \right) \;. \label{Eq:emcurrent}
\end{split}
\end{equation}
Here $Q_u$, $Q_d$ and $Q_e$ are the electric charges of the up-type quarks, down-type quarks and charged leptons. The weak neutral current is given by
\begin{equation}
 J^\mu_{\mathrm{NC}} = \frac{1}{2} \bar{U}'_L \gamma^\mu U'_L - \frac{1}{2} \bar{D}'_L \gamma^\mu D'_L + \frac{1}{2}\bar{N}'_L \gamma^\mu N'_L - \frac{1}{2} \bar{E}'_L \gamma^\mu E'_L - \sin^2 \theta_W J^\mu_{\mathrm{em}} \;, \label{Eq:nccurrent}
\end{equation}
and the weak charged current by
\begin{equation}
 J^\mu_{CC} = \bar{U}'_L \gamma^\mu D'_L + N'_L \gamma^\mu E'_L \;. \label{Eq:cccurrent}
\end{equation}
It should be noted that the currents are diagonal in flavour space.

Now we turn to the Yukawa couplings, i.e.\ the couplings of the fermions to the Higgs field
\begin{equation}
 \mathcal{L}_{\mathrm{Yuk}} = - Y_e^{ij} \bar{L}'^i H E'^j_R - Y_d^{ij} \bar{Q}'^i H D'^j_R + Y_u^{ij} \bar{Q}'^i \epsilon H^* U'^j_R + \mathrm{h.c.} \;,
\end{equation}
where $\epsilon$ is the totally antisymmetric tensor with $\epsilon_{12} = -1$ acting in the $SU(2)_L$ space and $i,j = 1,2,3$ is the family index. The Yukawa couplings are three complex three-by-three matrices. If the Higgs field develops a vev the Yukawa matrices generate mass matrices for the fermions of the form $M_f = Y_f v/\sqrt{2}$, where $f=u,d,e$. If the SM would include right-handed neutrinos, mass terms for neutrinos could be generated in the same way but it is yet unclear if these kind of mass terms, the so-called Dirac mass terms, are the correct description for neutrino masses. Neutrinos could as well be Majorana particles for which a mass term with different properties can be constructed.

\begin{table}
\centering
\begin{tabular}{cc}
\toprule
Observable & Experimental Value \\ \midrule
$m_u$ in MeV & $1.22^{+0.48}_{-0.40}$ \\
$m_c$ in GeV & $0.59 \pm 0.08$ \\
$m_t$ in GeV & $162.9 \pm 2.8$ \\
\midrule
$m_d$ in MeV & $2.76^{+1.19}_{-1.14}$ \\
$m_s$ in MeV & $52 \pm 15$ \\
$m_b$ in GeV & $2.79 \pm 0.09$ \\
\midrule
$m_e$ in MeV & $0.48529$ \\
$m_\mu$ in MeV & $102.47$ \\
$m_\tau$ in MeV & $1742.2 \pm 0.2$ \\
\midrule
$\theta^{\mathrm{CKM}}_{12}$ & $0.2257^{+0.0009}_{-0.0010}$ \\[0.3pc]
$\theta^{\mathrm{CKM}}_{13}$ & $0.00359^{+0.00020}_{-0.00019}$ \\[0.3pc]
$\theta^{\mathrm{CKM}}_{23}$ & $0.0415^{+0.0011}_{-0.0012}$ \\[0.3pc]
$\delta_{\mathrm{CKM}}$ & $1.2023^{+0.0786}_{-0.0431}$ \\
\bottomrule
\end{tabular}
\caption[SM Fermion Masses and Mixing Angles]{Running masses of the SM fermions in the $\overline{\mathrm{MS}}$ renormalisation scheme at the top-scale $m_t(m_t) = 162.9$~GeV taken from \protect\cite{Xing:2007fb} and the CKM mixing angles and phase at the weak scale extracted from \protect\cite{Amsler:2008zzb}. We do not give the experimental errors for the electron and the muon mass since they are negligibly small. \label{Tab:fermionmassesandmixings} }
\end{table}

Within the SM there is no reason for the Yukawa couplings to be diagonal. In the quark sector it is experimentally proven that at least one of the Yukawa matrices is non-diagonal since otherwise there would be no quark-mixing. Therefore the following unitary rotations in flavour space are defined
\begin{equation}
\begin{split}
 U'_L = V_{u_L} U_L \;, \;\; U'_R = V_{u_R} U_R \;&,\;\; D'_L = V_{d_L} D_L \;, \;\; D'_R = V_{d_R} D_R \;, \\
 N'_L = V_{\nu_L} N_L \;&, \;\;  E'_L = V_{e_L} E_L \;, \;\; E'_R = V_{e_R} E_R \;.
\end{split}
\end{equation}
With these choices we can diagonalise all Yukawa and mass matrices respectively via
\begin{align}
 V_{u_L} M_u V^\dagger_{u_R} &= \mathrm{diag}(m_u, m_c, m_t) \;, \\
 V_{d_L} M_d V^\dagger_{d_R} &= \mathrm{diag}(m_d, m_s, m_b) \;, \\
 V_{e_L} M_e V^\dagger_{e_R} &= \mathrm{diag}(m_e, m_\mu, m_\tau) \;.
\end{align}
Experimental values for the fermion masses at the top-scale $m_t(m_t)$ are collected in Tab.~\ref{Tab:fermionmassesandmixings}.

The electromagnetic and the neutral current in Eqs.\ \eqref{Eq:emcurrent} and \eqref{Eq:nccurrent} are invariant under these field redefinitions. This is not the case for the charged current
\begin{equation}
 J^\mu_{CC} = \bar{U}_L \gamma^\mu V_{u_L} V^\dagger_{d_L} D_L + N_L \gamma^\mu V_{\nu_L} V^\dagger_{E_L} E_L \;. 
\end{equation}
There is no mass matrix for the neutrinos and therefore we can choose $V_{\nu_L} = V_{e_L}$ such that there is no lepton mixing in the gauge interactions. This is different for the quarks. $V_{u_L}$ and $V_{d_L}$ are already fixed by the diagonalisation conditions. The matrix $V_{u_L} V^\dagger_{d_L}$ is the Cabibbo--Kobayashi--Maskawa (CKM) matrix $V_{\mathrm{CKM}}$ \cite{Cabibbo:1963yz, *Kobayashi:1973fv}, see also App.~\ref{App:CKM}.

We count now the number of free parameters of the CKM matrix for $n$ generations. A unitary $n \times n$ matrix has $n^2$ real parameters. By rephasing the quark fields $(2n -1)$ phases can be absorbed. As observables $n(n-1)/2$ angles and $(n-1)(n-2)/2$ phases remain. In the SM, where $n$ is equal to three, there are three real mixing angles and one phase. Throughout this thesis we use the PDG parameterisation \cite{Amsler:2008zzb}
\begin{equation}
V_{\mathrm{CKM}} 
= \begin{pmatrix} c_{12} c_{13} & s_{12} c_{13} & s_{13} \e^{-\ci \delta_{\mathrm{CKM}}} \\ - s_{12} c_{23} - c_{12} s_{23} s_{13} \e^{\ci \delta_{\mathrm{CKM}}} & c_{12} c_{23} - s_{12} s_{23} s_{13} \e^{\ci \delta_{\mathrm{CKM}}} & s_{23} c_{13} \\ s_{12} s_{23} - c_{12} c_{23} s_{13} \e^{\ci \delta_{\mathrm{CKM}}} & -c_{12} s_{23} - s_{12} c_{23} s_{13} \e^{\ci \delta_{\mathrm{CKM}}} & c_{23} c_{13} \end{pmatrix} \;,
\end{equation}
where $s_{ij}$ and $c_{ij}$ are abbreviations for $\sin \theta^{\mathrm{CKM}}_{ij}$ and $\cos \theta^{\mathrm{CKM}}_{ij}$. The three angles $\theta^{\mathrm{CKM}}_{12}$, $\theta^{\mathrm{CKM}}_{13}$ and $\theta^{\mathrm{CKM}}_{23}$ are the CKM angles and $\delta_{\mathrm{CKM}}$ is the CKM phase. Experimental values for the mixing angles and the CKM phase are collected in Tab.\ \ref{Tab:fermionmassesandmixings}. The CKM phase is of special phenomenological importance because it induces violation of the CP symmetry, the symmetry between matter and antimatter.

\section{Open Questions in the Standard Model}

In this section we discuss some of the open questions in the SM. Although the SM describes nature to a very high accuracy there are some open issues. One of the big open questions is the unification of the SM with a theory of gravity into a theory of quantum gravity. We do not address this question within this thesis. Also, in the following, we mention only those problems which are of special importance for this work.

\subsection{Neutrino Masses}

The question of how to embed neutrino masses into the SM was already mentioned several times before. Neutrinos are massive, see, for example, the reviews \cite{Altarelli:2010gt, Camilleri:2008zz, *Dore:2008dp, King:2007nw, *King:2003jb, *Mohapatra:2005wg, *Mohapatra:2006gs, *Albright:2009cn}, whereas in the SM there is no mass term for the neutrinos and they are strictly massless.

One might think that it is straightforward to give neutrinos a mass by including right-handed neutrinos and writing down a Dirac mass term as it works out in the same way for all the other fermions. But this is not the only possible way to add a mass term for neutrinos. If neutrinos are Majorana particles, which means that they are their own antiparticles, then there is also the possibility to write down a mass term which has different properties. One consequence of the Majorana nature of neutrinos would be the possibility of neutrinoless double beta decay. Another consequence for Majorana neutrinos is that in the leptonic mixing matrix, the Pontecorvo--Maki--Nakagawa--Sakata (PMNS) matrix \cite{Pontecorvo:1957qd, *Maki:1962mu}, two additional CP violating phases appear which cannot be absorbed in the lepton fields.

If neutrinos are Majorana particles the smallness of the neutrino masses could be elegantly explained by the seesaw mechanism \cite{Minkowski:1977sc, *GellMann:1980vs, *Yanagida:1979as, *Glashow:1979nm, *Mohapatra:1980yp, *Schechter:1981cv}. To explain this, we assume for the moment that there is only one left-handed neutrino and we add a right-handed neutrino to our theory, then the neutrinos can have a Dirac mass $m_D$, which couples left-handed to right-handed neutrinos, and a Majorana mass $m_R$, which couples right-handed to right-handed neutrinos. In this simplest extension there is no direct renormalisable mass term for the left-handed neutrinos allowed due to the $U(1)_Y$ symmetry. Such a term could nevertheless be introduced by including, for example, scalar $SU(2)_L$ triplets. Different ways of implementing the seesaw mechanism are classified by their additional field content to generate neutrino masses. In the type I seesaw mechanism extra gauge singlet fermions are added, which are commonly called right-handed neutrinos. Models with an additional scalar $SU(2)_L$ triplet are often referred to as type II seesaw mechanism. These are the only types of seesaw mechanism relevant for this work.

Let us go back to the mechanism itself. If the masses $m_D$ and $m_R$ fulfil the relation $m_D \ll m_R$ we have two mass eigenstates. One is the heavy mass eigenstate $N$ which primarily consists of the original right-handed neutrino with a mass $m_N \approx m_R$ and one is the light eigenstate $\nu$ which consists primarily of the left-handed neutrino with $m_\nu \approx m_D^2/m_R$. This mechanism can be extended straight-forwardly to three generations.

Although $m_D$ can be expected to be of the order of the other fermion masses, the masses of the light neutrinos are still small due to the suppression by the large masses $m_R$. In the context of left-right symmetric extensions of the SM this can be easily implemented because heavy right-handed neutrinos are necessary ingredients of these theories.

\subsection{The Hierarchy Problem} \label{Sec:hierarchyproblem}

In the SM the Higgs boson mass is not protected by a symmetry. Therefore it can have large radiative corrections. In a naive renormalisation scheme, where at a high scale $\Lambda$ the loop momenta are cut off, the radiative corrections to the Higgs boson mass are given by
\begin{equation}
\delta m_H^2 = \frac{\Lambda^2}{16 \pi^2} c_0 \log \frac{\Lambda}{\mu_R} \;, \label{Eq:Higgsmasscorrection}
\end{equation}
 to leading order where $c_0 = 3/(2v^2) (m_H^2 + 2 M_W^2 + M_Z^2 - 4 m_t^2)^2$. Roughly speaking the Higgs boson mass is proportional to the cutoff scale as long as there is no severe amount of fine-tuning between the bare Higgs boson mass and the loop corrections. If the SM is valid up to the Planck scale at which a theory of quantum gravity has to be invoked, the Higgs boson mass would also be expected to be of the order of the Planck scale. Electroweak precision data \cite{Amsler:2008zzb, Alcaraz:2009jr} however prefers a Higgs boson with a mass around the EW scale. Without introducing new physics the SM therefore has a large amount of fine tuning because the hierarchy  between the EW scale and the Planck scale is not protected by any symmetry. The most prominent solution to this problem is supersymmetry which is introduced in the next chapter.

\subsection{Charge Quantisation and Anomaly Cancellation}

\begin{figure}
\centering
\includegraphics{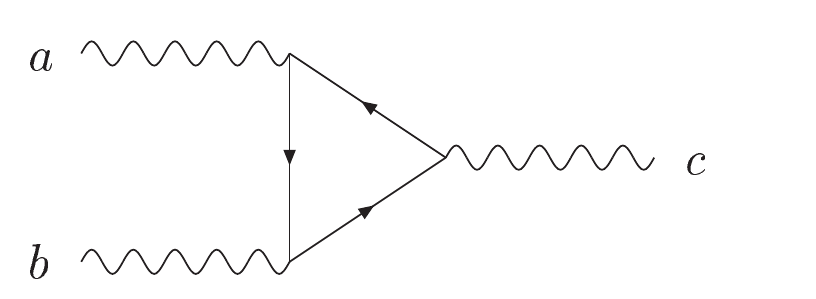}
\caption[Anomaly Feynman Diagram]{Triangle diagram contributing to an anomaly with external currents $j^a$, $j^b$ and $j^c$ and chiral fermions running in the loop. \label{Fig:Anomaly}}
\end{figure}

From the mathematical point of view the charges of a $U(1)$ gauge group do not have to be quantised. Therefore it is quite astonishing that the charges of the $U(1)_Y$ respectively $U(1)_{\mathrm{em}}$ group in the SM are quantised in such a way, that the hydrogen atom is neutral to very high precision. In the SM the lepton and the quark sector appear to be very distinct from each other so that it is very surprising that the sum of the three proton valence quark charges are just the negative of the electron charge. This can be explained in the context of GUTs and is explained in Ch.~\ref{Ch:GUTs}.

In a chiral gauge theory like the SM, so called anomalies appear. A symmetry is called anomalous if the symmetry is broken by radiative corrections although the tree-level Lagrangian respects this symmetry. In terms of Feynman diagrams the existence of an anomaly can be calculated by evaluating the diagram in Fig.~\ref{Fig:Anomaly}. The result is proportional to
\begin{equation}
\mathcal{A}^{abc} = \mathrm{Tr} [t^a \{ t^b,t^c \}] \;, \label{Eq:SMAnomaly}
\end{equation}
where the $t^a$ are the group generators belonging to the external currents in the triangle graph. Every possible configuration of external currents has to be checked.

In the SM the anomaly $\mathcal{A}^{abc}$ vanishes as it should be to keep the gauge symmetries unbroken. However, this happens only due to the fact that the anomalies from the loops with quarks and leptons cancel. This is a surprising result and seems to be connected to charge quantisation since there we need a similar cancellation between the electric charges of the leptons and the quarks. Nevertheless, charge quantisation and anomaly cancellation are related to each other but not necessarily quite the same. For example in a $SU(5)$ embedding of the SM, charge quantisation follows directly from the structure of the gauge group but the necessary anomaly cancellation between different representations is still accidental within $SU(5)$. Nevertheless, it can be shown that $SO(10)$ for example, another well motivated candidate for a GUT group, is free of gauge anomalies.

\subsection{The Flavour Puzzle}

\begin{figure}
\centering
\includegraphics[scale=0.7]{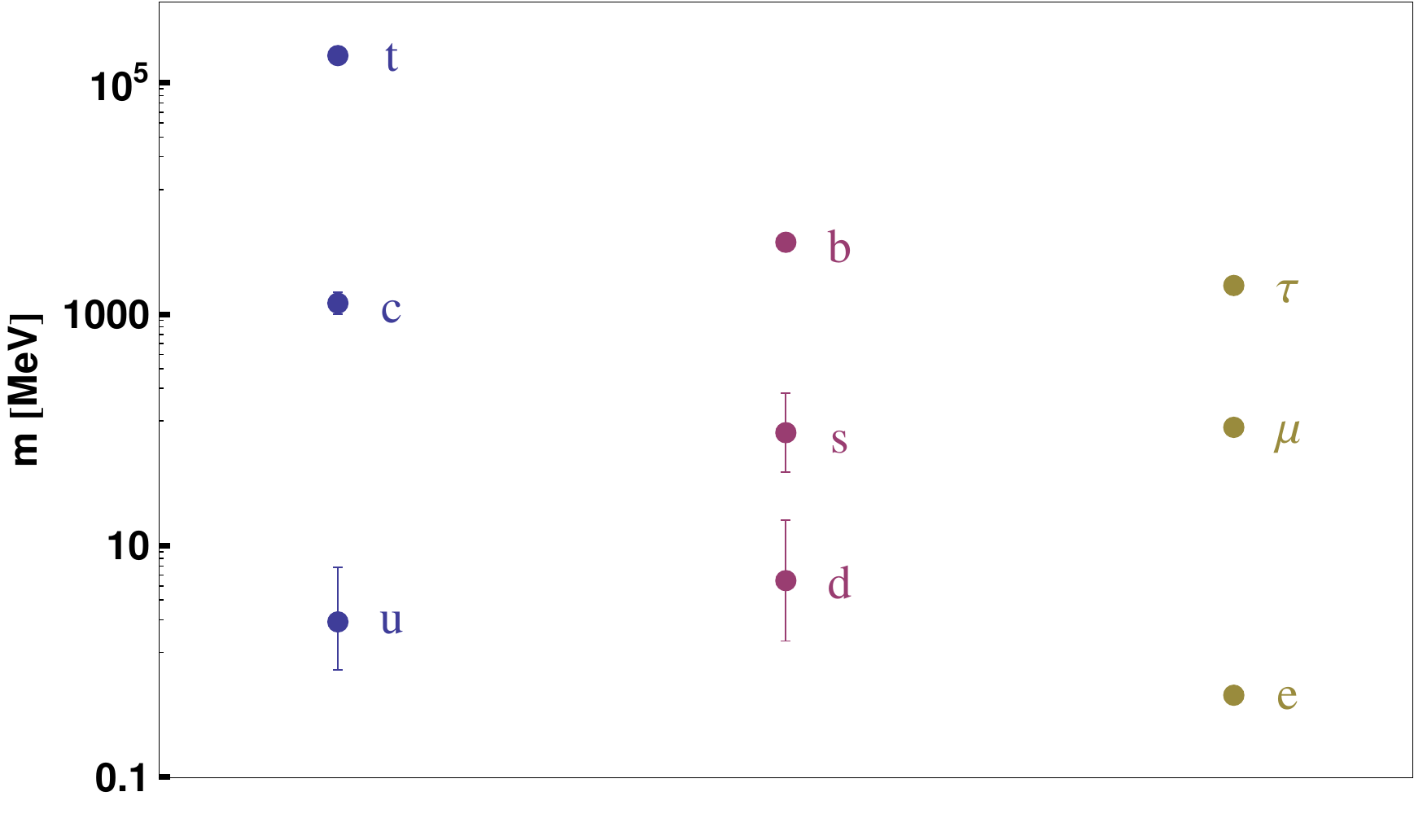}
\caption[Pattern of Fermion Masses]{Pattern of fermion masses in the SM. The 1 $\sigma$ errors are multiplied with a factor of three. \label{Fig:quarkmasses}}
\end{figure}

The SM has 19 free parameters from which 13 are related to the flavour sector. That seems unsatisfactory for a fundamental physical theory. Also the pattern of the fermion masses in the SM, see Fig.~\ref{Fig:quarkmasses}, seems to demand a deeper understanding. For example, the masses of the up-type quarks differ by a factor of approximately one thousand from each other whereas the down-type quark masses differ by approximately a factor of one hundred from each other.

This becomes even worse by including neutrino masses and mixings. This introduces three additional masses, three additional mixing angles and, depending on the nature of the neutrinos, i.e.\ Dirac or Majorana, one or three additional CP violating phases. The hierarchy between the masses gets even stronger. It is known that the sum of the neutrino masses cannot be larger than about 0.6~eV \cite{Jarosik:2010iu}. Even if the neutrino masses are close to this upper bound the top quark would still be about $10^{11}$ times heavier.

However, this is not the only strange pattern in the flavour sector. The mixing angles in the quark and the lepton sector are very different from each other. While the mixing angles are all small in the quark sector in the lepton sector two mixing angles are known to be large while the third one is very small. One of the leptonic mixing angles is even close to maximal.

Although the SM (respectively an extension with right-handed neutrinos) can in principle describe these patterns by imposing that the masses and mixing angles are just as we observe them, maybe there is a more fundamental theory which elegantly describes the flavour patterns. For example, the PMNS matrix can be described in terms of discrete symmetries which may be interpreted as a hint towards an underlying family symmetry.

%% file: kap_03_SUSY.tex
\chapter{Supersymmetry} \label{Ch:SUSY}

In this chapter we want to review some fundamental concepts of supersymmetry (SUSY) based on \cite{Martin:1997ns, Aitchison:2007fn, *Bailin:1994qt, Trenkel:2009phd}. We start with a short motivation for supersymmetric field theories in Sec.~\ref{Sec:SUSYMotivation} and then discuss some basic formal aspects of SUSY necessary for this thesis in Sec.~\ref{Sec:SUSYFormal}. We end this chapter with an introduction to the MSSM in Sec.~\ref{Sec:SUSYMSSM} and a brief discussion of SUSY breaking in Sec.~\ref{Sec:SUSYbreaking}.

\section{Motivation} \label{Sec:SUSYMotivation}

The initial motivation for introducing supersymmetry in particle physics was the realisation that despite a \emph{no-go} theorem by Coleman and Mandula \cite{Coleman:1967ad} the $S$-matrix can have symmetries beyond the internal symmetries and the Poincar\'{e} symmetry. This \emph{no-go} theorem can be circumvented by introducing symmetries whose algebras fulfil \emph{anti-}commutation relations instead of commutation relations \cite{Haag:1974qh}. This class of symmetries is called supersymmetries. But it turned out that supersymmetric field theories have other nice properties of which we give now three prominent examples.

Probably the most prominent feature of SUSY is the solution of the hierarchy problem, see Sec.~\ref{Sec:hierarchyproblem}. In a supersymmetric version of the SM, where SUSY is unbroken, the radiative corrections to the Higgs mass are exactly cancelled by diagrams with SUSY partners of the SM particle fields in the loops. Even in a field theory with softly broken SUSY the radiative corrections to the Higgs boson mass are under control as long as the scale of the SUSY particle masses is not too far above the TeV scale \cite{Witten:1981nf, *Kaul:1981hi}. Therefore the prospects of finding SUSY at the LHC is very high, see, e.g.\  \cite{Cassel:2010px, Ellis:2008di, Abdallah:2009zz, *Bechtle:2009zz, of:2009qj, *Lungu:2009nh, Germer:2010vn, Ehrenfeld:2009rt}.

\begin{figure}
\centering
\includegraphics[scale=0.55]{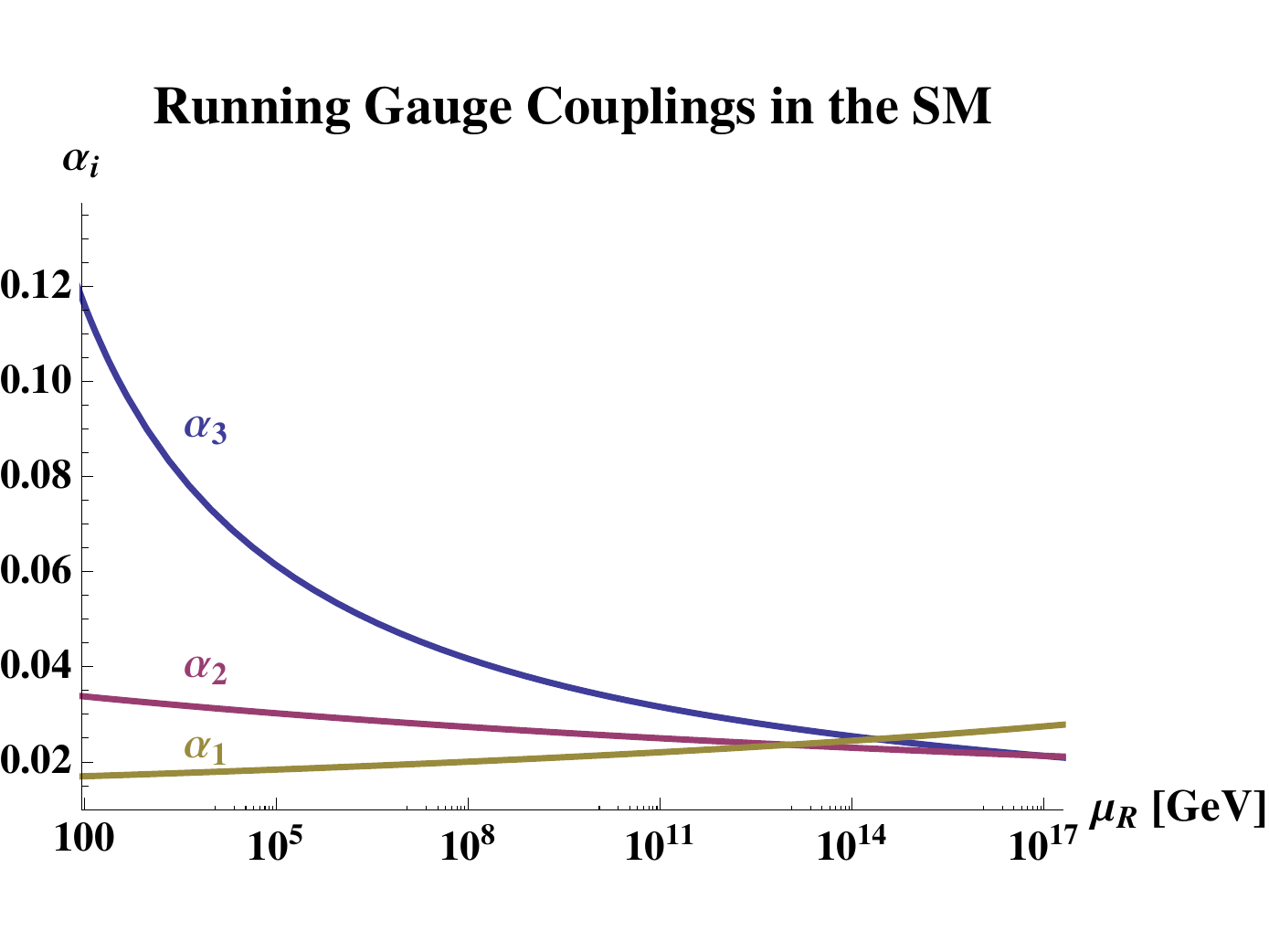} \hspace{0.3cm}
\includegraphics[scale=0.55]{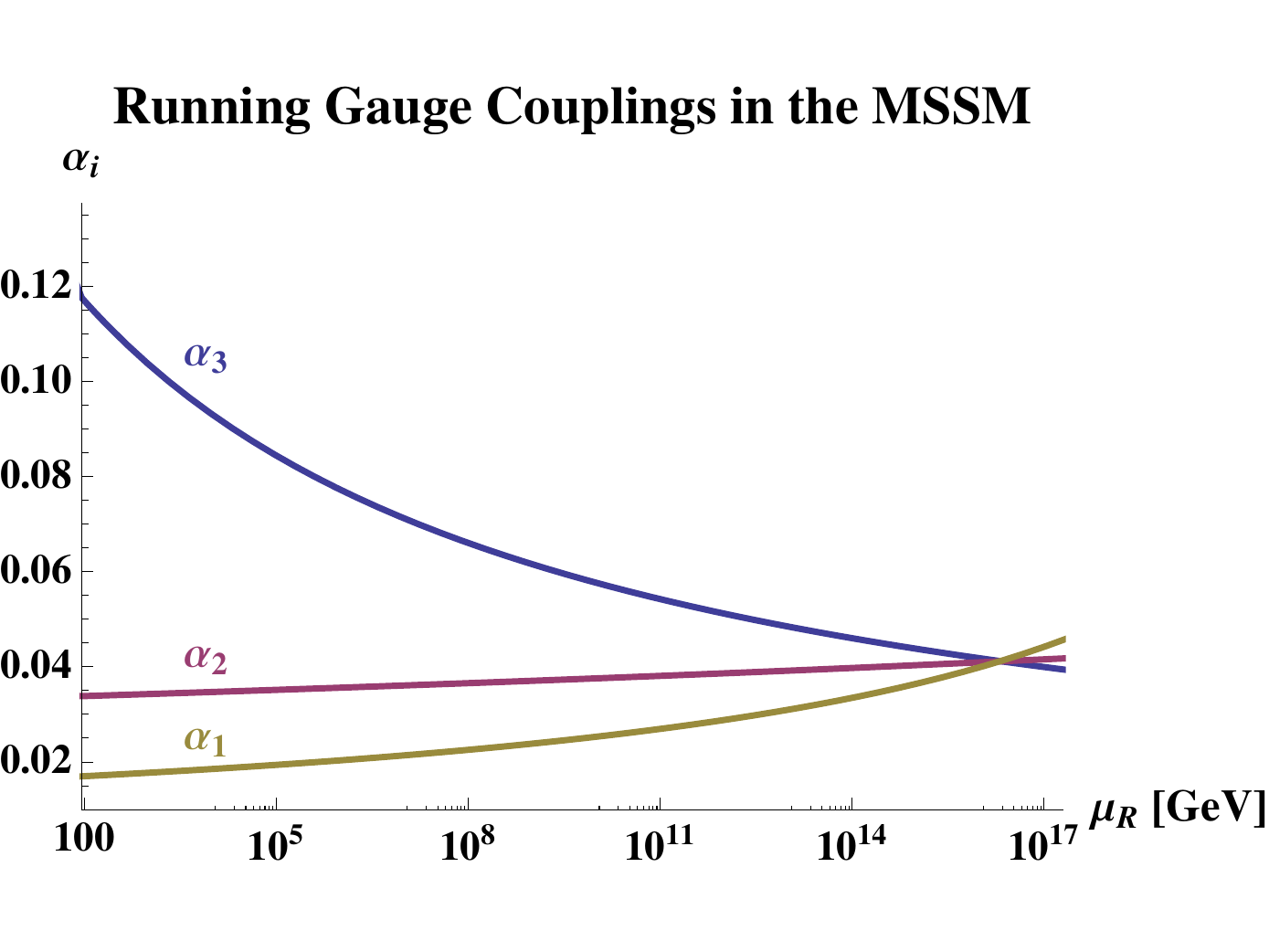}
\caption[Running Gauge Couplings]{Running of the gauge couplings on one-loop level in the SM (left) and the MSSM (right). The SUSY scale $M_{\mathrm{SUSY}}$ was set to $M_Z \approx 90$~GeV. \label{Fig:gaugeunification}}
\end{figure}

The second motivation which concerns us is gauge coupling unification. The sizes of the three gauge couplings in the SM depend on the energy scale at which they are probed. While they only come close to each other in the SM at a scale of roughly $10^{14}$~GeV the gauge couplings almost perfectly unify at a scale of roughly $10^{16}$~GeV in the MSSM \cite{Amaldi:1991cn, *deBoer:1994dg}, see also Fig.\ \ref{Fig:gaugeunification}. Unification of gauge couplings is an essential ingredient for GUTs where the three SM gauge groups are unified to one simple group. From that point of view SUSY and GUTs seem to fit together quite well.

The third argument often invoked in favour of SUSY is that within SUSY models in which the lightest SUSY particle (LSP) is stable on cosmological time scales this particle is a viable candidate for dark matter. This is true, for example, in $\mathcal{R}$-parity conserving SUSY models like the MSSM. There, the LSP is stable and if it carries neither colour nor electric charge, like a neutralino or the gravitino, it is a good candidate for dark matter, for a recent review see \cite{Ellis:2010kf}. Nevertheless, this third argument in favour of SUSY plays only a minor role in this thesis since our main focus is not on cosmological aspects.

\section{Some Formal Aspects of Supersymmetry} \label{Sec:SUSYFormal}

In the following we discuss some basic formal properties of SUSY acting as a foundation for the rest of the work. 

\subsection{The Supersymmetry Algebra} \label{Sec:SUSYalgebra}

Since the SUSY algebra is based on anticommutation relations the generators of the SUSY algebra have to be fermionic operators $Q_\alpha$ and their conjugate $\bar{Q}_{\dot{\alpha}}$ \cite{Wess:1973kz, *Volkov:1973ix, Wess:1974tw}. The generators of the SUSY algebra $Q_\alpha$ and $\bar{Q}_{\dot{\alpha}}$ are Weyl spinors, see App.~\ref{App:Weyl}. Their algebra is given by
\begin{equation}
\begin{split}
 \{ Q_\alpha, Q_\beta \} &= \{ \bar{Q}_{\dot{\alpha}} , \bar{Q}_{\dot{\beta}} \} =0 \;, \\
 \{ Q_\alpha, \bar{Q}_{\dot{\beta}} \} &= 2 (\sigma^\mu)_{\alpha \dot{\beta}} P_\mu \;. 
\end{split} \label{Eq:SUSYalgebra}
\end{equation}
We are working here in $\mathcal{N}=1$ SUSY where there is only one type of SUSY generator.

Besides commuting with the generators of $G_{\mathrm{SM}}$ the commutation relations for $Q_\alpha$ and $\bar{Q}_{\dot{\alpha}}$ with the generators of the Poincar\'{e} group are the following
\begin{equation}
\begin{split}
 [Q_\alpha, P_\mu] &= [\bar{Q}_{\dot{\alpha}},P_\mu] =0 \;, \\
 [Q_\alpha,M_{\mu \nu}] &= - \frac{1}{2} {(\sigma_{\mu \nu})_\alpha}^\beta Q_\beta \;,\\
 [\bar{Q}_{\dot{\alpha}}, M_{\mu \nu}] &= - \frac{1}{2} {(\bar{\sigma}_{\mu \nu})_{\dot{\alpha}} }^{\dot{\beta}} \bar{Q}_{\dot{\beta}} \;.
\end{split}
\end{equation}
The matrices $\sigma^\mu$, $\sigma^{\mu \nu}$ and $\bar{\sigma}^{\mu \nu}$ are defined in Eqs.~\eqref{Eq:sigmamu} and \eqref{Eq:sigmamunu}. $P^\mu$ are the generators for space-time translations and $M_{\mu \nu}$ are the generators for Lorentz transformations.

In Eq.~\eqref{Eq:SUSYalgebra} the connection between SUSY and gravity is insinuated. Indeed by changing to \emph{local} SUSY transformations, gravity can be incorporated in to what is called Supergravity (SUGRA) \cite{Freedman:1976xh, *Deser:1976eh}. Nevertheless, we work here only with global SUSY transformations which only induce constant translations in Minkowski space.

Although we seem to restrict ourselves here to some very special SUSY transformations (global, $\mathcal{N}=1$ SUSY), these transformations are most relevant for phenomenological applications and already give gauge coupling unification and a solution to the hierarchy problem. However, introducing extended SUSY with $\mathcal{N} > 1$ in four dimensions leads to conceptual problems since this does not allow for chiral fermions and parity violation to the observed amount \cite{Sohnius:1985qm}.

In the application of SUSY to particle physics every fermionic state is related to a bosonic state, or vice versa, via
\begin{equation}
 Q_\alpha |\mathrm{fermion}\rangle = |\mathrm{boson}\rangle \quad \mathrm{or} \quad Q_\alpha |\mathrm{boson}\rangle = |\mathrm{fermion}\rangle \;.
\end{equation}
This already follows from the conservation of angular momentum. Since the SUSY generator is a spinor it carries fractional spin and therefore changes the statistic of a given state if it is applied on that state.

A particle $f$ and its superpartner $\tilde{f}$ must have the same mass if SUSY is unbroken:
\begin{equation}
\begin{split}
m_{\tilde{f}}^2 | \tilde{f} \rangle = P^2 | \tilde{f} \rangle = P^2 Q | f \rangle = Q P^2 |f\rangle = Q m_f^2 |f\rangle = m_f^2 |\tilde{f}\rangle \;,
\end{split}
\end{equation}
where we have used the fact that the SUSY generators commute with the momentum operator $P$. Identical masses for a particle and its superpartner are not in agreement with experiment. The superpartner of, e.g.\  the electron was not found so far in experiment. That means that if SUSY is a symmetry of nature it has to be broken. We come back to this point in Sec.~\ref{Sec:SUSYbreaking} but before we discuss an elegant way of describing fields and their superpartners in terms of superfields.

\subsection{Superspace and Superfields} \label{Sec:Superfields}

The description of SUSY via superfields in superspace is an elegant way to handle SUSY transformations and writing down SUSY invariant Lagrangians. Superspace extends the four-dimensional Minkowski space with two additional Grassmann valued dimensions, for a discussion of Grassmann variables see App.~\ref{App:Grassmann}.

Every point in this space has the same number of bosonic and fermionic degrees of freedom (dof) and is given by the supercoordinate $X=(x^\mu, \theta_{\alpha}, \bar{\theta}^{\dot{\alpha}})$ where $x^\mu$ are the usual four coordinates in Minkowski space and $\theta_\alpha$ and $\bar{\theta}^{\dot{\alpha}}$ are additional Grassmann valued coordinates.

A superfield $\Phi=\Phi(x,\theta,\bar{\theta})$ in superspace contains all bosonic and fermionic components of a given supermultiplet. Since Grassmann numbers are nilpotent we can expand the superfield in the Grassmann variables and the result is a finite series
\begin{equation}
\begin{split}
\Phi(x,\theta,\bar{\theta}) &= \phi(x) + \theta \psi(x) + \bar{\theta} \bar{\chi}(x) + \theta \theta F(x) + \bar{\theta} \bar{\theta} H(x) + \theta \sigma^\mu \bar{\theta} A_\mu(x) \\
& + (\theta \theta) \bar{\theta} \bar{\lambda}(x) +(\bar{\theta} \bar{\theta}) \theta \xi(x) +(\theta \theta) (\bar{\theta} \bar{\theta}) D(x) \label{Eq:superfield} \;,
\end{split}
\end{equation}
where $\phi, F, H, D$ are complex scalar fields, $A_\mu$ is a complex vector field and $\psi, \bar{\chi}, \bar{\lambda}, \xi$ are two-component Weyl spinor fields. The four complex scalars and the complex vector field give 16 bosonic dof while the four Weyl spinors give 16 fermionic dof. These are more dof than we need for describing fields with a spin not greater than one. Imposing (SUSY) covariant constraints on a superfield reduces the number of dof and we end up with an irreducible representation of SUSY where the  redundant field components are removed.

Since Grassmann variables anticommute we can use them to rewrite the SUSY algebra in Eq.~\eqref{Eq:SUSYalgebra} in terms of commutation relations
\begin{equation}
\begin{split}
[\theta Q, \theta Q] &= [\bar{\theta} \bar{Q}, \bar{\theta} \bar{Q}]  = 0 \;, \\
[\theta Q, \bar{\theta} \bar{Q}] &= 2 \theta \sigma^\mu \bar{\theta} P_\mu \;,\\
[\theta Q, P_\mu] &= [\bar{\theta} \bar{Q}, P_\mu] = 0 \;.
\end{split}
\end{equation}
We know from the theory of Lie algebras how to construct the transformation of the corresponding symmetry from such relations. Hence, we obtain for a group element of a (global) SUSY transformation
\begin{equation}
S(y_\mu, \xi^\alpha, \bar{\xi}_{\dot{\alpha}}) = \mathrm{exp} \left( -\ci (\xi^\alpha Q_\alpha + \bar{\xi}_{\dot{\alpha}} \bar{Q}^{\dot{\alpha}} + y_\mu P^\mu )  \right) \;,
\end{equation}
where $y_\mu$ is a Minkowski variable and $\xi$ and $\bar{\xi}$ are Grassmann variables. This transformation can be applied to the superfield $\Phi$
\begin{equation}
S(y_\mu,\xi,\bar{\xi}) \Phi(x_\mu,\theta,\bar{\theta}) = \Phi(x_\mu + y_\mu + \ci\, \xi \sigma_\mu \bar{\theta} - \ci\, \theta \sigma_\mu \bar{\xi}, \xi + \theta, \bar{\xi} + \bar{\theta}) \;.
\end{equation}
If we expand the above results in $\xi$ and $\bar{\xi}$ we get the infinitesimal SUSY transformation $\delta_S (\xi,\bar{\xi})$ acting on a superfield as
\begin{equation}
\begin{split}
\delta_S (\xi, \bar{\xi}) \Phi(x,\theta,\bar{\theta}) &= - \ci [\xi Q + \bar{\xi} \bar{Q}] \Phi(x,\theta,\bar{\theta}) \\
&= \left[ \xi^\alpha \frac{\partial}{\partial \theta^\alpha} + \bar{\xi}_{\dot{\alpha}} \frac{\partial}{\partial \bar{\theta}_{\dot{\alpha}} }  + \ci (\xi \sigma_\mu \bar{\theta} - \theta \sigma_\mu \bar{\xi}) \frac{\partial}{\partial x_\mu}  \right]  \Phi(x,\theta,\bar{\theta}) \;,
\end{split}
\end{equation}
from which a linear representation of the SUSY generators can be derived
\begin{align}
Q_\alpha &= \ci\, \partial_\alpha  - (\sigma^\mu \bar{\theta})_\alpha  \partial_\mu \;, \\
\bar{Q}_{\dot{\alpha}} &= -\ci\, \bar{\partial}_{\dot{\alpha}} + (\theta \sigma^\mu)_{\dot{\alpha}} \partial_\mu \;, \\
P_\mu &= \ci\, \partial_\mu \;,
\end{align}
where we have made use of the abbreviations $\partial_\alpha = \partial/\partial \theta^{\alpha}$, $\bar{\partial}_{\dot{\alpha}} = \partial/\partial\bar{\theta}^{\dot{\alpha}} =- \epsilon_{\dot{\alpha} \dot{\beta}} \partial/\partial \bar{\theta}_{\dot{\beta}}$ and $\partial_\mu = \partial/\partial x^\mu$. Now we have everything together to define the SUSY covariant derivatives $D_\alpha$ and $\bar{D}_{\dot{\alpha}}$ in analogy to the covariant derivatives in gauge theories. The covariant derivatives have to be invariant under the SUSY transformations $\delta_S$, i.e.\ $[\xi Q + \bar{\xi} \bar{Q},D_\alpha] =0$, which can also be written in terms of anticommutation relations as
\begin{equation}
\{ D_\alpha, Q_\beta \} = \{D_\alpha, \bar{Q}_{\dot{\beta}} \} = \{ \bar{D}_{\dot{\alpha}}, Q_\beta \} = \{ \bar{D}_{\dot{\alpha}}, \bar{Q}_{\dot{\beta}} \} = 0 \;.
\end{equation}
Using all of the above equations, explicit expressions for $D_\alpha$ and $\bar{D}_{\dot{\alpha}}$ can be derived
\begin{equation}
\begin{split}
D_\alpha &= \ci\, \partial_\alpha + (\sigma^\mu \bar{\theta})_\alpha \partial_\mu \;, \\
\bar{D}_{\dot{\alpha}} &= -\ci\, \bar{\partial}_{\dot{\alpha}} - (\theta \sigma^\mu)_{\dot{\alpha}} \partial_\mu \;, \\
\{ D_\alpha, \bar{D}_{\dot{\beta}} \} &= -2 (\sigma^\mu)_{\alpha \dot{\beta}} P_\mu \;.
\end{split}
\end{equation}
Having this definitions at hand we can define the superfield constraints for the fields we are interested in
\begin{align}
\bar{D}_{\dot{\alpha}} \Phi &= 0 &\rightarrow && &\text{left-handed chiral superfield,} \label{Eq:LHsuperfield} \\
D_{\alpha} \Phi &= 0 &\rightarrow && &\text{right-handed anti-chiral superfield,} \label{Eq:RHsuperfield }\\
\Phi &= \Phi^\dagger &\rightarrow && &\text{vector superfield.} \label{Eq:Vsuperfield}
\end{align}
A product of two chiral superfields is again a chiral superfield whereas the product of a chiral and a conjugated chiral superfield gives a vector superfield.

Solving Eq.~\eqref{Eq:LHsuperfield} a left-handed chiral superfield $\Phi_L$ can be written as
\begin{equation}
\begin{split}
\Phi_L(x,\theta,\bar{\theta}) &= \phi(x) + \sqrt{2} \theta \psi(x) + \theta \theta F(x) + \ci \, \theta \sigma^\mu \bar{\theta} \partial_\mu \phi(x) \\
& - \frac{\ci}{\sqrt{2}} (\theta \theta) \left( \partial_\mu \psi(x) \sigma^\mu \bar{\theta} \right) - \frac{1}{4} (\theta \theta) (\bar{\theta} \bar{\theta}) \partial^\mu \partial_\mu \phi(x) \;.
\end{split}
\end{equation}
Comparing the left-handed chiral superfield with a general superfield, see Eq.~\eqref{Eq:superfield}, we see that the $\bar{\theta}$ and the $\bar{\theta} \bar{\theta}$ components do not enter and the other components are partially related to each other. $\Phi_L$ has four fermionic dof from the two-component Weyl spinor $\psi(x)$ and four bosonic dof from the two complex scalar fields $\phi(x)$ and $F(x)$. So in general a chiral superfield still has the same number of bosonic and fermionic dof. Hence it is guaranteed that the SUSY algebra closes.

Nevertheless, there is a subtlety. If the superfield is put on-shell the dof of the spinor field $\psi$ reduces to two. In this case the equations of motion of the auxiliary field $F$ become trivial ($F=F^*=0$) and thus the field $F$ can be eliminated from the superfield.

The behaviour of the component fields under infinitesimal SUSY transformations is
\begin{equation}
\begin{split}
\delta_S (\xi, \bar{\xi}) \phi(x) &= \sqrt{2} \xi \psi(x) \,, \\
\delta_S (\xi, \bar{\xi}) \psi_\alpha (x) &= \sqrt{2} F(x) \xi_\alpha - \sqrt{2} \left( \sigma^\mu \bar{\xi} \right)_\alpha \partial_\mu \phi(x) \;, \\
\delta_S (\xi, \bar{\xi}) F(x) &= \partial_\mu \left( \ci \sqrt{2} \psi(x) \sigma^\mu \bar{\xi} \right) \;. \label{Eq:superLHtrafo}
\end{split} 
\end{equation}
Here we want to stress two points. First of all we see here explicitly that via a SUSY transformation, fermionic dof are transformed into bosonic dof and vice versa. Secondly the SUSY transformation of the auxiliary field $F$ is a total derivative and hence does not play any role in the action of a physical system.

In complete analogy,the relations for right-handed chiral superfields can be derived via $\Phi_R = \overline{\Phi_L} = (\Phi_L)^\dagger$. The (non-Abelian) SUSY gauge transformations with generators $T^a$ and gauge coupling $g$ transform a chiral superfield as
\begin{equation}
\Phi_L \rightarrow \e^{-2\, \ci \, g \Lambda} \Phi_L \quad \text{and} \quad \overline{\Phi_L} \rightarrow \overline{\Phi_L} \e^{2\, \ci \, g \bar{\Lambda} } \;, \label{Eq:gaugechiraltrafo}
\end{equation}
where $\Lambda = \Lambda^a T^a$ and $\Lambda^a(x,\theta,\bar{\theta})$ are chiral superfields.

In Eq.~\eqref{Eq:Vsuperfield} we defined the constraint for a vector superfield $V(x,\theta,\bar{\theta})$ which gives in component notation
\begin{equation}
\begin{split}
V(x,\theta,\bar{\theta}) &= C(x) + \ci \, \theta \chi(x) - \ci \, \bar{\theta} \bar{\chi}(x) + \theta \sigma^\mu \bar{\theta} A_\mu(x) \\
&+ \frac{\ci}{2} \theta \theta \left[ M(x) + \ci \, N(x) \right] - \frac{\ci}{2} (\bar{\theta} \bar{\theta}) \left[ M(x) - \ci \, N(x) \right] \\
&+ \ci (\theta \theta) \bar{\theta} \left[ \bar{\lambda}(x)  + \frac{\ci}{2} \bar{\sigma}^{\mu} \partial_\mu \chi(x) \right] - \ci (\bar{\theta} \bar{\theta}) \theta \left[ \lambda(x) + \frac{\ci}{2} \sigma^\mu \partial_\mu \bar{\chi}(x) \right] \\
&+ \frac{1}{2} (\theta \theta) (\bar{\theta}\bar{\theta}) \left[ D(x) - \frac{1}{2} \partial^\mu \partial_\mu C(x) \right] \;,
\end{split}
\end{equation}
where $C,M,N,D$ are real scalar fields, $\chi, \lambda$ are complex Weyl spinors and $A_\mu$ is a real spin-one vector field. Off-shell, the eight bosonic dof match again the eight fermionic dof. We also still have the freedom to choose a gauge. Accordingly we choose the Wess--Zumino gauge \cite{Wess:1974tw} where $C=M=N=0$ and $\chi=0$. Then we end up with the gauge field $A_\mu$, the gaugino $\lambda$, its fermionic superpartner, and the bosonic auxiliary field $D$.

The field $D$ plays an analogous role for vector superfields as $F$ did for the chiral superfields. Off-shell it is needed to match the bosonic and fermionic dof and close the SUSY algebra whereas on-shell it can be eliminated by its equations of motion.

The behaviour of the component fields under infinitesimal SUSY transformations is
\begin{equation}
\begin{split}
\delta_S (\xi, \bar{\xi}) A^\mu &= \ci \left( \xi \sigma^\mu \bar{\lambda} - \lambda \sigma^\mu \bar{\xi} \right)  - \partial^\mu (\xi \chi + \bar{\xi} \bar{\chi}) \;, \\
\delta_S (\xi, \bar{\xi}) \lambda_\alpha &= -\ci \, D \, \xi_\alpha - \frac{1}{2} {(\sigma^\mu \bar{\sigma}^\nu)_\alpha }^\beta \xi_\beta (\partial_\mu A_\nu - \partial_\nu A_\mu) \;, \\
\delta_S (\xi, \bar{\xi}) D &= \partial_\mu \left( - \xi \sigma^\mu \bar{\lambda} + \lambda \sigma^\mu \bar{\xi} \right) \;. \label{Eq:superVtrafo}
\end{split}
\end{equation}
Again we see that SUSY transforms bosonic into fermionic dof and vice versa and the SUSY transformation of the auxiliary field $D$ is a total derivative.

Under a non-Abelian SUSY gauge transformation the vector superfield $V$ transforms as
\begin{equation}
\e^{2 g V} \rightarrow \e^{-2 \, \ci \, g \bar{\Lambda}} \e^{2 g V} \e^{2 \, \ci \, g \Lambda} \;, \label{Eq:gaugeVtrafo}
\end{equation}
where $V=V^a T^a$. For Abelian gauge groups this simplifies to
\begin{equation}
V \rightarrow V + \ci (\Lambda- \bar{\Lambda}) \,.
\end{equation}

Now we have defined all necessary superfields and their transformations and we can continue by building SUSY invariant Lagrangians.

\subsection{Supersymmetric Lagrangians}

In this section we want to discuss how to construct Lagrangian densities $\mathcal{L}(x)$ invariant under SUSY and gauge transformations out of chiral and vector superfields.

To be more concrete the action has to be invariant under infinitesimal SUSY transformations
\begin{equation}
\delta_S \int \dd^4 x \, \mathcal{L}(x) = 0 \;.
\end{equation}
Therefore it is sufficient if $\mathcal{L}(x)$ changes only up to a total space-time derivative under SUSY transformations. From Eqs.~\eqref{Eq:superLHtrafo} and \eqref{Eq:superVtrafo} we know that the $F$-terms of chiral and the $D$-terms of vector superfields transform as total derivatives under SUSY transformations and therefore a SUSY invariant action can be constructed from them. We define
\begin{equation}
\mathcal{L}(x) \equiv \mathcal{L}_F + \mathcal{L}_D = \int \dd^2 \theta \, \mathcal{L}_f + \int \dd^2 \theta \, \dd^2 \bar{\theta} \, \mathcal{L}_d + \mathrm{h.c.,}
\end{equation}
where in $\mathcal{L}_F$ only the $F$-terms of $\mathcal{L}_f$ appear whereas in $\mathcal{L}_D$ only the $D$-terms of $\mathcal{L}_d$ appear due to the Grassmann nature of $\theta$ and $\bar{\theta}$, cf.\ App.~\ref{App:Grassmann}.

We start with the discussion of $\mathcal{L}_f$ which is an analytic function of chiral superfields and therefore a chiral superfield itself. $\mathcal{L}_f$ can be given in terms of the gauge invariant \emph{holomorphic} superpotential $\mathcal{W}$,
\begin{equation}
 \mathcal{L}_f = \mathcal{W}(\{\Phi_i\}) = \sum_i a_i \Phi_i + \frac{1}{2} \sum_{ij} m_{ij} \Phi_i \Phi_j + \frac{1}{3!} \sum_{ijk} \lambda_{ijk} \Phi_i \Phi_j \Phi_k \;, \label{Eq:Superpotential}
\end{equation}
where all $\Phi_i$ are left-chiral superfields and the couplings $m_{ij}$, $\lambda_{ijk}$ are totally symmetric under the interchange of $i,j,k$. We do not want to spoil the renormalisability of our theory for which reason only terms maximally trilinear in the superfields appear in the superpotential. The $F$-term of the superpotential is given in terms of the component fields $\phi_i$, $\psi_i$ and $F_i$
\begin{align}
 \int \dd^2 \theta \, \mathcal{W}(\{ \Phi_i \}) &= \sum_i a_i F_i + \sum_{ij} m_{ij} \left( \phi_i F_j - \frac{1}{2} \psi_i \psi_j \right) + \sum_{ijk} \frac{\lambda_{ijk}}{2} \left( \phi_i \phi_j F_k - \phi_i \psi_j \psi_k \right) \nonumber\\
&= \sum_j \frac{\partial \mathcal{W}(\phi)}{\partial \phi_j} F_j - \frac{1}{2} \sum_{jk} \frac{\partial^2 \mathcal{W}(\phi)}{\partial \phi_j \partial \phi_k} \psi_j \psi_k \;,
\end{align}
where in the last line the superpotential is understood to be a function of only the scalar fields $\phi_i$. This provides us with mass terms for the fermions and Yukawa-type interactions.

Now we have a closer look on $\mathcal{L}_D$. This provides us with kinetic terms for scalars and fermions since they are of the form $\bar{\Phi} \Phi$. To be more concrete the gauge invariance of the SUSY Lagrangian demands that
\begin{equation}
 \mathcal{L}_d = \bar{\Phi} \, \e^{2gV} \Phi \;,
\end{equation}
cf.\ Eqs.~\eqref{Eq:gaugechiraltrafo} and \eqref{Eq:gaugeVtrafo}. We also have to replace everywhere the usual derivative $\partial_\mu$ with the gauge covariant derivative $D_\mu$
\begin{equation}
 \partial_\mu \rightarrow D_\mu = \partial_\mu + \ci \, g  A_\mu^a T^a \;,
\end{equation}
where $T^a$ are the generators of the gauge transformations and $A^a_\mu$ are the vector components of a general vector superfield. Since we now know the structure of $\mathcal{L}_d$ we can write down $\mathcal{L}_D$ in component notation
\begin{equation}
\begin{split}
 \mathcal{L}_D &= \sum_i \int \dd^2 \theta \, \dd^2 \bar{\theta} \, \bar{\Phi}_i \, \e^{2 g V} \Phi_i \\
 &= \sum_i \left[ D_\mu \phi_i D^\mu \phi_i^* + \ci \, \bar{\psi}_i \bar{\sigma}^\mu D_\mu \psi_i - \sqrt{2} g \left( \bar{\psi}_i \bar{\lambda} \psi_i + \psi_i^* \lambda \psi_i \right) + g \phi_i^* T^a D^a \psi_i + F_i^* F_i \right] \;,
\end{split} \label{Eq:LSUSYD}
\end{equation}
which involves, besides kinetic terms for the scalars and fermions, interaction terms of the scalars and fermions with the gauge boson fields and ``SUSY-gauge interactions'' involving gauginos.

Now we are only lacking kinetic terms for gauginos and gauge bosons and couplings between gauge bosons and gauginos although they are allowed by gauge invariance. These terms are added to our theory by introducing the additional part $\mathcal{L}_{\mathrm{kin}}$,
\begin{equation}
 \mathcal{L}_{\mathrm{kin}} = \frac{1}{16 g^2} \mathrm{Tr} (W_\alpha W^\alpha) \;,
\end{equation}
to the Lagrangian where $W_\alpha$ are field strength tensors defined as
\begin{equation}
 W_\alpha \equiv \frac{1}{4} \bar{D} \bar{D} \, \e^{-2 g V} D_\alpha \e^{2 g V} \;. \label{Eq:SUSYfieldstrengthtensors}
\end{equation}
For Abelian gauge groups, Eq.~\eqref{Eq:SUSYfieldstrengthtensors} simplifies to
\begin{equation}
 W_\alpha = \frac{g}{2} \bar{D} \bar{D} D_\alpha V \;.
\end{equation}
It can be easily shown that $W_\alpha W^\alpha$ is gauge invariant. $W_\alpha$ is a chiral superfield and hence also $W_\alpha W^\alpha$ is chiral. Therefore $\mathcal{L}_{\mathrm{kin}}$ is a chiral superfield and the $F$-component transform as a total derivative under SUSY transformations and can be added to the $F$-term of the SUSY Lagrangian
\begin{equation}
 \frac{1}{16 g^2} \int \dd^2 \theta \, \left[ \mathrm{Tr}(W_\alpha W^\alpha) + \mathrm{h.c.} \right] = - \frac{1}{4} F_{\mu \nu}^a F^{\mu \nu \, a} + \ci \, \bar{\lambda}^a \bar{\sigma}^\mu (D_\mu \lambda)^a + \frac{1}{2} D^a D^a \;,
\end{equation}
with the usual field strength tensors
\begin{equation}
 F_{\mu \nu}^a = \partial_\mu A_\nu^a - \partial_\nu A_\mu^a + g f^{abc} A_\mu^b A_\nu^c \;,
\end{equation}
where $f^{abc}$ are the gauge group structure constants.

In summary the final SUSY Lagrangian $F$-term reads
\begin{equation}
\begin{split}
 \mathcal{L}_F &= \int \dd^2 \theta \, \mathcal{L}_f + \mathrm{h.c.} \\
 &= \int \dd^2 \theta  \left[\mathcal{W} + \overline{\mathcal{W}} \right] + \frac{1}{16 g^2} \int \dd^2 \theta \, \left[ \mathrm{Tr}(W_\alpha W^\alpha) + \mathrm{h.c.} \right] \;, \label{Eq:LSUSYF}
\end{split}
\end{equation}
and the complete SUSY Lagrangian is given by $\mathcal{L} = \mathcal{L}_F + \mathcal{L}_D$ where $\mathcal{L}_D$ is given by Eq.~\eqref{Eq:LSUSYD} and $\mathcal{L}_F$ is given by Eq.~\eqref{Eq:LSUSYF}.

It is interesting to note that if we expand the SUSY Lagrangian in component fields the auxiliary fields $F$ and $D$ do not obtain any kinetic terms as we already said before. We can define from them the scalar potential $\mathcal{V}$,
\begin{equation}
 \mathcal{V} \equiv \sum_i \left( - F_i^* F_i - \frac{\partial \mathcal{W}}{\partial \phi_i} F_i - \frac{ \partial \overline{\mathcal{W}}(\phi^*) }{\partial \phi_i^*} F_i^* \right) + \frac{1}{2} \sum_a D^a D^a \;,
\end{equation}
where $F_i$ is the $F$-component of the superfield $\Phi_i$ and $D^a$ is the $D$-component of the vector superfield $V^a$. Since the auxiliary fields have no kinetic terms their Euler--Lagrange equations read
\begin{equation}
 \frac{\partial \mathcal{L}}{\partial F} = \frac{\partial \mathcal{L}}{\partial D} = 0 \;.
\end{equation}
With the help of these equations we can eliminate $F$ and $D$ from the Lagrangian. The scalar potential is then
\begin{equation}
 \mathcal{V} = \sum_i F_i^* F_i + \frac{1}{2} \sum_a (D^a)^2 = \sum_i \frac{\partial \mathcal{W}(\phi)}{\partial \phi_i} \frac{\partial \mathcal{\overline{W}}(\phi^*)}{\partial \phi_i^*} + \frac{1}{2} \sum_l g_l^2 \sum_a \left( \sum_i \phi_i^* T_l^a \phi_i \right)^2 \label{Eq:ScalarPotential} \;,
\end{equation}
where $l$ sums over the gauge groups of the theory with the corresponding gauge coupling $g_l$ and generators $T_l^a$. The scalar potential is a sum of absolute values squared and therefore positive for every field configuration.

\section{The Minimal Supersymmetric Extension of the Standard Model} \label{Sec:SUSYMSSM}

In this section we want to discuss the minimal supersymmetric extension of the SM (MSSM) \cite{Nilles:1983ge, Haber:1984rc, Barbieri:1987xf}. This SUSY extension of the SM is minimal in the sense that it introduces the least possible amount of new fields without spoiling the solution to the hierarchy problem. It is based on the following principles:
\begin{itemize}
 \item The SM is extended by $\mathcal{N} = 1$ SUSY.
 \item The MSSM is invariant under the SM gauge group.
 \item The SUSY breaking scale is near the EW scale.
 \item $\mathcal{R}$-parity is conserved.
\end{itemize}
The first two principles are related to the requirement of minimality and lead to the field content as described in Sec.~\ref{Sec:MSSMfields}. The third principle is needed to guarantee that the MSSM solves the hierarchy problem. SUSY breaking is discussed in Sec.~\ref{Sec:SUSYbreaking} and in the MSSM leads to the parameters as described in Sec.~\ref{Sec:MSSMparameters}. $\mathcal{R}$-parity conservation is required to forbid fast proton decay and is discussed in Sec.~\ref{Sec:Rparity}.

\subsection{Field Content of the MSSM} \label{Sec:MSSMfields}

\renewcommand{\arraystretch}{1.3}
\begin{table}
\centering
\begin{tabular}{ccccc}
\toprule
Superfield & Label & Bosonic Part & Fermionic Part & Quantum Numbers  \\ \midrule
chiral & $Q$ & $\tilde{q}_L = (\tilde{u}_L,\tilde{d}_L)^T$ & $q_L = (u_L, d_L)^T$ & $\left(\mathbf{3},\mathbf{2},+\frac{1}{3}\right)$ \\
chiral & $\bar{U}$ & $\tilde{u}_R^*$ & $\bar{u}_R$ & $\left(\mathbf{\overline{3}},\mathbf{1},-\frac{4}{3}\right)$  \\
chiral & $\bar{D}$ & $\tilde{d}_R^*$ & $\bar{d}_R$ & $\left(\mathbf{\overline{3}},\mathbf{1},+\frac{2}{3}\right)$  \\  \midrule
chiral & $L$ & $\tilde{l}_L = (\tilde{\nu}_L,\tilde{e}_L)^T$ & $l_L = (\nu_L, e_L)^T$ & $\left(\mathbf{1},\mathbf{2}, -1\right)$ \\
chiral & $\bar{E}$ & $\tilde{e}_R^*$ & $\bar{e}_R$ & $\left(\mathbf{1},\mathbf{1},+2\right)$  \\ \midrule
chiral & $H_u$ & $h_u = (h_u^+, h_u^0)^T$ & $\tilde{h}_u = (\tilde{h}_u^+,\tilde{h}_u^0)^T$ & $\left(\mathbf{1},\mathbf{2}, +1\right)$ \\
chiral & $H_d$ & $h_d = (h_d^0, h_d^-)^T$ & $\tilde{h}_d = (\tilde{h}_d^0,\tilde{h}_d^-)^T$ & $\left(\mathbf{1},\mathbf{2}, -1\right)$ \\ \midrule
vector & $G^a$ & $g^a$ & $\tilde{g}^a$ & $\left(\mathbf{8},\mathbf{1},  0\right)$ \\
vector & $W^i$ & $W^i$ & $\tilde{W}^i$ & $\left(\mathbf{1},\mathbf{3},  0\right)$ \\
vector & $B$ & $B$ & $\tilde{B}$ & $\left(\mathbf{1},\mathbf{1},  0\right)$ \\
\bottomrule
\end{tabular}
\caption[MSSM Field Content]{Field content of the MSSM. The superfields are labelled with a capital letter. The SUSY partners of the SM fields (cf.\ Tab.~\ref{Tab:SMfields}) are denoted with a tilde and the subscripts $L$ and $R$ of the scalar SUSY fields refer to the chirality of the corresponding fermionic partner. We have suppressed family and colour indices for the chiral matter fields to streamline notation. The index $a=1,\ldots,8$ enumerates the vector superfields of $SU(3)_C$ and the index $i=1, 2, 3$ enumerates the vector superfields of $SU(2)_L$. \label{Tab:MSSMfields}}
\end{table}
\renewcommand{\arraystretch}{1.0}

In principle the MSSM field content is already fixed by using the SM gauge group and restricting ourselves to $\mathcal{N} = 1$ SUSY and the requirement of not introducing unnecessary additional fields. Although the field content can be derived straightforwardly we want to discuss the fields in a little bit more detail in the following.

Since the generators of the SM gauge group commute with the SUSY generators the fields in each supermultiplet have to have the same quantum numbers. Within the SM there are no pairs of fermions and bosons having the same gauge quantum numbers apart from the lepton and the Higgs doublet. There was a proposal that they are contained in the same supermultiplet \cite{Fayet:1976et} but this would result, apart from missing anomaly cancellation, in lepton-number violation and a neutrino mass incompatible with experimental data. Therefore  SUSY partners of both doublets have to be added to the field content of the SM.

All fermions in the SM are chiral and hence have to belong to a chiral superfield. So for every fermion an additional complex scalar is added in the MSSM with the same quantum numbers. The naming scheme is that in front of the name of the fermion a `s' as abbreviation for scalar is added. So the scalar partner of the $\tau$-lepton is called `stau'. The scalar partner of the $b$-quark is called sbottom and so on. The generic term for the scalar partners of the fermions is correspondingly sfermions. The label of the SUSY partner of a fermion $f$ is denoted with a tilde $\tilde{f}$.

The gauge bosons of the SM are vector fields and therefore they also belong to vector superfields. The fermionic partners therein are called gauginos. In more detail the SUSY partner of the gluon $g$ is called gluino $\tilde{g}$, the SUSY partners of the $W^i$-bosons Winos $\tilde{W}^i$ and the SUSY partner of the $B$-boson is called Bino $\tilde{B}$. After EW symmetry breaking the electrically neutral gauginos are Majorana particles whereas the electrically charged gauginos are combined to form Dirac particles.

The Higgs sector of the MSSM cannot be minimally extended by putting the SM Higgs into a chiral superfield. The superpotential has to be holomorphic. Thus there may not appear a field and its conjugate at the same time in the superpotential. But in the SM the up-type quarks couple to the Higgs field while the charged leptons and the down-type quarks couple to its conjugate. If we want to give all fermions a mass via the Higgs mechanism we therefore have to add another Higgs doublet. This has the consequence that in the MSSM five physical Higgs bosons appear in the particle spectrum and not only one as in the SM. These two Higgs doublets are included in two chiral superfields with opposite hypercharge. The fermionic partners of the Higgs bosons are called Higgsinos. Since the chiral Higgs fields are conjugated to each other their gauge anomaly contributions cancel which is an additional nice feature of the MSSM.

We have collected the complete field content of the MSSM in Tab.~\ref{Tab:MSSMfields}.

\subsection{$\mathcal{R}$-Parity} \label{Sec:Rparity}

The superpotential from Eq.~\eqref{Eq:Superpotential} has to respect the SM gauge group and with the field content of the MSSM, see Tab.~\ref{Tab:MSSMfields}, the most general superpotential doing this can be decomposed as \cite{Sakai:1981pk, Weinberg:1981wj}
\begin{equation}
 \mathcal{W} = \mathcal{W}_R + \mathcal{W}_{\slashed{\mathcal{R}}} \;,
\end{equation}
where the two parts, written in terms of the MSSM superfields, are
\begin{align}
 \mathcal{W}_{\mathcal{R}} &= (Y_u)_{ij} Q_i \epsilon H_u \bar{U}_j - (Y_e)_{ij} L_i \epsilon H_d \bar{E}_j - (Y_d)_{ij} Q_i \epsilon H_d \bar{D}_j + \mu H_d \epsilon H_u \;, \label{Eq:WRconserving} \\
  \mathcal{W}_{\slashed{\mathcal{R}}} &= \frac{1}{2} \lambda_{ijk} L_i \epsilon L_j \bar{E}_k + \lambda'_{ijk} L_i \epsilon Q_j \bar{D}_k + \kappa_i L_i \epsilon H_u  + \frac{1}{2} \lambda''_{ijk} \bar{D}_i \bar{D}_j \bar{U}_k   \;, \label{Eq:WRviolating}
\end{align}
where we have suppressed gauge indices. The $\epsilon$ tensors are inserted to contract the $SU(2)_L$ indices. Due to gauge invariance $\lambda_{ijk}$ and $\lambda''_{ijk}$ are antisymmetric in the first two indices. Factors of $1/2$ are introduced to avoid double counting in scattering amplitudes.

The superpotential $\mathcal{W}_{\mathcal{R}}$ is necessary since it induces fermion Yukawa couplings and masses. The last term, the Higgs $\mu$ term, is necessary as well to generate successful electroweak symmetry breaking.

The first three terms in Eq.~\eqref{Eq:WRviolating} violate lepton number while the last term violates baryon number. The combination of baryon and lepton number violating operators gives rise to rapid proton decay \cite{Smirnov:1996bg, *Bhattacharyya:1998bx, *Barbier:2004ez}. Such operators should hence be strongly suppressed since there are severe bounds on proton decay, see, e.g.\  \cite{SuperK:2009gd, Raby:2008pd}.

In the SM such operators are absent because there baryon and lepton number are accidentally preserved and hence proton decay is no problem within the SM. In the MSSM an additional symmetry has to be invoked. In principle there are three discrete symmetries which are consistent with an underlying anomaly-free $U(1)$ gauge theory. Global symmetries are broken by quantum gravity effects and therefore a gaugeable discrete $U(1)$ subgroup is desirable  \cite{Dreiner:2005rd, Ibanez:1991pr, *Ibanez:1991hv, Mohapatra:2007vd}. In the MSSM  $\mathcal{R}$-parity \cite{Farrar:1978xj} or equivalently matter parity is invoked and we do not discuss the other two possibilities, proton-hexality and baryon-triality.

By introducing $\mathcal{R}$-parity to each particle a multiplicative discrete quantum number $P_{\mathcal{R}}$ is assigned where
\begin{equation}
P_{\mathcal{R}} = (-1)^{3(B-L) + 2s}
\end{equation}
with $B$ being the baryon number, $L$ the lepton number and $s$ the spin of the given particle. Since the spin enters here the different components of a superfield have different $\mathcal{R}$-parities. Our definition is such that the SM fields have even $\mathcal{R}$-parity whereas SUSY particles have odd $\mathcal{R}$-parity. Imposing this symmetry forbids the superpotential $\mathcal{W}_{\slashed{\mathcal{R}}}$ from Eq.~\eqref{Eq:WRviolating} as desired.

\subsection{MSSM Spectrum and Soft SUSY Parameters} \label{Sec:MSSMparameters}

In this section we want to discuss briefly the new parameters introduced by (softly broken) SUSY and the resulting spectrum.

If SUSY was unbroken it would be a very economic extension of the SM. In the MSSM without the SUSY breaking terms only one additional parameter is added. The new parameter is the $\mu$ parameter that mixes the two Higgs superfields in the superpotential, see \eqref{Eq:WRconserving}. Although this is very economic, the question arises why this parameter is of the order of the EW scale to generate correct EW symmetry breaking and not of the GUT or the Planck scale. This is the so-called $\mu$-problem \cite{Hall:1983iz, *Kim:1983dt, *Inoue:1985cw, *Anselm:1986um, *Giudice:1988yz} which can be solved by introducing an additional symmetry and a gauge singlet which breaks this symmetry and generates $\mu$ spontaneously. Those kinds of models are called NMSSM, see, e.g.\  \cite{Fayet:1974pd, *Nilles:1982dy, *Frere:1983ag, *Derendinger:1983bz, *Greene:1986th, *Ellis:1986mq, *Durand:1988rg, *Drees:1988fc, *Ellis:1988er, *Elliott:1994ht, *King:1995vk}. Nevertheless, we want to stick to the MSSM here, where $\mu$ is simply assumed to be of the right scale. 

If SUSY is broken the MSSM is not so economic anymore. Indeed there are about 100 soft SUSY breaking parameters. Those parameters are called soft because they do not spoil the solution to the hierarchy problem. So the number of additional parameters in the MSSM is restricted by the demand of not reintroducing quadratic divergences in the theory. In the following we classify the soft SUSY breaking parameters. For a more extensive treatment, see, e.g.\  the review \cite{Chung:2003fi}.

First of all there are mass terms for the gauginos
\begin{equation}
-\frac{1}{2} (M_1 \tilde{B} \tilde{B} + M_2 \tilde{W} \tilde{W} + M_3 \tilde{g} \tilde{g}) + \mathrm{h.c.} \;,
\end{equation}
where we have suppressed gauge indices. Secondly there are the Higgs mass parameters
\begin{equation}
- m_{h_u}^2 h_u^\dagger h_u - m_{h_d}^2 h_d^\dagger h_d - (b h_u \epsilon h_d + \mathrm{h.c.}) \;.
\end{equation}
These mass terms together with the $F$- and $D$-terms generate the whole Higgs potential which breaks EW symmetry. Furthermore there are the three-by-three sfermion mass matrices before EW symmetry breaking
\begin{equation}
-\tilde{q}_L^\dagger m_{\tilde{Q}}^2 \tilde{q}_L -\tilde{u}_R^* m_{\tilde{U}}^2 \tilde{u} _R -\tilde{d}_R^* m_{\tilde{D}}^2 \tilde{d}_R -\tilde{l}_L^\dagger m_{\tilde{L}}^2 \tilde{l}_L -\tilde{e}_R^* m_{\tilde{E}}^2 \tilde{e}_R \;,
\end{equation}
and finally the trilinear couplings
\begin{equation}
- (A_U \tilde{q}_L h_u \tilde{u}_R^* - A_D \tilde{q}_L h_d \tilde{d}_R^* - A_E \tilde{l}_L h_d \tilde{e}_R^*) + \mathrm{h.c.} \,.
\end{equation}
The sfermion mass matrices and the trilinear couplings are matrices in family space which can give rise to additional flavour transitions. Their structures are thereby severely constrained by phenomenology. But to this point within the MSSM itself, their structure is not restricted, which gives rise to the SUSY flavour problem. The SUSY flavour problem is an extension of the SM flavour problem in the sense that the MSSM contains even more flavourful quantities than the SM whose patterns have to be explained.

The soft parameters also introduce a lot of additional complex phases which result in electric dipole moments that are severely constrained by phenomenology as well. The open question of the phases in the MSSM is the SUSY CP problem.

We now summarise briefly the MSSM spectrum. First of all, in addition to the SM spectrum there are the sfermions. For every fermion in the SM we have two scalar partners appearing in the spectrum. After EW symmetry breaking the left- and right-handed scalars mix and give two different mass eigenstates per fermion. So there are all in all twelve squarks and nine sleptons. Since there are no right-handed neutrinos in the SM there are also no right-handed sfermions in the MSSM.

The gluinos, the superpartners of the gluon, are their own mass eigenstate while the other gauginos mix with the Higgsinos after EW symmetry breaking. The electrically charged Winos mix with the electrically charged Higgsino components and give the two charginos. The electrically neutral Wino, the Bino and the neutral Higgsinos form four mass eigenstates called the neutralinos. The lightest neutralino is also a natural dark matter candidate as long as $\mathcal{R}$-parity is conserved and the lightest neutralino is the LSP.

We would also like to restate here that the MSSM is a two Higgs doublet model and therefore in the MSSM there are five physical Higgs mass eigenstates. If the Higgs potential preserves the CP symmetry there are two CP even mass eigenstates, one CP odd mass eigenstate and two oppositely charged Higgs bosons with the same mass.

\section{Supersymmetry Breaking} \label{Sec:SUSYbreaking}

We know that SUSY, even if it is realised in nature, has to be broken. In this section we therefore discuss briefly how SUSY can be broken in general and then we give three explicit, popular examples of SUSY breaking schemes.

\subsection{General Aspects of SUSY Breaking}

We start our discussion of SUSY breaking with some general aspects. If SUSY is spontaneously broken the vacuum state $\vert 0 \rangle$ is no longer invariant under SUSY transformations $Q_\alpha \vert 0 \rangle \neq 0$ and $\bar{Q}_{\dot{\alpha}} \vert 0 \rangle \neq 0$ where $Q_\alpha$ and $\bar{Q}_{\dot{\alpha}}$ are the SUSY generators, cf.\ \eqref{Eq:SUSYalgebra}. The SUSY algebra relates the Hamiltonian $\mathcal{H}$ to the SUSY generators via
\begin{equation}
 \mathcal{H} = P^0 = \frac{1}{4} ( Q_1 \bar{Q}_1 + \bar{Q}_1 Q_1 + Q_2 \bar{Q}_2 + \bar{Q}_2 Q_2 ) \;.
\end{equation}
Therefore in broken global SUSY the vacuum energy has to be strictly positive
\begin{equation}
 \langle 0 \vert \mathcal{H} \vert 0 \rangle = \frac{1}{4} \left( \left\vert Q_1 \vert 0 \rangle \right\vert^2 + \left\vert \bar{Q}_1 \vert 0 \rangle \right\vert^2 + \left\vert Q_2 \vert 0 \rangle \right\vert^2 + \left\vert \bar{Q}_2 \vert 0 \rangle \right\vert^2 \right) > 0 \;.
\end{equation}
Furthermore, neglecting space-time dependent effects and fermion condensates the vev of the Hamiltonian $\langle 0 \vert \mathcal{H} \vert 0 \rangle$ is equal to the vev of the scalar potential $\langle 0 \vert \mathcal{V} \vert 0 \rangle$ which is generated by the $F$- and $D$-terms, see Eq.~\eqref{Eq:ScalarPotential}. That tells us that SUSY is broken if at least one of the auxiliary field components has a non-vanishing vev for all possible field configurations. Such a vev generates mass terms for the SUSY partners of a given theory. In spontaneously broken SUSY theories a sum rule for the tree-level squared masses of the particles can be derived, see, e.g.\  \cite{Martin:1997ns},
\begin{equation}
 \mathrm{STr}(m^2) \equiv \sum_j (-1)^j (2j+1) \mathrm{Tr}(m_j^2)  = 0 \;, \label{Eq:SUSYMassSumRule}
\end{equation}
where the sum is taken over all particles with spin $j$ and the last equality is valid in theories with non-anomalous $U(1)$ gauge symmetries, as the $U(1)_Y$ in the MSSM.

If we would assume a family symmetry or simply conservation of individual lepton numbers the sum rule \eqref{Eq:SUSYMassSumRule} decomposes into different blocks and one obtains, for example, for the (s)electrons
\begin{equation}
 m_{\tilde{e}_1}^2 + m_{\tilde{e}_2}^2 = 2 m_e^2 \;,
\end{equation}
which is in conflict with experiment since there are no charged scalars in this mass regime. If we would impose lepton flavour violation as it is suggested by neutrino oscillations it is still very difficult to obtain viable SUSY partner masses in agreement with phenomenological data from the lepton masses and lepton flavour violation. Therefore, a different scenario is proposed where SUSY is broken in a \emph{hidden sector} and the breaking is then mediated to the visible sector via (flavour-blind) interactions or radiative corrections. Before we discuss three of these breaking scenarios we want to discuss briefly how to give the $F$- or the $D$-terms a vev.

In the MSSM neither the $F$- nor the $D$-terms can acquire a vev. One simple example to give a vev to the $F$-terms is the model proposed by O'Raifeartaigh \cite{O'Raifeartaigh:1975pr}. The proposed superpotential with three superfields is
\begin{equation}
 \mathcal{W} = -k \Phi_1 + m \Phi_2 \Phi_3  + \frac{y}{2} \Phi_1 \Phi_3^2 \;,
\end{equation}
which is only allowed if $\Phi_1$ is a gauge singlet. From this superpotential the scalar potential with the corresponding $F$-terms can be derived
\begin{equation}
\begin{split}
 \mathcal{V} &= \vert F_1\vert^2 + \vert F_2\vert^2 + \vert F_3\vert^2 \;, \\
 F_1 &= k - \frac{y}{2} {\phi_3^*}^2 \;, \quad F_2 = - m \phi_3^* \;, \quad F_3 = - m\phi_2^* - y \phi_1^* \phi_3^* \;.
\end{split}
\end{equation}
Here $F_1=0$ and $F_2=0$ cannot be fulfilled simultaneously, so one of the $F$-terms acquires a vev and SUSY is spontaneously broken.

The $D$-terms of a $U(1)$ symmetry can obtain a vev due to the fact that for $U(1)$ symmetries a term linear in the auxiliary field $D$ can be introduced into the scalar potential
\begin{equation}
 \mathcal{V} = \kappa D - \frac{1}{2} D^2 - g D \sum_i q_i \vert \phi_i \vert^2 \;,
\end{equation}
where $\kappa$ is a constant with units of $[\text{mass}]^2$ and $q_i$ are the charges of the scalar fields $\phi_i$ under the $U(1)$ transformation. The term linear in $D$ is called Fayet--Iliopoulos term since Fayet and Iliopoulos proposed this kind of SUSY breaking \cite{Fayet:1974jb}. The presence of this linear term changes the equations of motions to
\begin{equation}
 D = \kappa - g \sum_i q_i \vert \phi_i \vert^2 \;.
\end{equation}
If we assume that all the scalar fields $\phi_i$ have $F$-term masses $m_i$
\begin{equation}
 \mathcal{V} = \sum_i \vert m_i \vert^2 \vert \phi_i \vert^2 + \frac{1}{2} ( \kappa - g \sum_i q_i \vert \phi_i \vert^2 )^2 \;,
\end{equation}
and $\vert m_i \vert^2 > g q_i \kappa$ for each $i$ we have a SUSY breaking minimum which preserves the $U(1)$ symmetry. In the MSSM one could try to use the $U(1)_Y$ group for SUSY breaking but this does not work out since squarks and sleptons have no superpotential mass terms and in this case the minimum of the superpotential in the above equation would break $SU(3)_C$ or $U(1)_{\mathrm{em}}$ which is excluded by experiment.

We now have an idea how SUSY could be broken and we know that this cannot happen at tree-level in the visible sector. In the following we want to discuss three scenarios for the mediation of SUSY breaking which lead to special structures for the soft SUSY parameters. Indeed this is what is usually done due to our missing understanding of SUSY breaking. The soft SUSY breaking terms which are compatible with the solution to the hierarchy problem are parameterised in terms of a few parameters motivated by a certain SUSY breaking scenario. Here, we discuss only three common scenarios used in this work.

\subsection{Minimal Anomaly Mediated Supersymmetry Breaking} \label{Sec:AMSB}

The anomaly mediated SUSY breaking scenario (AMSB) is based on the assumption of extra dimensions. If there are extra dimensions and SUSY is broken on a separate brane this breaking is mediated to the visible world via the superconformal anomaly \cite{Randall:1998uk, *Giudice:1998xp, *Gherghetta:1999sw}.

The parameter $m_{3/2}$, the vev of the auxiliary field in the supergravity multiplet, determines the overall mass scale of the SUSY particle masses. However, in the simplest AMSB model the sleptons are tachyonic. To resolve this issue, in the minimal AMSB scenario (mAMSB) an additional universal scalar soft mass $m_0$ is introduced. The spectrum is completely determined by the four parameters $m_{3/2}$, $m_0$, $\tan \beta$ and $\mu$. The two parameters $\mu$, the Higgs superfield mixing parameter, and $\tan \beta$, the ratio of the Higgs vevs, are needed to describe EWSB completely. Although $\mu$ is in principle a free parameter, the modulus of $\mu$ can be calculated from conditions for successful EWSB if all other parameters of the Higgs potential are known. The sign of $\mu$ stays undetermined and usually is not counted as a free parameter since it has only two possible values and as we  discuss later in more detail a negative $\mu$ parameter is disfavoured by measurements of the anomalous magnetic moment of the muon $(g-2)_\mu$, see Sec.~\ref{Sec:g2mu}. So in the end we are left with only the three parameters $m_{3/2}$, $m_0$ and $\tan \beta$.

Explicitly, the boundary conditions at the GUT scale for $a=1,2,3$ in mAMSB are given by 
\begin{equation}
\begin{split}
 M_a (M_{{\mathrm{GUT}}}) & =  \frac{\beta(g_a)}{g_a} m_{3/2} \;, \\
 A_y (M_{{\mathrm{GUT}}}) & =  -\frac{\beta_y}{y}  m_{3/2} \;, \\
 \tilde{m}_{\tilde{f}}^2 (M_{{\mathrm{GUT}}}) & =  - \frac{1}{4} \left[
                \beta(g_a) \frac{\partial \gamma_{\tilde{f}}}{\partial g_a}
                + \beta_y \frac{\partial \gamma_{\tilde{f}}}{\partial y}
                \right] m_{3/2}^2 + m_0^2 \;,
\end{split}
\end{equation}
where $M_a$ are the gaugino masses, $A_y$ the trilinear couplings, $\tilde{m}_{\tilde f}$ the sfermion soft mass parameters, $\beta(g_a)$ are the $\beta$ functions of the gauge couplings $g_a$, $\beta_y$ the $\beta$ function of the Yukawa coupling $y$ and the $\gamma_{\tilde{f}}$ is the anomalous dimension of the superfield $\tilde{f}$.

\subsection{Minimal Gauge Mediated Supersymmetry Breaking} \label{Sec:GMSB}

In the minimal gauge mediated SUSY breaking scenario (mGMSB) \cite{Dine:1993yw, *Dine:1994vc, *Dine:1995ag, *Ambrosanio:1997rv, *Giudice:1998bp} SUSY breaking is mediated via gauge interactions. The messenger fields are assumed to be complete five-dimensional representations of $SU(5)$. This choice does not spoil gauge coupling unification. The SUSY breaking sector is coupled on one-loop level to the gauginos and on two-loop level to the remaining SUSY fields. The trilinear couplings arise at two-loop order and are additionally suppressed by a factor of $\alpha_a/4\pi$, $a=1,2,3$, and hence are negligibly small. Since SUSY breaking is mediated via gauge interactions the soft scalar masses are predicted to be universal at $\Lambda$.

In these scenarios  the SUSY spectrum depends on six parameters: the messenger mass $m_{\mathrm{mess}}$, the number of $\mathbf{5} \oplus \overline{\mathbf{5}}$ messenger fields $n_5$, the soft SUSY breaking mass scale $\Lambda$, the constant $c_{\mathrm{grav}}$ needed to calculate the gravitino mass, $\tan \beta$ and the sign of $\mu$. The modulus of $\mu$ can be calculated from the conditions for successful EWSB as in the mAMSB case. Later on we set $c_{\mathrm{grav}} = 1$ without loss of generality, since we do not investigate observables depending on the gravitino mass. Thus we are left with only four free parameters in this case.

The universal boundary conditions are applied at the messenger scale and read for the gaugino masses $M_a$, $a=1,2,3$ and the sfermion soft mass parameters $\tilde{m}_{\tilde f}$
\begin{equation}
\begin{split}
 M_a (m_{\mathrm{mess}}) & =  \frac{g^2_a}{16 \pi^2} n_5 \Lambda
        \tilde{g} \left( \frac{\Lambda}{m_{\mathrm{mess}}} \right) \;, \\
 \tilde{m}_{\tilde f} (m_{\mathrm{mess}}) & =  2 \Lambda^2 \sum_a 
        \left( \frac{g_a^2}{16 \pi^2} \right)^2 C_a n_5 
        \tilde{f} \left( \frac{\Lambda}{m_{\mathrm{mess}}} \right) \;,
\end{split}
\end{equation}
where $C_a$ is the quadratic Casimir invariant of the MSSM scalar field in question and
\begin{equation}
\begin{split}
 \tilde{g}(x) & =  \frac{1}{x^2} \left[ (1+x)\ln(1+x) +
                (1-x) \ln (1-x) \right]  \; ,\\
 \tilde{f}(x) & =  \frac{1+x}{x^2} \left[ \ln(1+x)
                -2 \mathrm{Li}_2 \left(\frac{x}{1+x}\right)
                +\frac{1}{2} \mathrm{Li}_2\left(\frac{2x}{1+x}\right)
                \right] + (x \rightarrow - x) \; ,
\end{split}
\end{equation}
are functions appearing in the calculation of the loop diagrams.

\subsection{Constrained Minimal Supersymmetric Standard Model} \label{Sec:CMSSM}

The constrained MSSM (CMSSM) is a SUSY breaking scenario inspired by supergravity (SUGRA) theories where SUSY breaking is mediated via gravitational interactions \cite{Nilles:1982ik, *Nilles:1982xx, Chamseddine:1982jx, *Barbieri:1982eh}. To avoid large flavour changing neutral currents (FCNC) usually the assumption is made that the couplings of the SUSY breaking sector to the visible sector is universal in flavour space. This kind of model is called minimal supergravity (mSUGRA) since a minimal choice for the K\"ahler potential is used \cite{Chamseddine:1982jx, *Barbieri:1982eh}. There is a subtle difference between CMSSM and mSUGRA concerning the gravitino mass which shall not bother us here since we always assume that the gravitino is not the LSP and for the other parts it is irrelevant.

The boundary conditions for the soft SUSY breaking parameters at the GUT scale are then
\begin{equation}
\begin{split}
 M_a (M_{{\mathrm{GUT}}}) & =  m_{1/2} \;, \\
 A_Y (M_{{\mathrm{GUT}}}) & =  A_0 Y \;, \\
 \tilde{m}_{\tilde{f}}^2 (M_{{\mathrm{GUT}}}) & =  m_0^2 \;,
\end{split}
\end{equation}
where $M_a$, $a=1,2,3$, are the gaugino masses, $A_Y$ are the trilinear couplings which are assumed to be proportional to the particular Yukawa matrices $Y$ and $\tilde{m}_{\tilde{f}}$ are the sfermion masses. All in all in the CMSSM we have therefore four parameters fully describing the MSSM spectrum. They are $m_{1/2}$, $A_0$, $m_0$ and $\tan \beta$. The absolute value of $\mu$ is determined from the conditions for EWSB and the sign of $\mu$ is set to be positive due to the constraint from the anomalous magnetic moment of the muon.

%% file: kap_04_GUTs.tex
\chapter{Grand Unified Theories} \label{Ch:GUTs}

In this chapter we discuss the idea of Grand Unified Theories (GUTs) on the basis of $SU(5)$ gauge theory which is the smallest simple gauge group containing the SM gauge group and being compatible with the field content of the SM. This GUT is of special importance for this work since later on we construct explicit flavour models within $SU(5)$.

The guiding principle behind GUTs is the quest for unification of forces. The SM knows three gauge interactions which can be described in terms of symmetry groups. Therefore it seems compelling to look for a larger group which contains the SM gauge group and hence describes the SM gauge interactions in terms of one simple group. Besides unification of forces, GUTs also shed some light on the questions of neutrino masses, charge quantisation and anomaly cancellation. Very recently it was even shown how to embed the inflaton into a GUT representation \cite{Antusch:2010va}.

The first step towards a GUT was the unification of colour and lepton number in the Pati--Salam (PS) model \cite{Pati:1973uk} with the gauge group $SU(4)_C \times SU(2)_L \times SU(2)_R$. Later on also simple groups were proposed, like $SU(5)$ \cite{Georgi:1974sy}, $SO(10)$ \cite{Georgi:1974my, *Fritzsch:1974nn} or $E_6$ \cite{Gursey:1975ki}. Here we want to discuss the relevant features of GUTs via the example of $SU(5)$. Our treatment of GUTs is based on the presentations in \cite{Cheng:1985bj, *Ross:1985ai, *Georgi:1982jb}.

We start with an introduction to $SU(5)$ and describe the field content in Sec.~\ref{Sec:SU5Intro}. Afterwards we discuss how electric charge is quantised in Sec.~\ref{Sec:GUTChargeQuantisation},  the unification of gauge couplings in Sec.~\ref{Sec:CouplingUnification} and proton decay in Sec.~\ref{Sec:ProtonDecay}. We end this chapter with a brief summary about symmetry breaking and fermion masses within $SU(5)$ in Sec.~\ref{Sec:GUTMasses}.

\section[Introduction to $SU(5)$ and Field Content]{Introduction to $\boldsymbol{SU(5)}$ and Field Content} \label{Sec:SU5Intro}

A general representation of $SU(5)$ can be written as a tensor under $SU(5)$, for more details, see App.~\ref{App:SUNReps}. The transformation matrices $U$ are in this case
\begin{equation}
[U]^i_m = [\mathrm{exp} (\ci \, \alpha^a \lambda^a/2)]^i_m \;,
\end{equation}
where the indices $i, m = 1, \ldots, 5$ and $a=0,\ldots,23$. The $\alpha^a$ are the transformation parameters and the $\lambda_a$  are the generators of $SU(5)$ (The group $SU(N)$ has $N^2-1$ generators).  The generators of $SU(5)$ are Hermitian, traceless $5 \times 5$ matrices with the normalisation $\mathrm{Tr}(\lambda^a \lambda^b) = 2 \delta^{ab}$, for example
\begin{equation}
\lambda^3 = \begin{pmatrix} 0 & & & & \\  & 0 & & & \\  & & 0 & & \\  & & & 1 & \\  & & & & -1 \\ \end{pmatrix} \quad \text{and} \quad \lambda^0 =  \frac{1}{\sqrt{15}}  \begin{pmatrix} 2 & & & & \\  & 2 & & & \\  & & 2 & & \\  & & & -3 & \\  & & & & -3 \\ \end{pmatrix} \;. \label{Eq:SU5matrices}
\end{equation}
The $SU(3)_C  \times SU(2)_L$ decomposition of a given representation is given by identifying the first three components of an $SU(5)$ index $i=1,2,3$ as colour indices and the last two components of an $SU(5)$ index $i=4,5$ as weak isospin indices. Below we label for convenience colour indices with Greek letters, $\alpha$, $\beta$, $\ldots$, and weak isospin indices with Latin letters, $r$, $s$, $\ldots$.

Taking a closer look at the SM field content, cf.\ Tab.~\ref{Tab:SMfields}, we see that all the left-handed matter fields fit nicely into the two representations
\begin{equation}
\begin{split}
F &= \overline{\mathbf{5}} = (\overline{\mathbf{3}},\mathbf{1}) + (\mathbf{1},\overline{\mathbf{2}}) =  \begin{pmatrix}
         d_R^{c} & d_B^{c} & d_G^{c} & e &-\nu
          \end{pmatrix}_L \:, \\
T &= \mathbf{10} =  (\overline{\mathbf{3}},\mathbf{1}) + (\mathbf{3},\mathbf{2}) + (\mathbf{3},\mathbf{1})
        = \frac{1}{\sqrt{2}}
           \begin{pmatrix}
           0 & -u_G^{c} & u_B^{c} & -u_{R} & -d_{R} \\
           u_G^{c} & 0 & -u_R^{c} & -u_{B} & -d_{B} \\
           -u_B^{c} & u_R^{c} & 0 & -u_{G} & -d_{G} \\
           u_{R} & u_{B} & u_{G} & 0 & -e^c \\
           d_{R} & d_{B} & d_{G} & e^c & 0
           \end{pmatrix}_L \:.
\end{split}
\end{equation}
where we have given the decomposition under $SU(3)_C \times SU(2)_L$, see also \cite{Slansky:1981yr}. The lower indices $R$, $B$ and $G$ denote the quark colours and the index $L$ denotes that $F$ and $T$ are left-handed. The hypercharge of the fields is discussed in the next section.

The gauge bosons of the SM are all unified into the twenty-four-dimensional adjoint representation of $SU(5)$ $A^i_j$ which can be decomposed as
\begin{equation}
A^i_j =\mathbf{24} = (\mathbf{8},\mathbf{1}) + (\mathbf{1},\mathbf{3}) + (\mathbf{1},\mathbf{1}) + (\mathbf{3},\mathbf{2}) +(\overline{\mathbf{3}},\mathbf{2}) \;.
\end{equation}
Using the index-labelling convention as described above
we can identify the SM gauge bosons as
\begin{itemize}
\item $A^\alpha_\beta = (\mathbf{8},\mathbf{1})$ are the gluons of $SU(3)_C$.
\item $A^r_s = (\mathbf{1},\mathbf{3})$ are the $W$ bosons of $SU(2)_L$.
\item $ \sqrt{3/20} A^r_r - \sqrt{1/15} A^\alpha_\alpha  = (\mathbf{1},\mathbf{1})$ is the $U(1)_Y$ B-field.
\end{itemize}
Then twelve of the gauge bosons in $A^i_j$ are still unassigned. They carry both $SU(3)_C$ and $SU(2)_L$ indices
\begin{equation}
A_\alpha^r = (\mathbf{3},\mathbf{2}) \quad \text{and} \quad A_r^\alpha  = (\overline{\mathbf{3}},\mathbf{2}) \;,
\end{equation}
which are commonly denoted as $X$ and $Y$ gauge bosons
\begin{equation}
A_\alpha^r = (X_\alpha,Y_\alpha) \quad \text{and} \quad A_r^\alpha  = \begin{pmatrix} X^\alpha \\ Y^\alpha \end{pmatrix}\;.
\end{equation}
In the next section we derive the operator for electromagnetic charge, but already here we note that the new gauge bosons have the electric charges
\begin{equation}
Q_X = - \frac{4}{3} \quad \text{and} \quad Q_Y  = -\frac{1}{3} \;.
\end{equation}

The Higgs fields are discussed in Sec.~\ref{Sec:GUTMasses} where we discuss symmetry breaking.

\section{Charge Quantisation} \label{Sec:GUTChargeQuantisation}

If the SM gauge group is unified into one simple non-Abelian group, electric charge has to be quantised since the eigenvalues of the generators of a simple non-Abelian group are discrete. Electric charge $Q$ is an additive quantum number and therefore must be a linear combination of  diagonal generators. $SU(5)$ is a Lie group of rank four and therefore also has four diagonal generators. $Q$ has to commute with the generators of $SU(3)_C$ and then only two possibilities remain, namely $T^3$ and $T^0$ with $T^a = \lambda^a/2$. $T^3$ is the diagonal generator of weak isospin and hence we have
\begin{equation}
Q = I^3 + \frac{Y}{2} = T^3 + c \, T^0 \;.
\end{equation}
The coefficient $c$ can be determined by comparing the eigenvalues of $Y$ with the eigenvalues of $T^0$ and we obtain
\begin{equation}
c = - \sqrt{\frac{5}{3}} \;.
\end{equation}
For the fundamental representation this yields the correct electric charge,
\begin{equation}
Q(\bar{F}_i) = \begin{pmatrix} - \frac{1}{3} & & & & \\ & - \frac{1}{3} & & & \\  & & - \frac{1}{3} & & \\  & & & 1 & \\   & & & & 0  \\   \end{pmatrix} = Q_i \, \delta_{ij} \;. \label{Eq:SU5Q}
\end{equation}
The conjugate representation $\overline{\mathbf{5}}$ has the opposite charge $Q(F^i) = - Q_i \, \delta_{ij}$. For a rank two tensor $\psi$ the electric charge can be calculated via
\begin{equation}
\begin{split}
Q(\psi_{ij}) &= Q_i + Q_j \;,\\
Q(\psi_i^j) &= Q_i - Q_j \;.
\end{split}
\end{equation}
It can be checked that for the ten-dimensional fermion representation $T$ or the twenty-four-dimensional adjoint representation $A^i_j$ the charges of the fields are correctly reproduced as well. For higher-rank tensors the above formula can be generalised straightforwardly.

It is interesting to note that in GUTs a relation between colour and electromagnetic charge emerges. Since the generators of $SU(N)$ groups are traceless we get the relation
\begin{equation}
N_C \, Q_d + Q_{e^c} = 0 \;,
\end{equation}
from Eq.~\eqref{Eq:SU5Q} where $N_C$ is the number of colours. In $SU(3)_C$ there are exactly three colours which enforces that quarks carry 1/3 of the charge of leptons. Within $SU(5)$ (and other GUTs) the pattern of gauge quantum numbers hence seems to be a lot less arbitrary than in the SM.

We note that charge quantisation emerges here only since all generators of charge are part of a non-Abelian gauge group. If electric charge is a linear combination of generators where one generators belongs to a $U(1)$ factor as it is the case, for example, in flipped $SU(5)$ with the gauge group $SU(5) \times U(1)$ charge does not have to be quantised anymore. Nevertheless charge quantisation is a salient feature in a wide class of models as long as neutrinos are Majorana particles \cite{Babu:1989ex, *Babu:1989tq}. In our concrete flavour models in Chs.~\ref{Ch:GUTImplications} and \ref{Ch:Model} we always assume that neutrinos are Majorana particles.

We now turn to anomaly freedom in $SU(5)$. In $SU(N)$ gauge theories the anomaly from Eq.~\eqref{Eq:SMAnomaly} simplifies to
\begin{equation}
\mathcal{A}^{abc} = \mathrm{Tr} [t^a \{ t^b,t^c \}] = \frac{1}{2} \, A(R) \, d^{abc}\;,
\end{equation}
where $R$ is the considered representation and $d^{abc}$ are the symmetric structure constants of the gauge group. $A(R)$ is independent of the generators and therefore we use $t^a = t^b = t^c = Q$ to calculate the ratio of the anomalies of the matter content
\begin{equation}
\frac{A(F)}{A(T)} = \frac{\mathrm{Tr} \, Q^3 (F)}{\mathrm{Tr} \, Q^3 (T)} = \frac{3 (1/3)^2 + (-1)^3 + 0^3}{3(-2/3)^3 + 3(2/3)^3 + 3(-1/3)^3 + 1^3} = -1 \;.
\end{equation}
Hence, by the extension of the gauge group to $SU(5)$ the theory remains free of anomalies. Nevertheless, this anomaly cancellation still appears accidental. However, on the level of $SO(10)$ this cancellation is lifted to an intrinsic feature.

\section{Unification of Gauge Couplings} \label{Sec:CouplingUnification}

In GUTs the gauge couplings have to unify to match the covariant derivatives
\begin{equation}
\begin{split}
D^{\mathrm{SM}}_\mu &= \partial_\mu + \ci \, g_3 \sum_{a=1}^8 g_\mu^a T^a  + \ci \,  g_2 \sum_{r=1}^3 W_\mu^r I^r + \ci \, g' B_\mu \frac{Y}{2} \;, \\
D^{SU(5)}_\mu &= \partial_\mu  + \ci \, g_5 \sum_{a=0}^{23} A_\mu^a \frac{\lambda^a}{2} \;,
\end{split}
\end{equation}
where $g_3$, $g_2$ and $g'$ are the SM gauge couplings, $g_\mu^a$, $W_\mu^r$ and $B_\mu$ are the SM gauge fields with their generators $T^a$, $I^r$ and $Y/2$. Correspondingly $g_5$ is the $SU(5)$ gauge coupling with the gauge fields $A_\mu^a$ and generators $\lambda^a/2$.

As we have already discussed in the last section hypercharge is not correctly normalised. The matching to a generator of $SU(5)$ gives
\begin{equation}
Y = -\sqrt{\frac{5}{3}} \lambda^0 \quad \text{and} \quad g' = - \sqrt{\frac{3}{5}} g_1 \;,
\end{equation}
where $g_1$ is the correctly normalised gauge coupling of the $U(1)_Y$ group. With the help of this definition the two covariant derivatives of the SM and $SU(5)$ can be matched if the gauge couplings unify,
\begin{equation}
g_1=g_2=g_3=g_5 \;. \label{Eq:UnificationCondition}
\end{equation}
This relation obviously is not true at the weak scale, but gauge couplings run and at a high scale this relation can be satisfied to a high accuracy, see Fig.~\ref{Fig:gaugeunification}. As we have already discussed in Ch.~\ref{Ch:SUSY} this unification works out much better in the MSSM. That is  one of the reasons why we mainly discuss SUSY GUTs in this thesis. The extension of non-SUSY GUTs to SUSY GUTs is straightforward since the SUSY generators commute with the generators of internal symmetry groups. All fields  are extended to superfields and we have to introduce the EWSB Higgs doublets in pairs. This is discussed in the next section.

We now give the RGEs for the gauge couplings and discuss shortly how to determine the GUT scale from them, the scale at which the gauge couplings unify. The RGEs at one-loop level are given by
\begin{equation}
16 \pi^2 \mu_R \frac{\dd g_i}{\dd \mu_R } = b_i g_i^3 \;,
\end{equation}
where $\mu_R$ is the renormalisation scale and $i=1,2,3$ enumerates the gauge group. The $b_i$ are the $\beta$-function coefficients and have the values $(b_3,b_2,b_1) = (-7, -19/6,41/10)$ in the SM and $(b_3,b_2,b_1) = (-3, 1, 33/5)$ in the MSSM. The analytic solution of the above differential equation is
\begin{equation}
\frac{1}{g_i ^2 (\mu_R)} = \frac{1}{g_i^2(\mu_0)}  + \frac{b_i}{8 \pi^2 } \ln \frac{\mu_R}{\mu_0} \;,
\end{equation}
where $\mu_0$ is the boundary scale which we set to $M_Z$ since the gauge couplings at this scale are well determined. Setting two of the GUT scale gauge couplings equal allows us to determine the GUT scale from that equation. For the non-SUSY case this would be roughly $10^{14}$~GeV but we are not interested in this case since the gauge couplings within the SM unify poorly. In the MSSM case the GUT scale $M_{\mathrm{GUT}}$ is approximately $2 \times 10^{16}$~GeV, a value we use in the rest of this thesis as the definition for the GUT scale.

From the unification condition \eqref{Eq:UnificationCondition} we can now predict  the third gauge coupling. For the fine structure constant at the GUT scale we have $\alpha_5 (M_{\mathrm{GUT}}) \approx 1/24$ from which the third gauge coupling at the GUT scale can be determined and via its RGE evolution we can compare its low energy value with the experimental data. As mentioned above in the SM the gauge couplings unify poorly and hence the prediction there is also bad in contrast to the MSSM where the gauge couplings unify, or in other words, the prediction for the third gauge coupling is good.

\section{Proton Decay} \label{Sec:ProtonDecay}

\begin{figure}
\centering
\includegraphics[scale=0.7]{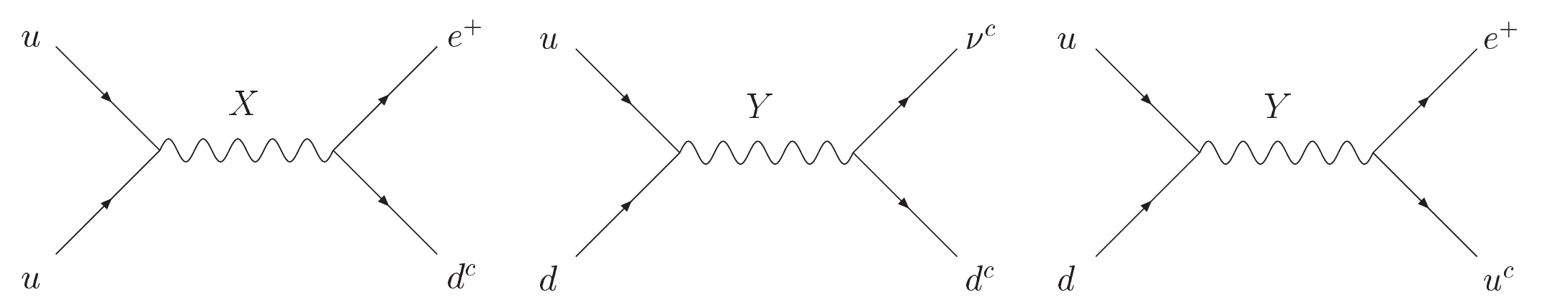}
\caption[Proton Decay by Gauge Boson Exchange]{Feynman diagrams contributing to proton decay. After integrating out the heavy gauge bosons $X$ and $Y$ they give effective dimension six operators mediating proton decay which are the dominant contribution in non-SUSY $SU(5)$. \label{Fig:NonSUSYpdecay}}
\end{figure}

\begin{figure}
\centering
\includegraphics[scale=0.7]{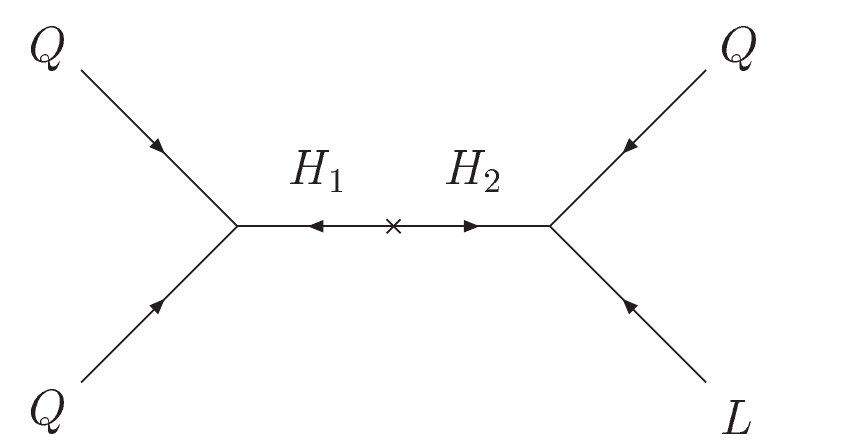}
\caption[Proton Decay by Triplet Exchange]{Supergraph contributing to proton decay in SUSY GUTs. The cross denotes a mass insertion. After going to component notation and integrating out the heavy Higgsino triplets it gives effective dimension five operators mediating proton decay in SUSY $SU(5)$. \label{Fig:SUSYpdecay}}
\end{figure}

We do not focus in this thesis on proton decay but, nevertheless, it is one important prediction of GUTs and the discovery of proton decay is often claimed to be a \emph{smoking gun} signature of GUTs although for example in $\mathcal{R}$-parity violating SUSY models the proton also is not stable. The proton is even unstable in the SM due to sphaleron processes but they are so strongly suppressed that proton decay in the SM is far beyond experimental reach.

In non-SUSY GUTs the main contribution to proton decay stems from effective dimension six operators generated by the exchange of the heavy gauge bosons $X$ and $Y$, depicted in Fig.~\ref{Fig:NonSUSYpdecay}. In SUSY GUTs with $\mathcal{R}$-parity conservation the main contribution to proton decay is mediated by the exchange of heavy triplets, depicted in Fig.~\ref{Fig:SUSYpdecay}. Closely related to proton decay in SUSY GUTs is hence the doublet-triplet-splitting problem. If the triplets could be made arbitrarily heavy, proton decay could be strongly suppressed.  The question is how to make the triplet components of the five-dimensional Higgs representations heavy while keeping the doublet components light.

We do not discuss these issues and possible solutions here in detail, for more information see, e.g.\  \cite{Senjanovic:2009kr, Raby:2008pd}, or for a recent concrete $SO(10)$ model compatible with actual bounds on proton decay, see \cite{Babu:2010ej}. Throughout this thesis we simply assume that some mechanism is at work which resolves these issues since our main focus is on the flavour sector.

\section{Symmetry Breaking and Fermion Masses}  \label{Sec:GUTMasses}

The $SU(5)$ can be spontaneously broken similar as the $SU(2)_L \times U(1)_Y$ is broken in the SM. In the simplest case the necessary Higgs field is in a twenty-four-dimensional adjoint representation of $SU(5)$ which breaks the GUT group to the SM group provided the vev is proportional to $\lambda^0$, cf.\ Eq.~\eqref{Eq:SU5matrices}. The vev is obviously invariant under the SM gauge group. The additional gauge bosons $X$ and $Y$ become massive by the spontaneous breakdown of $SU(5)$ and have GUT scale masses if the vev of the twenty-four is of the order of the GUT scale.

In the simplest case the SM symmetry breaking is triggered by a five-dimensional representation of $SU(5)$ or in the SUSY case by a pair of five-dimensional representations conjugated to each other. For the masses of the down-type quarks and charged leptons, a conjugated five-dimensional representation is needed which we call $\bar{H}_5$. The doublet component then acquires an EW scale vev and the operators generating the masses for the down-type quarks and charged leptons have the structure $y F T \bar{H}_5$ with $y$ being the Yukawa coupling. Since such operators give simultaneously masses to down-type quarks and charged leptons the ratio of the respective Yukawa couplings is fixed at the GUT scale. In this case the Yukawa couplings are even exactly equal. This relation is only approximately true for the third generation and Yukawa coupling ratios at the GUT scale are one of the main topics of this thesis and are discussed in detail later on. The up-type quark masses can be generated by an operator of the form $y T T H_5$ where again $y$ is the Yukawa coupling.

Here, it becomes clear why flavour model building in the context of (SUSY) GUTs is appealing. In $SU(5)$ we only have two types of flavours ($F$ and $T$) appearing in three copies, one for each generation. This is much more economic than in the SM. In $SO(10)$, all matter fields of one generation are unified into one single sixteen-dimensional representation and thus there appears only one type of flavour in three copies which is even more economic.

%% file: kap_05_NewGUTRelations.tex
\chapter[New GUT Predictions for Yukawa Coupling Ratios]{New GUT Predictions for\\ Yukawa Coupling Ratios} \label{Ch:GUTYukawaRelations}

In this chapter we discuss GUT predictions for Yukawa coupling ratios based on \cite{Antusch:2009gu}. We focus on $SU(5)$ \cite{Georgi:1974sy} and PS \cite{Pati:1973uk} GUTs which we assume to be embedded in a $SO(10)$ GUT \cite{Georgi:1974my, *Fritzsch:1974nn}. We do not work out this embedding explicitly but we use it as a motivation for the possible Higgs field content. We want to emphasise here that, although the focus of this thesis is on SUSY GUTs, the results of this chapter rely on the underlying group structure only and are therefore also valid in non-SUSY GUTs.

Within unified theories Yukawa couplings emerge from operators involving the unified fermion representations as well as Higgs fields in GUT representations where one of the fields has to contain an EW  Higgs. Thus in general, each such operator generates Yukawa couplings for different types of fermions, for example for down-type quarks as well as for charged leptons, which are related to each other by the group theoretical Clebsch--Gordan (CG) factors from GUT symmetry breaking.

\section{Conditions for Predictions} \label{sec:Conditions}

Now we clarify under which conditions such relations lead to observable predictions for quark and lepton masses.

The first condition results in simple relations between entries of the Yukawa matrices and the quark and charged lepton masses and it demands that the Yukawa matrices in the flavour basis are hierarchical and dominated by the diagonal elements. This situation is approximately realised in many approaches to unified model building, but only regarding the second and third generation. In other words we claim that the masses of the second generation of quarks and charged leptons are directly related to the 2-2 entries of the Yukawa matrices and the masses of the third generation to the 3-3 entries. For the masses of the first generation of fermions this condition is often violated and the relation to the elements of the Yukawa matrices depends on additional assumptions, e.g.\  whether there is a texture zero in the 1-1 entry of the Yukawa matrices, see, e.g.\  \cite{Gatto:1968ss}. For the phenomenological analysis later on we therefore focus mainly on the second and third generation.

The second condition is that there is only one operator dominating the relevant elements of the Yukawa matrices. This requirement is necessary because if, for instance, two operators would contribute with similar strength, the resulting prediction would be an intermediate value and it would be difficult to disentangle the fundamental relations and respective operators.

In this chapter we assume that these two conditions are satisfied sufficiently well.

\section[$b-\tau$ Unification and Georgi--Jarlskog Relations]{$\boldsymbol{b-\tau}$ Unification and Georgi--Jarlskog Relations} \label{Sec:ExampleRelations}

There are two examples of quark and lepton mass relations at the GUT scale which are ubiquitous in many classes of unified models of flavour. These are third partial family Yukawa unification, or $b$-$\tau$ unification, and the so-called Georgi--Jarlskog (GJ) relations \cite{Georgi:1979df}, i.e.\ $y_\mu/y_s=3$ and $y_e/y_d=1/3$.

In the following, we review them briefly in the context of $SU(5)$ GUTs. In $SU(5)$ all the SM matter fields are contained in the representations $F^i$ and $T^i$, where $i$ is the family index, see Ch.~\ref{Ch:GUTs} and App.~\ref{App:SUNReps} for more details. If the Yukawa matrix 3-3 entries for down-type quarks and charged leptons are generated by an operator of the form $F^3 T^3 \bar{H}_5$ where the five-dimensional $\bar{H}_5$ contains an EW Higgs $SU(2)_L$ doublet, then
it is easy to see that the resulting prediction is $y_b/y_\tau = 1$, i.e.\  $b$-$\tau$  Yukawa coupling unification. On the other hand, if the relevant 2-2 entries of the Yukawa matrices are generated by the operator $F^2 T^2 \bar{H}_{45}$ with the EW Higgs field contained in the 45-dimensional $SU(5)$ representation $\bar{H}_{45}$ then $y_\mu / y_s = -3$ is predicted. This results from the fact that the $\bar{H}_{45}$ can be written as a traceless tensor and for the trace to vanish the factor of $-3$ for the charged leptons has to compensate the colour factor of 3 for the quarks. In the following we choose the Yukawa couplings to be positive, which is the reason for referring to the GJ relation as $y_\mu / y_s = 3$. The second GJ relation $y_e/y_d = 1/3$ emerges from a special assumption about the Yukawa matrix texture where the 1-1 entry is zero.

In the following we derive in addition to $b$-$\tau$ unification and the GJ relations various alternative relations between quark and lepton Yukawa couplings emerging from higher-dimensional operators in unified theories.

\section{New GUT Predictions}

\begin{figure}
 \centering
 \includegraphics[scale=0.9]{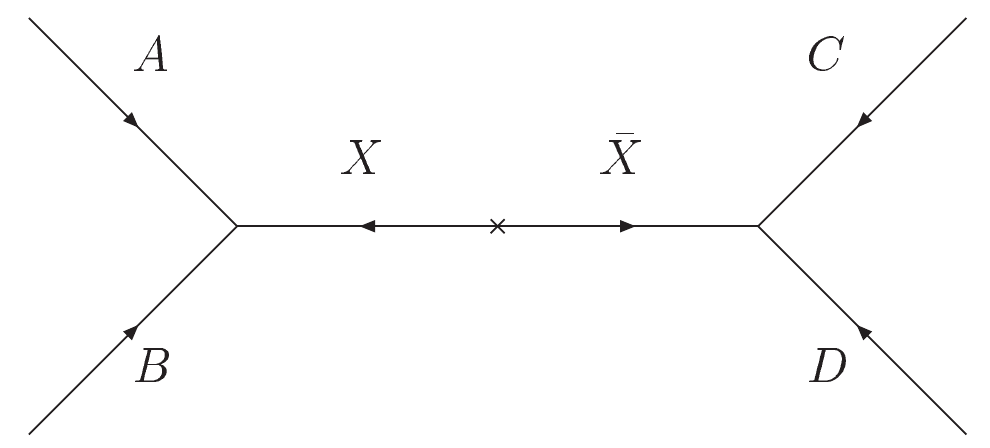}
 \caption[Feynman Graph Generating New GUT Relations]{Feynman graph with heavy messenger fields $X$ and $\bar{X}$. When the messenger fields are effectively integrated out of the theory below their mass scales, higher-dimensional operators are generated which can lead to GUT relations between quark and lepton Yukawa couplings. \label{Fig:Messenger}}
\end{figure}

When the conditions specified in Sec.~\ref{sec:Conditions} are satisfied, the predicted relations between quark and lepton Yukawa couplings at the GUT scale mainly depend on the specific operator which dominates the relevant entries of the Yukawa matrices. The simplest types of operators in this context are the renormalisable ones, for example the operators mentioned above which lead to $b$-$\tau$ unification and the GJ relation for the second generation. The different predictions result from different Higgs representations which contain the EW Higgs(es).

The procedure to obtain possible predictions for Yukawa coupling ratios in these cases is the following: The operators include two matter fields and one Higgs field. For the matter fields, the common matter representations of the unified theories are chosen. Doing so, the possible representations of the Higgs field are fixed by the condition that the operator has to be a gauge singlet after contracting all gauge indices and that the Higgs field has to include the usual SM (MSSM) Higgs(es). Explicit expressions for the Higgs vevs are given later.

New possibilities, in addition to the Yukawa coupling ratios of 1 and $-3$ can arise in particular when effective, higher-dimensional operators are taken into account. In many unified flavour models using family symmetries, the renormalisable dimension-four operators are forbidden by that symmetry and the Yukawa couplings are generated from higher-dimensional operators in the effective theory limit. Later on we give explicit examples which use the relations proposed in this chapter. Non-renormalisable operators are typically generated from integrating out messenger fields $X$ and $\bar{X}$, cf.\ Fig.~\ref{Fig:Messenger}.
The (super)fields $A$, $B$, $C$ and $D$ can be either matter or Higgs fields. In total the effective operator has to contain two matter fields, one Higgs field breaking EW symmetry and one Higgs field with a GUT scale vev. The latter is only allowed to break the unified gauge symmetry but not the EW symmetry. At low energies, the Yukawa operators of the SM (MSSM) are realised with some of the Yukawa couplings related to each other due to the  underlying unified group structure.

We restrict ourselves to messenger fields and GUT scale Higgs fields which are included in the common $SO(10)$ representations, i.e.\ $\mathbf{10}$, $\mathbf{16}$, $\mathbf{45}$, $\mathbf{54}$, $\mathbf{120}$, $\mathbf{126}$ and $\mathbf{210}$ and their conjugates under $SO(10)$. With these restrictions, we cover the cases of most GUT models based on $SO(10)$ broken to the SM gauge group via PS or $SU(5)$. For a list of used representations and their symmetry properties in terms of Young tableaux see App.~\ref{App:RepList}. In the next subsections we derive the results for the cases of (SO(10) broken to the SM via) $SU(5)$ or PS. A summary of the results is contained in Tabs.~\ref{Tab:SU5Relations} and \ref{Tab:PSRelations}.

\subsection[Predictions from $SU(5)$ Unification]{Predictions from $\boldsymbol{SU(5)}$ Unification} \label{sec:SU5}

\begin{table}
\begin{center}
\begin{tabular}{cccc} \toprule
($A$, $B$) & ($C$, $D$) & $X$ & $y_e/y_d$ \\ \midrule
($F$, $T$) & ($\bar{H}_5$, $H_{24}$) & $\overline{\mathbf{5}}$ & 1 \\
($F$, $T$) & ($\bar{H}_5$, $H_{24}$) & $\overline{\mathbf{45}}$ & -3 \\
($F$, $T$) & ($\bar{H}_5$, $H_{75}$) & $\overline{\mathbf{45}}$ & -3 \\
($F$, $T$) & ($\bar{H}_{45}$, $H_{24}$) & $\overline{\mathbf{5}}$ & 1 \\
($F$, $T$) & ($\bar{H}_{45}$, $H_{24}$) & $\overline{\mathbf{45}}_1$ & -3 \\
($F$, $T$) & ($\bar{H}_{45}$, $H_{24}$) & $\overline{\mathbf{45}}_2$ & - \\
($F$, $T$) & ($\bar{H}_{45}$, $H_{75}$) & $\overline{\mathbf{5}}$ & 1 \\
($F$, $T$) & ($\bar{H}_{45}$, $H_{75}$) & $\overline{\mathbf{45}}_1$ & -3 \\
($F$, $T$) & ($\bar{H}_{45}$, $H_{75}$) & $\overline{\mathbf{45}}_2$ & - \\
($F$, $\bar{H}_5$) & ($T$, $H_{24}$) & $\mathbf{10}$ & 6 \\
($F$, $\bar{H}_5$) & ($T$, $H_{24}$) & $\mathbf{15}$ & 0 \\
($F$, $\bar{H}_5$) & ($T$, $H_{75}$) & $\mathbf{10}$ & -3 \\
($F$, $\bar{H}_{45}$) & ($T$, $H_{24}$) & $\mathbf{10}$ & -18 \\
($F$, $\bar{H}_{45}$) & ($T$, $H_{24}$) & $\mathbf{40}$ & 0 \\
($F$, $\bar{H}_{45}$) & ($T$, $H_{75}$) & $\mathbf{10}$ & 9 \\
($F$, $\bar{H}_{45}$) & ($T$, $H_{75}$) & $\mathbf{40}$ & 0 \\
($F$, $H_{24}$) & ($T$, $\bar{H}_5$) & $\mathbf{5}$ & -3/2 \\
($F$, $H_{24}$) & ($T$, $\bar{H}_5$) & $\mathbf{45}$ & 3/2 \\
($F$, $H_{75}$) & ($T$, $\bar{H}_5$) & $\mathbf{45}$ & -3 \\
($F$, $H_{24}$) & ($T$, $\bar{H}_{45}$) & $\mathbf{5}$ & 9/2 \\
($F$, $H_{24}$) & ($T$, $\bar{H}_{45}$) & $\mathbf{45}$ & -1/2 \\
($F$, $H_{75}$) & ($T$, $\bar{H}_{45}$) & $\mathbf{45}$ & 1 \\
($F$, $H_{75}$) & ($T$, $\bar{H}_{45}$) & $\mathbf{50}$ & 0 \\
\bottomrule
\end{tabular}
\end{center}
\caption[Dimension-five $SU(5)$ Yukawa Coupling Relations]{
Dimension-five operators within $SU(5)$ unification and resulting predictions for the GUT scale ratios $y_e/y_d$, where $e$ and $d$ stand for any charged lepton and down-type quark of the same generation. $A, B, C, D$ and $X$ correspond to the fields in the Feynman graph in Fig.~\ref{Fig:Messenger} which generates the dimension-five operator after integrating out the heavy messenger fields. The index of the Higgs fields $H$ gives the dimension of their representation. If the messenger representation $X$ has an index, there is more than one way to combine the fields $A$ and $B$ or $C$ and $D$ to form this representation leading to possibly different predicted ratios $y_e/y_d$. A dash means that $y_d$ is zero. At the stage of $SU(5)$ unification the dimension-five operators predict no relation between the up-type quark Yukawa couplings and any other Yukawa couplings. \label{Tab:SU5Messenger}}
\end{table}

\begin{table}
\begin{center}
\begin{tabular}{cc}\toprule
Operator Dimension &	$y_e/y_d$ 	\\ 
\midrule 4	&	1	\\
	&	-3	\\
\midrule 5	&	-1/2	\\
	&	1	\\
	&	$\pm$3/2	\\
	&	-3	\\
	&	9/2	\\
	&	6	\\
	&	9	\\
	&	-18	\\
\bottomrule
\end{tabular}
\end{center}
\caption[Summary of $SU(5)$ Yukawa Coupling Relations]{
Summary of possible $SU(5)$ predictions for the GUT scale ratios $y_e/y_d$, where $e$ and $d$ stand for any charged lepton and down-type quark of the same generation. 
\label{Tab:SU5Relations}}
\end{table}

As mentioned before, we perform our analysis at the stage of $SU(5)$ or PS unification for simplicity. However, what we have in mind is a possible embedding into $SO(10)$ GUTs. In GUTs based on the unifying gauge group $SU(5)$, the fermions of the SM are embedded in the GUT representations $F^i$ and $T^i$, see Ch.~\ref{Ch:GUTs}, where $i=1,2,3$ is the family index. The $F^i$ and $T^i$, plus an extra SM singlet, form the matter representations $\mathbf{16}^i$ of $SO(10)$.

The commonly used GUT representations for the Higgs fields in $SU(5)$ are $\bar{H}_5$, $H_{24}$ and $\bar{H}_{45}$. Their notation and vevs are specified as
\begin{align}
 \left( \bar{H}_5 \right)^a = \overline{\mathbf{5}}^{\, a}  & \;, \quad  \langle \left( \bar{H}_5 \right)^5 \rangle = v_5 \;,\\
\left(H_{24}\right)^a_b = \mathbf{24}^{\, a}_{\, b} &\;, \quad  \langle \left(H_{24}\right)^a_b \rangle = v_{24} (2 \delta^a_\alpha \delta_b^\alpha - 3 \delta^a_\beta \delta_b^\beta)\;, \label{Eq:H24vev} \\
\left(\bar{H}_{45}\right)^{ab}_c = -\left(\bar{H}_{45}\right)^{ba}_{c} = \overline{\mathbf{45}}^{\, ab}_{\, c} &\;,  \quad\langle \left(\bar{H}_{45}\right)^{i5}_j \rangle = v_{45} \left(\delta^i_j - 4 \delta^{i4} \delta_{j4}\right)\;,
\end{align}
where $a,b=1,\ldots,5$, $\alpha=1,2,3$, $\beta=4,5$ and $i,j=1,\ldots,4$. The vevs $v_5$ and $v_{45}$ are assumed to be of the EW scale whereas $v_{24}$ is of the order of the GUT scale. $H_{24}$ breaks $SU(5)$ down to $G_{\mathrm{SM}}$. For the determination of the vevs of the GUT-breaking Higgs fields we have neglected the influence of the vevs of the Higgs fields which break the EW symmetry, which provides a very good approximation.

In addition, we also consider the 75-dimensional Higgs representation $H_{75}$. $H_{24}$ and $H_{75}$ are the only nontrivial representations which are included in the common $SO(10)$ representations and have a SM singlet component that can obtain a GUT scale vev without breaking the SM gauge symmetries. The vev of $H_{75}$ is constructed via its Young tableau, see App.~\ref{App:SUNReps}, from the vev of $H_{24}$ which preserves the SM gauge group.

At dimension four, only operators containing $\bar{H}_5$ and $\bar{H}_{45}$ can generate Yukawa couplings for the down-type quarks and charged leptons. The $\bar{H}_5$ gives $b$-$\tau$ unification and the $\bar{H}_{45}$ the GJ relation already mentioned in Sec.~\ref{Sec:ExampleRelations}. For the construction of the dimension-five operators we can add an additional $H_{24}$ or a $H_{75}$ to the dimension-four operators. All possible resulting combinations of external and messenger fields are listed in Tab.~\ref{Tab:SU5Messenger}, including the corresponding Yukawa coupling ratio. If the messenger representation in the table has an index, there is more than one way to combine the fields $A$ and $B$ or $C$ and $D$ to form this representation.

The resulting relations are collected in Tab.~\ref{Tab:SU5Relations}. Since the operators do not relate the up-type quark Yukawa couplings to any other Yukawa couplings, we only present the predicted ratio $y_e/y_d$, where $e$ and $d$ stand for any charged
lepton and down-type quark of the same generation. Higher-dimensional operators involving the Higgs representation $H_{24}$ have also been considered in \cite{Duque:2008ah}. The possible CG factor $3/2$ is mentioned there as well, however it has not been postulated as a GUT prediction. The CG factor of $3/2$ was also already claimed in \cite{Ellis:1979fg} as a correction term to make the relations $y_d/y_s$ and $y_e/y_\mu$ viable in $SU(5)$ but it was not used there as a postulated Yukawa coupling relation as well.

To illustrate how the relations from dimension five operators are generated, let us discuss the operator leading to the new prediction $y_e/y_d = 9/2$. Using the notation of Fig.~\ref{Fig:Messenger} we can assign $A=F$, $B=H_{24}$, $C=T$ and $D=\bar{H}_{45}$. At the left vertex $F$ and $H_{24}$ are combined to a $\overline{\mathbf{5}}$ to couple to the messenger field $X=\mathbf{5}$. From the vev of $H_{24}$ the down-type quarks are multiplied by a factor of 2 and the leptons by a factor of $-3$, cf.~Eq.~\eqref{Eq:H24vev}. At the right vertex $T$ and $\bar{H}_{45}$ are combined to form a $\mathbf{5}$. Since $\bar{H}_{45}$ is traceless, this, similar to the GJ relation, leads to an additional relative factor of $-3$ for the down-type quarks compared to the charged leptons. In combination, this gives the relative factor of 9/2.

\subsection{Predictions from Pati--Salam Unification} \label{Sec:PS}

\begin{table}
\begin{center}
\begin{tabular}{cccc} \toprule
($A$, $B$) & ($C$, $D$) & $X$ & $(y_e/y_d,y_u/y_d,y_\nu/y_u)$ \\ \midrule
($R$, 
$\phi$) & 
($\bar{R}$, 
$h$) & 
$(\mathbf{\overline{4}},\mathbf{\overline{2}},\mathbf{\overline{3}})$ & (1, 1, 1) \\
($R$, 
$\bar{R}$) & 
($h$, 
$\hat{\phi}$) & 
$(\mathbf{15},\mathbf{\overline{2}},\mathbf{2})$ & (-3, 1, -3) \\
($R$, 
$\bar{R}$) & 
($h$, 
$\tilde{\phi}$) & 
$(\mathbf{15},\mathbf{\overline{2}},\mathbf{2})$ & (-3, 1, -3) \\
($R$, 
$\bar{R}$) & 
($\tilde{h}$, 
$\phi$) & 
$(\mathbf{15},\mathbf{\overline{2}},\mathbf{2})$ & (-3, 1, -3) \\
($R$, 
$\bar{R}$) & 
($\tilde{h}$, 
$\hat{\phi}$) & 
$(\mathbf{1},\mathbf{\overline{2}},\mathbf{2})$ & (1, 1, 1) \\
($R$, 
$\bar{R}$) & 
($\tilde{h}$, 
$\hat{\phi}$) & 
$(\mathbf{15}_1,\mathbf{\overline{2}},\mathbf{2})$ & (-3, 1, -3) \\
($R$, 
$\bar{R}$) & 
($\tilde{h}$, 
$\hat{\phi}$) & 
$(\mathbf{15}_2,\mathbf{\overline{2}},\mathbf{2})$ & (-3, 1, -3) \\
($R$, 
$\tilde{h}$) & 
($\bar{R}$, 
$\hat{\phi}$) & 
$(\mathbf{\overline{4}},\mathbf{1},\mathbf{\overline{2}})$ & (9, 1, 9) \\
($\bar{R}$, 
$\tilde{h}$) & 
($R$, 
$\hat{\phi}$) & 
$(\mathbf{4},\mathbf{2},\mathbf{1})$ & (9, 1, 9) \\
($R$, 
$\bar{R}$) & 
($\tilde{h}$, 
$\tilde{\phi}$) & 
$(\mathbf{1},\mathbf{\overline{2}},\mathbf{2})$ & (1, 1, 1) \\
($R$, 
$\bar{R}$) & 
($\tilde{h}$, 
$\tilde{\phi}$) & 
$(\mathbf{15},\mathbf{\overline{2}},\mathbf{2})_1$ & (-3, 1, -3) \\
($R$, 
$\bar{R}$) & 
($\tilde{h}$, 
$\tilde{\phi}$) & 
$(\mathbf{15},\mathbf{\overline{2}},\mathbf{2})_2$ & (-3, 1, -3) \\
($R$, 
$\tilde{h}$) & 
($\bar{R}$, 
$\tilde{\phi}$) & 
$(\mathbf{\overline{4}},\mathbf{1},\mathbf{\overline{2}})$ & (9, 1, 9) \\
($\bar{R}$, 
$\tilde{h}$) & 
($R$, 
$\tilde{\phi}$) & 
$(\mathbf{4},\mathbf{2},\mathbf{1})$ & (9, 1, 9) \\
\bottomrule
\end{tabular}
\end{center}
\caption[Dimension-five PS Yukawa Coupling Relations]{
Dimension-five operators within PS unification and resulting predictions for the GUT scale ratios $y_e/y_d$, $y_u/y_d$ and $y_\nu/y_u$, where $\nu$, $e$, $d$ and $u$ stand for any neutrino, charged lepton, down-type and up-type quark of the same generation. $A, B, C, D$ and $X$ correspond to the fields in the Feynman graph in Fig.~\ref{Fig:Messenger} which generates the dimension-five operator after integrating out the heavy messenger field. For a description of the fields and their representation see Sec.~\ref{Sec:PS}. If the messenger representation $X$ has an index, there is more than one way to combine the fields $A$ and $B$ or $C$ and $D$ to form this representation leading to possibly different predicted ratios.  \label{Tab:PSMessenger}}
\end{table}

We now turn to the case of classes of $SO(10)$ GUTs where the breaking to the SM proceeds via the PS breaking chain (Remember: $G_{\mathrm{PS}} = SU(4)_C \times SU(2)_L \times SU(2)_R$).  At the stage of PS unified theories, the fermions of the SM are embedded in the representations  $(\mathbf{4},\mathbf{2},\mathbf{1}) $ and $(\mathbf{\overline{4}},\mathbf{1},\mathbf{\overline{2}})$ of the PS gauge group as
\begin{align}
  R^{i}_{\alpha a} & =  (\mathbf{4},\mathbf{2},\mathbf{1})^i =
         \begin{pmatrix}
         u_L^R & u_L^B & u_L^G & \nu_L \\
         d_L^R & d_L^B & d_L^G & e_L^-
         \end{pmatrix}^i , \\
  \bar{R}^{i \alpha x} & = 
         (\mathbf{\overline{4}},\mathbf{1},\mathbf{\overline{2}})^i
       = \begin{pmatrix}
         \bar{d}_R^R & \bar{d}_R^B & \bar{d}_R^G & e_R^+ \\
         \bar{u}_R^R & \bar{u}_R^B & \bar{u}_R^G & \bar{\nu}_R
         \end{pmatrix}^i ,
\end{align}
where $\alpha = 1,\ldots,4$ is an $SU(4)_C$ index, $a,x = 1,2$ are $SU(2)_{L,R}$ indices and $i=1,2,3$ is a family index. The fields in $R^i$ form $SU(2)_L$ doublets and the fields in $\bar{R}^i$ 
$SU(2)_L$ singlets as indicated by the index $L$ and $R$. The MSSM Higgs $SU(2)_L$ doublets $h_u$ and $h_d$ are contained in the bi-doublet representation
\begin{equation}
  h_x^a = (\mathbf{1},\overline{\mathbf{2}},\mathbf{2}) = \begin{pmatrix}
         h_u^+ & h_d^0 \\
         h_u^0 & h_d^-
         \end{pmatrix}.
\end{equation}
It acquires the vevs $v_u$ and $v_d$ in the $h_u^0$ and $h_d^0$ directions, respectively, which break the EW symmetry. The breaking of the PS gauge symmetry to the SM can be achieved with the Higgs representations
\begin{align}
  H_{\alpha b} &= (\mathbf{4},\mathbf{1},\mathbf{2}) = \begin{pmatrix}
         u_H^R & u_H^B & u_H^G & \nu_H \\
         d_H^R & d_H^B & d_H^G & e_H^-
         \end{pmatrix}, \\
  \bar{H}^{\alpha x} &= (\overline{\mathbf{4}},\mathbf{1},\overline{\mathbf{2}}) = \begin{pmatrix}
         \bar{d}_H^R & \bar{d}_H^B & \bar{d}_H^G & e_H^+ \\
         \bar{u}_H^R & \bar{u}_H^B & \bar{u}_H^G & \bar{\nu}_H
         \end{pmatrix},
\end{align}
obtaining GUT scale vevs in the neutrino directions $\langle {\nu}_H \rangle$ and $\langle {\bar{\nu}}_H \rangle$.

Alternative to the bi-doublet and the quartets, other representations can contain the SM (MSSM) Higgs fields or can break the PS group to the SM. For example, the PS representation $\tilde{h}=(\mathbf{15},\overline{\mathbf{2}},\mathbf{2})$ can contain Higgs $SU(2)_L$ doublets which can develop an EW scale vev. This representation leads to the GJ relation in PS: The fifteen-dimensional representation can be written as a traceless tensor and the relative factor of $-3$ for the leptons comes in to compensate the number of quark colours. 

The PS group is left-right symmetric and hence there are also Yukawa coupling relations for the up-type quark and neutrino sector. For the neutrino Yukawa couplings, dimension four operators with $h$ ($\tilde{h}$) lead to the relation $y_\nu/y_u = 1$ ($y_\nu/y_u = -3$). 

Furthermore, the PS Higgs representations $\phi = (\mathbf{1},\mathbf{1},\mathbf{3})$, $\hat{\phi}=(\mathbf{15},\mathbf{1}, \mathbf{1})$ and $\tilde{\phi}=(\mathbf{15},\mathbf{1},\mathbf{3})$ can arise from the common $SO(10)$ representations and they have singlet components which can develop a GUT scale vev. Their inclusion in the effective operators which generate the Yukawa couplings can lead to new relations for the GUT scale Yukawa coupling ratios. We note that there are other fields like $SU(4)_C$ sextets or complete singlets common in PS models which we do not consider here explicitly, since they do not lead to new predictions. 

In Tab.~\ref{Tab:PSMessenger} we have listed the possible combinations of external and messenger fields which can appear in the diagram of Fig.~\ref{Fig:Messenger}.  The results for the GUT scale Yukawa ratios $y_e/y_d$ and $y_u/y_d$, where $e$, $d$ and $u$ stand for any charged lepton, down-type and up-type quark of the same generation, are presented in Tab.~\ref{Tab:PSRelations}.  Furthermore, we also list the results for certain dimension-six operators from\cite{Allanach:1996hz}, for which only the fields  $R$, $\bar{R}$, $h$, $H$ and $\bar{H}$ are taken into account.

\begin{table}
\begin{center}
\begin{tabular}{cc} \toprule
Operator Dimension &	($y_e/y_d$, $y_u/y_d$) \\
\midrule 4	&	(1, 1)     \\
	&	(-3, 1)     \\
\midrule 5	&	(1, 1)     \\
	&	(-3, 1)     \\
	&	(9, 1)     \\
\midrule 6	&	(0, 1/2)   \\
	&	(0, $\pm$1)     \\
	&	(0, 2)     \\
	&	(3/4, 0)   \\
	&	(3/4, 1/2) \\
	&	(3/4, $\pm$1)   \\
	&	(3/4, 2)   \\
	&	(1, 0)     \\
	&	(1, 1/2)   \\
	&	(1, $\pm$1)     \\
	&	(1, 2)     \\
	&	(2, 0)     \\
	&	(2, 1/2)   \\
	&	(2, $\pm$1)     \\
	&	(2, 2)     \\
	&	(-3, 0)     \\
	&	(-3, 1/2)   \\
	&	(-3, $\pm$1)     \\
	&	(-3, 2)     \\

\bottomrule
\end{tabular}
\end{center}
\caption[Summary of PS Yukawa Coupling Relations]{Summary of possible PS predictions for the GUT scale ratios $y_e/y_d$ and  $y_u/y_d$, where  $e$, $d$ and $u$ stand for any charged lepton, down-type quark and up-type quark of the same generation.  The predictions from certain dimension-six operators, taken from \protect\cite{Allanach:1996hz}, are also included. Predictions for $y_\nu/y_u$ can be read off from Tab.~\ref{Tab:PSMessenger} respectively looked up in \protect\cite{Allanach:1996hz}.
 \label{Tab:PSRelations}}
\end{table}

%% file: kap_06_SUSYThresholdCorrections.tex
\chapter[Supersymmetric Threshold Corrections to the Yukawa Couplings]{Supersymmetric Threshold Corrections\\ to the Yukawa Couplings} \label{Ch:SUSYThresholdCorrections}

In this chapter we discuss supersymmetric threshold corrections to the Yukawa couplings based on \cite{Antusch:2008tf}. For medium or large $\tan \beta$ in the MSSM it is well known that supersymmetric one-loop corrections to the Yukawa couplings are enhanced and that they therefore can have a large impact on the GUT scale Yukawa couplings and trigger Yukawa coupling unification like $y_t = y_b = y_\tau$  \cite{Hall:1993gn, Blazek:1995nv, Carena:1994bv, *Hempfling:1993kv, Bagger:1996ei, King:2000vp}. Here we discuss the corrections in the EW unbroken phase, since this is a good approximation and the results can be given in a compact analytic form. The discussion of their implementation in the RG running and the resulting GUT scale Yukawa couplings is postponed to the next chapter.

We start with a description of the corrections within an effective field theory approach in Sec.~\ref{Sec:EFTThresholds} where it is shown how these corrections come in and in Sec.~\ref{Sec:SUSYThresholdResults} we give explicit expressions and numerical values for the SUSY threshold corrections in the EW unbroken phase which can be used as simple estimates in model building.

\section{Effective Field Theory Approach} \label{Sec:EFTThresholds}

In the effective field theory approach, field degrees of freedom which cannot be excited can be integrated out of the theory and then give, for example, new vertices. In the following, we want to discuss one example, which is quite important for the Yukawa couplings in SUSY models with large $\tan \beta$  \cite{Hall:1993gn, Blazek:1995nv, Carena:1994bv, *Hempfling:1993kv}.

\begin{figure}
\centering
\includegraphics[scale=0.9]{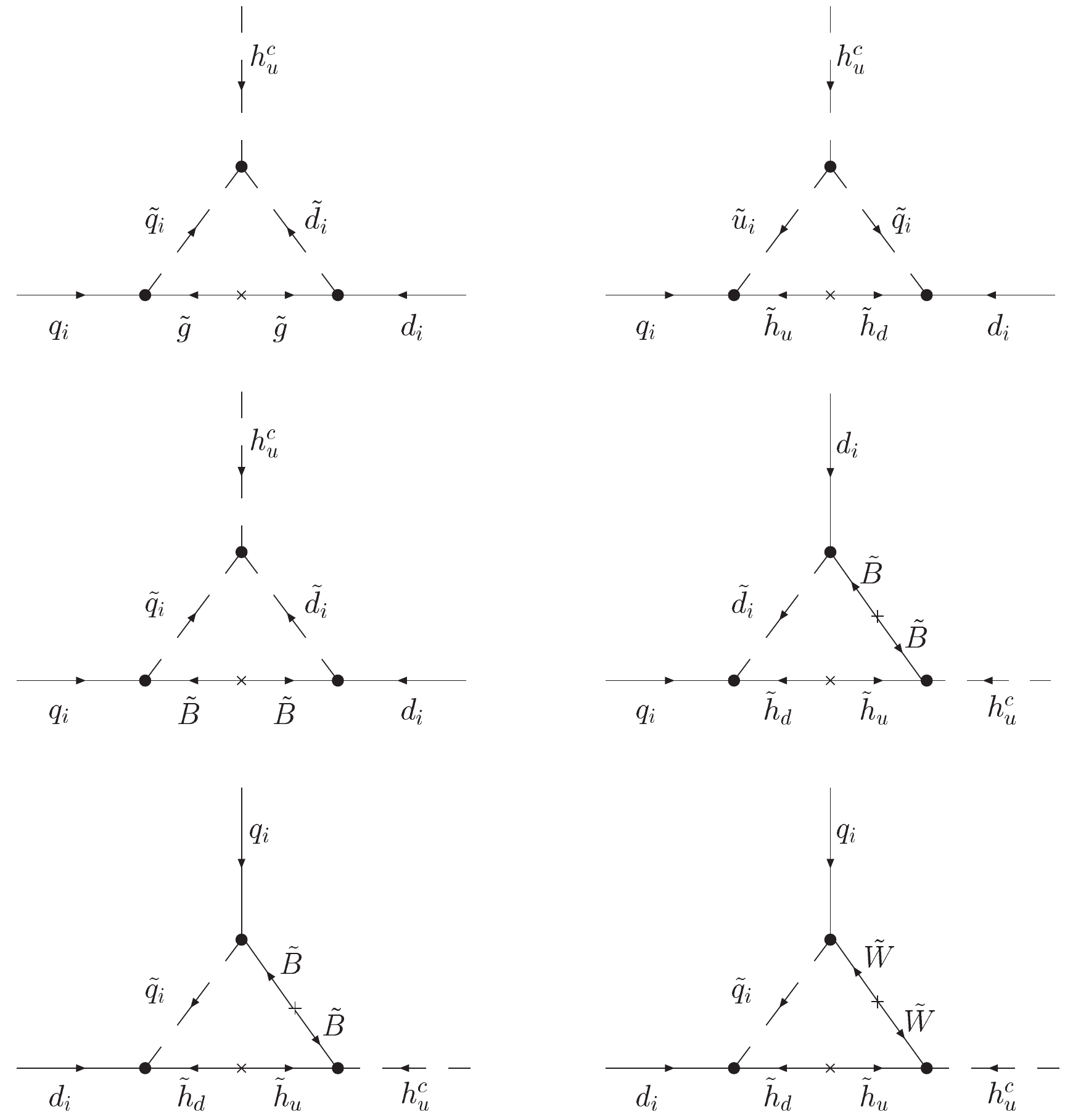}
\caption[Down-Type Quark Yukawa SUSY Threshold Corrections]{Feynman diagrams contributing to the SUSY threshold corrections to down-type quark Yukawa couplings in the EW unbroken phase. \label{Fig:QuarkThresholds}}
\end{figure}

\begin{figure}
\centering
\includegraphics[scale=0.9]{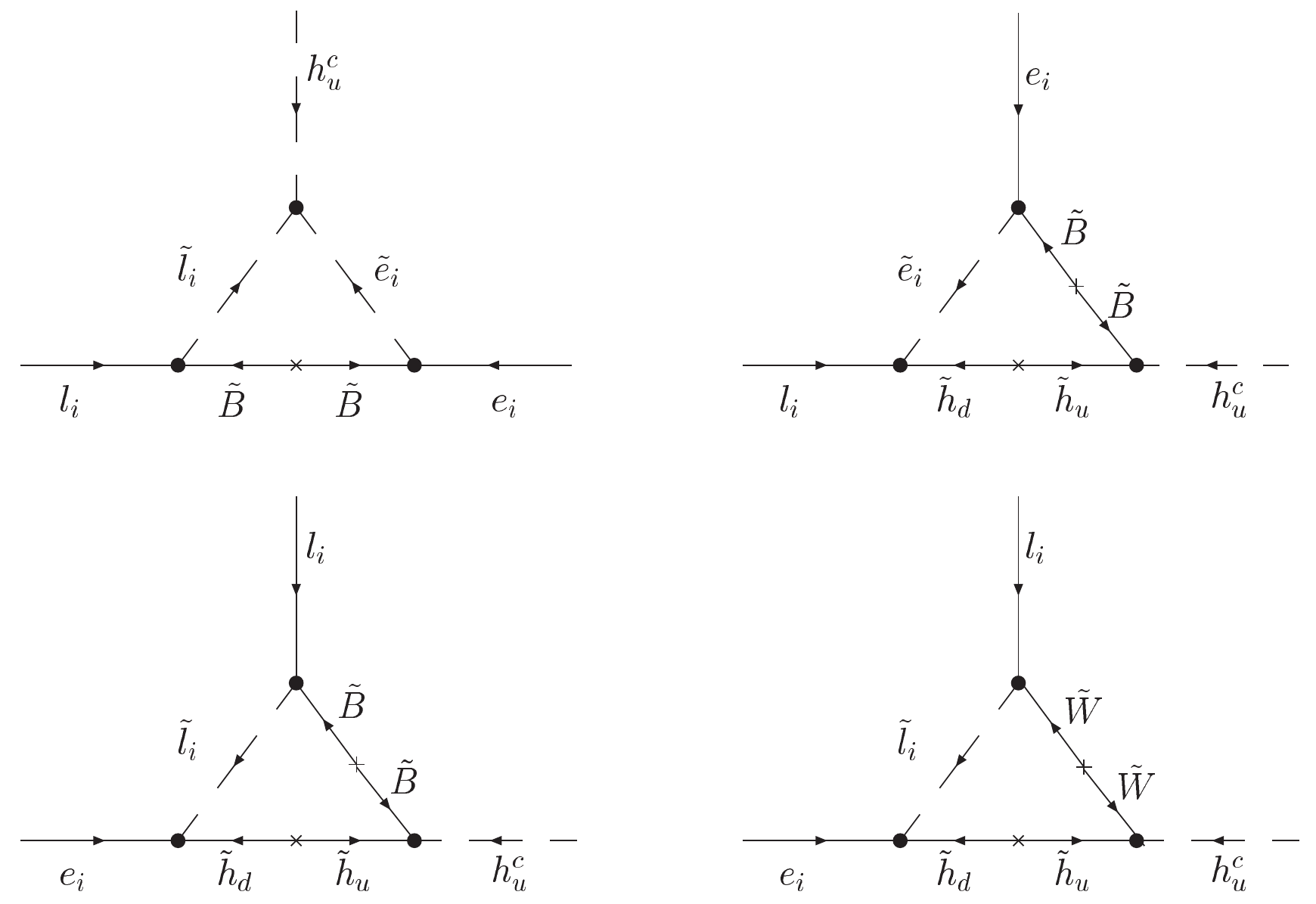}
\caption[Charged Lepton Yukawa SUSY Threshold Corrections]{Feynman diagrams contributing to the SUSY threshold corrections to charged lepton Yukawa couplings in the EW unbroken phase. \label{Fig:LeptonThresholds}}
\end{figure}

As we have discussed in Ch.~\ref{Ch:SUSY} in the MSSM, on tree-level, the down-type quarks and charged leptons couple only to the down-type Higgs field $h_d$ while the up-type quarks couple only to the up-type Higgs field $h_u$. This statement is not true anymore on one-loop level. On one-loop level, the down-type quarks and charged leptons couple to the up-type Higgs $h_u$ as well via the diagrams shown in Figs.~\ref{Fig:QuarkThresholds} and \ref{Fig:LeptonThresholds}. Similar diagrams apply to the up-type quarks generating a coupling of them to the down-type Higgs field $h_d$ which give negligible corrections as we discuss later on. The diagrams  are shown for the case of broken SUSY but unbroken EW symmetry. In this chapter we use this approximation for simplicity reasons. It turns out that this is quite good \cite{Buras:2002vd} and later on we use a more precise approach, which, however, gives less insight about parameter dependencies.

In the limit of heavy superpartner particle masses, the relevant couplings for, i.e.\ the $d$-quarks, are
\begin{equation}
\begin{split}
\mathcal{L}_{\mathrm{Yuk}} &= y_d h_d \epsilon q_L \bar{d}_R - \delta \tilde{y}_d h_u^c \epsilon q_L \bar{d}_R + \mathrm{h.c.} \\
& \stackrel{\mathrm{EWSB}}{\longrightarrow} -(y_d v_d + \delta \tilde{y}_d v_d) \bar{d}_R d_L - y_d h_d^0 \bar{d}_R d_L - \delta \tilde{y}_d h_u^{0 *} \bar{d}_R d_L + \mathrm{h.c.} \;,
\end{split}
\end{equation}
where $y_d$ is the tree-level MSSM $d$-quark Yukawa coupling and $\delta \tilde{y}_d$ the Yukawa coupling to the \emph{wrong} Higgs field $h_u$ generated effectively. Hence, after EWSB the $d$-quark mass has contributions from both Higgs fields
\begin{equation}
m_d = y_d v_d + \delta \tilde{y}_d v_u = m_d^{(0)} \left( 1 + \eta_d \tan \beta \right) \;,
\end{equation}
where we have used that  $m_d^{(0)} = y_d v_d$, $\eta_d = \delta \tilde{y}_d/y_d$ and $\tan \beta = v_u/v_d$. Here, we can see that the SUSY threshold corrections encoded in $\eta_d$ are more important for larger $\tan \beta$, since the one-loop suppression in $\eta_d$ can be partially cancelled in this case.

In our analysis we have included these corrections in form of a matching relation between the SM Yukawa couplings $y_i^{\mathrm{SM}}$ and the MSSM Yukawa couplings $y_i^{\mathrm{MSSM}}$ at the SUSY scale $M_{\mathrm{SUSY}}$:
\begin{equation}
y_i^{\mathrm{MSSM}} = \frac{y_i^{\mathrm{SM}}}{\cos \beta (1 + \eta_i \tan \beta)} \;, \label{Eq:YukawaMatchingRelation1}
\end{equation}
for $i=e,\mu,\tau,d,s$ and $b$. This matching relation concerns only the charged leptons and down-type quarks. In the matching relation for the up-type quarks $\cot \beta$ instead of $\tan \beta$ appears. For large $\tan \beta$ this induces a suppression in addition to the one-loop suppression. That is the reason for not including the SUSY threshold corrections for the up-type quarks and simply using the tree-level matching relation
\begin{equation}
y_i^{\mathrm{MSSM}} = \frac{y_i^{\mathrm{SM}}}{\sin \beta} \;, \label{Eq:YukawaMatchingRelation2}
\end{equation}
for $i=u,c$ and $t$. We have not included neutrino Yukawa couplings due to the lack of right-handed neutrinos in the MSSM but they would fulfil the same matching relation as the up-type quarks. The SM model Yukawa couplings $y_i^{\mathrm{SM}}$ can be calculated via $y_i^{\mathrm{SM}} = \bar{m}_i/v$, where $\bar{m}_i$ are the $\overline{\text{MS}}$-masses of the fermions and $v$ is the vev of the SM Higgs. Note that since the calculation in this and the next chapter is one-loop, we can use $\overline{\text{MS}}$-quantities as well as $\overline{\text{DR}}$-quantities. In the following we drop the MSSM-label for the Yukawa couplings and imply $y \equiv y^{\mathrm{MSSM}}$.

For large $\tan \beta$, the correction to the Yukawa couplings $\eta_i \tan \beta$ can become quite sizeable. In principle, it can even happen that the one-loop corrections are larger than the tree-level couplings. In this case the down-type quark and charged lepton masses would predominantly be generated by the coupling to the up-type Higgs field $h_u$. For that reason higher order calculations seem to be necessary. However, it can be shown that this relative $\tan \beta$-enhancement is a one-loop effect and does not repeat in higher orders \cite{Carena:1999py}.

\section{Formulas and Numerical Values} \label{Sec:SUSYThresholdResults}

In this section we give explicit formulas for the quantities $\eta_i$ in Eq.~\eqref{Eq:YukawaMatchingRelation1}, which in addition to $\tan \beta$ govern the size of the SUSY threshold corrections. The formulas are exactly valid for the case of unbroken EW symmetry, which is appropriate for the situation that the sparticle masses are in the TeV range and that we apply the matching conditions at these energies above the EW scale $M_{\mathrm{EW}}$. There are also expressions for the full corrections \cite{Pierce:1996zz} which we use later on. Nevertheless, here and in the next chapter we use the simplified Ansatz because it offers more insight into the parameter dependencies.

Turning to the corrections for the down-type quarks we can decompose the corrections as $\eta_i = \eta_i^G + \eta_i^B + \eta_i^W + \eta^y \delta_{ib}$, with \cite{Freitas:2007dp}
\begin{align} \label{Eq:expq}
 \eta_i^G & =  -\frac{2 \alpha_S}{3 \pi} \frac{\mu}{M_3} H_2(u_{\tilde{Q}_i},u_{\tilde{d}_i}) \; , \\
\label{Eq:expq_bino}
 \eta_i^B & =  \frac{1}{16 \pi^2} \left[ \frac{{g'}^2}{6} \frac{M_1}{\mu} \left( H_2(v_{\tilde{Q}_i}, x_1) + 2 H_2(v_{\tilde{d}_i}, x_1) \right) + \frac{{g'}^2}{9} \frac{\mu}{M_1} H_2(w_{\tilde{Q}_i},w_{\tilde{d}_i}) \right] \;, \\
 \eta_i^W & =  \frac{1}{16 \pi^2} \frac{3 g_2^2}{2} \frac{M_2}{\mu} H_2(v_{\tilde{Q}_i}, x_2) \; , \\
 \eta^y &= - \frac{y_t^2}{16 \pi^2} \frac{A_t}{\mu} H_2(v_{\tilde{Q}_3}, v_{\tilde{u}_3}) \; ,
\label{Eq:expq_end}
\end{align}
where $u_{\tilde{f}} = m_{\tilde{f}}^2/M_3^2$, $v_{\tilde{f}} = m_{\tilde{f}}^2/\mu^2$, $w_{\tilde{f}} = m_{\tilde{f}}^2/M_1^2$, $x_1 = M_1^2/\mu^2$ and $x_2 = M_2^2/\mu^2$ for $i=d,s,b$ and all mass parameters are assumed to be real. The correction $\eta^y$ is only relevant for the $b$-quarks due to the strong hierarchy of the quark Yukawa couplings. 
The function $H_2$ is defined as
\begin{equation}
H_2(x,y) = \frac{x \ln x}{(1-x)(x-y)} + \frac{y \ln y}{(1-y)(y-x)} \;.
\end{equation}
Note that $H_2$ is negative for positive $x$ and $y$ and $|H_2|$ is maximal if its arguments are minimal, and vice versa.

The corrections for the charged leptons stem from diagrams similar to the ones for the quarks and are shown in Fig.~\ref{Fig:LeptonThresholds}. One difference between the corrections for quarks and charged leptons is of course that the SUSY QCD loop contributions $\eta_i^G$ are absent. Another difference concerns the contributions $\eta_i^B$ with binos in the loops, where due to the different hypercharge of the charged leptons the prefactors for these contributions are changed. In the last term in Eq.~\eqref{Eq:expl_bino} this causes an enhancement by a factor of $-9$ compared to the corresponding term in the quark sector in Eq.~\eqref{Eq:expq_bino}. The contribution from the diagrams with winos $\eta_i^W$, on the other hand, is equal for quarks and leptons. A further difference between the corrections for quarks and charged leptons is that in the seesaw framework, which we consider later on, the $\tau$-lepton Yukawa coupling does not have a relevant correction of the $\eta^y$-type because the corresponding vertex correction is suppressed by the heavy mass scale of the right-handed neutrinos. For the corrections for the charged leptons we find
\begin{align} \label{Eq:expl_bino}
 \eta_i^B & =  \frac{1}{16 \pi^2} \left[ \frac{{g'}^2}{2} \frac{M_1}{\mu} \left(- H_2(v_{\tilde{L}_i}, x_1) + 2 H_2(v_{\tilde{e}_i}, x_1) \right) - {g'}^2 \frac{\mu}{M_1} H_2(w_{\tilde{L}_i},w_{\tilde{e}_i}) \right] , \\
 \eta_i^W & =  \frac{1}{16 \pi^2} \frac{3 g_2^2}{2} \frac{M_2}{\mu} H_2(v_{\tilde{L}_i}, x_2) \; , \label{Eq:expl_end}
\end{align}
for $i=e,\mu,\tau$.

\begin{table}
\begin{center}
\begin{tabular}{cccc}\toprule
SUSY Parameter in TeV & Case $g_+$ & Case $g_-$ & Case $a$ \\ \midrule
$m_{\tilde{f}}$		& [0.5, 1.5]  & [0.5, 1.5] & [0.5, 1.5]  \\ 
$M_1$ 			& [0.5, 1]    & [0.5, 1]   & [1.65, 3.3] \\ 
$M_2$ 			& [1, 2]      & [1, 2]     & [0.5, 1]  \\ 
$M_3$ 			& [3, 6]      & [3, 6]     & [-9, -4.5]  \\
$\mu$ 			& 0.5      & -0.5    &  0.5  \\
$A_t$ 			& $\pm$1  & $\pm$1 &  $\pm$1  \\
$M_{\mathrm{SUSY}}$ 	& 1  & 1 &  1  \\ \bottomrule
\end{tabular}
\end{center}
\caption[Example SUSY Parameter Ranges]{Example ranges of SUSY parameters at the matching scale $M_{\mathrm{SUSY}}$ used in our analyses in Chs.~\ref{Ch:SUSYThresholdCorrections} and \ref{Ch:AFirstGlance}. The choices of gaugino masses in the cases $g_\pm$ are inspired by universal gaugino masses at the GUT scale and in case $a$  by anomaly-mediated SUSY breaking. We can therefore use the low-energy approximations $M_1:M_2:M_3 = 1:2:6$ for the cases $g_\pm$ and $M_1:M_2:M_3 = 3.3:1:-9$ for case $a$ as constraints. \label{Tab:SUSYParameters}}
\end{table}

To give a list of numerical values of the SUSY threshold corrections at the matching scale $M_{\mathrm{SUSY}}$ we have to specify the soft SUSY parameters and the $\mu$ parameter entering the SUSY threshold corrections $\eta_i$. This is usually the aim of SUSY breaking schemes, where the low scale soft SUSY breaking parameters are calculated from only a handful of parameters, see also Sec.~\ref{Sec:SUSYbreaking}. This is not what we do here. Here, we simply impose values for the SUSY parameters at the scale $M_{\mathrm{SUSY}}$.

All parameters should be around the TeV scale because we do not want to spoil the solution to the hierarchy problem and since otherwise the approximation of having only one matching scale fails. Nevertheless, to make the parameter space manageable, we have used low energy relations between the gaugino mass parameters. For the  cases $g_+$ and $g_-$ we have used the relation $M_1:M_2:M_3 = 1:2:6$ which is inspired by CMSSM scenarios, see Sec.~\ref{Sec:CMSSM} for more details and references. For case $a$, we have used the relation $M_1:M_2:M_3 = 3.3:1:-9$ which is inspired by AMSB scenarios, see Sec.~\ref{Sec:AMSB} for more details and references. Note that in case $a$ the gluino mass parameter has a negative sign which switches the sign of the SUSY QCD contribution $\eta^G_i$ in comparison to the case $g_+$.

We only introduce relations for the gaugino mass parameters. For the sfermion mass parameters $m_{\tilde{f}}$ and the $\mu$ and $A_t$ parameters we do not apply restrictions from a specific model of SUSY breaking. The example ranges are chosen such that we can neglect effects which are suppressed by $M_{\mathrm{EW}}/M_{\mathrm{SUSY}}$, like mixing between left- and right-handed sfermions, and such that our approach of one-step matching is justified to a good approximation.  We note that left-right mixing effects may nevertheless be important \cite{Buras:2002vd}, for instance in so-called inverted scalar mass hierarchy scenarios \cite{Blazek:2001sb}. 
We also note that we do not specify here the remaining SUSY parameters which do not enter the formulas for the threshold corrections and hence we cannot apply various relevant phenomenological constraints on the spectrum. We extend our approach to a more phenomenological one in the next chapter. The  SUSY parameter ranges we have used in this analysis are listed in Tab.~\ref{Tab:SUSYParameters}.

\begin{table}
\begin{center}
\begin{tabular}{lccc}\toprule
		SUSY Threshold Correction &   Case $g_+$   & Case $g_-$ & Case $a$ \\ \midrule
$\eta_i^G$ in $10^{-3}$ 		&   [3.52, 9.31] & [-9.31, -3.52] & [-7.85, -3.10] \\[0.1pc]
$\eta_i^B$ for quarks in $10^{-3}$ 	& [-0.31, -0.08] &   [0.08, 0.31] & [-0.30, -0.16] \\[0.1pc]
$\eta_i^B$ for leptons in $10^{-3}$ 	& [-0.18,  0.30] & [-0.30,  0.18] &  [0.01,  0.30]  \\[0.1pc]  
$\eta_i^W$ in $10^{-3}$ 		& [-2.21, -0.98] &   [0.98, 2.21] & [-2.21, -0.72] \\[0.1pc] 
$\eta^y/\mathrm{sign}(A_t)$  in $10^{-3}$  		&  [0.84, 4.65] & [-4.65, -0.84] & [0.84, 4.65] \\ \bottomrule
\end{tabular}
\end{center}
\caption[Values for SUSY Threshold Corrections]{Ranges for the various contributions to the SUSY threshold corrections corresponding to the example ranges of SUSY parameters in Tab.~\ref{Tab:SUSYParameters} for $i=d,s,b$ and $i=e,\mu,\tau$, respectively. For the charged leptons there is no contribution $\eta_i^G$ and $\eta^y$. The ranges for $\eta_i^W$ are the same for quarks and leptons. \label{Tab:ThresholdValues} }
\end{table}

The numerical values for the SUSY threshold corrections, the main result of this chapter, are listed in Tab.~\ref{Tab:ThresholdValues}. First of all, we see that the $\eta_i^G$ and the $\eta^y$ corrections give the largest contributions. The EW corrections are smaller due to the smaller gauge couplings but they can still be in the 10\% region of the leading corrections. Especially for the charged leptons, the EW corrections are important since they are the only corrections. For the down-type quarks, the threshold corrections can make up to 50\% and therefore cannot be neglected at all as long as $\tan \beta$ is large.

In the next chapter, we include these threshold corrections into the RGE running of the Yukawa couplings and study their effects on the GUT scale Yukawa couplings and their ratios.

%% file: kap_07_FirstGlance.tex
\chapter[Yukawa Couplings at the GUT Scale: A First Glance]{Yukawa Couplings at the GUT Scale:\\ A First Glance}  \label{Ch:AFirstGlance}

The main purpose of this chapter based on \cite{Antusch:2008tf} is to include SUSY threshold corrections in the renormalisation group evolution of the Yukawa couplings and calculate the possible GUT scale ranges for quark and charged lepton Yukawa couplings as well as for the ratios $y_e/y_d$, $y_\mu/y_s$, $y_\tau/y_b$ and $y_t/y_b$, which are important input parameters for GUT model building. Despite the possible importance of SUSY threshold effects, in studies which interpolate the running of the Yukawa couplings to the GUT scale, these effects are typically ignored \cite{Fusaoka:1998vc,Xing:2007fb}. 

Nevertheless, the effects of SUSY threshold corrections on the possibility of third family Yukawa unification $y_t = y_b = y_\tau$ and also on the less restrictive possibility $y_b = y_\tau$, emerging in $SU(5)$ GUTs, has been extensively studied in the literature, see, e.g.\ ~\cite{Hall:1993gn, Blazek:1995nv, Carena:1994bv, *Hempfling:1993kv, Bagger:1996ei, King:2000vp}. Furthermore, recent studies \cite{Altmannshofer:2008vr} have addressed the phenomenological viability of this relation and have pointed out that under certain assumptions on the soft breaking parameters at the GUT scale, $y_t = y_b = y_\tau$ may be already quite challenged by experimental data from B physics.

Regarding $y_\mu/y_s$, compared to \cite{Ross:2007az} we include additional corrections from electroweak loops with binos and winos for quarks and charged leptons, which, as we have shown in the last chapter, can have significant impact. Furthermore, instead of trying to fit the GJ relations by a sparticle spectrum, our aim is to analyse which alternative GUT scale relations may be possible and whether the GJ relation lies within the projected GUT scale ranges.

We start our discussion of possible GUT scale Yukawa couplings in Sec.~\ref{Sec:RGERunning} with a description of the semianalytic approach we have used to include the SUSY threshold corrections in the RG evolution. This approach is quite easy to implement by using the ranges for the SUSY threshold corrections we derived in the last chapter and collected in Tab.~\ref{Tab:ThresholdValues}. Afterwards we discuss the dependencies of the GUT scale Yukawa couplings on the soft SUSY parameters and possible additional right-handed neutrino thresholds. We end this chapter with a discussion of the impact of the whole sparticle spectrum on the GUT scale Yukawa couplings and ratios summarised in Fig.~\ref{Fig:Scatter} and Tabs.~\ref{Tab:Ratfinal} and \ref{Tab:Yukfinal}.

\section{Renormalisation Group Running from the EW to the GUT Scale} \label{Sec:RGERunning}

The calculation of Yukawa couplings of quarks and charged leptons at the GUT scale can be accomplished by solving the corresponding renormalisation group equations (RGEs) from the low-energy to the GUT scale. For this we use the  REAP package introduced in \cite{Antusch:2005gp}, where also a summary of the relevant RGEs can be found.

For our analysis, we take as input values the running quark and lepton masses at the top scale $m_t (m_t)$, which have been calculated in \cite{Xing:2007fb} with recent experimental values for the low-energy quark and charged lepton masses, and evolve them first to the SUSY scale $M_{\mathrm{SUSY}}$ using the SM RGEs.

At the SUSY scale we match the SM with the MSSM to obtain the running $\overline{\text{MS}}$ Yukawa couplings via Eqs.~\eqref{Eq:YukawaMatchingRelation1} and \eqref{Eq:YukawaMatchingRelation2} in our numerical approach. Since we consider one-loop running only, we can neglect issues of scheme dependence such as transformations from $\overline{\text{MS}}$ to $\overline{\text{DR}}$ quantities. Two-loop running (and scheme-dependent) effects are small compared to the $\tan \beta$-enhanced threshold corrections and can be neglected. For the semianalytic approach we have neglected the SUSY threshold corrections at the SUSY scale for a first approximation and not taken them into account until the GUT scale which turns out to be a good approximation.

As next step, we solve the RGEs from the SUSY scale to $M_{\mathrm{GUT}}$ taking into account possible intermediate right-handed neutrino thresholds as discussed in \cite{Antusch:2002rr}. For our numerical calculations we use REAP, which solves the complete set of one-loop RGEs and automatically includes right-handed neutrino thresholds.
We comment on possible effects of right-handed neutrino thresholds depending on additional degrees of freedom in seesaw models, in Sec.~\ref{Sec:RHnus}. If not stated otherwise, they are ignored in our remaining analysis.

We note that there are SUSY scenarios which may lead to corrections to our approach in this chapter of one-step matching at the SUSY scale in the EW unbroken phase. For example, if the sparticle spectrum is light, effects of EWSB may have to be taken into account for the calculation of the SUSY threshold corrections. Another example is the possibility of having a split sparticle spectrum, in which case matching at one scale would be a bad approximation. When we present explicit examples in the following, we choose parameters where our assumptions are justified to a good approximation.

\section{Semianalytic Approach} \label{Sec:SemianalyticalApproach}

The ranges for the $\eta_i$ from Tab.~\ref{Tab:ThresholdValues} can be used to obtain analytic estimates for the ratios of the Yukawa couplings at the GUT scale. For example, in leading order the GUT scale ratio $y_e/y_d$ is given by
\begin{equation}
 \frac{y_e(M_{\mathrm{GUT}})}{y_d(M_{\mathrm{GUT}})} \approx \frac{\hat{y}_e(M_{\mathrm{GUT}})}{\hat{y}_d(M_{\mathrm{GUT}})} \:\frac{1 + \eta_d \tan \beta}{1 + \eta_e \tan \beta} = \frac{\hat{y}_e(M_{\mathrm{GUT}})}{\hat{y}_d(M_{\mathrm{GUT}})} \left( 1 + \left(\eta_d - \eta_e \right) \tan \beta \right) + \mathcal{O}(\eta_e^2 \tan^2 \beta) \;,
\end{equation}
where $\hat{y}(M_{\mathrm{GUT}})$ denotes the Yukawa couplings at the GUT scale without SUSY threshold corrections included. We use the analogous formula for the second generation. For the ratio $y_t/y_b$ keep in mind that $\eta_t$ is set to zero since this threshold is suppressed by $\tan \beta$.
Later on we compare these estimates with the numerical results for the same ranges of MSSM parameters, cf.~Tab.~\ref{Tab:SError}. For the values of the masses and Yukawa couplings at the GUT scale, we take the values calculated with REAP setting all SUSY threshold corrections to zero and using the best-fit values for the fermion masses as low-energy input. These values for the Yukawa couplings are collected in Tab.~\ref{Tab:YukawaNoThresholds}. In the following we refer to the case without SUSY threshold corrections as case 0.

\begin{table}
\centering
\begin{tabular}{cccc}\toprule
 $\tan \beta$&  $\hat{y}_e$ in $10^{-4}$ & $\hat{y}_\mu$ in $10^{-2}$ & $\hat{y}_\tau$  \\ \midrule
$30$ & 0.62 & 1.31 & 0.23 \\ 
$40$ & 0.88 & 1.85 & 0.34 \\ 
$50$ & 1.21 & 2.55 & 0.51 \\ \bottomrule
\multicolumn{1}{c}{}
\end{tabular}

\begin{tabular}{cccc}\toprule
 $\tan \beta$ &  $\hat{y}_d$ in $10^{-4}$ & $\hat{y}_s$ in $10^{-2}$ & $\hat{y}_b$  \\ \midrule
$30$ & 1.57 & 0.30 & 0.18 \\ 
$40$ & 2.22 & 0.43 & 0.26 \\ 
$50$ & 3.06 & 0.59 & 0.39 \\ \bottomrule
\multicolumn{1}{c}{}
\end{tabular}

\begin{tabular}{cccc}\toprule
 $\tan \beta$ &  $\hat{y}_u$ in $10^{-6}$ & $\hat{y}_c$ in $10^{-4}$ & $\hat{y}_t$ \\ \midrule
$30$ & 2.73 & 1.33 & 0.49 \\ 
$40$ & 2.75 & 1.34 & 0.50 \\ 
$50$ & 2.77 & 1.35 & 0.52 \\ \bottomrule
\end{tabular}

\caption[GUT Scale Yukawa Couplings without SUSY Threshold Corrections]{Best-fit values for the Yukawa couplings at the GUT scale without SUSY threshold corrections (case 0) for $M_{\mathrm{SUSY}} = 1$~TeV and different values of $\tan \beta$.\label{Tab:YukawaNoThresholds} }
\end{table}

\begin{table}
\centering
\begin{tabular}{ccccc}\toprule
			& Case 0 & Case $g_+$ & Case $g_-$  & Case $a$ \\ \midrule
 $y_e/y_d$ 	& 0.39 & [0.35, 0.64] & [0.15, 0.44] & [0.16, 0.44] \\ 
 $y_\mu/y_s$ 	& 4.35 & [3.83, 7.01] & [1.69, 4.87] & [1.81, 4.85] \\ 
 $y_\tau/y_b$ 	& 1.32 & [1.16, 2.13] & $\leq 1.48$  & $\leq 1.47$ \\ 
 $y_t/y_b$ 	& 1.93 & [1.65, 2.92] & $\leq 2.21$  & $\leq 1.98$ \\ \bottomrule
\end{tabular}
\caption[Semianalytic Estimates for the GUT Scale Yukawa Coupling Ratios]{Semianalytic estimates for the ranges of the Yukawa coupling ratios at the GUT scale corresponding to the example ranges of SUSY parameters in Tab.~\ref{Tab:SUSYParameters} for $\tan \beta = 40$. Case 0 refers to the case without SUSY threshold corrections. For the ranges involving $y_b$, the lower boundaries depend on the cut we had to introduce in order to keep $y_b$ perturbative up to $M_{\mathrm{GUT}}$ and thus have been omitted. \label{Tab:GJRAnalytical}}
\end{table}

The results of these estimates are collected in Tab.~\ref{Tab:GJRAnalytical}. We note that these estimates are naive in the sense that we have combined the maximal and minimal values of each of the contributions to $\eta_i$, neglecting possible correlations between them. For example, we do not account for the effect that the QCD corrections become large if $M_3$ and thus the gaugino masses are large and the sfermion masses small, whereas the wino corrections become large if the gaugino masses and the sfermion masses are small. However, as we see later, the estimates nevertheless work pretty well. Another effect we can immediately see from the analytic estimates is that $y_b$ can become non-perturbatively large if $\eta_b \tan \beta$ is absolutely large but negative. In fact, it can even occur that $\eta_i \tan \beta \approx -1$ if $\tan \beta$ is sufficiently large, which spoils the perturbative expansion in $y_b$. Whenever non-perturbative values of $y_b$ occur, we only give the upper boundaries of the ranges $y_\tau/y_b$ and $y_t/y_b$ and the lower boundary for $y_b$ itself in the result tables.

The naive estimates already suggest that with SUSY thresholds included, a wide range of GUT scale values of down-type quark and charged Yukawa couplings could be realised. Taking a preliminary look at the predicted ratio for $y_\mu/y_s$, the naive estimates suggest that with the SUSY parameters of case $g_+$, the GUT scale value of $y_\mu/y_s$ is typically significantly larger than the GJ relation of $y_\mu/y_s = 3$. On the other hand, the scenarios $g_-$ and $a$ are well compatible with the GJ relation. Beyond the GJ relation, the naive estimates also imply that with SUSY thresholds included, a large variety of GUT model predictions for these ratios might be compatible with the low-energy data on quark and lepton masses. A full numerical analysis for the example SUSY parameter ranges $g_\pm$ and $a$ is presented at the end of this chapter.

\section{Parameter Dependencies} \label{Sec:ParameterDependencies}

In the following we discuss the dependencies of the GUT scale Yukawa coupling ratios on the SUSY parameters. We start with the discussion of the $\mu$ parameter and $\tan \beta$ which are important parameters for EWSB. Afterwards we continue with $M_{\mathrm{SUSY}}$ and $A_t$. The SUSY scale should be around the TeV scale in order not to spoil the solution to the hierarchy problem. For the trilinear coupling $A_t$ there is no such constraint and in CMSSM the trilinear couplings are essentially free parameters. Especially the sign can be chosen which makes this parameter very interesting since it can enhance or suppress the SUSY threshold corrections depending on the sign. After a brief discussion of possible right-handed neutrino threshold effects we turn to the discussion of the impact of the whole sparticle spectrum summarised in Fig.~\ref{Fig:Scatter} and Tabs.~\ref{Tab:Ratfinal} and \ref{Tab:Yukfinal}.

\subsection[Dependence on $\mu$ and $\tan \beta$]{Dependence on $\boldsymbol{\mu}$ and $\boldsymbol{\tan \beta}$}

Before we proceed with the numerical analysis for the example ranges of MSSM para\-meters of Tab.~\ref{Tab:SUSYParameters}, we discuss the dependence on $\mu$ and $\tan \beta$, which have been kept fixed so far. The dependence on $\mu$ is rather important, because all corrections are proportional to $\mu$ or $1/\mu$. The parameter $\mu$ therefore gives the overall sign of the corrections and determines if the Yukawa couplings are enhanced or reduced by the SUSY threshold effects. In addition, $\tan \beta$ is very important due to the fact that the threshold corrections are almost linear in $\tan \beta$ and because for successful complete third family Yukawa unification we need a large value of $\tan \beta$.

\begin{figure}
 \centering
 \includegraphics[scale=0.6]{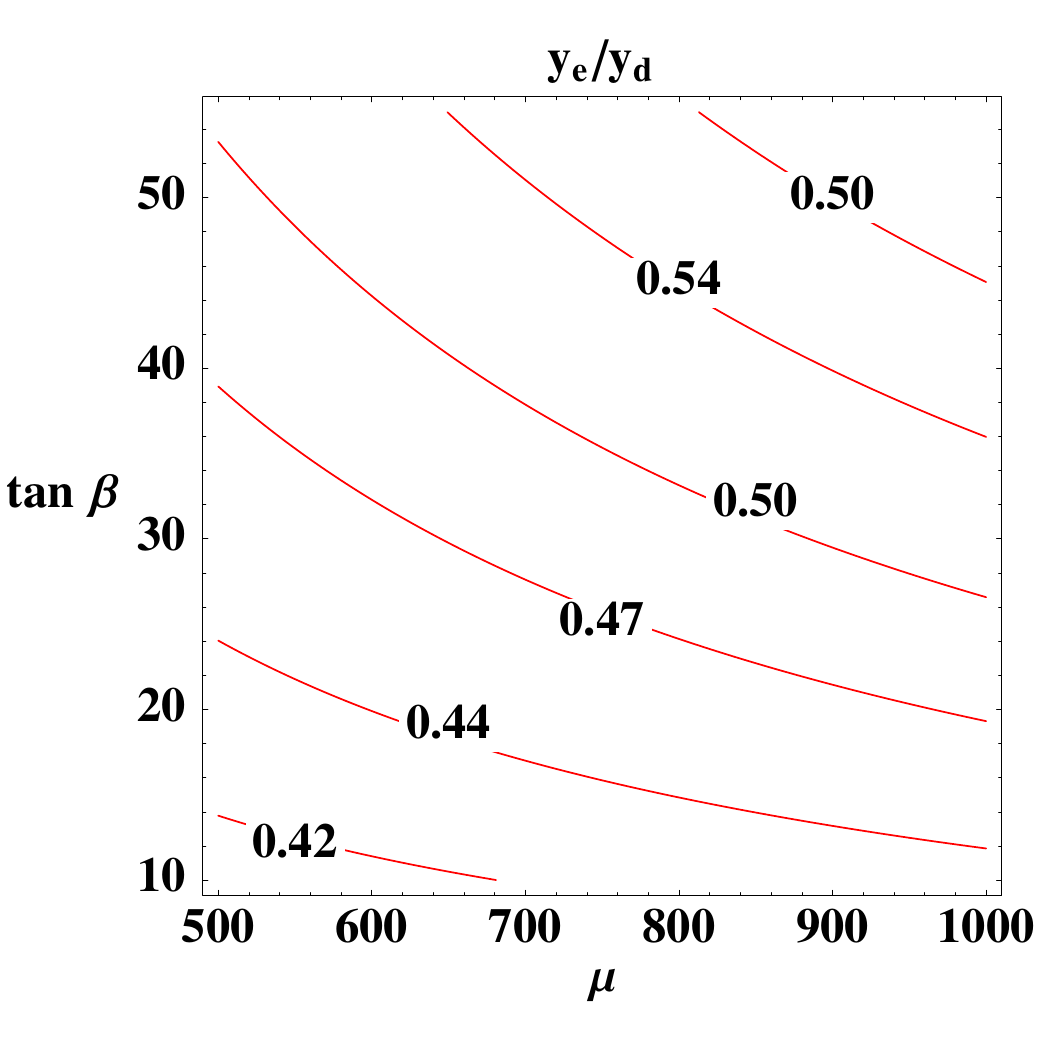} \hspace{1cm}
 \includegraphics[scale=0.6]{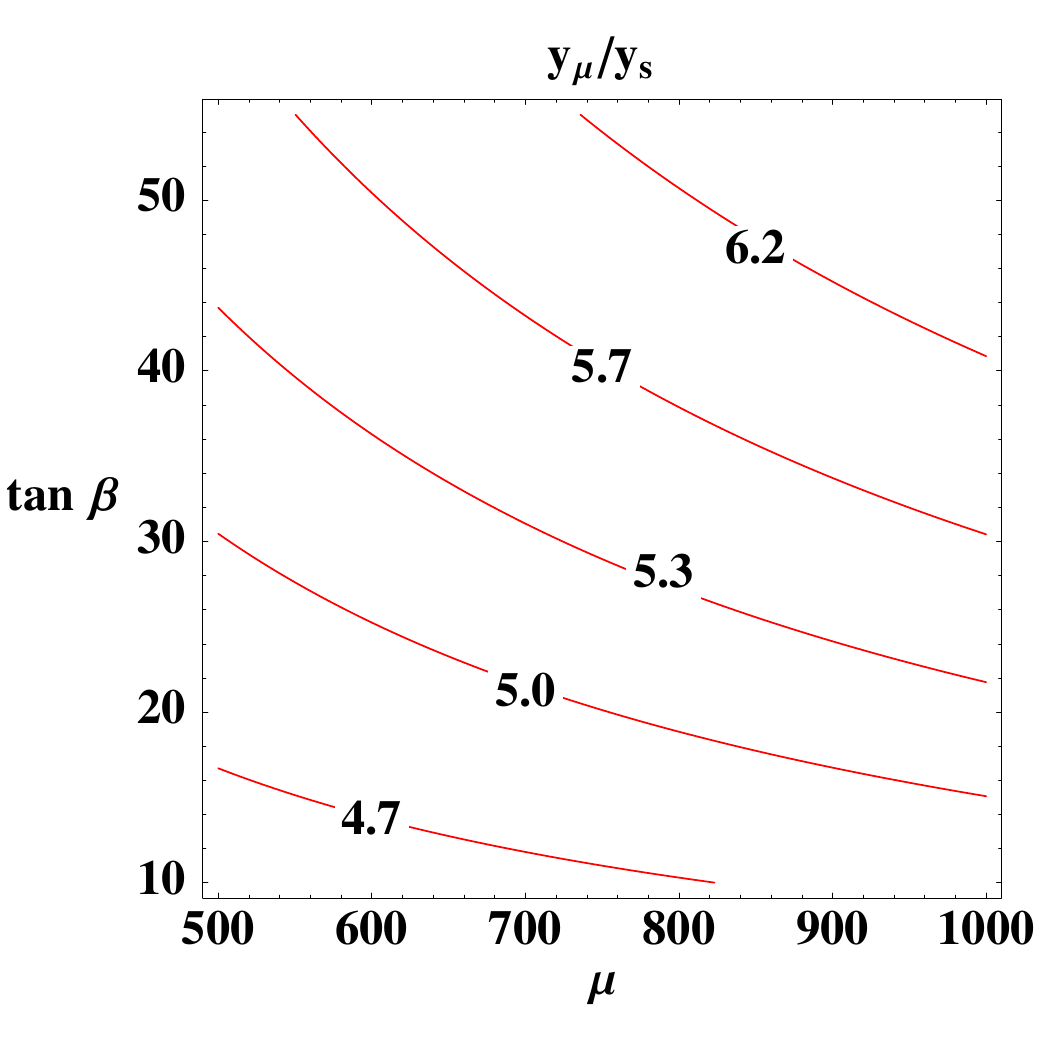}
 \includegraphics[scale=0.6]{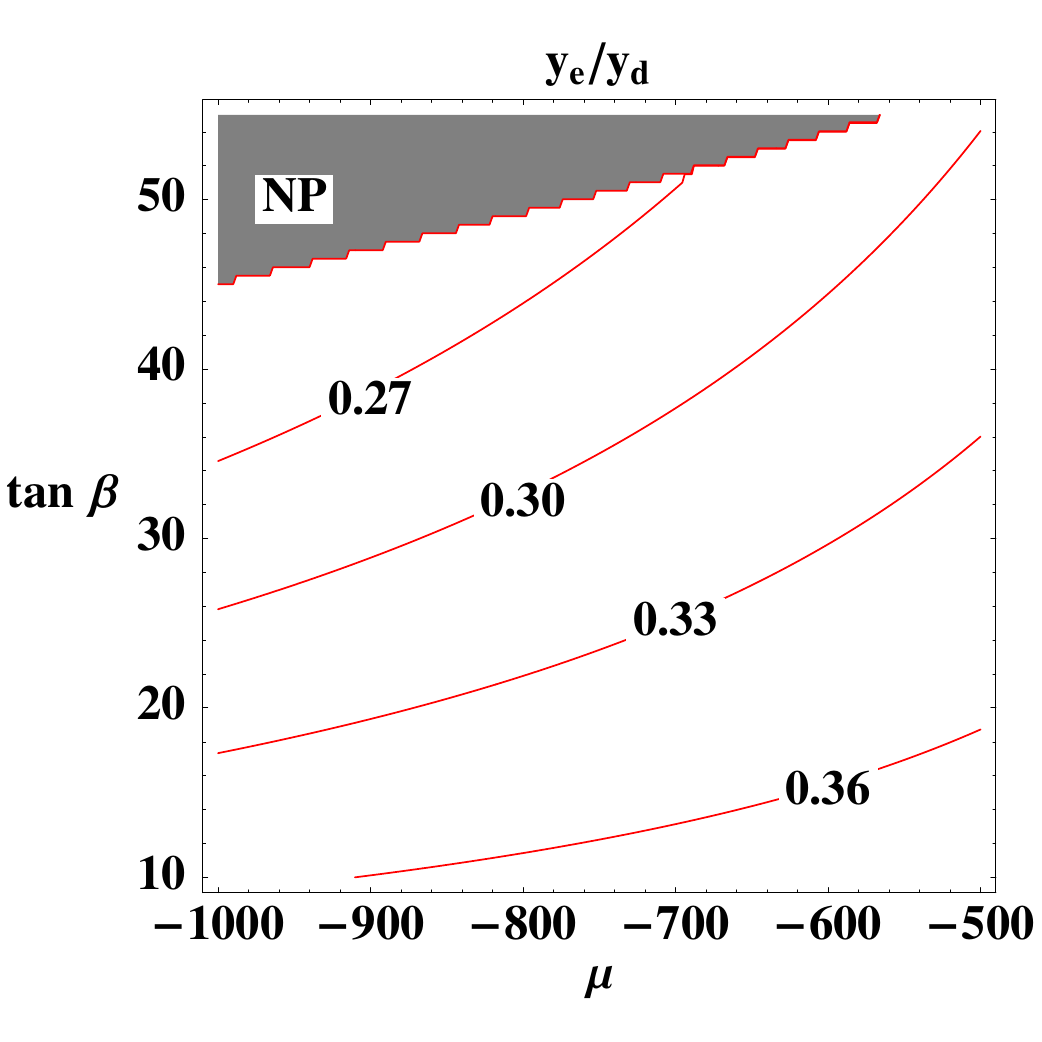} \hspace{1cm}
 \includegraphics[scale=0.6]{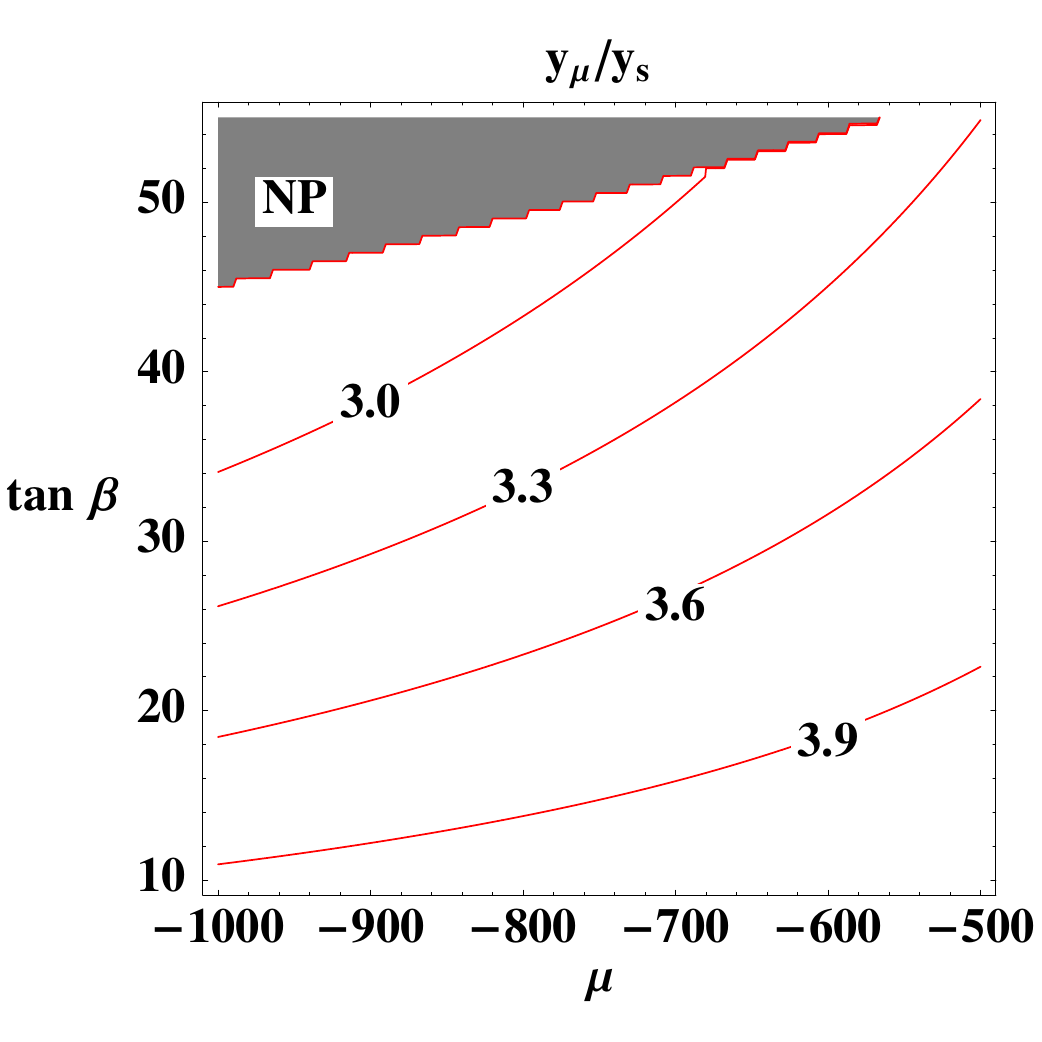}
 \includegraphics[scale=0.6]{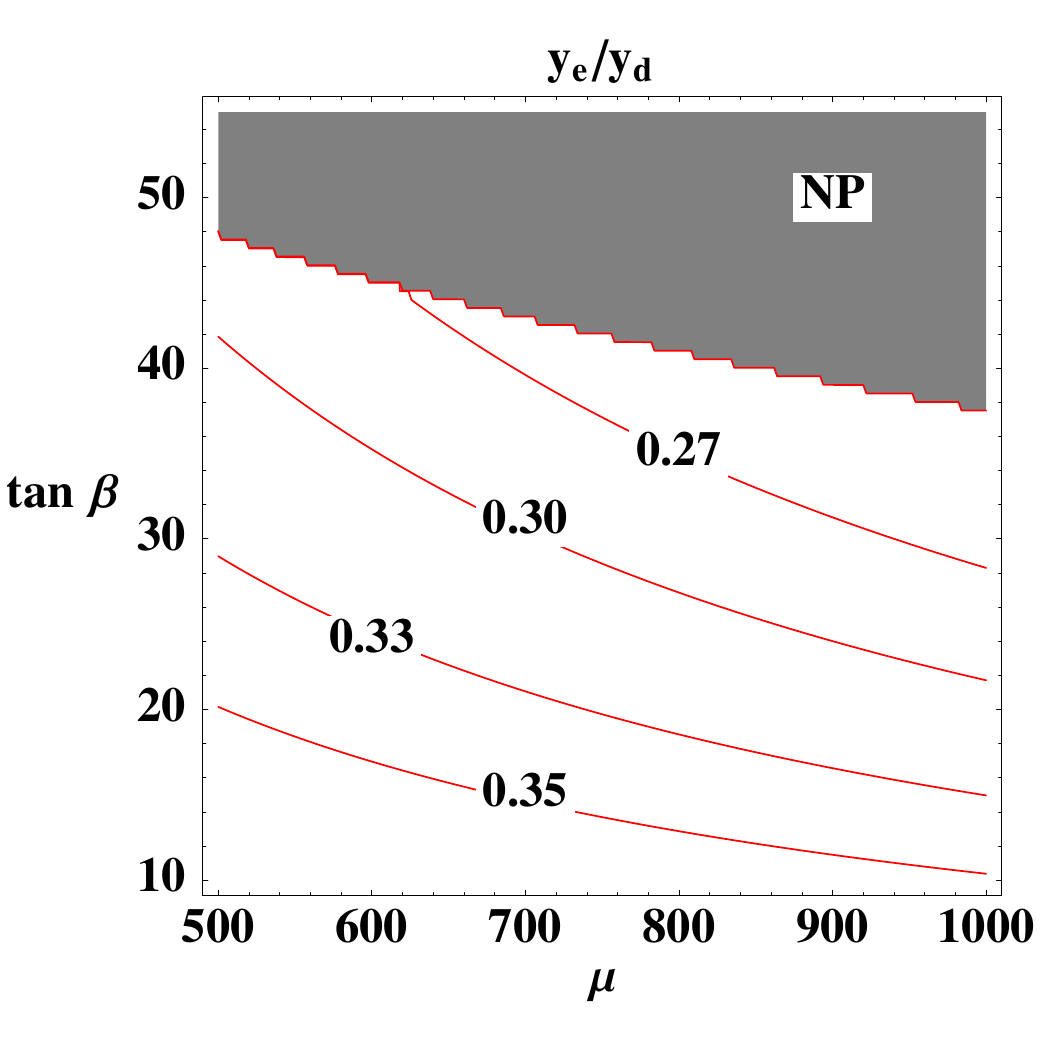} \hspace{1cm}
 \includegraphics[scale=0.6]{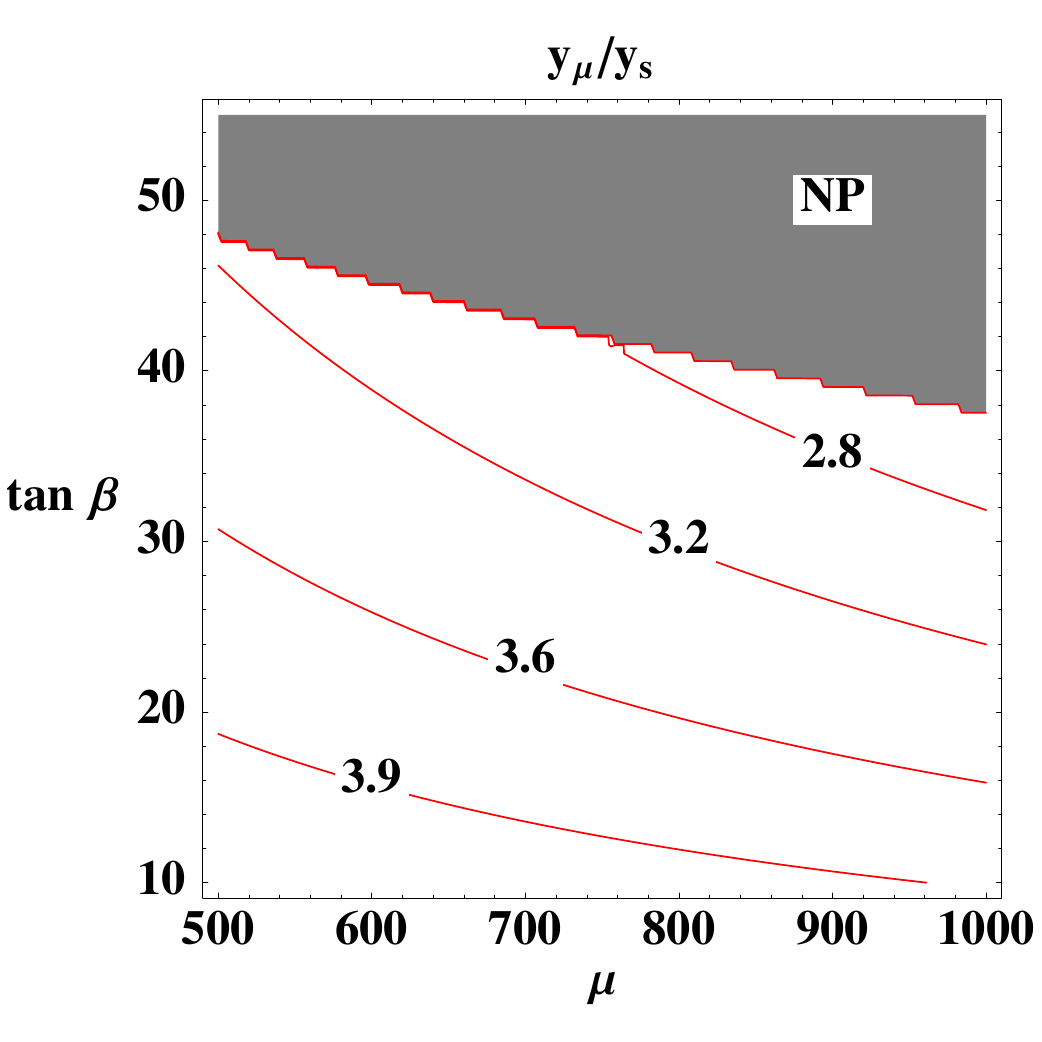}
 \caption[Dependence of $y_e/y_d$ and $y_\mu/y_s$ at the GUT Scale on $\mu$ and $\tan \beta$]{Contour plots for the GUT scale ratios $y_e/y_d$ (left side) and $y_\mu/y_s$ (right side) for $M_3>0$, $\mu>0$ (first row), $M_3>0$, $\mu<0$ (second row) and $M_3<0$, $\mu>0$ (third row) in the $\mu$-$\tan \beta$ plane. In the grey areas labelled with NP the value of $y_b$ becomes non-perturbatively large. \label{Fig:Contour12}}
\end{figure}

\begin{figure}
 \centering
 \includegraphics[scale=0.6]{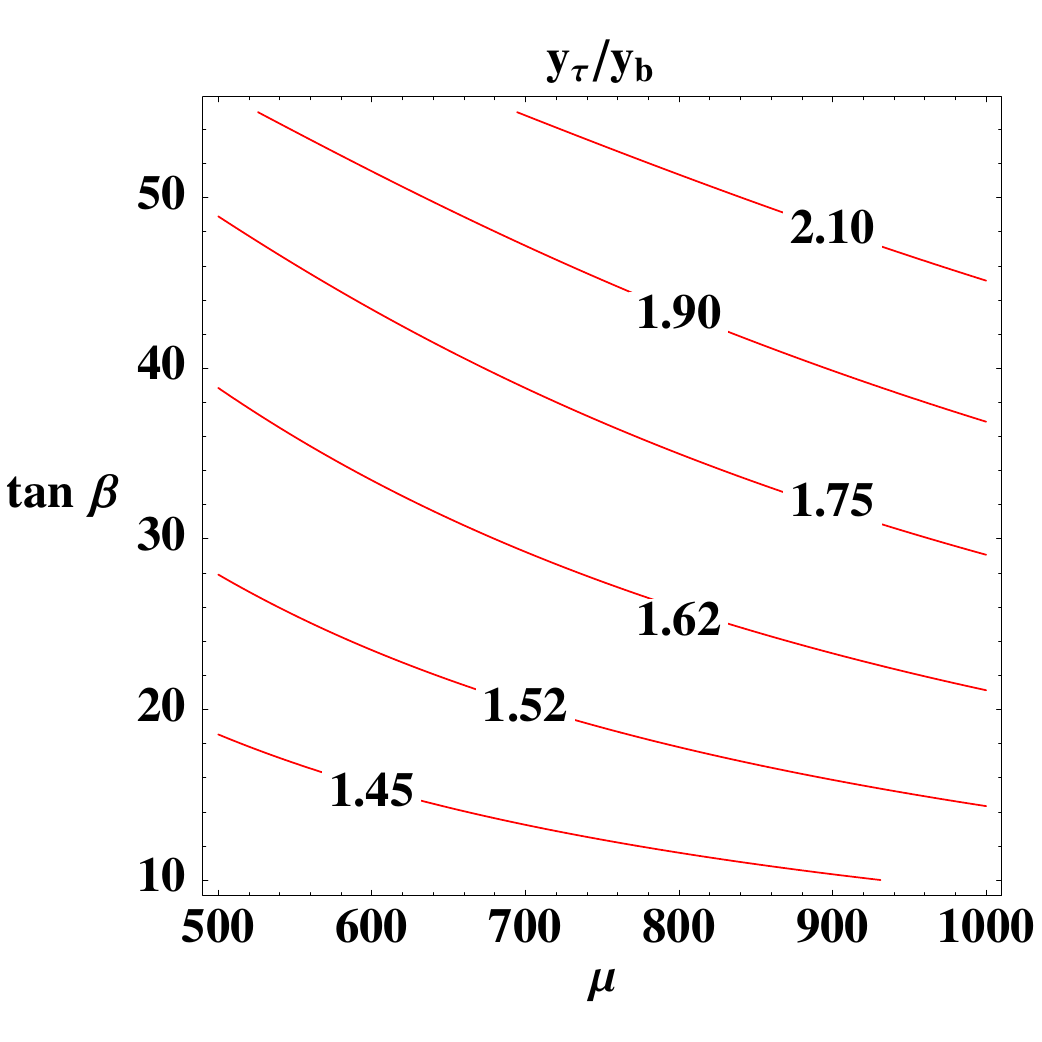} \hspace{1cm}
 \includegraphics[scale=0.6]{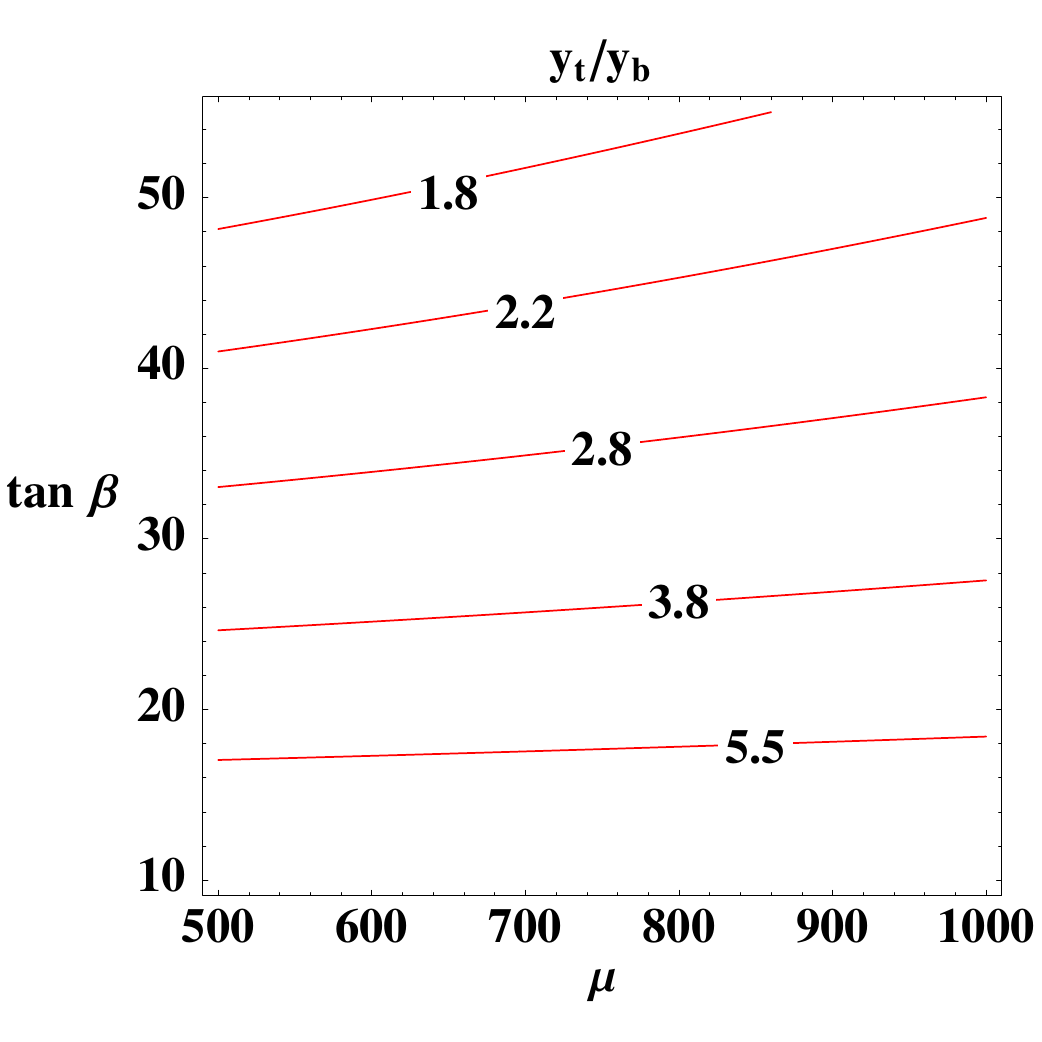}
 \includegraphics[scale=0.6]{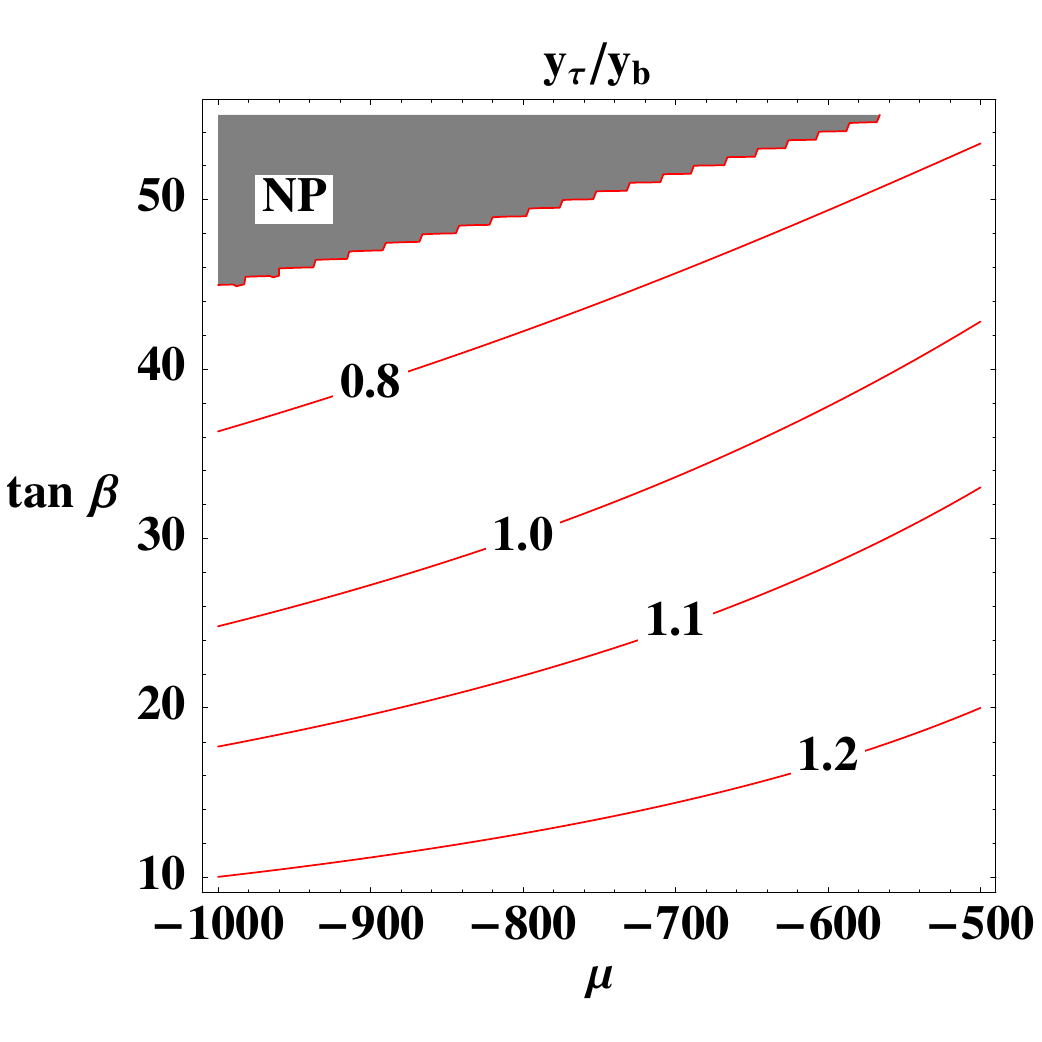} \hspace{1cm}
 \includegraphics[scale=0.6]{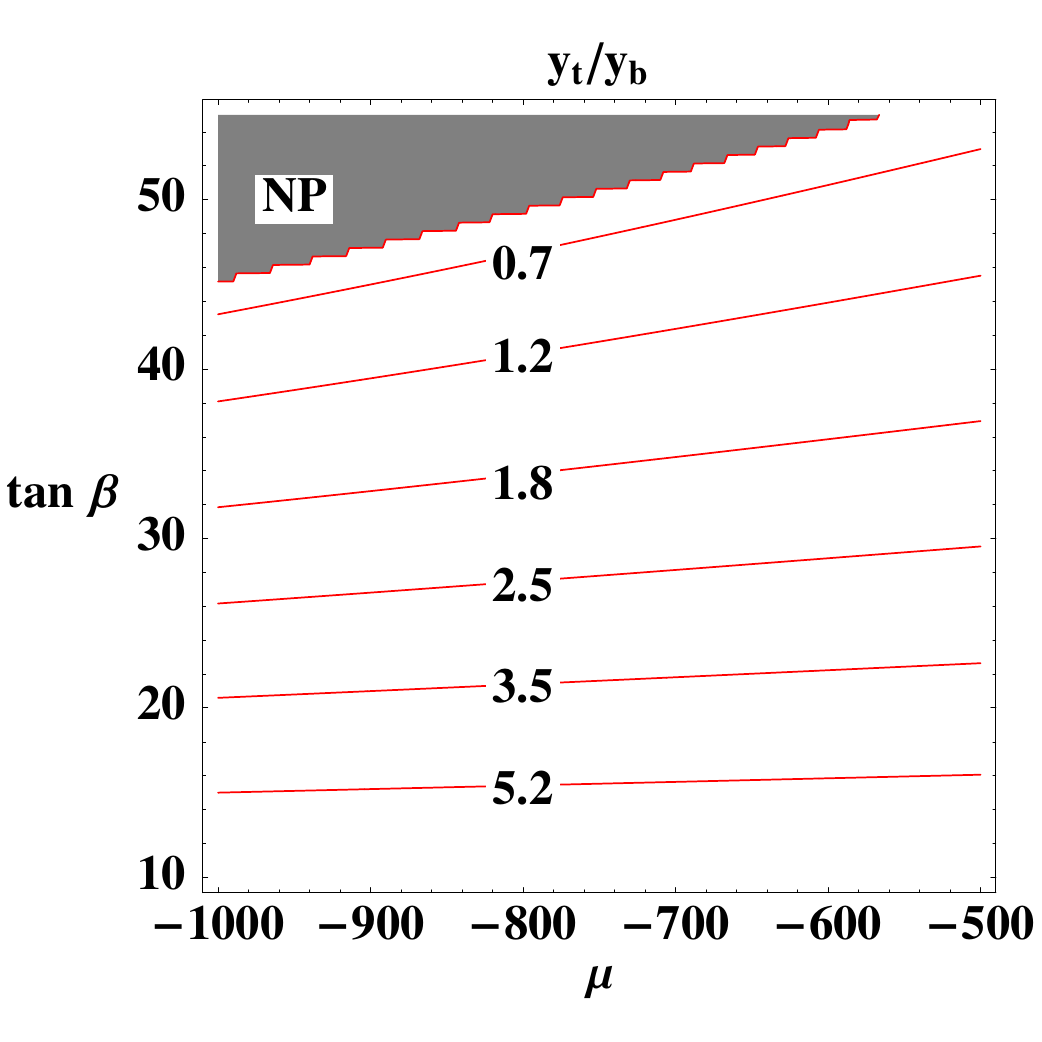}
 \includegraphics[scale=0.6]{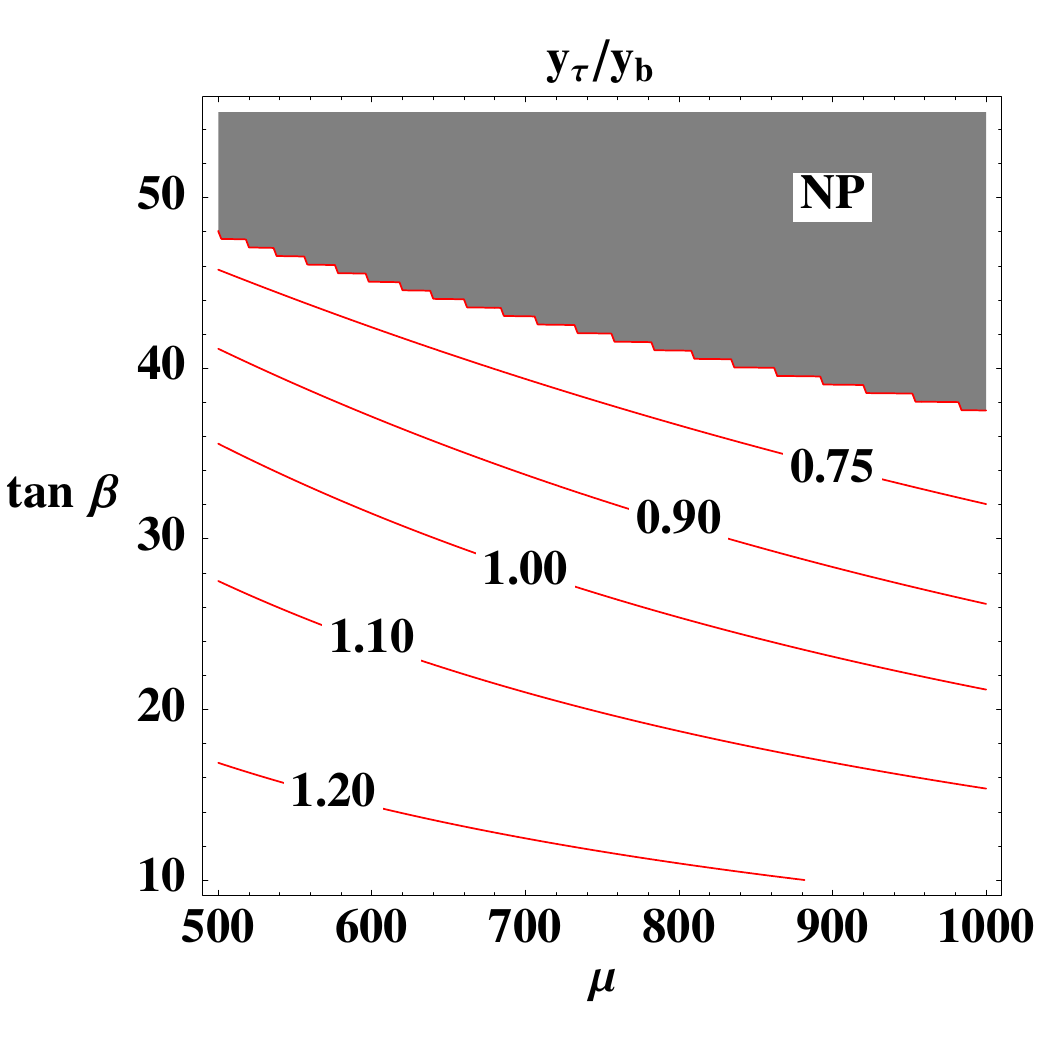} \hspace{1cm}
 \includegraphics[scale=0.6]{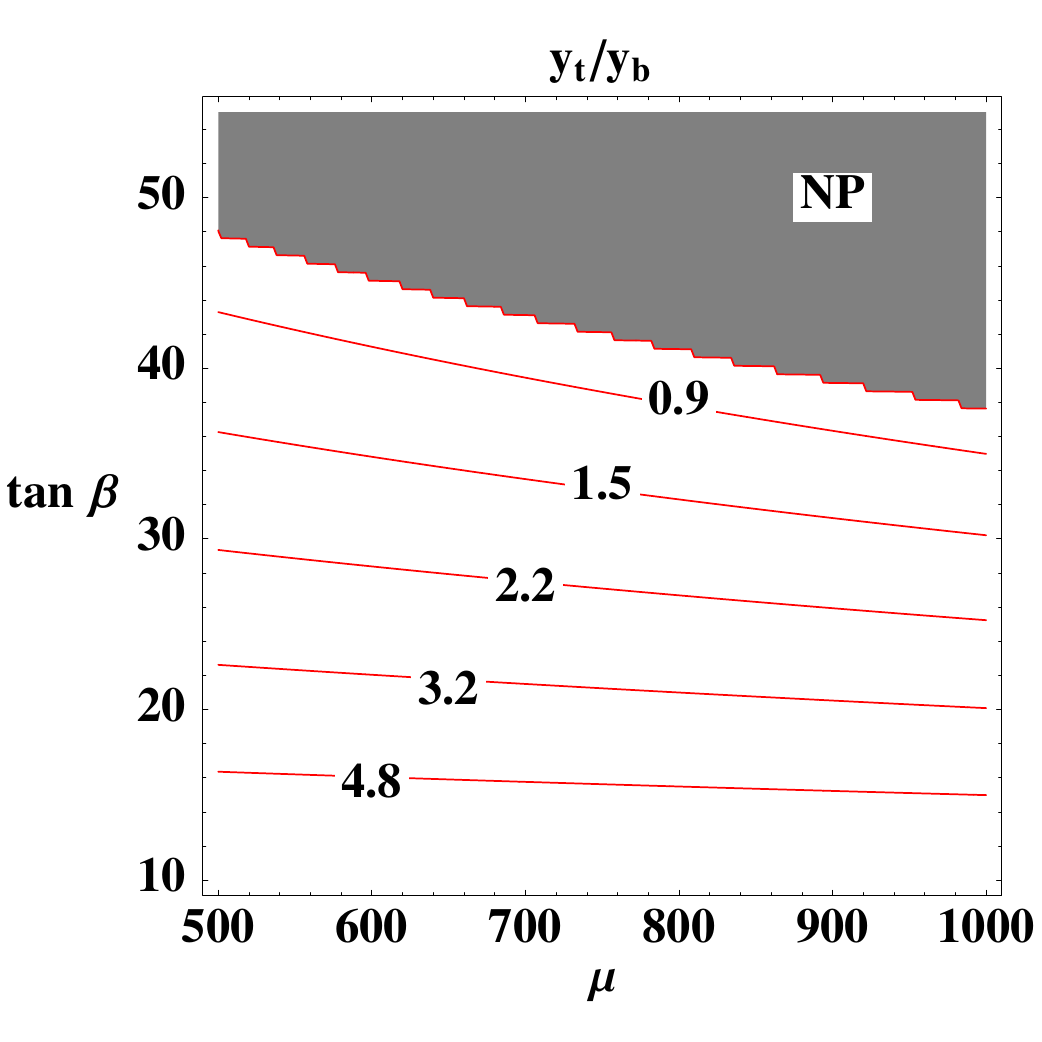}
 \caption[Dependence of $y_\tau/y_b$ and $y_t/y_b$ at the GUT Scale on $\mu$ and $\tan \beta$]{Contour plots for the GUT scale ratios $y_\tau/y_b$ (left side) and $y_t/y_b$ (right side) for $M_3>0$, $\mu>0$ (first row), $M_3>0$, $\mu<0$ (second row) and $M_3<0$, $\mu>0$ (third row) in the $\mu$-$\tan \beta$ plane. In the grey areas labelled with NP the value of $y_b$ becomes non-perturbatively large. \label{Fig:Contour34}}
\end{figure}

To isolate the effects of these parameters, we have turned off the right-handed neutrino 
thresholds, put $A_t$ to zero and all the other soft SUSY breaking parameters and the SUSY scale to $1$ TeV with both signs allowed for $M_3$ but with positive $M_1$ and $M_2$. In Figs.~\ref{Fig:Contour12} and \ref{Fig:Contour34} the numerical results are presented as contour plots in the $\mu$-$\tan \beta$ plane for the four ratios $y_e/y_d$, $y_\mu/y_s$, $y_\tau/y_b$ and $y_t/y_b$ for different combinations of the signs of $\mu$ and $M_3$. 

There are several interesting points we would like to remark: First, the overall dependence in the plots illustrates the anticipated behaviour from the fact that the leading contribution from gluino loops is proportional to $\mu$ and that the overall size of the corrections is proportional to $\tan \beta$. They also illustrate that for $\mu M_3 < 0$ (second and third row in the figures) the total corrections enhance the down-type quark Yukawa couplings leading to more stringent restrictions for the possible values of $\tan \beta$ from perturbativity of $y_b$ up to the GUT scale. On the other hand, for $\mu > 0$ and $M_3 > 0$ (first row in the figures) the total corrections lower the down-type quark Yukawa couplings and in principle larger values for $\tan \beta$ are possible.

Second, interesting conclusions can also be drawn from comparing the second to the third row of the figures. From the leading SUSY QCD contribution which is invariant under a simultaneous change of sign in $\mu$ and $M_3$, one might expect that the plots in the second and third row look very similar if understood as results for $|\mu|$. Differences are entirely  induced by the contributions from wino and bino loops, since we have chosen $A_t = 0$. Inspection of the numerical results show significant differences for $\mu < 0$ and $M_3 > 0$ and $\mu > 0$ and $M_3 < 0$, which confirms that the EW contributions are indeed important and cannot be ignored as we have already concluded from our estimates for the SUSY threshold corrections in the last chapter.

\subsection[Dependence on $M_{\mathrm{SUSY}}$ and $A_t$]{Dependence on $\boldsymbol{M_{\mathrm{SUSY}}}$ and $\boldsymbol{A_t}$}

The GUT scale values of the quark and lepton Yukawa couplings also depend on $M_\mathrm{SUSY}$ and $A_t$. While the correction to the bottom Yukawa coupling can be significant as can be seen from Tab.~\ref{Tab:ThresholdValues}, the effects on the down and strange quark Yukawa couplings are quite weak since they  only stem from indirect effects (modified RG evolution) due to the change of the bottom Yukawa coupling.

We have also looked at the dependence on $M_\mathrm{SUSY}$ by fixing all other parameters and varying only $M_\mathrm{SUSY}$. We find that changing $M_\mathrm{SUSY}$ can have some effect on the GUT scale value of the Yukawa couplings due to the difference in the RGEs between SM and MSSM; however this effect is typically much smaller than the uncertainty induced by the quark mass errors and the sparticle spectrum, and it nearly cancels out when we consider Yukawa coupling ratios. We have therefore fixed $M_\mathrm{SUSY}$ to $1$~TeV in our numerical examples. 

\subsection{Right-handed Neutrino Threshold Effects} \label{Sec:RHnus}

For analysing the possible dependence on threshold effects from the right-handed neutrino sector, we have taken the three examples for SUSY parameters from our analysis on the $\mu$ and $\tan \beta$ dependence from the previous chapter and fixed $\mu$ to $\pm 0.5$ TeV and $\tan \beta$ to 40. For these example parameter points we have investigated the effects for the three different scenarios of sequential dominance  \cite{King:1998jw, King:1999cm, *King:1999mb, *Antusch:2004gf, *Antusch:2010tf, King:2002nf, King:2005bj} also used as examples in \cite{Antusch:2007dj}. With the largest neutrino Yukawa coupling being ${\cal O}(1)$, we found that the deviations are typically smaller than 5~\%, which is small compared to the SUSY threshold effects in our examples and also compared to the uncertainties induced by the present quark mass errors, especially for the first and second generation.  

\subsection{Impact of the Sparticle Spectrum} \label{Sec:ImpactSparticles}

As already stated in the previous chapter, the GUT scale values of the quark and lepton Yukawa couplings strongly depend on the sparticle spectrum. We analyse now its impact numerically in more detail. Because of the large number of relevant parameters, we do not attempt to discuss each of them separately, but rather make a parameter scan.

For our scan, we take the three example ranges of SUSY parameters used for our analytic estimates for the SUSY threshold corrections in Ch.~\ref{Ch:SUSYThresholdCorrections}, which are listed explicitly in Tab.~\ref{Tab:SUSYParameters}. In addition, for $\tan \beta$ we assume a range from 30 to 50. The sfermion mass parameters and $A_t$ are scanned with a step size of 1 TeV including $A_t = 0$, the mass of the lightest gaugino was changed with a step size of 0.5 TeV and $\tan \beta$ is scanned  with a step size of 10. 

\begin{figure}
   \centering
  \includegraphics[scale=0.54]{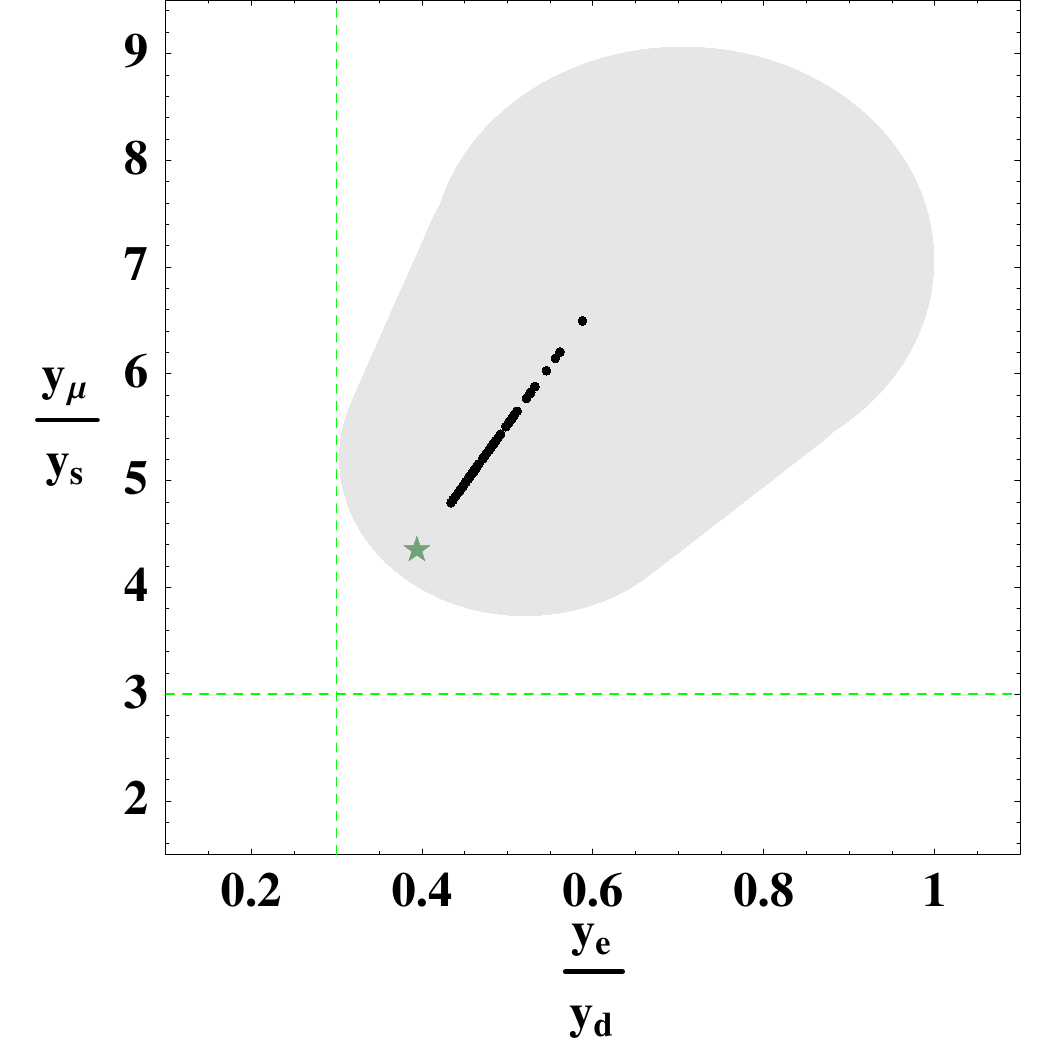} \hspace{1cm} \vspace{0.2cm}
  \includegraphics[scale=0.54]{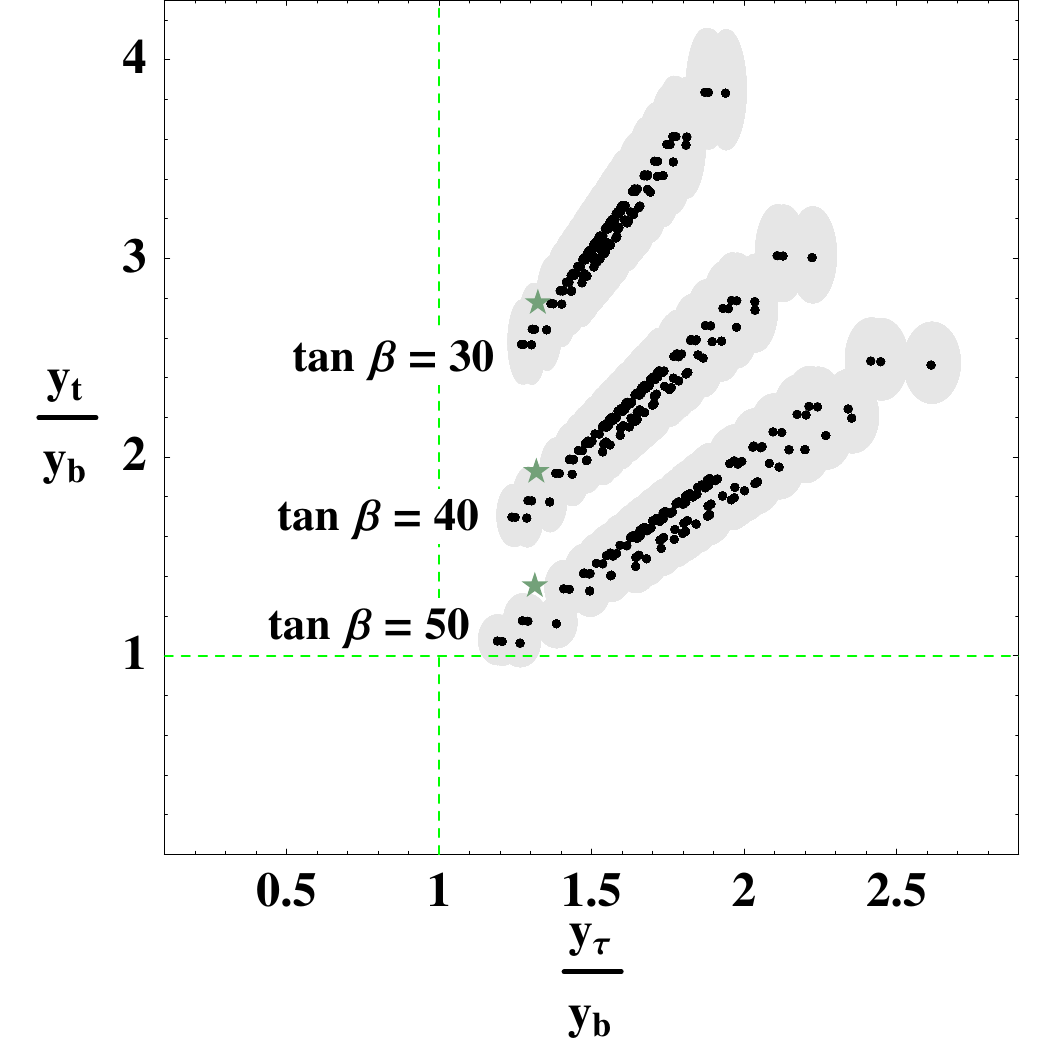} 
  \includegraphics[scale=0.54]{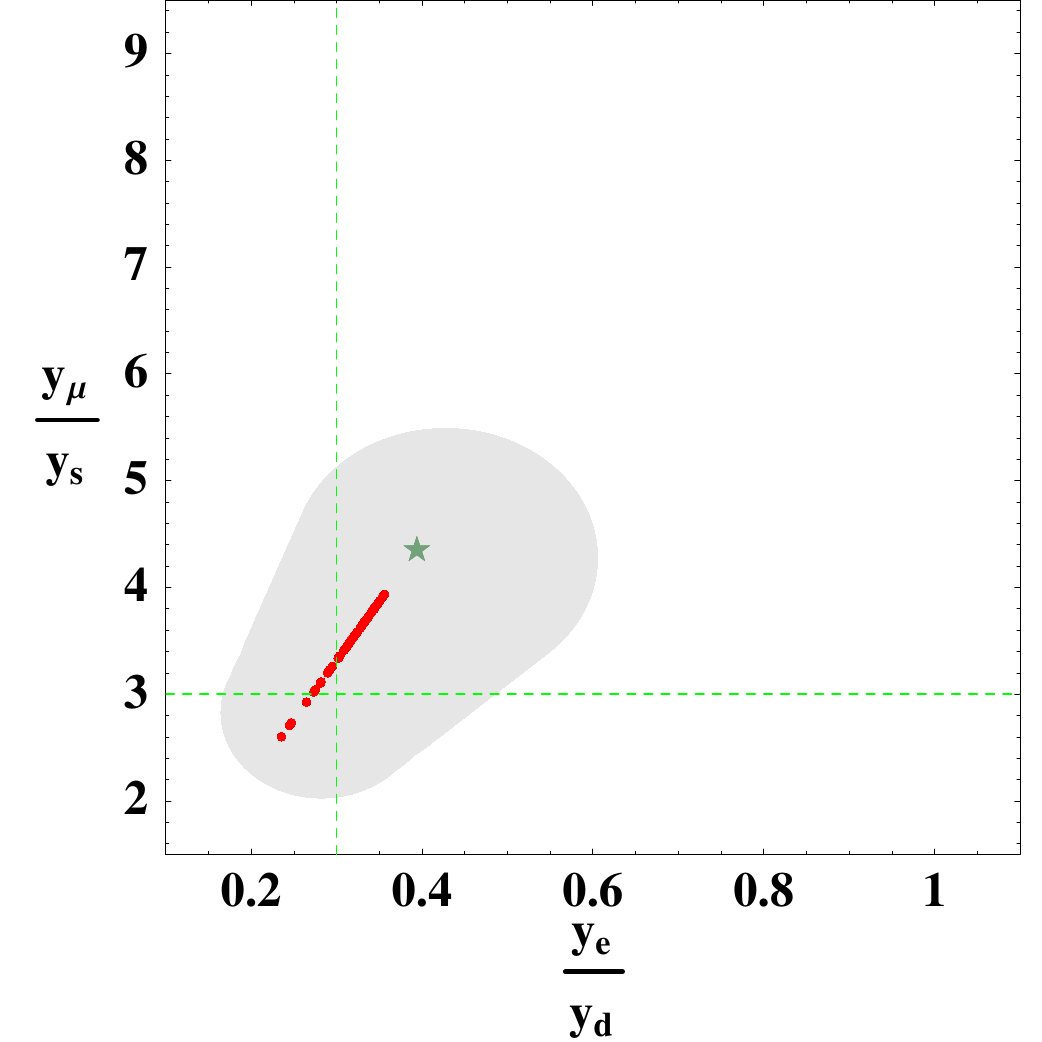} \hspace{1cm} \vspace{0.2cm}
  \includegraphics[scale=0.54]{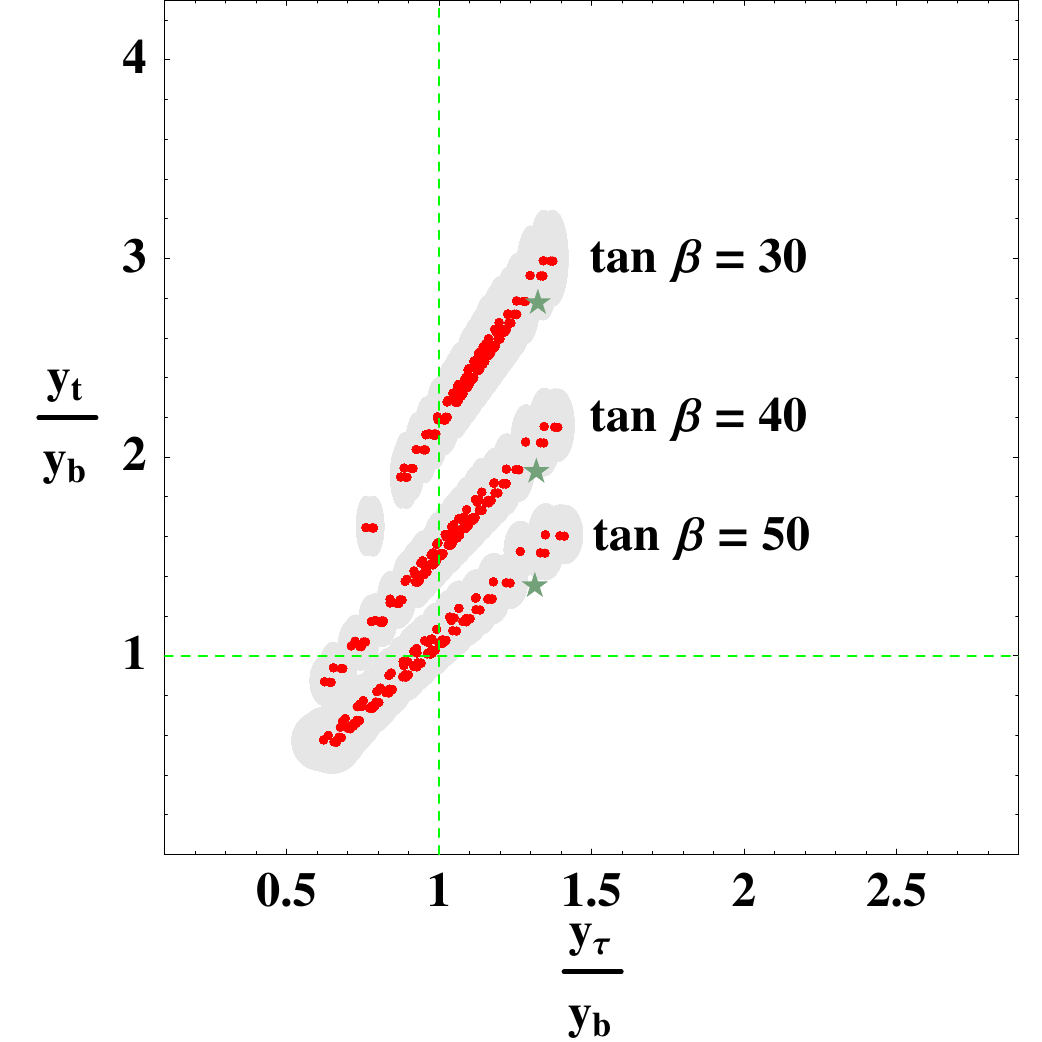}
  \includegraphics[scale=0.54]{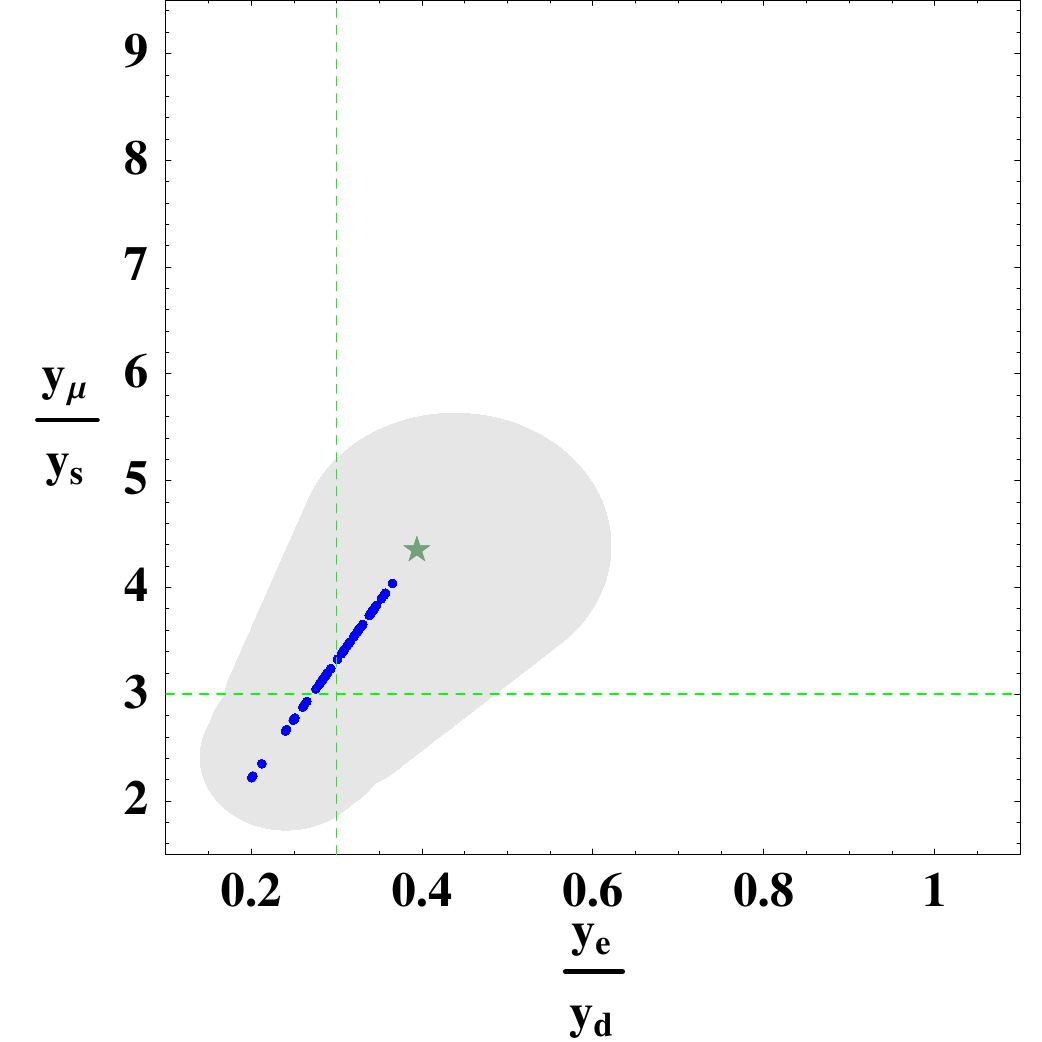} \hspace{1cm}
  \includegraphics[scale=0.54]{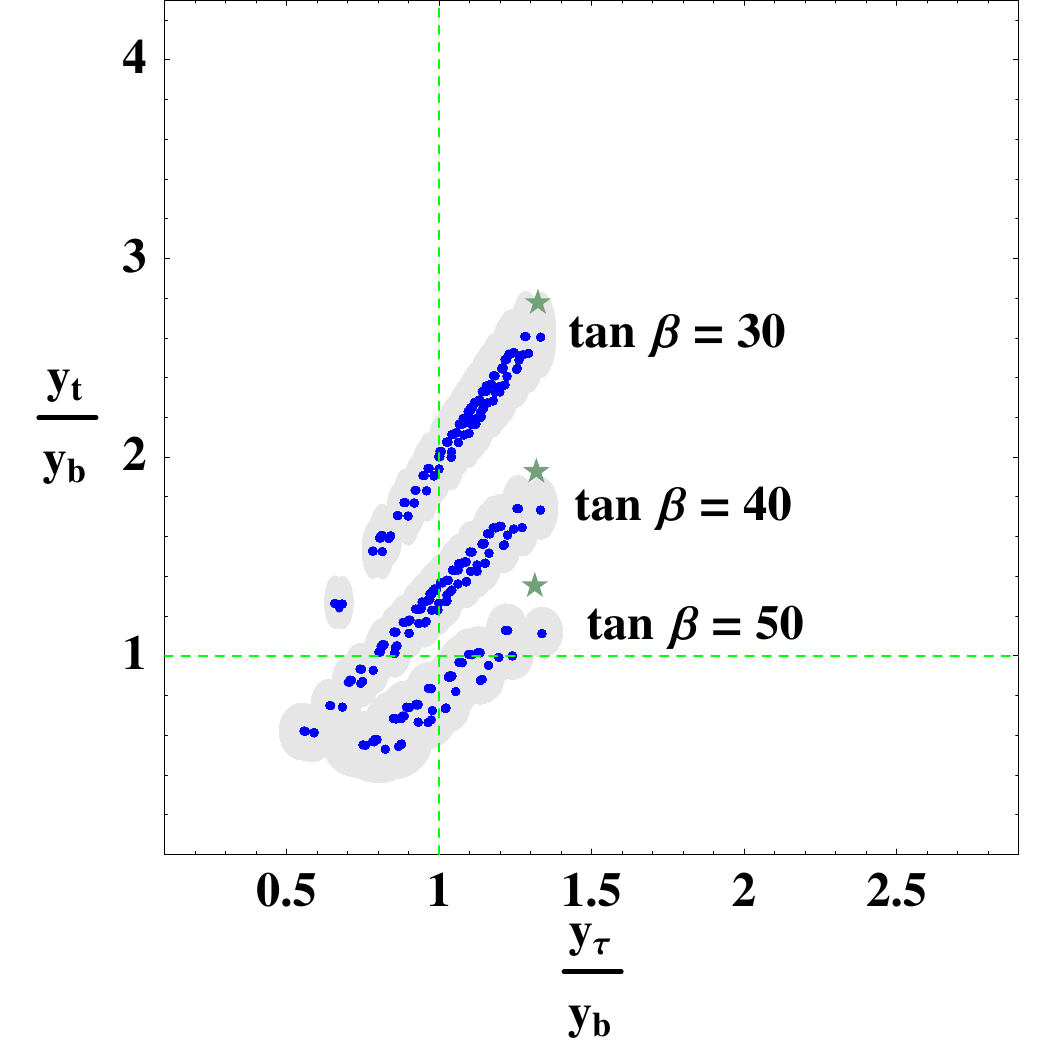}
   \caption[Scatter Plots of the GUT Scale Yukawa Coupling Ratios]{Scatter plots illustrating the ranges of GUT scale ratios $y_e/y_d$, $y_\mu/y_s$,  $y_\tau/y_b$ and $y_t/y_b$ corresponding to the example scan ranges of SUSY parameters in Tab.~\ref{Tab:SUSYParameters} including $A_t = 0$  for $\tan \beta = 30$, 40 and 50 (first row: case $g_+$, second row: case $g_-$, third row: case $a$). The green dashed lines in the left plots correspond to the GJ relations. In the right plots they indicate the ratios $y_\tau/y_b=1$ and $y_t/y_b=1$. The green stars correspond to case 0 (no SUSY threshold corrections). The point-bands in the plots on the right correspond to the different values of $\tan \beta$. 
The grey areas correspond to the (1$\sigma$) quark mass errors for each shown data point.    \label{Fig:Scatter}}   
\end{figure}

Although this seems to be a rather coarse scan, we note that we have discussed in Ch.~\ref{Ch:SUSYThresholdCorrections} that the extremal values of the SUSY threshold corrections correspond to the extremal values of the SUSY parameters. Therefore, increasing the number of points would not lead to enlarged ranges for the GUT scale quantities but only lead to more dense point clouds in the plots in Fig.~\ref{Fig:Scatter}. 

The parameter points for which $y_b$ becomes non-perturbative have been dropped from our analysis. We note that for simplicity we have introduced a slightly more restrictive cut and included only parameter points where $y_b < 1$ at $M_\mathrm{SUSY}$ which ensures perturbativity up to $M_\mathrm{GUT}$ but eventually removes a few allowed parameter points with large but still perturbative $y_b$. 

Furthermore, we have put $M_{\mathrm{SUSY}}$ to be 1~TeV. The masses of the first two sfermion generations have been assumed identical, which is inspired by universal high scale boundary conditions for sfermions. We note that a small mass splitting does not change the conclusions from this plot. However, a large mass splitting can reduce the threshold effects due to a reduced value of the function $|H_2|$ in the formulas \eqref{Eq:expq} to \eqref{Eq:expl_end}.  

The results of our parameter scans are presented as scatter plots in Fig.~\ref{Fig:Scatter}.
The grey areas correspond to the 1$\sigma$ quark mass errors for each shown data point. 
In the first column, $y_s$ and $y_d$ have been varied, and  in the second column $y_b$ and $y_t$, using best-fit values for the remaining fermion masses. In comparison to the quark mass errors, the charged lepton mass errors are negligible. In Fig.~\ref{Fig:Scatter} only those data points are shown where $y_b$, including its $1\sigma$ error, stays perturbative. 

For comparison we have also included the best-fit values, which would be obtained without SUSY threshold effects. It can be seen that, with SUSY threshold corrections included, the shown ratios for almost all parameter points are significantly shifted to larger values for the case $g_+$ and to smaller values for the cases $g_-$ and $a$.

The plots also reveal some interesting features, which we discuss now. For example we notice that small values for $y_\mu/y_s$ are correlated (nearly linearly) to small values of $y_e/y_d$. This correlation is connected to our assumption that the sfermion masses of the first two families are assumed to be identical. Nevertheless, because of the quark mass errors, this correlation is somewhat smeared.

Looking next at the third family relations $y_\tau/y_b$ versus $y_t/y_b$, we find that there is an additional $\tan \beta$ dependence, which is due to the fact that the relation of top mass to Yukawa coupling differs from the relations for down-type quarks and charged leptons by a factor of $\tan \beta$. For all three cases we can therefore distinguish three bands, which correspond to the three values of $\tan \beta$ in our parameter scan. The scans show that for the cases $g_-$ and $a$ it is in principle possible to obtain third family Yukawa unification for $\tan \beta \approx 50$ (in fact for $\tan \beta$ somewhat below $50$ in case $a$), whereas for $g_+$ we found that it could not be exactly realised. Although $y_b \approx y_t$ can be achieved for $\tan \beta = 50$, $A_t = -1$~TeV, $\mu = 0.5$~TeV, light gaugino masses, $m_{\tilde{d}_3} = 1.5$ TeV and $m_{\tilde{Q}_3} = m_{\tilde{u}_3} = 0.5$~TeV, we found $y_\tau/y_b \gtrsim 1.1$ in the considered parameter range. 

As a first glance on the new relations for the second generation we see that the ratio $y_\mu/y_s = 9/2$ can be achieved for all three scenarios and $y_\mu/y_s = 6$ is easily achieved within case $g_+$ while it is slightly above the edge of the 1$\sigma$ area for the other cases. For the third generation the relation $y_\tau/y_b = 3/2$ from $SU(5)$ for case $g_+$ seems possible which is again slightly above the 1$\sigma$ ranges for the other cases.  For the cases $g_-$ and $a$ the relation $y_t = 2 y_b = 2 y_\tau$ from PS seems to be possible. Nevertheless, these considerations should be  studied in more detail as we do it in the next chapter. There, the whole SUSY spectrum is specified in terms of SUSY breaking schemes and various phenomenological constraints are applied. This is important since a constraint for the anomalous magnetic moment of the muon $(g-2)_\mu$, for example, would strongly disfavour the case $g_-$.

The ranges for the quark and lepton Yukawa couplings and their mentioned ratios at the GUT scale are presented in Tab.~\ref{Tab:SError} without quark mass errors. They can be compared to the results of our semianalytic treatment  given in Tab.~\ref{Tab:GJRAnalytical}. Comparing the two tables, one can see that the ``mean values'' of the ranges agree well for the first two generations, however the extremal values are somewhat different. This is no surprise since in our naive estimates we have generically overestimated the ranges of the $\eta_i$ since we have ignored possible correlations between the corrections. For the third generation, the ranges (boundaries) are slightly shifted. This effect is caused by the modified RG running with SUSY threshold corrections included. For the first two generations this effect is smaller due to the smaller Yukawa couplings. 

While the quark mass errors are still ignored in Tabs.~\ref{Tab:GJRAnalytical} and \ref{Tab:SError}, they are included in our final results listed in Tabs.~\ref{Tab:Ratfinal} and \ref{Tab:Yukfinal}. Here, we have varied all the quark mass errors simultaneously. Comparing case 0 (without SUSY threshold corrections) to the cases $g_\pm$ and $a$, we see that the ranges for all types of Yukawa couplings, for down-type quarks, charged leptons and up-type quarks for all three generations can be significantly affected by the SUSY threshold corrections. We note that for the up-type quarks, the changes are indirect in the sense that they are induced by modified RG running mainly due to the corrected $b$-quark and $\tau$-lepton Yukawa couplings.   

\begin{table}
\centering
\begin{tabular}{ccccc}
\toprule 				& Case 0 & Case $g_+$ & Case $g_-$ & Case $a$ \\ \midrule
$y_e/y_d$ 		& 0.39 & [0.45, 0.55] & [0.26, 0.34] & [0.24, 0.36] \\ 
$y_\mu/y_s$ 		& 4.35 & [4.95, 6.03] & [2.92, 3.80] & [2.65, 3.92] \\ 
$y_\tau/y_b$ 		& 1.32 & [1.23, 2.23] & $\leq 1.40$  & $\leq 1.35$  \\ 
$y_t/y_b$ 		& 1.93 & [1.67, 3.03] & $\leq 2.17$  & $\leq 1.76$  \\ \midrule
$y_e$ in $10^{-4}$	& 0.88 & [0.87, 0.97] & [0.80, 1.09] & [0.92, 1.44] \\
$y_\mu$ in $10^{-2}$	& 1.85 & [1.83, 2.06] & [1.68, 2.31] & [1.95, 3.05] \\
$y_\tau$			& 0.34 & [0.34, 0.39] & [0.31, 0.44] & [0.36, 0.60] \\ \midrule
$y_d$ in $10^{-4}$	& 2.22 & [1.66, 2.10] & [2.39, 3.80] & [2.73, 5.73] \\
$y_s$ in $10^{-2}$	& 0.43 & [0.32, 0.40] & [0.46, 0.73] & [0.52, 1.10] \\
$y_b$			& 0.26 & [0.16, 0.30] & $\geq 0.23$  & $\geq 0.29$  \\ \midrule
$y_u$ in $10^{-6}$	& 2.75 & [2.73, 2.76] & [2.74, 2.83] & [2.75, 2.87] \\
$y_c$ in $10^{-3}$	& 1.34 & [1.33, 1.35] & [1.34, 1.38] & [1.34, 1.40] \\
$y_t$			& 0.50 & [0.48, 0.51] & [0.50, 0.58] & [0.50, 0.62] \\ \bottomrule
\end{tabular}
\caption[GUT Scale Yukawa Couplings and Ratios without Quark Mass Errors]{Ranges for the GUT scale Yukawa couplings and ratios corresponding to the example ranges of SUSY parameters in Tab.~\ref{Tab:SUSYParameters}, including additionally $A_t = 0$ from our numerical analysis with $\tan \beta = 40$. Case 0 refers to the case without SUSY threshold corrections. Quark mass errors are not yet included. The results can be compared with the semianalytic estimates of Tab.~\ref{Tab:GJRAnalytical}. Our final results, which include the experimental mass errors, are given in Tabs.~\ref{Tab:Ratfinal} and \ref{Tab:Yukfinal}. Where $y_b$ becomes non-perturbatively large we have given only the boundary for which $y_b$ stays perturbative up to the GUT scale. \label{Tab:SError}}
\end{table}

\begin{table}
\centering
\begin{tabular}{cccccc}
\toprule  	$\tan \beta$		& Ratio 	& Case 0 & Case $g_+$ & Case $g_-$ & Case $a$ \\ \midrule
 $30$ 	& $y_e/y_d$	& [0.28, 0.67] & [0.30, 0.86] & [0.21, 0.61] & [0.20, 0.62] \\
 		& $y_\mu/y_s$	& [3.39, 6.07] & [3.73, 7.79] & [2.54, 5.49] & [2.40, 5.63] \\
 		& $y_\tau/y_b$	& [1.27, 1.38] & [1.20, 2.02] & [0.71, 1.43] & [0.60, 1.39] \\
 		& $y_t/y_b$	& [2.56, 3.02] & [2.36, 4.19] & [1.50, 3.28] & [1.14, 2.87] \\ \midrule
 $40$ 	& $y_e/y_d$	& [0.28, 0.67] & [0.31, 0.93] & [0.19, 0.59] & [0.17, 0.60] \\
		& $y_\mu/y_s$	& [3.39, 6.07] & [3.85, 8.41] & [2.28, 5.30] & [2.07, 5.47] \\
		& $y_\tau/y_b$	& [1.26, 1.38] & [1.16, 2.32] & $\leq 1.46$  & $\leq 1.41$  \\
		& $y_t/y_b$	& [1.77, 2.11] & [1.55, 3.31] & $\leq 2.38$  & $\leq 1.94$  \\ \midrule
 $50$ 	& $y_e/y_d$	& [0.28, 0.67] & [0.32, 1.00] & [0.16, 0.57] & [0.14, 0.59] \\
		& $y_\mu/y_s$	& [3.39, 6.07] & [3.98, 9.06] & [2.02, 5.12] & [1.72, 5.31] \\
		& $y_\tau/y_b$	& [1.25, 1.38] & [1.08, 2.73] & $\leq 1.49$  & $\leq 1.43$  \\
		& $y_t/y_b$	& [1.22, 1.50] & [0.94, 2.74] & $\leq 1.81$  & $\leq 1.31$  \\ \midrule
\end{tabular}
\caption[GUT Scale Yukawa Coupling Ratios with Quark Mass Errors Included]{Ranges for the GUT scale ratios $y_e/y_d$, $y_\mu/y_s$,  $y_\tau/y_b$ and $y_t/y_b$ corresponding to the example ranges of SUSY parameters $g_+$, $g_-$ and $a$ defined in Tab.~\ref{Tab:SUSYParameters} including $A_t = 0$. The results have been extracted from the parameter scan with $\tan \beta = 30$, $40$ and $50$, where in addition to the SUSY threshold corrections the present experimental errors for the quark masses have been included. Case 0 refers to the case without SUSY threshold corrections. Where $y_b$ becomes non-perturbatively large we have given only the boundary for which $y_b$ stays perturbative up to the GUT scale.
\label{Tab:Ratfinal}}
\end{table}

\begin{table}
\centering
\begin{tabular}{cccccc}
\toprule 		$\tan \beta$		& Yukawa Coupling		& Case 0       & Case $g_+$ & Case $g_-$ & Case $a$ \\ \midrule
 $30$ 	& $y_e$ in $10^{-4}$	& 0.62         & [0.62, 0.67] & [0.58, 0.66] & [0.63, 0.79] \\
		& $y_\mu$ in $10^{-2}$	& [1.30, 1.32] & [1.32, 1.41] & [1.22, 1.40] & [1.34, 1.66] \\
		& $y_\tau$		& 0.23         & [0.23, 0.25] & [0.22, 0.25] & [0.24, 0.30] \\ \cmidrule{2-6}
		& $y_d$ in $10^{-4}$	& [0.92, 2.26] & [0.75, 2.14] & [0.98, 3.11] & [1.06, 3.89] \\
		& $y_s$ in $10^{-2}$	& [0.21, 0.39] & [0.17, 0.37] & [0.23, 0.53] & [0.25, 0.67] \\
		& $y_b$			& [0.17, 0.18] & [0.12, 0.20] & [0.16, 0.34] & [0.18, 0.48] \\ \cmidrule{2-6}
		& $y_u$ in $10^{-6}$	& [1.79, 3.88] & [1.78, 3.89] & [1.78, 3.93] & [1.79, 3.98] \\
		& $y_c$ in $10^{-3}$	& [1.14, 1.54] & [1.13, 1.54] & [1.14, 1.56] & [1.14, 1.58] \\
		& $y_t$			& [0.46, 0.51] & [0.46, 0.52] & [0.46, 0.54] & [0.46, 0.57] \\ \midrule
 $40$ 	& $y_e$ in $10^{-4}$	& [0.87, 0.88] & [0.86, 0.99] & [0.79, 1.62] & [0.91, 1.77] \\
		& $y_\mu$ in $10^{-2}$	& [1.83, 1.87] & [1.82, 2.08] & [1.67, 3.43] & [1.93, 3.74] \\
		& $y_\tau$		& [0.34, 0.35] & [0.34, 0.39] & [0.30, 0.67] & [0.36, 0.76] \\ \cmidrule{2-6}
		& $y_d$ in $10^{-4}$	& [1.30, 3.21] & [0.97, 3.05] & [1.40, 8.41] & [1.59, 9.81] \\
		& $y_s$ in $10^{-2}$	& [0.30, 0.55] & [0.23, 0.52] & [0.33, 1.44] & [0.37, 1.69] \\
		& $y_b$			& [0.25, 0.27] & [0.15, 0.33] & $\geq 0.22$  & $\geq 0.27$  \\ \cmidrule{2-6}
		& $y_u$ in $10^{-6}$	& [1.79, 3.91] & [1.78, 3.92] & [1.79, 4.07] & [1.80, 4.15] \\
		& $y_c$ in $10^{-3}$	& [1.14, 1.55] & [1.14, 1.56] & [1.14, 1.62] & [1.14, 1.65] \\
		& $y_t$			& [0.47, 0.53] & [0.46, 0.54] & [0.47, 0.65] & [0.48, 0.72] \\ \midrule
$50$ 	& $y_e$ in $10^{-4}$	& [1.18, 1.23] & [1.13, 1.53] & [1.04, 2.31] & [1.30, 3.80] \\
		& $y_\mu$ in $10^{-2}$	& [2.50, 2.60] & [2.39, 3.24] & [2.19, 4.89] & [2.74, 8.04] \\
		& $y_\tau$		& [0.50, 0.52] & [0.47, 0.69] & [0.42, 1.07] & [0.56, 2.20] \\ \cmidrule{2-6}
		& $y_d$ in $10^{-4}$	& [1.77, 4.46] & [1.20, 4.56] & [1.91, 12.22]& [2.36, 22.68]\\
		& $y_s$ in $10^{-2}$	& [0.41, 0.77] & [0.28, 0.78] & [0.44, 2.10] & [0.55, 3.90] \\
		& $y_b$			& [0.36, 0.42] & [0.19, 0.60] & $\geq 0.30$  & $\geq 0.43$  \\ \cmidrule{2-6}
		& $y_u$ in $10^{-6}$	& [1.81, 3.95] & [1.79, 4.00] & [1.80, 4.18] & [1.81, 4.21] \\
		& $y_c$ in $10^{-3}$	& [1.15, 1.57] & [1.14, 1.59] & [1.15, 1.66] & [1.16, 1.67] \\
		& $y_t$			& [0.49, 0.56] & [0.46, 0.59] & [0.48, 0.78] & [0.50, 0.86] \\ \bottomrule
\end{tabular}
\caption[GUT Scale Yukawa Couplings with Quark Mass Errors Included]{Ranges for the GUT scale values of the Yukawa couplings corresponding to the example ranges of SUSY parameters $g_+$, $g_-$ and $a$ defined in Tab.~\ref{Tab:SUSYParameters} including $A_t = 0$. The results have been extracted from the parameter scan with $\tan \beta = 30$, $40$ and $50$, where in addition to the SUSY threshold corrections the present experimental errors for the quark masses have been included. Case 0 refers to the case without SUSY threshold corrections. Where $y_b$ becomes non-perturbatively large we have given only the boundary for which $y_b$ stays perturbative up to the GUT scale. \label{Tab:Yukfinal}}
\end{table}

%% file: kap_08_Pheno.tex
\chapter[Yukawa Couplings at the GUT Scale: A Phenomenological Approach]{Yukawa Couplings at the GUT Scale:\\ A Phenomenological Approach}  \label{Ch:Pheno}

When testing the predictions of SUSY GUTs for quark and lepton Yukawa coupling ratios $y_e/y_d$, $y_\mu/y_s$, $y_\tau/y_b$ and $y_t/y_b$ the $\tan \beta$-enhanced SUSY threshold corrections are of particular importance, as has been emphasised in the last two chapters, see also  \cite{Ross:2007az,Antusch:2008tf}.  In this chapter based on \cite{Antusch:2009gu} we extend the analysis of the preceding chapter and analyse which ratios between quark and lepton Yukawa couplings can be realised at the GUT scale when phenomenological constraints are taken into account.  For explicitness, we consider three minimal, but characteristic SUSY breaking scenarios, namely mAMSB, mGMSB and CMSSM, which provide boundary conditions for the soft SUSY parameters, as is already briefly reviewed in Sec.~\ref{Sec:SUSYbreaking}. After RG evolution to low energies and including SUSY threshold corrections, see Sec.~\ref{Sec:NumericalProcedure}, we apply the phenomenological constraints described in Sec.~\ref{Sec:ExperimentalConstraints}.

Performing the above-described analysis, we arrive at phenomenologically allowed GUT scale ratios within the considered parameter ranges of the three SUSY breaking scenarios mAMSB, mGMSB and CMSSM. These results are independent of any underlying GUT model. Finally, in Sec.~\ref{Sec:TheoryvsExperiment}, we compare them with the GUT predictions considered in Ch.~\ref{Ch:GUTYukawaRelations}.

\section{SUSY Breaking Parameters} \label{Sec:ScanRanges}

SUSY, if realised in nature, obviously has to be broken in order to be consistent with the experimental non-observation of sparticles so far. To keep SUSY as a solution to the hierarchy problem this breaking should be soft as it is discussed in Ch.~\ref{Ch:SUSY}. 

In this thesis we consider three common and characteristic examples for supersymmetry breaking scenarios, namely mAMSB \cite{Randall:1998uk, *Giudice:1998xp, *Gherghetta:1999sw}, mGMSB \cite{Dine:1993yw, *Dine:1994vc, *Dine:1995ag, *Ambrosanio:1997rv, *Giudice:1998bp} and CMSSM \cite{Nilles:1982ik, *Nilles:1982xx, Chamseddine:1982jx, *Barbieri:1982eh} which provide boundary conditions for the soft SUSY breaking parameters at high energies. Since we have already discussed these three SUSY breaking schemes in Sec.~\ref{Sec:SUSYbreaking} we give in Tab.~\ref{Tab:SUSYParameterRanges} only the ranges and stepwidths for the soft SUSY breaking parameters which we have used in our numerical scan. In all schemes we choose the sign of $\mu$ to be positive in order to improve consistency with the experimental results on $(g-2)_\mu$, as is discussed in Sec.~\ref{Sec:g2mu}. 
The absolute value of $\mu$ is determined numerically to achieve successful electroweak symmetry breaking.

We note that we have not included neutrino masses in our analysis since we focus on Yukawa coupling ratios for charged fermions and right-handed (s)neutrinos are also not included in the minimal SUSY breaking scenarios mAMSB \cite{Randall:1998uk, *Giudice:1998xp, *Gherghetta:1999sw}, mGMSB \cite{Dine:1993yw, *Dine:1994vc, *Dine:1995ag, *Ambrosanio:1997rv, *Giudice:1998bp} and CMSSM  \cite{Nilles:1982ik, *Nilles:1982xx, Chamseddine:1982jx, *Barbieri:1982eh}. 

\begin{table}
\begin{center}
\begin{tabular}{cccc}\toprule
mAMSB Parameter 		& Minimum & Maximum & Stepwidth \\ \midrule
$m_0$ in TeV 	& 0   & 3   & 0.1 \\ 
$m_{3/2}$ in TeV	& 20  & 200 & 10  \\ 
$\tan \beta$ 	& 20  & 60  & 2.5  \\ 
\bottomrule
\end{tabular}

\vspace{1pc}

\begin{tabular}{cccc}\toprule
mGMSB Parameter 			& Minimum & Maximum & Stepwidth \\ \midrule
 $n_5$		& 1   & 5   & 1 \\
 $\Lambda$ in TeV		& 10  & 200 & 20  \\
 $m_{\mathrm{mess}}$
	& $1.01 \Lambda$ & $10^5 \Lambda$ & $10^4 \Lambda$ \\
 $c_{\mathrm{grav}}$	& 1  & 1 & - \\
 $\tan \beta$		& 20  & 60  & 2 \\
\bottomrule
\end{tabular}

\vspace{1pc}

\begin{tabular}{cccc}\toprule
CMSSM Parameter 		& Minimum & Maximum & Stepwidth \\ \midrule
 $m_0$ in TeV	& 0  & 3 & 0.2  \\
 $m_{1/2}$ in TeV	& 0  & 3 & 0.2 \\
 $A_0$ in TeV	& -3  & 3 & 1.5 \\
 $\tan \beta$	& 20  & 60 & 5  \\
\bottomrule
\end{tabular}
\end{center}
\caption[SUSY Parameter Ranges for the Numerical Scan]{Parameter ranges and stepwidths used in our numerical scan
         for the mAMSB, mGMSB and CMSSM scenario (from top to bottom). \label{Tab:SUSYParameterRanges}}
\end{table}

\section{Numerical Procedure} \label{Sec:NumericalProcedure}

Using the soft breaking parameters specified in the last section as high scale boundary conditions, the MSSM parameters are run to low energies using a modified version of SoftSUSY 2.0.18 \cite{Allanach:2001kg} which we have used for calculating the spectrum. SoftSUSY runs in loops to achieve consistency with high scale boundary conditions as well as with low scale input, thereby determining $|\mu|$. From SoftSUSY we read out the Yukawa couplings of the quarks and charged leptons at the GUT scale. Our modifications to the SoftSUSY code are the following:
\begin{itemize}

\item In SoftSUSY 2.0.18, the threshold corrections are included as self-energy corrections to the fermion masses, but only for the third generation. We have included the SUSY threshold corrections for the first two generations, using mainly the formulas of \cite{Pierce:1996zz}. The large logs appearing in the formulas in \cite{Pierce:1996zz} are already resummed in the gauge couplings and therefore are not included anymore, see also \cite{Allanach:2001kg}. For the first two generations we have set the external momenta of the fermions to zero. This provides a very good approximation since corrections are of the 
order of $m_f/M_{\mathrm{SUSY}}$, where $m_f$ is the mass of the corresponding (light) fermion and $M_{\mathrm{SUSY}}$ is the mass scale of the SUSY particles involved in the loops. We have also updated the experimental data on the quark masses according to \cite{Xing:2007fb}, see also Tab.~\ref{Tab:fermionmassesandmixings}. 

\item We have furthermore modified SoftSUSY 2.0.18 to include left-right mixing for the first two families, which was set to zero. The left-right sfermion mixing angle $\theta_{\tilde{f}}$ is defined (at tree-level) as
\begin{equation} \label{Eq:SfermionMixingAngle}
\sin ( 2 \theta_{\tilde{f}} ) = \frac{2 m_f (A_f - \mu \tan \beta)}
                                {m_{\tilde{f}_1}^2 - m_{\tilde{f}_2}^2},
\end{equation}
where $f=e$, $\mu$, $\tau$, $d$, $s$ and $b$. $A_f$ is the corresponding trilinear coupling and $m_{\tilde{f}_{1/2}}^2$ are the corresponding mass eigenvalues of the sfermion mass matrix. For our study it was necessary to include left-right mixing for all families since we found that for some parameter points it is not negligible.  For example, in the mAMSB scenario for $m_0 = 500$~GeV, $m_{3/2} = 20$~TeV and $\tan \beta = 30$ we obtain $\theta_{\tilde{s}}  \approx 0.58$ and $\theta_{\tilde{b}} \approx 0.35$. This large mixing can be understood from the fact that the splitting between the sfermion mass eigenstates in this example is mainly driven by the mass of the corresponding fermion. Then both, the numerator and the denominator of Eq.~\eqref{Eq:SfermionMixingAngle}, are small, leading to a sizeable mixing. 
 
\item Some of the points in our parameter scan are already excluded by SoftSUSY and are not displayed in our results. This happens for example if the spectrum contains tachyons or if it is not possible to achieve  successful EWSB, see the SoftSUSY manual \cite{Allanach:2001kg}. In addition, we have also made SoftSUSY reject parameter points where the calculated SUSY threshold corrections are so large that the
perturbative expansion is spoiled. 
\end{itemize}  

Regarding the calculation of some of the experimental constraints described in detail in the next section, we have exported the SUSY spectra calculated from SoftSUSY to micrOMEGAs 2.2 CPC \cite{Belanger:2001fz, *Belanger:2004yn, *Belanger:2006is, *Belanger:2008sj} using the SLHA \cite{Skands:2003cj} interface.

\section{Experimental Constraints} \label{Sec:ExperimentalConstraints}

We turn now to the discussion of the experimental constraints used in our analysis. We briefly discuss below each experimental constraint and give references for further details.

\subsection{Direct Detection}

The LEP experiments have searched for SUSY particles with negative results \cite{Amsler:2008zzb}. In our analysis we exclude parameter points with a chargino or slepton (sneutrino and charged slepton) lighter than the LEP bounds.

We have not applied the LEP bound for the Higgs boson mass which holds only in the SM (or approximately for a SM-like Higgs). However, for almost all parameter points which pass the remaining constraints we have checked that the lightest CP-even Higgs boson was heavier than the LEP bound and for the other parameter points it was still above 105 GeV. For these points there may be some tension with the LEP data. However, for the outcome of our study it makes no difference if they are included or excluded.

\subsection{Electroweak Precision Observables}

We have furthermore included constraints from electroweak precision observables (EWPO) such as the W boson mass $M_W$ and the effective leptonic weak mixing angle $\sin^2 \theta_{\mathrm{eff}}$. These observables are known to a high accuracy from LEP and Tevatron.

In \cite{Aaltonen:2007ps} a combined world result for the W boson mass of
\begin{equation}
 M_W = 80.429 \pm 0.039 \; \mathrm{GeV}
\end{equation}
is given and in \cite{ALEPH:2005ema, *Alcaraz:2006mx} the up-to-date experimental result for the effective leptonic weak mixing angle is listed as
\begin{equation}
 \sin^2 \theta_{\mathrm{eff}} = 0.23153 \pm 0.00016 \;.
\end{equation}
By applying these results as a constraint we demand that the theoretical predictions for a given parameter point calculated by our modified SoftSUSY version lie within the above given $1\sigma$ errors.

\subsection[$\mathrm{BR}(b \rightarrow s \gamma)$]{$\boldsymbol{\mathrm{BR}(b \rightarrow s \gamma)}$}

The decay $b \rightarrow s \gamma$ occurs in the SM as well as in the MSSM at one loop level, which makes it very interesting as a probe of physics beyond the SM. The present experimental value for
$\mathrm{BR}(b \rightarrow s \gamma)$, released by the Heavy Flavour Averaging Group (HFAG), is \cite{HFAG, *Barate:1998vz, *Chen:2001fja, *Koppenburg:2004fz, *Abe:2001hk, *Aubert:2002pb}
\begin{equation}
 \mathrm{BR}(b \rightarrow s \gamma) 
   = \left( 3.55 \pm 0.24^{+0.09}_{-0.10} \pm 0.03 \right)
     \times 10^{-4} \; ,
\end{equation}
where the first error is the combined statistical and uncorrelated systematic uncertainty, and the other two errors are correlated systematic theoretical uncertainties and corrections respectively.

We evaluate $\mathrm{BR}(b \rightarrow s \gamma)$ for our data points using micrOMEGAs \cite{Belanger:2001fz, *Belanger:2004yn, *Belanger:2006is, *Belanger:2008sj} and exclude the data points which do not lie within the interval $\left( 3.55^{+0.36}_{-0.37} \right) \times 10^{-4}$. For our analysis we simply  sum the errors to define the allowed region.

\subsection[$\mathrm{BR}(B_s \rightarrow \mu^+ \mu^-)$]{$\boldsymbol{\mathrm{BR}(B_s \rightarrow \mu^+ \mu^-)}$}

The present experimental upper limit on $\mathrm{BR}(B_s \rightarrow \mu^+ \mu^-)$ from the Fermilab Tevatron collider is $5.8 \times 10^{-8}$ at the 95 \% C.L. \cite{Aaltonen:2007kv}. The SM prediction for this branching ratio is $\left( 3.4 \pm 0.5 \right) \times 10^{-9}$ \cite{Buchalla:1993bv, *Misiak:1999yg, *Buchalla:1998ba, *Buras:2003td}, leaving some room for a possible large SUSY contribution. We have calculated this contribution using the micrOMEGAs package. We impose the constraint that the SUSY contribution does not exceed the experimental bound minus the lower limit of the SM contributions.

An approximate formula for the SUSY corrections to $\mathrm{BR}(B_s \rightarrow \mu^+ \mu^-)$ is \cite{Buras:2002vd, Isidori:2001fv}
\begin{equation}
\begin{split}
\mathrm{BR}(B_s \rightarrow \mu^+ \mu^-) \simeq \;  & 3.5 \times 10^{-5}
\left[ \frac{\tan \beta}{50} \right]^6 \left[ \frac{\tau_{B_s}}{1.5 \, \mathrm{ps}} \right]
 \left[ \frac{F_{B_s}}{230 \, \mathrm{MeV}} \right]^2 \left[ \frac{|V_{ts}|}{0.04} \right]^2
  \\
  & \times  \frac{\bar{m}_t^4}{M_A^4}
   \frac{(16 \pi^2 \epsilon_Y)}{(1+ \tilde{\epsilon}_3 \tan \beta)^2(1+ \epsilon_0 \tan \beta)^2}
\end{split}   
\end{equation}
where $\bar{m}_t$ is the running top mass and $\tilde{\epsilon}_3 = \epsilon_0 + y_t^2 \epsilon_Y$. The full expressions for $\epsilon_0$ and $\epsilon_Y$ can be found in \cite{Buras:2002vd, Isidori:2001fv}. The branching ratio is proportional to $\tan^6 \beta$ as well as to $\epsilon_Y$, which in turn is proportional to the trilinear coupling of the stops. This means that large $\tan \beta$ and a large trilinear coupling pushes the branching ratio to larger values whereas a heavier CP-odd Higgs boson can suppress the branching ratio.

\subsection{Anomalous Magnetic Moment of the Muon} \label{Sec:g2mu}

The results for the anomalous magnetic moment of the muon $(g-2)_\mu$ or for the parameter $a_\mu = 1/2 \, (g-2)_\mu $  respectively are still not completely settled. In particular there is some tension between the preliminary $\tau$ data from BELLE \cite{Hayashii:2005ih} and the $e^+ e^-$ data \cite{Davier:2007ua} for the hadronic contributions, for a review see, e.g.\  \cite{Stockinger:2006zn, *Stockinger:2007pe, *Zhang:2008pka}. With the $e^+ e^-$ data for the hadronic contributions and the final result of the Brookhaven E821 experiment \cite{Bennett:2004pv, *Bennett:2006fi} the difference between the experiment and the theoretical SM prediction is
\begin{equation}
 a_\mu^{\mathrm{exp}} - a_\mu^{\mathrm{theo}}
   = (27.5 \pm 8.4) \times 10^{-10}
\end{equation}
equivalent to a 3.3$\sigma$ deviation. Three other recent evaluations yield slightly different numbers \cite{Miller:2007kk, *Jegerlehner:2007xe, *Hagiwara:2006jt}. Because of the discrepancies between the electron and the $\tau$ data and the slight differences in the theoretical predictions we only
use as constraint that the SUSY contributions to $(g-2)_\mu$ have the right sign to make $a_\mu^{\mathrm{exp}} - a_\mu^{\mathrm{theo}}$ smaller and that they are not too large, $0 \leq a_\mu \leq 35.9 \times 10^{-10}$.

For the calculation of $(g-2)_\mu$ we use micrOMEGAs which has implemented the formulas from \cite{Martin:2001st}. There is also an approximate formula given in \cite{Stockinger:2006zn, *Stockinger:2007pe, *Zhang:2008pka} for the case that all SUSY parameters are set to $M_{\mathrm{SUSY}}$, $\mathrm{sgn} (M_1) = \mathrm{sgn} (M_2)$ and all parameters are real:
\begin{equation}
\delta a_\mu^{SUSY} \approx 13 \tan \beta \:\mathrm{sgn} (\mu M_{1,2})
  \left( \frac{100 \, \mathrm{GeV}}{M_{\mathrm{SUSY}}} \right)^2 10^{-10} \;.
\end{equation}
From this formula we already see that large values of $\tan \beta$ can lead to conflicts with experimental observations if also the SUSY scale is not too large. The anomalous magnetic moment receives also larger corrections for smaller smuon and muon-sneutrino masses and larger neutralino and chargino masses. Furthermore, we can also see the dependence on the sign of $\mu$. For example, our constraints exclude a negative $\mu$ if both $M_1$ and $M_2$ are positive.

\subsection{Dark Matter} \label{Sec:CDM}

In the $\mathcal{R}$-parity conserving MSSM  the LSP provides an interesting candidate for the dark matter particle. It may be the lightest neutralino, but may alternatively be the gravitino. The WMAP Collaboration, after five years of data taking, has released $\Omega_m h^2 = 0.1143 \pm 0.0034$ for the dark matter density in the Universe \cite{Hinshaw:2008kr}.

If one makes the assumption of a \emph{standard} cosmological evolution as well as that dark matter predominantly consists of the lightest neutralino, this would imply rather strong constraints on the parameter space of SUSY models. However, other particles may contribute to dark matter in addition to a neutralino LSP, which relaxes this bound to the requirement that the relic density of the neutralino, which we require to be the LSP, should not exceed the dark matter observed by WMAP.  

We discuss this relaxed bound separately in the following, since it may be taken as a possible constraint under additional assumptions. However, since it can be avoided if, for instance, the cosmological evolution is non-standard or if a small amount of $\mathcal{R}$-parity violation is introduced, we do not include it in our final results. Furthermore, in mGMSB the gravitino is the LSP and its relic density depends on its mass, which we treat as a free parameter in this setup such that no constraint can be applied.

\section{Allowed Quark and Lepton Mass Ratios at the GUT Scale} \label{Sec:Results}

Performing the numerical scan over the parameter ranges for the SUSY breaking scenarios specified in Sec.~\ref{Sec:ScanRanges}, we obtain as final results the scatter plots with allowed GUT scale values for the quark and lepton Yukawa coupling ratios of interest shown in Fig.~\ref{Fig:FinalResults}. For each of the parameter points, corresponding to specific boundary conditions for the SUSY breaking parameters at high energies, we apply the experimental constraints from direct searches, EWPO, $BR(B_s \rightarrow \mu^+ \mu^-)$, $BR(b \rightarrow s \gamma)$ and $(g-2)_\mu$ described in Sec.~\ref{Sec:ExperimentalConstraints}. Detailed Plots for each experimental constraint separately are given in App.~\ref{App:Plots}.

Values shown in black are consistent with the applied constraints, whereas dots in red mark parameter points which are excluded. We note that there is no one-to-one correspondence between the soft SUSY breaking parameters and the Yukawa coupling ratios. Therefore it can happen that allowed and forbidden parameter points give the same result and that black dots overlap with red ones. The grey regions around the black dots indicate the allowed ratios when the experimental 1$\sigma$ errors on the quark masses are included. The other lines and dots correspond to possible GUT predictions and are discussed in Sec.~\ref{Sec:TheoryvsExperiment} in more detail. We now discuss the impact of the experimental constraints in the considered SUSY breaking scenarios.

\subsection{mAMSB} \label{Sec:Constraints_mAMSB}

The first row in Fig.~\ref{Fig:FinalResults} shows the combined results for mAMSB. For the considered mAMSB parameter range, see Tab.~\ref{Tab:SUSYParameterRanges}, we can see from the left plot that, with quark mass errors included, $y_\mu /y_s$ in the range from 2.48 to 5.72 and  $y_e / y_d$  in the range from 0.21 to 0.65 are possible. The right plot shows that for $y_\tau / y_b$ values in the range from 0.98 to 1.30 and for $y_t / y_b$ in the range from 1.37 to 4.78 are allowed. Compared to the yellow squads indicating the values calculated without taking the SUSY threshold corrections into account, we see that all ratios are reduced. As discussed in the previous chapter and in \cite{Ross:2007az,Antusch:2008tf}, the reason for this is that the sign of the dominant $\tan \beta$-enhanced correction parameter $\eta_i^G$ is negative for negative gluino mass $M_3$ when $\mu$ is positive, which enhances the down-type Yukawa couplings at the SUSY scale and finally lowers the possible values of the ratios at $M_{\mathrm{GUT}}$.  Large SUSY threshold corrections, and thus lower values of the GUT scale ratios correspond to large $\tan \beta$. The plots also show that there is a strong correlation between $y_\mu / y_s$ and $y_e / y_d$, which stems from the fact that the masses of the first two sfermion generations are very similar.

One can see from the plots how the phenomenological constraints restrict the possible effects of the threshold corrections on the GUT scale ratios.  First of all, a sparticle spectrum free of tachyons already excludes values of $m_0$ below about 200 GeV. Furthermore, we found that large values of $\tan \beta$ above 50 did not lead to a viable spectrum.  These parameter points were rejected by the numerics and are not displayed in Fig.~\ref{Fig:FinalResults}. In the parameter range we considered, the strongest constraint was $b \rightarrow s \gamma$, which disfavours large values of $\tan \beta$. In mAMSB, EWPO also provide a significant constraint and disfavour large values of $\tan \beta$. Compared to $b \rightarrow s \gamma$ and EWPO, the limits from $B_s \rightarrow \mu^+ \mu^-$ and $(g-2)_\mu$ are much less constraining, cf.\ Fig.~\ref{Fig:Plots_mAMSB}. Including all constraints the minimal allowed $m_0$ rises to about 600~GeV and the maximal $\tan \beta$ reduces to about 45.

Finally, under the assumption that the neutralino is the LSP, stable due to $\mathcal{R}$-parity, and that the evolution of the universe is standard up to temperatures where the LSP freezes out, the LSP relic density could be used as an additional constraint. In particular the parameter points which lead to a LSP relic density larger than the dark matter density or where the LSP is charged would be excluded. The impact which this constraint would have is shown in Fig.~\ref{Fig:Plots_mAMSB}. The consequence would be that only a small region where the threshold corrections are comparatively small would remain allowed.

\subsection{mGMSB}

The combined results for mGMSB are shown in the second row in Fig.~\ref{Fig:FinalResults}. Compared to the case of mAMSB and following the arguments of Sec.~\ref{Sec:Constraints_mAMSB}, positive $M_3$ with positive $\mu$ leads to a positive threshold correction parameter $\eta_i^G$ which lowers the down-type Yukawa couplings and consequently enlarges the Yukawa coupling ratios compared to the case without threshold effects included. For the considered mGMSB parameter range, see Tab.~\ref{Tab:SUSYParameterRanges}, we can see from the left plot that, with quark mass errors included, $y_\mu / y_s$ in the range from 3.62 to 7.69 and  $y_e / y_d$  in the range from 0.30 to 0.87 are possible. The right plot shows that for $y_\tau / y_b$ values in the range from 1.35 to 2.09 and for $y_t / y_b$ in the range from 1.01 to 5.26 are allowed.

Turning to the individual experimental constraints, in mGMSB with the parameter range specified in Tab.~\ref{Tab:SUSYParameterRanges} all applied constraints lead to a significant reduction of the possible GUT scale ratios. As in mAMSB, the strongest constraint comes from $b \rightarrow s \gamma$, followed by EWPO and $(g-2)_\mu$ and finally by limits from direct searches and $B_s \rightarrow \mu^+ \mu^-$, cf.\ Fig.~\ref{Fig:Plots_mGMSB}.  We note that due to the correlation between $y_\mu / y_s$ and $y_e / y_d$ many parameter points lead to the same ratio which means that the dots would lie on top of each other. If at least one of the parameter points is consistent with the phenomenological constraints, the ratio is shown in black. Dark matter constraints are not discussed since the gravitino is generically the LSP in GMSB and the gravitino mass essentially represents a free parameter in our setup.

\subsection{CMSSM}

In the CMSSM, as in mGMSB, with positive $M_3$ and $\mu$ the SUSY threshold corrections tend to reduce the down-type Yukawa couplings and consequently enlarge the Yukawa coupling ratios at the GUT scale. The combined results for CMSSM  are shown in the third row of Fig.~\ref{Fig:FinalResults}. For the CMSSM parameter ranges specified in Tab.~\ref{Tab:SUSYParameterRanges} we find that, with quark mass errors included, $y_\mu / y_s$ can be in the range from 3.44 to 7.73  and  $y_e / y_d$ in the range from 0.29 to 0.87. The right plot shows that for $y_\tau / y_b$ values in the range from 1.28 to 2.10 and for $y_t / y_b$ in the range from 0.97 to 5.71 are allowed.

Fig.~\ref{Fig:Plots_CMSSM} shows  the impact of the individual constraints. The main consequence regarding the allowed GUT scale ratios is that points where the SUSY threshold corrections tend to reduce the GUT scale ratios are excluded. This is in agreement with \cite{Altmannshofer:2008vr}, where it has been argued that third family Yukawa coupling unification within the inverted scalar mass hierarchy scenario \cite{Blazek:2001sb} requires a region of parameter space where $A_0 \approx - 2 m_0$ and $\mu, m_{1/2} \ll m_0$ and that this inevitably leads to conflicts with bounds on, e.g.\  $B_s \rightarrow \mu^+ \mu^-$ because of the large trilinear coupling. We note that we have not focused on this specific correlation between the parameters which explains why we have only relatively few (excluded) parameter points which are close to third family Yukawa unification.

We see there also the constraints coming from the requirement that the neutralino relic density does not exceed the observed dark matter density, under the assumptions that the neutralino is the stable LSP and that the cosmic history is standard. We find that the impact of this constraint would be that a certain region with large $\tan \beta$ would be favoured. However, we would like to note that there are comparatively thin parameter space regions which lead to a viable neutralino relic density, i.e.\ the so-called funnel and coannihilation regions. Since our parameter space is comparatively coarse, we cannot exclude that we have missed viable parameter points in these thin regions. Such points could lead to additional possibilities for allowed GUT scale ratios. The few points with larger $y_t/y_b$, i.e.\ smaller $\tan \beta$, belong to these thin parameter space regions. Due to this and the fact that the dark matter constraint applies only under additional assumptions, this constraint is not included in the final results.

\section{Allowed GUT Scale Ratios Compared to Theory Predictions} \label{Sec:TheoryvsExperiment}

\begin{figure}
 \centering
 \includegraphics[scale=0.56]{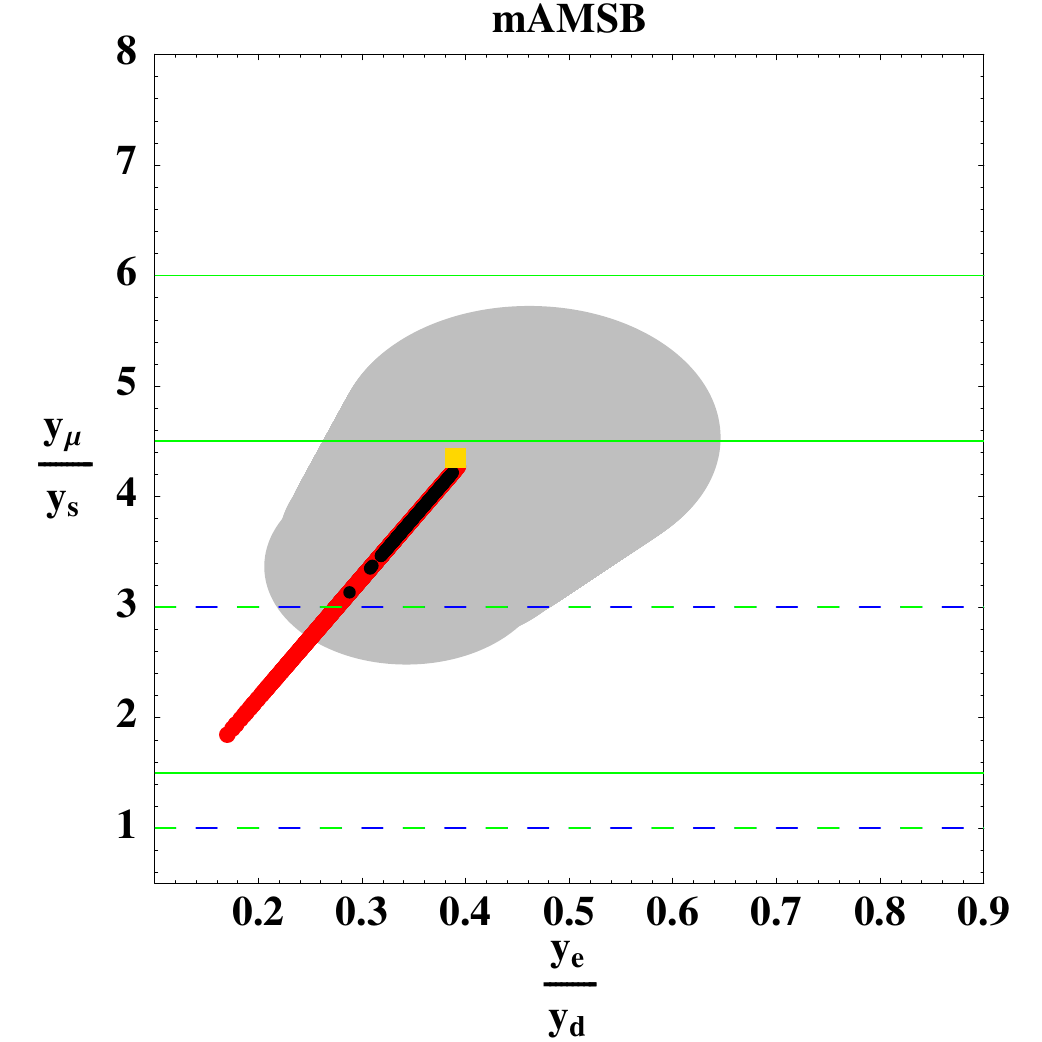} \hspace{0.4cm}
 \includegraphics[scale=0.56]{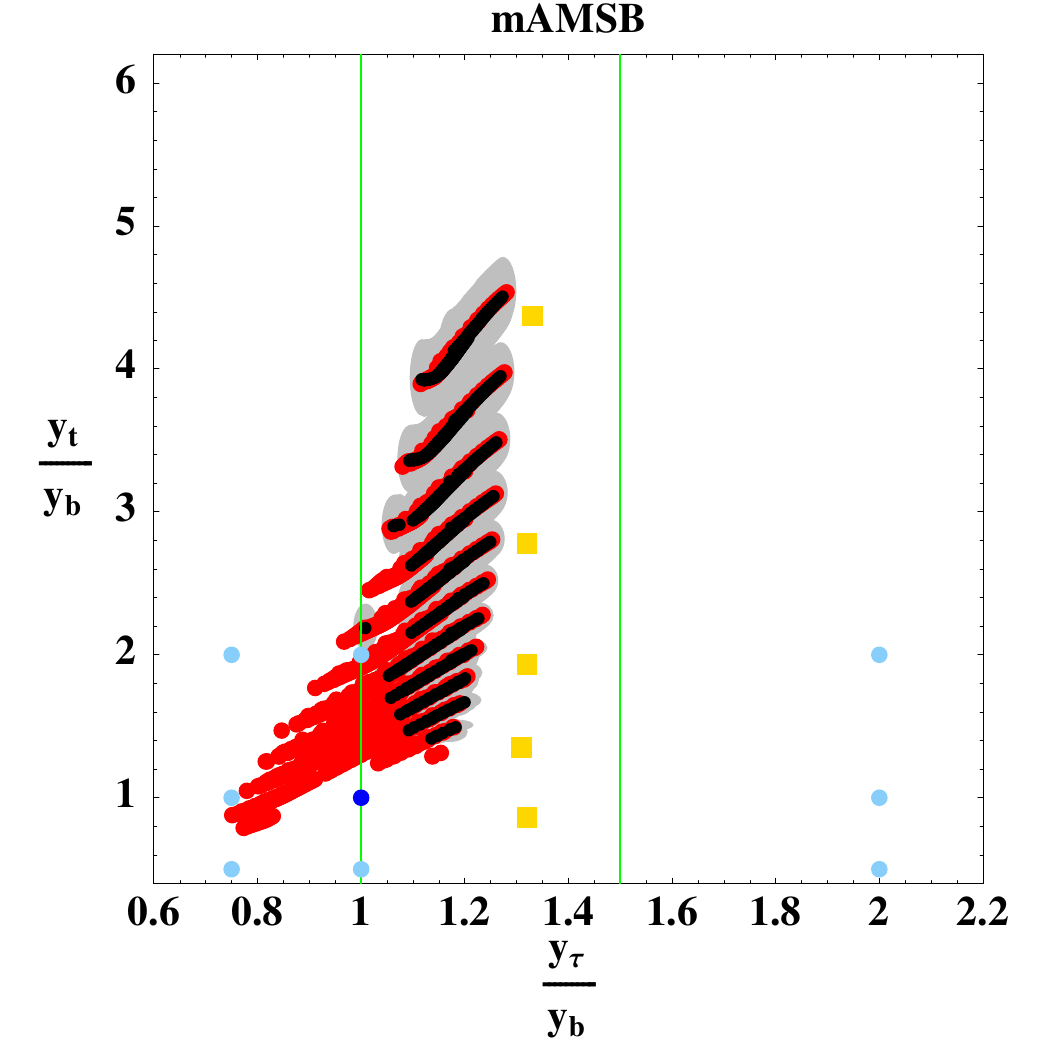}
 \includegraphics[scale=0.56]{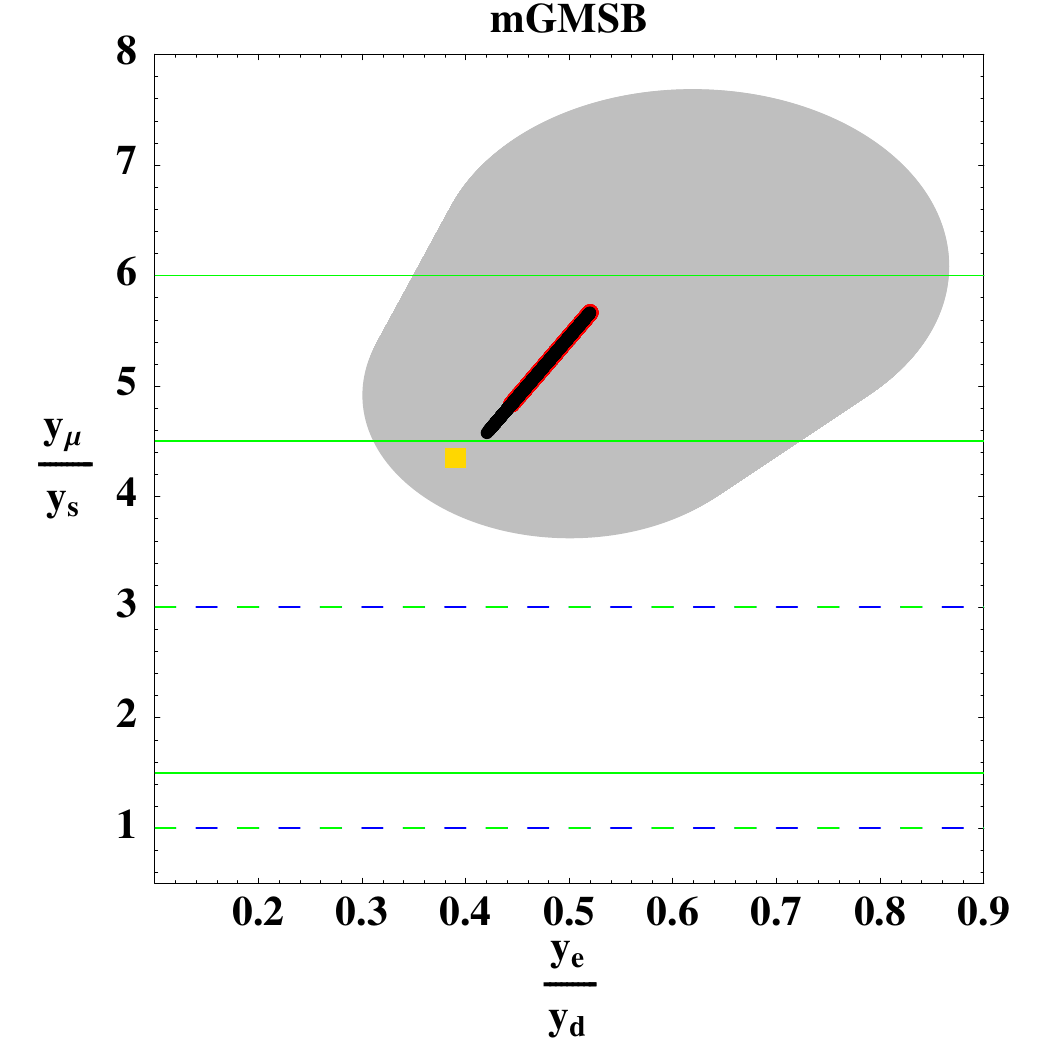} \hspace{0.4cm}
 \includegraphics[scale=0.56]{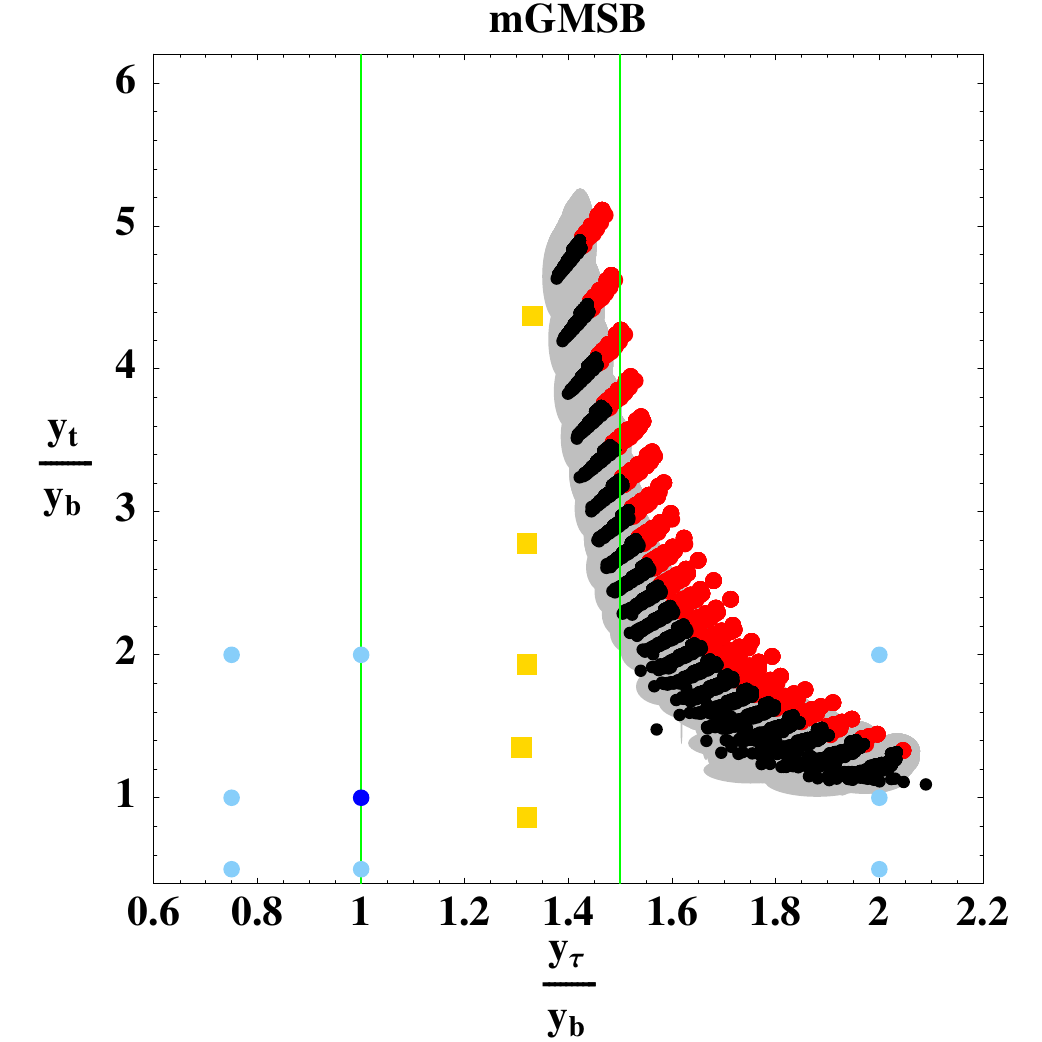}
 \includegraphics[scale=0.56]{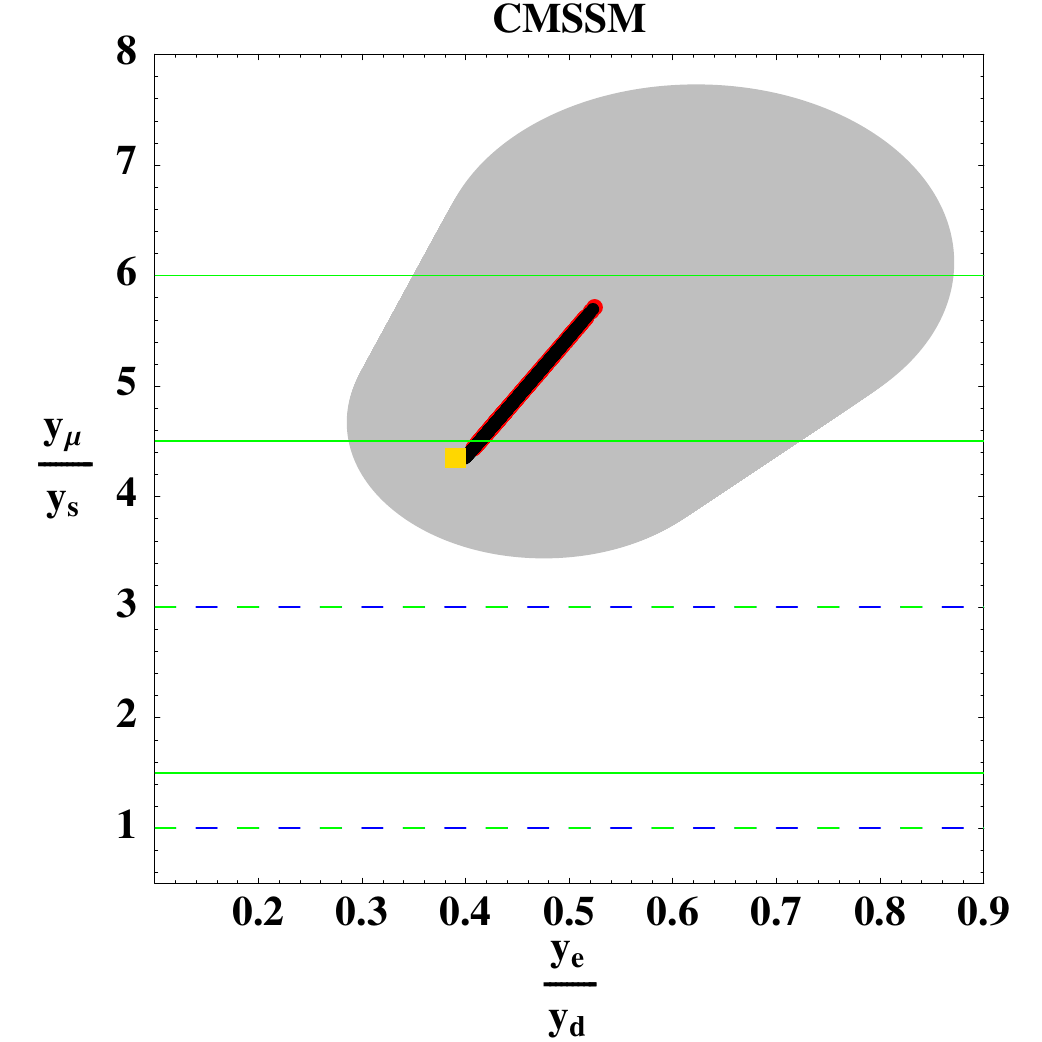} \hspace{0.4cm}
 \includegraphics[scale=0.56]{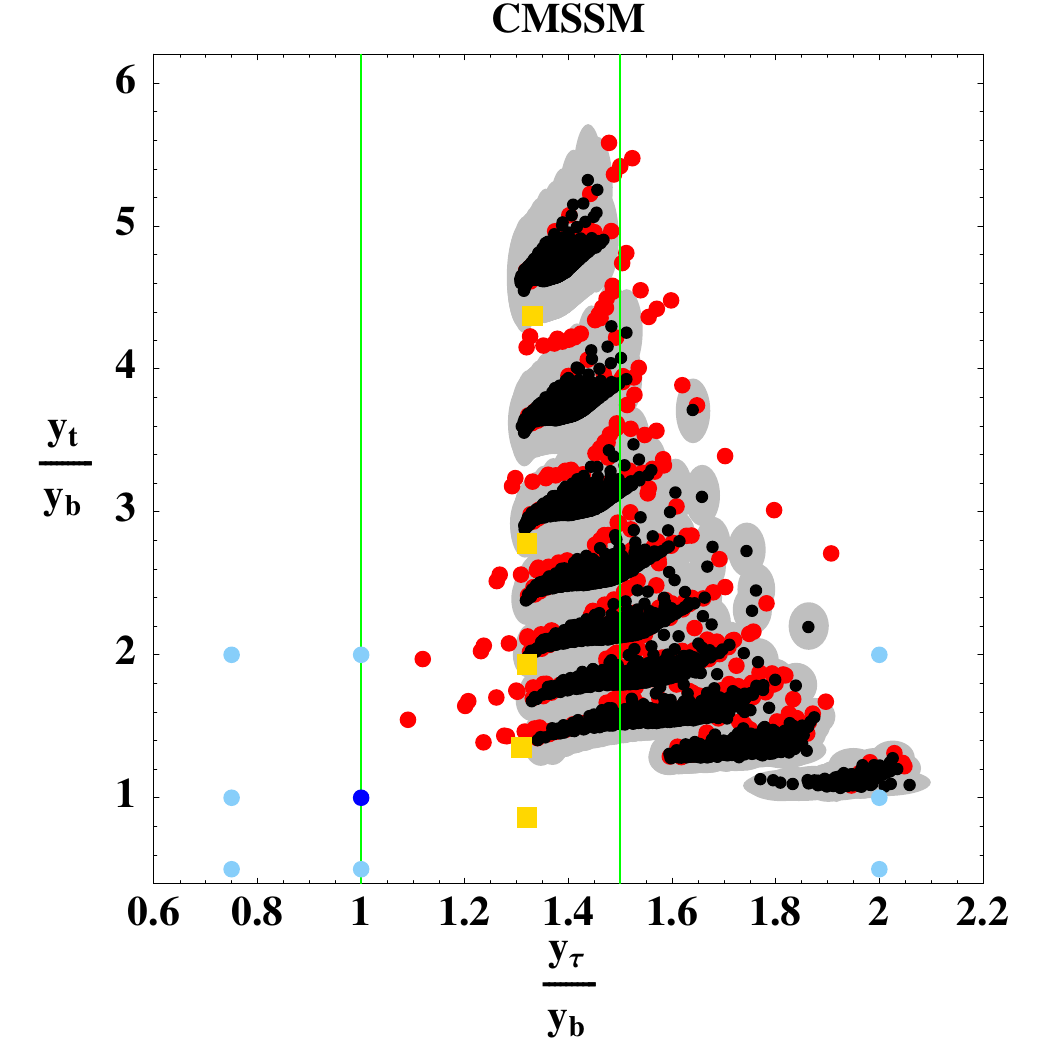}
 \caption[Final Results of the Phenomenological Scan]{Final results for mAMSB, mGMSB and CMSSM. 
          The (red) black points are the (excluded) allowed points after
          applying the constraints. The grey regions indicate the uncertainties 
	  from experimental quark mass errors. 
	  The green lines are predictions from $SU(5)$, the dashed lines from $SU(5)$ and PS 
	  and the (light) blue points from PS (dimension-six operators). 
	  The yellow squads are the GUT scale Yukawa ratios without including
          SUSY threshold corrections for
          $\tan \beta = 20$, $30$, $40$, $50$ and $60$ from top to bottom.
   \label{Fig:FinalResults}}
\end{figure}

As discussed in the previous section, within mAMSB, mGMSB and CMSSM only certain ranges of GUT scale ratios $y_e/y_d$, $y_\mu/y_s$, $y_\tau/y_b$ and $y_t/y_b$ are allowed when phenomenological constraints from electroweak precision observables, $B$ physics, $(g-2)_\mu$ and mass limits on sparticles are taken into account. In this section we compare these ranges with the possible predictions for these ratios from unified theories. Fig.~\ref{Fig:FinalResults} contains our final results. The red dots correspond to parameter points which are excluded by phenomenological constraints, while the black dots are allowed with grey regions indicating the experimental 1$\sigma$ errors on the quark masses.

The possible theory predictions discussed in Ch.~\ref{Ch:GUTYukawaRelations} are shown in Fig.~\ref{Fig:FinalResults} as green and blue lines and dots. We note that for Yukawa coupling ratios only the modulus of the ratio is relevant, since a sign only corresponds to a global phase redefinition. We display therefore in the following always the modulus of the predicted ratios. The different colours have the following meaning: Green straight lines denote the predictions from $SU(5)$ GUTs and the dashed lines in green and blue denote the predictions which can arise in PS and $SU(5)$ unification. For the third family the dark blue points denote the predictions from operators up to dimension five in PS unification, whereas the light blue points denote predictions which can arise from certain dimension-six operators.

We have collected the results of this section also in Tab.~\ref{Tab:TheoryvsExperiment} where a checkmark (cross) denotes (in)compatibility with our phenomenological scan described in the previous sections. In the following we want to discuss this in more detail.

\subsection{GUT Predictions in the mAMSB Scenario}

From Fig.~\ref{Fig:FinalResults} we see that mAMSB is the only considered scenario where the GJ relation $y_\mu/y_s = 3$ is allowed. Its realisation requires intermediate $\tan \beta$ (around 30) and a comparatively heavy sparticle spectrum corresponding to $m_0$ above about 1 TeV and $m_{3/2}$ above about 100 TeV. Interestingly, this parameter region would also be compatible (with quark mass errors included) with the second GJ relation $y_e/y_d = 1/3$, which arises in the presence of a texture zero in the 1-1 elements of the Yukawa matrices and under the assumption that they are symmetric.   

In addition to the GJ relation, mAMSB is also compatible with the ratio $y_\mu/y_s = 9/2$. This ratio arises in all scenarios whenever the SUSY threshold corrections are comparatively small, for instance if $\tan \beta$ is small such that there is no $\tan \beta$ enhancement. In Fig.~\ref{Fig:FinalResults} the yellow squad shows the GUT scale ratios which would result when the SUSY threshold corrections were ignored. In the absence of SUSY threshold corrections a value close to $y_\mu/y_s = 9/2$ would result as well.

Regarding the third generation we find that third family Yukawa unification $y_t = y_b = y_\tau$ is not compatible with mAMSB. The parameter points which come close to this relation are all excluded because either the spectrum contains tachyons and/or because it is not possible to achieve successful electroweak symmetry breaking there. Partial third family Yukawa unification $y_\tau/y_b = 1$ turns out to be possible.  Interestingly, $y_\tau/y_b = 1$ is realised in combination with $y_t/y_b = 2$. Both relations can emerge simultaneously from a dimension-six operator within PS unified theories. 

The GUT predictions $y_t = 2 y_b = 2 y_\tau$ and $y_\mu/y_s = 3$ can be realised for the same region of parameter space where $\tan \beta$ is intermediate and the sparticle spectrum is rather heavy.  We would like to note that including the dark matter constraint would exclude this parameter space region, see Fig.~\ref{Fig:Plots_mAMSB}. However, for example, in variants of mAMSB where a small amount of $\mathcal{R}$-parity violation is introduced or in non-standard cosmology, this constraint might be avoided.

\subsection{GUT Predictions in the mGMSB and CMSSM Scenarios}

The allowed GUT scale ranges within mGMSB and CMSSM differ significantly from the ranges in mAMSB. This is due to the fact that the sign of the generically dominant $\tan \beta$-enhanced SUSY QCD threshold correction is governed by $\mathrm{sgn}(\mu M_3)$ which is positive in mGMSB and CMSSM but negative in mAMSB.  It has turned out that mGMSB and CMSSM are in fact compatible with the same theory predictions. We therefore discuss both scenarios together in this subsection.

For mGMSB and CMSSM the GJ relation $y_\mu/y_s = 3$ is disfavoured. For small $\tan \beta$, i.e.\ small threshold corrections, both scenarios (and also mAMSB) are compatible with $y_\mu/y_s = 9/2$. In addition, for large $\tan \beta$, i.e.\ large SUSY threshold corrections, the theory prediction $y_\mu/y_s = 6$ can be compatible with phenomenological constraints. The GUT scale ratios $y_\mu/y_s = 9/2$ as well as $y_\mu/y_s = 6$ can be realised in $SU(5)$ GUTs, however, within our setup, not from the PS gauge group. 

Regarding the third generation we again find that third family Yukawa unification $y_t = y_b = y_\tau$ is incompatible. However, interesting alternative relations are compatible with data: One example is the GUT scale prediction $y_\tau/y_b = 3/2$ which arises in the context of $SU(5)$ GUTs. It can be realised for moderate values of $\tan \beta$, e.g.\  $\tan \beta \approx 25$ in CMSSM, while it would be disfavoured for large values of $\tan \beta$. We would like to remark that this region of parameter space is also consistent with the GUT prediction $y_\mu/y_s = 9/2$. 
For large $\tan \beta$, i.e., large SUSY threshold corrections, on the other hand, the relations  
$y_\tau/y_b = 2$ and $y_t/y_b = 1$ are allowed. Interestingly, the relation $ 2 y_t = 2 y_b = y_\tau$ can also emerge as a prediction from dimension-six operators within PS unified theories. 
The parameter space where $2 y_t = 2 y_b = y_\tau$ is realised additionally allows to realise the GUT relation $y_\mu/y_s = 6$. However, while $y_\mu/y_s = 6$ appears in $SU(5)$ the relation $2 y_t = 2 y_b = y_\tau$ can emerge from PS.  In our scan we found no parameter point in mGMSB and CMSSM where partial third family Yukawa unification $y_\tau/y_b = 1$ was compatible with experimental constraints. 

\begin{table}
\centering
\begin{tabular}{ccccc}\toprule
Prediction & Gauge Group & mAMSB & mGMSB & CMSSM \\ \midrule
$y_\mu/y_s = 3$      & $SU(5)$, PS & $\checkmark$ &  $\color{red} \times$ &  $\color{red}\times$  \\
$y_\tau/y_b = 1$    & $SU(5)$, PS & $\checkmark$ &  $\color{red}\times$ &  $\color{red}\times$  \\
$y_\mu/y_s = 6$      & $SU(5)$     &  $\color{red}\times$ & $\checkmark$ & $\checkmark$  \\
$y_\mu/y_s = 9/2$    & $SU(5)$     & $\checkmark$ & $\checkmark$ & $\checkmark$  \\
$y_\tau/y_b = 3/2$   & $SU(5)$     &  $\color{red}\times$ & $\checkmark$ & $\checkmark$  \\
$y_t = y_b = y_\tau$ & PS &  $\color{red}\times$ &  $\color{red}\times$ &  $\color{red}\times$ \\
$2 y_t = 2 y_b = y_\tau$  & PS & $\color{red}\times$ & $\checkmark$ & $\checkmark$  \\
$y_t = 2 y_b = 2 y _\tau$ & PS & $\checkmark$ &  $\color{red}\times$ &  $\color{red}\times$ \\
\bottomrule
\end{tabular}
\caption[GUT Scale Ratios Compared to Theory Predictions]{Comparison of GUT scale predictions for Yukawa coupling ratios to phenomenological data. A checkmark (cross) denotes (in)compatibility within the phenomenological constraints, see also Fig.~\ref{Fig:FinalResults}. For the origin of the predictions see Ch.~\ref{Ch:GUTYukawaRelations}. \label{Tab:TheoryvsExperiment}}
\end{table}

\subsection{Comparison with the Previous Approach}

In the last chapter, see also \cite{Antusch:2008tf}, the impact of the $\tan \beta$-enhanced SUSY threshold corrections on all down-type quark and charged lepton Yukawa couplings has been analysed numerically and analytically. For this purpose the threshold corrections have been treated in the EW unbroken phase. The possible ranges for the GUT scale values of the Yukawa couplings and their ratios have been calculated for three example ranges of low energy SUSY parameters and are collected in Fig.~\ref{Fig:Scatter} and Tabs.~\ref{Tab:Ratfinal}  and \ref{Tab:Yukfinal}.

Compared to the previous approach our results are in good qualitative agreement, cf. Fig.~\ref{Fig:Scatter} where the results are presented in a similar way as in Fig.~\ref{Fig:FinalResults}. The example SUSY parameter range $a$ was inspired by anomaly mediated SUSY breaking and the SUSY parameter ranges $g_+$ (and $g_-$) by scenarios with gaugino unification and $\mu >0$ ($\mu < 0$). Quantitatively there are, nevertheless, differences, which are larger for the third family. For example for $\tan \beta = 30$ in the mAMSB case we find, before applying experimental constraints, $y_\mu / y_s$ can be in the range 2.41--5.73, whereas in case $a$ we found the very similar range 2.40--5.63. On the other hand, for the ratio $y_\tau/y_b$ we find an allowed range of 0.94--1.28 within mAMSB compared to 0.60--1.39 for the example SUSY parameter range $a$. However, since in the present study we are considering explicit SUSY breaking scenarios at high energy resulting in different low energy SUSY spectra, there is no reason to expect perfect quantitative agreement. 

The main difference from our previous approach is of course that the consideration of explicit SUSY breaking scenarios allows to take phenomenological constraints into account. Their restrictions on the allowed GUT scale ratios depend on the explicit minimal SUSY breaking scenario. However, we expect that some consequences are also characteristic for variants of the considered schemes. For example, it has turned out that there is a certain tension between realising GUT predictions which require large SUSY threshold corrections and the experimental constraints which basically restrict the effects of SUSY loops to the observables. It has also turned out that, contrary to claims in \cite{Ross:2007az,Antusch:2008tf}, it may be  challenging to realise third family Yukawa unification in AMSB-like SUSY breaking scenarios.

%% file: kap_09_GUTImplications.tex
\chapter[Implications for GUT Model Building and a Concrete Application]{Implications for GUT Model Building\\ and a Concrete Application} \label{Ch:GUTImplications}

In the previous chapters  we have seen that comparatively wide ranges of Yukawa coupling ratios are allowed at the GUT scale, if possible SUSY threshold corrections are taken into account. In turn some of these are compatible with our newly proposed ratios from Ch.~\ref{Ch:GUTYukawaRelations}. From today's perspective, since we do not have any experimental confirmation for low-energy SUSY and thus no knowledge of the sparticle spectrum, this can have a large impact for GUT model building as we discuss in  Sec.~\ref{Sec:Implications}.

Subsequently, we turn to a concrete application, a $SO(3) \times SU(5)$ GUT flavour model in Sec.~\ref{Sec:ModelI}, based on \cite{Antusch:2010xx} (work in progress). The model predicts TB neutrino mixing after the $SO(3)$ family symmetry is broken as an indirect result of the assumed \emph{neutrino flavon} alignment via constrained sequential dominance (CSD)  \cite{King:1998jw, King:1999cm, *King:1999mb, *Antusch:2004gf, *Antusch:2010tf, King:2002nf, King:2005bj}. The neutrino flavons break the $SO(3)$ symmetry and are  assumed to be aligned along the columns of the TB mixing matrix, but quadratic combinations of the neutrino flavons respect the neutrino family symmetry accidentally as discussed in \cite{King:2009ap}. Further flavons are assumed to be misaligned compared to the neutrino flavons and together they are responsible for quark and charged lepton masses and quark mixing. We make a detailed fit to quark masses and mixing using the misaligned quark flavons and show that a simple ansatz for the phase of one of the misaligned quark flavons leads to successful quark CP violation. Thereby we take advantage of the proposed Yukawa coupling ratios $y_\mu / y_s = 6$ and $y_\tau / y_b = -3/2$ which are viable for large $\tan \beta$ as discussed in the preceding chapters. Interestingly, the phases in the quark and lepton sector have one common origin resulting in definite predictions for the effective neutrino mass $m_{ee}$ appearing in neutrinoless double beta decay in contrast to the general case where due to phase uncertainties no sharp prediction is possible \cite{Feruglio:2002af}.

\section{Implications for GUT Model Building} \label{Sec:Implications}

In the following, we discuss some of the implications of the preceding chapters on GUT model building. We start with a brief comparison of our findings with previous studies.

\subsection{Comparison with Previous Studies}

The viability of third family Yukawa unification $y_t = y_b = y_\tau$ and also the less restrictive possibility $y_b = y_\tau$ has been extensively studied in the literature, see, e.g.\ \cite{Carena:1994bv, *Hempfling:1993kv,Ross:2007az,Antusch:2008tf,Altmannshofer:2008vr,Antusch:2009gu,Bagger:1996ei,King:2000vp, Blazek:2002ta}. A recent study \cite{Altmannshofer:2008vr} has reconsidered the phenomenological viability of this relations and it has been pointed out that in a variant of the CMSSM with non-universal soft Higgs mass parameters the relation $y_t = y_b = y_\tau$ is quite challenged by the experimental data from B physics. In \cite{Ross:2007az} the authors find $b$-$\tau$ Yukawa coupling unification to be viable for small $\tan \beta = 1.3$ and large $\tan \beta = 38$. While for small $\tan \beta$ the SUSY threshold corrections are negligible, for large $\tan \beta$ they fit the threshold corrections without phenomenological viability check.

For the GJ relation, it has been shown in \cite{Ross:2007az,Antusch:2008tf,Antusch:2009gu}, that the threshold corrections can be consistent with the latest experimental data on quark masses. If the GJ relations are assumed at high energies, this can be understood as a constraint on the SUSY parameter space and points to scenarios with AMSB-like SUSY breaking with $M_3 < 0$ and $\mu > 0$ if phenomenological consistency with experimental results on $(g-2)_\mu$ is assumed to be restored by SUSY loops \cite{Stockinger:2006zn, *Stockinger:2007pe, *Zhang:2008pka}.

Another interesting aspect is that the newly proposed Yukawa coupling ratios at the GUT scale open up new possibilities for constructing SUSY GUT models to address the flavour problem.  One example for an application of such alternative GUT scale ratios $y_\mu/y_s$ and $y_e/y_d$ can be found in \cite{Antusch:2005ca}, where an approach has been presented to realise the phenomenologically successful  relation $\theta^{\mathrm{MNS}}_{12} + \theta^{\mathrm{CKM}}_{12} \approx \pi/4$, the so-called quark-lepton complementarity \cite{Raidal:2004iw, *Minakata:2004xt}, in unified theories. In this approach, the Yukawa matrices for the charged leptons and down-type quarks emerge from the identical higher-dimensional operators where quarks and leptons are unified in representations of the PS gauge group. After spontaneous breaking of PS to the SM gauge group, CG factors lead to different GUT scale values for the charged lepton and down-type quark Yukawa couplings. In the approach of \cite{Antusch:2005ca}, for example, $y_\mu/y_s = 2$ was postulated at $M_\mathrm{GUT}$ which we have however found to be challenged even within the mAMSB scenario. 

More generally, the assumption that the Yukawa matrices for the charged leptons and down-type quarks are generated from the same set of higher-dimensional operators in quark-lepton unified theories leads to a large variety of possible ratios $y_\mu/y_s$ and $y_e/y_d$ which correspond to different choices of operators and their associated CG factors. In Tab.~\ref{Tab:TheoryvsExperiment} we give a collection of possible CG factors in the context of $SU(5)$ or PS theories. Any of the viable combinations of CG factors which results in ratios $y_\mu/y_s$ and $y_e/y_d$ consistent with the phenomenological constraints are a priori interesting new options for GUT model building.

\subsection{GUT Scale Ratios for the First Fermion Generation}

As discussed in Ch.~\ref{Ch:GUTYukawaRelations}, predictions for the ratios between quark and charged lepton masses at the GUT scale can arise if two conditions are satisfied: a hierarchical structure of the Yukawa matrices and the situation that one single GUT operator dominates the relevant Yukawa matrix element.  The simplest case which can lead to predictions for the first generation of fermions is that the submatrix for the first and the second fermion generation is also hierarchical. Then the masses of the first fermion generation would be approximately determined by the diagonal elements (i.e.\ the 1-1 elements) of the corresponding Yukawa matrices and the phenomenologically allowed range for $y_e/y_d$ can directly be compared to the theory predictions in Tabs.~\ref{Tab:SU5Relations}  and \ref{Tab:PSRelations} of Ch.~\ref{Ch:GUTYukawaRelations}. The theory prediction $y_e/y_d = 1/2$, possible in $SU(5)$, or the relation $y_e/y_d = 3/4$ from PS unification would be compatible with the experimental constraints. 

In many GUT models of fermion masses and mixings, however, a different situation is realised: There, the Yukawa matrices are symmetric and have vanishing 1-1 entries, see, e.g.\  \cite{Gatto:1968ss}. In this case, the mass of the electron and down-type quark are inversely proportional to the masses of the second generation and, in addition, depend on the 1-2 entries of the Yukawa matrices which are equal to the 2-1 entries by assumption. More precisely, the prediction for the ratio $y_e/y_d$ is then given by
\begin{equation}
\frac{y_e}{y_d} = \frac{y_s}{y_\mu}\, \frac{(Y_e)^2_{12}}{(Y_d)^2_{12}} \;.
\end{equation} 
For ${(Y_e)_{12}}/{(Y_d)_{12}}=1$ and $y_\mu/y_s = 3$ we recover the second GJ relation $y_e/y_d=1/3$ which is consistent with our results when quark mass errors are included. Interestingly, it is possible to realise both relations within mAMSB. With ${(Y_e)_{12}}/{(Y_d)_{12}}=1$, no alternative GUT prediction for $y_\mu/y_s$ is consistent with the above assumptions, due to the strong correlation between $y_e/y_d$ and $y_\mu/y_s$ as shown in Fig.~\ref{Fig:FinalResults}. 

However, with a different CG factor relating $(Y_e)_{12}$ and $(Y_d)_{12}$, the alternative GUT predictions $y_\mu/y_s = 9/2$ and $y_\mu/y_s = 6$ can well be consistent with the assumption of symmetric Yukawa matrices with zero 1-1 elements: The relation $y_\mu/y_s = 9/2$ is consistent with $y_e/y_d = 1/2$, which would require ${(Y_e)_{12}}/{(Y_d)_{12}} \approx 3/2$. Similarly, $y_\mu/y_s = 6$ is consistent with $y_e/y_d = 3/2$, which would require ${(Y_e)_{12}}/{(Y_d)_{12}}\approx 2$.  Of course, when one of the above assumptions, i.e.\ symmetric Yukawa matrices and zero 1-1 elements, is dropped then there are more possibilities. For example, without vanishing 1-1 element  the relation ${(Y_e)_{12}}/{(Y_d)_{12}}= 1$ can well be compatible with $y_\mu/y_s = 9/2$ or $y_\mu/y_s = 6$.

\subsection{Charged Lepton Corrections to Neutrino Mixing Angles}

In many GUT models of fermion masses and mixings, characteristic predictions can arise for the neutrino mixing angles which are, however, perturbed by the mixing coming from the charged lepton sector, see, e.g.\  \cite{Antusch:2005kw, Antusch:2008yc}. One typical example is the leptonic mixing angle $\theta^{\mathrm{MNS}}_{13}$. In many models the 1-3 mixing from the neutrino sector is very small or even zero ($\theta_{13}^\nu = 0$). Nevertheless, a total lepton mixing $\theta^{\mathrm{MNS}}_{13}$ can be induced from the possible corrections caused by mixing in the charged lepton mass matrix and is then given by
\begin{equation}
\theta^{\mathrm{MNS}}_{13} \approx \frac{\theta_{12}^e}{\sqrt{2}}  \;,
\end{equation}   
where $\theta_{12}^e$ is the charged lepton 1-2 mixing angle for a hierarchical mass matrix given by $\theta_{12}^e \approx (Y_e)^2_{12}/(Y_e)^2_{22}$. Assuming, for instance,
\begin{equation}
\frac{(Y_e)^2_{12}}{(Y_d)^2_{12}} = 1 \quad \text{and} \quad \left| \frac{(Y_e)^2_{22}}{(Y_d)^2_{22}} \right| \approx y_\mu/y_s = 3 \;,
\end{equation}
we obtain
\begin{equation}
\theta^{\mathrm{MNS}}_{13} \approx \theta_{12}^d /(3\sqrt{2}) \;,
\end{equation}
where $\theta_{12}^d$ is the 1-2 mixing of the down-type quark mass matrix $Y_d$. Interestingly, in many GUT models $\theta_{12}^d$ is approximately equal to the Cabibbo angle $\theta_C \approx 13^\circ$, which under the above assumptions would yield $\theta^{\mathrm{MNS}}_{13} \approx 3^\circ$. This value emerges in many models as a prediction for the neutrino mixing $\theta^{\mathrm{MNS}}_{13}$, closely related to the GJ relation $y_\mu/y_s = 3$. 

In this context we would like to remark that the alternative GUT predictions $y_\mu/y_s = 9/2$ and $y_\mu/y_s = 6$ can lead to new predictions for the leptonic mixing angle $\theta^{\mathrm{MNS}}_{13}$, following the above chain of arguments. In particular, when $y_\mu/y_s = 9/2$ is realised in a unified model it could predict 
\begin{equation}
\theta^{\mathrm{MNS}}_{13}\approx 2\, \theta_C /(9\sqrt{2}) \approx 2^\circ\;.
\end{equation} 
Analogously, $y_\mu/y_s = 6$ could lead to the prediction 
\begin{equation}
\theta^{\mathrm{MNS}}_{13}\approx \theta_C /(6\sqrt{2}) \approx 1.5^\circ \;,
\end{equation}
 for the so far unmeasured leptonic mixing angle. Additional predictions are possible when the assumption $(Y_e)^2_{12}/(Y_d)^2_{12} = 1$ is replaced by a different group theoretical CG factor.

\section[A GUT Flavour Model for Large $\tan \beta$]{A GUT Flavour Model for Large $\boldsymbol{\tan \beta}$} \label{Sec:ModelI}

\sectionmark{A GUT Flavour Model for Large $\boldsymbol{\tan \beta}$}

In this section, we propose and describe a SUSY GUT model based on the unified $SU(5)$ gauge group as well as on the family symmetry $SO(3)$ amended by a product of $\mathbb{Z}_n$ symmetries, cf.\ Tab.~\ref{Tab:Symmetries1}. For a description of $\mathbb{Z}_n$ symmetries, see App.~\ref{App:Zn}. Note that it is always possible to replace any product of commuting discrete symmetries by a single Abelian group $U(1)$ with a suitable choice of charges for the fields, i.e.\ it is possible to replace the $\mathbb{Z}_3^2\times \mathbb{Z}_2^4$ symmetry by a single $U(1)$ symmetry with an appropriate choice of charges.

Indeed many models in the literature use an Abelian $U(1)$ symmetry rather than a product of $\mathbb{Z}_n$ symmetries to control the operators. Although this looks simpler, it should be remarked that first an $U(1)$ symmetry has infinitely many more group elements than any discrete symmetry, and second one must then confront the question of Goldstone bosons once the assumed global $U(1)$ symmetry is broken. If the auxiliary symmetry is gauged, one must further complicate the model by ensuring that it is anomaly free. Therefore, a discrete symmetry, even a large one, has definite advantages over a continuous $U(1)$ symmetry. Furthermore, large discrete symmetries are ubiquitous in string theory constructions.
Finally, the discrete symmetry used here is a rather simple one consisting of a product of $\mathbb{Z}_3$ and $\mathbb{Z}_2$ parity factors. Thus we regard this model as simple and attractive compared to other models invoking an $U(1)$ symmetry instead.

\subsection{Symmetries and Field Content}

\begin{table}
\begin{center}
\begin{tabular}{cccccccc}
\toprule
 & $SU(5)$ & $SO(3)$ & $\mathbb{Z}_2$ &  $\mathbb{Z}_6$ & $\mathbb{Z}_3$ & $\mathbb{Z}'_2$ & $\mathbb{Z}_2^{\mathrm{MP}}$ \\
\midrule
\multicolumn{8}{l}{Chiral Matter}  \\
\midrule
$F$ & $\mathbf{\bar{5}}$ & $\mathbf{3}$ & $+$ & 0 & 0 & $+$ & $-$ \\
$T_1$, $T_2$, $T_3$ & $\mathbf{10}$, $\mathbf{10}$, $\mathbf{10}$ & $\mathbf{1}$, $\mathbf{1}$, $\mathbf{1}$ & $+$, $+$, $-$  & 0, 2, 0  & 1, 0, 0 & $+$, $+$, $+$ & $-$, $-$, $-$ \\
$N_1$, $N_2$ & $\mathbf{1}$, $\mathbf{1}$ & $\mathbf{1}$, $\mathbf{1}$ & $+$, $+$  & 0, 2  & 1, 0 & $+$, $+$ & $-$, $-$ \\
\midrule
\multicolumn{8}{l}{Flavons \& Higgs Multiplets}  \\
\midrule
$\phi_{23}$, $\phi_{123}$, $\phi_3$ & $\mathbf{1}$, $\mathbf{1}$, $\mathbf{1}$ & $\mathbf{3}$, $\mathbf{3}$, $\mathbf{3}$ & $+$, $+$, $-$  & 0, 4, 0 & 2, 0, 0 & $-$, $-$, $-$ & $+$, $+$, $+$ \\
$\tilde{\phi}_{23}$ & $\mathbf{1}$ & $\mathbf{3}$ & $+$  & 1 & 0 & $-$ & $+$ \\
$H_5$, $\bar{H}_5$ & $\mathbf{5}$, $\mathbf{\overline{5}}$ & $\mathbf{1}$, $\mathbf{1}$ & $+$, $+$  & 0, 0  & 0, 0 & $+$, $+$ & $+$, $+$ \\
$H'_{5}$, $\bar{H}'_{5}$ & $\mathbf{5}$, $\mathbf{\overline{5}}$ & $\mathbf{1}$, $\mathbf{1}$  & $+$, $+$  & 3, 3 & 0, 0 & $+$, $+$ & $+$, $+$ \\
$H_{24}$ & $\mathbf{24}$ & $\mathbf{1}$ & $+$ & $0$ & $0$ & $-$ & $+$\\
\midrule
\multicolumn{8}{l}{Matter-like Messengers } \\
\midrule
$X_1$, $X_2$, $X_3$ & $\mathbf{5}$, $\mathbf{5}$, $\mathbf{5}$ & $\mathbf{1}$, $\mathbf{1}$, $\mathbf{1}$ & $+$, $+$, $-$  & 0, 2, 0  & 1, 0, 0 & $+$, $+$, $+$ & $-$, $-$, $-$ \\
$\bar{X}_1$, $\bar{X}_2$, $\bar{X}_3$ & $\mathbf{\bar{5}}$, $\mathbf{\bar{5}}$, $\mathbf{\bar{5}}$ & $\mathbf{1}$, $\mathbf{1}$, $\mathbf{1}$ & $+$, $+$, $-$  & 0, 4, 0 & 2, 0, 0 & $+$, $+$, $+$ & $-$, $-$, $-$ \\
$Y$, $\bar{Y}$ & $\mathbf{10}$, $\mathbf{\overline{10}}$ & $\mathbf{3}$, $\mathbf{3}$ & $+$, $+$  & 3, 3  & 0, 0 & $+$, $+$ & $-$, $-$ \\
$X'_1$, $X'_2$, $X'_3$ & $\mathbf{5}$, $\mathbf{5}$, $\mathbf{5}$ & $\mathbf{1}$, $\mathbf{1}$, $\mathbf{1}$ & $+$, $+$, $-$  & 0, 2, 0  & 1, 0, 0 & $-$, $-$, $-$ & $-$, $-$, $-$ \\
$\bar{X}'_1$, $\bar{X}'_2$, $\bar{X}'_3$ & $\mathbf{\bar{5}}$, $\mathbf{\bar{5}}$, $\mathbf{\bar{5}}$ & $\mathbf{1}$, $\mathbf{1}$, $\mathbf{1}$ & $+$, $+$, $-$  & 0, 4, 0 & 2, 0, 0 & $-$, $-$, $-$ & $-$, $-$, $-$ \\
$Y'$, $\bar{Y}'$ & $\mathbf{10}$, $\mathbf{\overline{10}}$ & $\mathbf{3}$, $\mathbf{3}$ & $+$, $+$  & 3, 3  & 0, 0 & $-$, $-$ & $-$, $-$ \\
\midrule
\multicolumn{8}{l}{Higgs-like Messengers }  \\
\midrule
$U_1$, $U_2$ & $\mathbf{5}$, $\mathbf{5}$ & $\mathbf{1}$, $\mathbf{1}$ & $+$, $+$  & 0, 2  & 1, 0 & $+$, $+$ & $+$, $+$ \\
$\bar{U}_1$, $\bar{U}_2$ & $\mathbf{\bar{5}}$, $\mathbf{\bar{5}}$ & $\mathbf{1}$, $\mathbf{1}$ & $+$, $+$  & 0, 4  & 2, 0 & $+$, $+$ & $+$, $+$ \\
$Z_1$, $Z_2$ & $\mathbf{1}$, $\mathbf{1}$ & $\mathbf{1}$, $\mathbf{1}$ & $+$, $+$ & 0, 4  & 2, 0 & $+$, $+$ & $+$, $+$ \\
$\bar{Z}_1$, $\bar{Z}_2$ & $\mathbf{1}$, $\mathbf{1}$ & $\mathbf{1}$, $\mathbf{1}$ & $+$, $+$ & 0, 2  & 1, 0 & $+$, $+$ & $+$, $+$ \\
\bottomrule
\end{tabular}
\end{center}
\caption[Representations and Charges in the $SO(3) \times SU(5)$ Flavour Model]{Representations and charges of the superfields. The subscript $i$ on the fields $T_i$, $N_i$, $U_i$, $\bar{U}_i$, $X_i$, $\bar{X}_i$, $X'_i$, $\bar{X}'_i$ and $Z_i$, $\bar{Z}_i$ is a family index. The flavon fields $\phi_i$, $\tilde{\phi}_{23}$ can be associated to a family via their charges under $\mathbb{Z}_2 \times \mathbb{Z}_6 \times \mathbb{Z}_3$.  The subscripts on the Higgs fields $H$, $\bar{H}$, $H'$ and $\bar{H}'$ denote the transformation properties under $SU(5)$. MP stands for matter parity. \label{Tab:Symmetries1}}
\end{table}

Let us start introducing the model by specifying the field content and the symmetries. The SM matter fields are contained in the fields $F_i$ and $T_i$, see Ch.~\ref{Ch:GUTs}, where $i$ is a family index.  In our model, we assume that the three generations $F_i$ form a triplet representation ${\bf 3}$ of a $SO(3)$ family symmetry while the three $T_i$ form singlets under $SO(3)$. We note that for this model every symmetry with real triplet representations is suitable. Nevertheless, we choose here the $SO(3)$ family symmetry. In the second model later on, we use an $A_4$ family symmetry instead, although there we only need real triplet representations as well. In the following, we suppress the $SO(3)$ indices. In addition to the matter fields $F_i$ and $T_i$, we consider two right-handed neutrinos, singlets under $SU(5)$ as well as under $SO(3)$, labelled by $N_1$ and $N_2$.

$SU(5)$ is spontaneously broken by the vev of the $H_{24}$ field, electroweak symmetry is broken by the vevs of the Higgs fields $H_5$, $\bar H_5$, $H'_{5}$, $\bar{H}'_{5}$ and $SO(3)$ is spontaneously broken by the vevs of the flavon fields, i.e.\ the family symmetry breaking Higgs fields $\phi_{123}$, $\phi_{23}$, $\phi_{3}$ and  $\tilde{\phi}_{23}$. We comment below on the specific directions in which we assume $SO(3)$ to be broken by the flavons.

Furthermore, we consider additional heavy messenger fields which, after effectively integrating them out of the theory, give rise to higher-dimensional operators generating the Yukawa coupling matrices as well as the mass matrix of the gauge singlet (right-handed) neutrinos $N_1$ and $N_2$.

The field content of our model as well as the symmetries are specified in Tab.~\ref{Tab:Symmetries1}. We would like to point out that we do not explicitly consider the full flavour and GUT Higgs sector of the model and just assume that the $SU(5)$ and $SO(3)$ breaking vevs are aligned in the desired directions of field space. We assume that in these sectors issues like doublet-triplet splitting are resolved. Without specifying these sectors, a reliable calculation of the proton decay rate is also beyond the scope of the present work.

\subsection[The $SO(3)\times SU(5)$ Symmetric Superpotential]{The $\boldsymbol{SO(3)\times SU(5)}$ Symmetric Superpotential}

\begin{figure}
\centering
\includegraphics[scale=0.72]{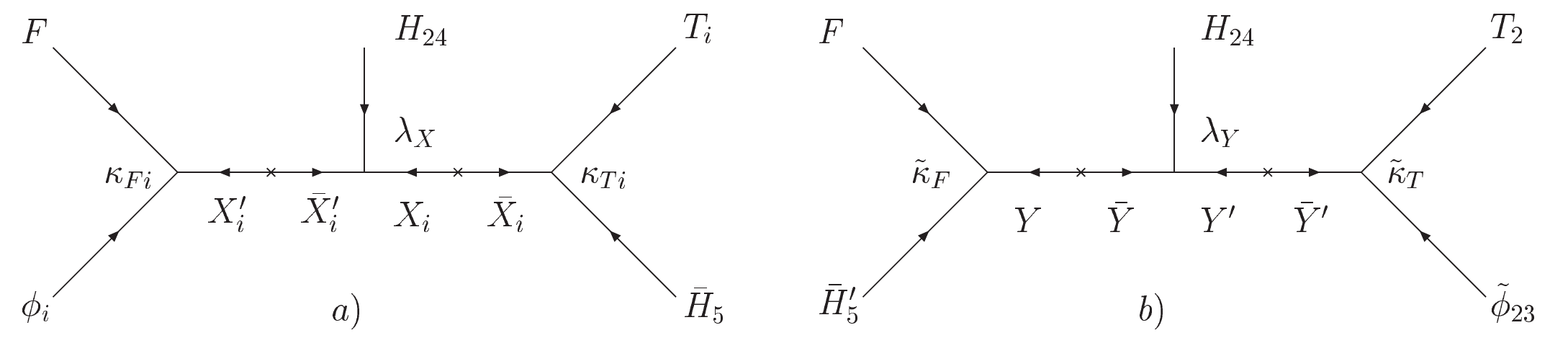}
\caption[Supergraphs for Down-type Fermion Masses in the $SO(3) \times SU(5)$ Model]{Supergraph diagrams inducing effective superpotential operators for the down-type quarks and charged leptons. \label{Fig:messenger1_d}}
\end{figure}

\begin{figure}
\centering
\includegraphics[scale=0.72]{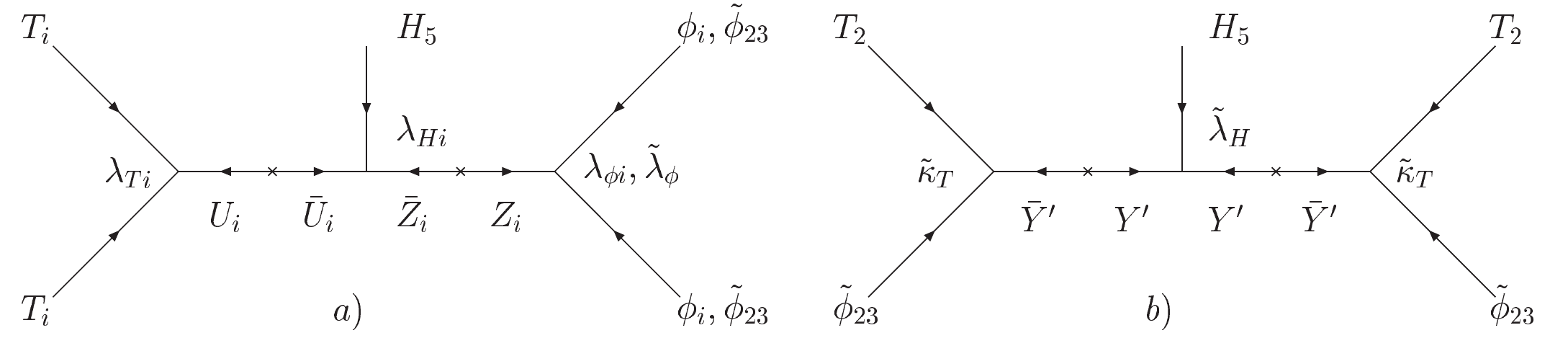}
\caption[Supergraphs for Up-type Quark Masses in the $SO(3) \times SU(5)$ Model]{Supergraph diagrams inducing effective superpotential operators for the up-type quarks. \label{Fig:messenger1_u}}
\end{figure}

\begin{figure}
\centering
\includegraphics[scale=0.72]{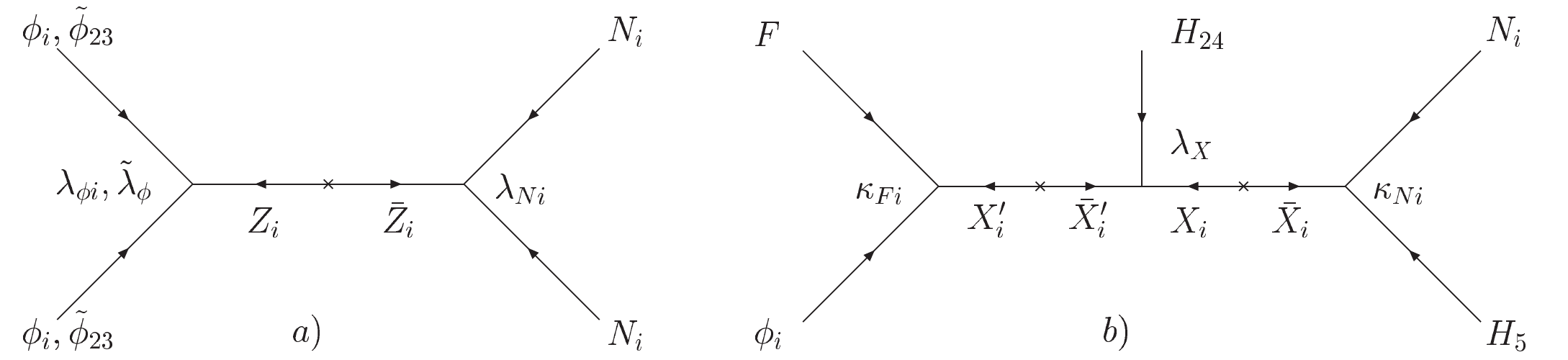}
\caption[Supergraphs for Neutrino Masses in the $SO(3) \times SU(5)$ Model]{Supergraph diagrams inducing effective superpotential operators for the neutrino sector. \label{Fig:messenger1_n}}
\end{figure}

With the field content and symmetries specified in Tab.~\ref{Tab:Symmetries1} the superpotential contains the following renormalisable terms:
\begin{align}
W_{H} &= \mu_{5} H_5 \bar{H}_5 + \mu'_{5} H'_{5} \bar{H}'_{5} + \mu_{24} H_{24}^2 \label{Eq:mu} \\
W_X &= \sum_i \left( M_{U_i} U_i \bar{U}_i + M_{X_i} X_i \bar{X}_i + M_{X'_i} X'_i \bar{X}'_i  + M_{Z_i} Z_i \bar{Z}_i \right) + M_Y Y \bar{Y} + M_{Y'} Y' \bar{Y}' \label{Eq:Xmass} \\
W_{\mathrm{int}} &=  \kappa_{Fi} F \phi_i X'_i + \kappa_{Ti} T_i \bar{H}_5 \bar{X}_i  + \tilde{\kappa}_{F} F \bar{H}'_5 Y + \tilde{\kappa}_{T} T_2 \tilde{\phi}_{23} \bar{Y}' + \kappa_{Ni} N_i H_5 \bar{X}_i \nonumber\\
& + \lambda_{\phi i} \phi_i^2 Z_i + \lambda_{Ni} N_i^2 \bar{Z}_i + \lambda_{Ti} T_i^2 U_i  + \lambda_{Hi} H \bar{U}_i \bar{Z}_i + \tilde{\lambda}_H Y' Y' H_5 \nonumber\\
& + \lambda_X H_{24} X_i \bar{X}'_i + \lambda_Y H_{24} Y' \bar{Y} + a_3 T_3^2 H_5 \label{Eq:Wint}  \;.
\end{align}
Integrating out the heavy messenger superfields denoted by $U$, $X$, $Y$ and $Z$, the Feynman diagrams in Figs.~\ref{Fig:messenger1_d}, \ref{Fig:messenger1_u} and \ref{Fig:messenger1_n}
 lead to the following effective non-renormalisable superpotential terms in the $SU(5)$ and $SO(3)$ unbroken phase:
\begin{align}
W_{Y_l} &= F \left( b_1 \frac{\phi_{23}}{M} T_1 + b_2  \frac{\phi_{123}}{M} T_2 +  b_3 \frac{\phi_{3}}{M} T_3 \right) \bar{H}_5 + \tilde{b}_2 F \frac{\tilde{\phi}_{23}}{M} T_2 \bar{H}'_{5} \;, \label{Eq:Yl} \\
W_{Y_u} &= \left( a_3 T_3^2 + a_2 \frac{\phi_{123}^2}{M^2} T_2^2  + \tilde{a}_2 \frac{\tilde{\phi}_{23}^2}{M^2} T_2^2 + a_1 \frac{\phi_{23}^2}{M^2} T_1^2  \right) H_5 \;, \label{Eq:Yu} \\
W_{Y_\nu} &=  F \left( a_{\nu_1} \frac{\phi_{23}}{M} N_1 + a_{\nu_2} \frac{\phi_{123}}{M} N_2 \right)H_5 \label{Eq:Ynu} \;, \\
W_{\nu}^{M_R} &= a_{R_1} \frac{\phi_{23}^2}{M^2} N_1^2 + \left( a_{R_2} \frac{\phi_{123}^2}{M^2} + \tilde{a}_{R_2} \frac{\tilde{\phi}_{23}^2}{M^2}\right) N_2^2 \;,  \label{Eq:MR}
\end{align}
where he have introduced an \emph{effective messenger scale} $M$.

After GUT symmetry breaking the $SU(2)_L$ doublet components from $H_5$ and $H'_{5}$ as well as $\bar{H}_5$ and $\bar{H}'_{5}$ respectively mix and only the light states acquire the $SU(2)_L$ breaking vevs which give the fermion masses. We parameterise the Higgs mixing with the mixing angles $\gamma$ and $\bar{\gamma}$ respectively as
\begin{equation}
\begin{split}
\begin{pmatrix} H_5 \\ H'_{5} \end{pmatrix} &= \begin{pmatrix} c_\gamma & - s_\gamma \\  s_\gamma & c_\gamma \end{pmatrix} \begin{pmatrix} H_l \\ H_{h} \end{pmatrix}  \;, \\
\begin{pmatrix} \bar{H}_{5} \\ \bar{H}'_{5} \end{pmatrix} &= \begin{pmatrix} c_{\bar{\gamma}} & - s_{\bar{\gamma}} \\  s_{\bar{\gamma}} & c_{\bar{\gamma}} \end{pmatrix} \begin{pmatrix} \bar{H}_l \\ \bar{H}_{h} \end{pmatrix}  \;,
\end{split}
\end{equation}
where we have used the common abbreviations $c_\gamma \equiv \cos \gamma$ and $s_\gamma \equiv \sin \gamma$ and similar for the other angle $\bar{\gamma}$. The light Higgs doublets are denoted with an index $l$ while the heavy Higgs doublets are denoted with an index $h$.

The effective couplings $a$ and $b$ appearing in the effective superpotential in Eqs.~\eqref{Eq:Yl}-\eqref{Eq:MR}
can be expressed in terms of the fundamental couplings from Eq.~\eqref{Eq:Wint}, the messenger masses from \eqref{Eq:Xmass},  the Higgs mixing angles and the effective messenger scale $M$ as
\begin{equation}
\label{Eq:bis}
b_i = \frac{\kappa_{Fi} \lambda_X \kappa_{Ti}}{c_{\bar{\gamma}}} \frac{M v_{24}}{M_{X_i} M_{X'_i}} \;, \quad \tilde{b}_2 = \frac{\tilde{\kappa}_{F} \lambda_Y \tilde{\kappa}_{T}}{s_{\bar{\gamma}}} \frac{M v_{24}}{M_Y M_{Y'} } \;,
\end{equation}
\begin{gather}
 \tilde{a}_2 =   \frac{s_{\bar{\gamma}}^2 }{c_\gamma}  \frac{M_{X_2}^2 M_{X'_2}^2}{v_{24}^2 M_{U_2} M_{Z_2}}  \frac{\lambda_{T2} \lambda_{H2} \tilde{\lambda}_{\phi} }{ \tilde{\kappa}_{F}^2 \lambda_X^2 \tilde{\kappa}_{T}^2 }  +  \frac{s_{\bar{\gamma}}^2 }{c_\gamma}  \frac{M_{X_2}^2 M_{X'_2}^2}{v_{24}^2 M_{Y}^2}  \frac{\tilde{\lambda}_{H}}{ \tilde{\kappa}_{F}^2 \lambda_Y^2} \;,\\
a_{i} = \frac{c_{\bar{\gamma}}^2 }{c_\gamma}  \frac{M_{X_i}^2 M_{X'_i}^2}{v_{24}^2 M_{U_i} M_{Z_i}}  \frac{\lambda_{Ti} \lambda_{Hi} \lambda_{\phi i} }{ \kappa_{Fi}^2 \lambda_X^2 \kappa_{Ti}^2 }  \;,
\end{gather}
\begin{equation}
a_{\nu_i} = \frac{c_{\bar{\gamma}}}{c_\gamma}  \frac{\kappa_{Ni} }{\kappa_{Ti}} \;,
\end{equation}
\begin{equation}
\label{Eq:aRs}
a_{R_i}= c_{\bar{\gamma}}^2 \frac{M_{X_i}^2 M_{X'_i}^2 }{v_{24}^2 M_{Z_i}} \frac{\lambda_{\phi i} \lambda_{Ni}}{\kappa_{Fi}^2 \lambda_X^2 \kappa_{Ti}^2} \;,\quad \tilde{a}_{R_2}= \frac{s_{\bar{\gamma}}^2}{3} \frac{M_{Y}^2 M_{Y'}^2}{v_{24}^2 M_{Z_2}} \frac{\tilde{\lambda}_{\phi} \lambda_{N2}}{\kappa_{F2}^2 \lambda_Y^2 \kappa_{T2}^2}   \;.
\end{equation}

\subsection{Vacuum Alignment}

In the following, we assume that the vevs of the $SO(3)$ breaking flavon fields point in the following directions in field space
\begin{equation} \label{Eq:vacuumalignment}
b_1 \frac{\langle \phi_{23} \rangle }{M}  = \begin{pmatrix}  0 \\  1\\  - 1  \end{pmatrix} \epsilon_{23} \;, \quad
b_2 \frac{\langle \phi_{123} \rangle }{M}  = \begin{pmatrix} 1 \\  1 \\ 1  \end{pmatrix} \epsilon_{123} \;, \quad
b_3 \frac{\langle \phi_{3} \rangle }{M}  = \begin{pmatrix}  0 \\  0\\  1  \end{pmatrix} \epsilon_{3} \;.
\end{equation}
These relations also define the quantities $\epsilon_{123}$, $\epsilon_{23}$ and $\epsilon_{3}$. The breaking of $SO(3)$ along the field directions of  $\phi_{123}$ and $\phi_{23}$ allows us to realise TB neutrino mixing via CSD \cite{King:1998jw, King:1999cm, *King:1999mb, *Antusch:2004gf, *Antusch:2010tf, King:2002nf, King:2005bj}. In the rest of this chapter, we assume that CP is only broken spontaneously by the vev of the flavon $\tilde{\phi}_{23}$. 

A priori, for the flavon $\tilde{\phi}_{23}$ one may suppose a less constrained alignment,
\begin{equation} \label{Eq:vacuumalignment2}
\tilde{b}_2 \frac{\langle \tilde\phi_{23} \rangle }{M}  = \begin{pmatrix}  0 \\  v \\  w  \end{pmatrix} \tilde{\epsilon}_{23} \;.
\end{equation}
However,  empirically we find that the numerical fit to quark masses and mixings, in particular quark CP violation, seems strongly to prefer that the vacuum alignment of the flavon $\tilde{\phi}_{23}$ has its second component along the imaginary direction. To simplify the results of the numerical fit, we shall restrict ourselves to the case:
\begin{equation}
v = -\ci \;. \label{Eq:phit23ansatz}
\end{equation}
In some future more ambitious theory, one may attempt to reproduce Eq.~\eqref{Eq:phit23ansatz} as a result of some
special vacuum alignment, but here we shall simply regard it as a special choice, or ansatz, which leads to a successful fit to quark CP violation.

\subsection{Numerical Fit to the SM Fermion Masses and Mixings}

We define our conventions for the Yukawa matrices such that the operators of the form $F T \phi \bar{H}$ and $T^2 \phi^2 H$ give the following Yukawa terms in the Lagrangian:
\begin{equation} \label{Eq:GutYukConvention}
\mathcal{L}_{\mathrm{Yuk}} = - (Y^*_d)_{ij} q_i \bar{d}_j h_d - (Y^*_e)_{ij} l_i \bar{e}_j h_d  - (Y_u^*)_{ij} q_i \bar{u}_j h_u + \mathrm{h.c.}\;,
\end{equation}
where the $SU(5)$ relation $Y_d = Y_e^T$ is fulfilled, if all CG factors are one. The convention we use here is the same as the one used by the Particle Data Group \cite{Amsler:2008zzb}.

\begin{table}
\begin{center}
\begin{tabular}{cc}
\toprule
Parameter & Fit Value \\
\midrule 
$\epsilon_{123}$ in $10^{-4}$ & $6.56$ \\
$\epsilon_{23}$ in $10^{-4}$ & $-5.74$ \\
$\tilde{\epsilon}_{23}$ in $10^{-3}$ & $-2.23$ \\
$\epsilon_3$ & $-0.155$ \\
$w$ & $3.21$ \\
\midrule
$\hat{y}_u$ in $10^{-6}$ & $2.87$\\
$\hat{y}_c$ in $10^{-3}$ & $1.39$\\
$\hat{y}_t$ & $0.526$\\
\midrule
$\eta_l \tan \beta$ in $10^{-2}$ & $-3.42$ \\
$\eta_q \tan \beta$ in $10^{-2}$ & $3.00$\\
$\eta_u \tan \beta$ in $10^{-2}$ & $10.5$\\
\bottomrule
\end{tabular}
\caption[Fitted Parameters for the $SO(3) \times SU(5)$ Model]{The model parameters for $\tan \beta = 30$ and $M_{\mathrm{SUSY}} = 500$~GeV from a fit to the experimental data.  \label{Tab:Parameters1}}
\end{center}
\end{table}

\begin{table}
\begin{center}
\begin{tabular}{cccc}
\toprule
Quantity (at $m_t(m_t)$) & Model & Experiment & Deviation  \\ \midrule
$y_\tau$ in $10^{-2}$ & 1.00 & 1.00 & $-0.0133$\% \\
$y_\mu$ in $10^{-4}$  & 5.89 & 5.89 & $0.0444$\% \\
$y_e$ in $10^{-6}$ & 2.79 & 2.79 & $0.0139$\% \\  \midrule
$y_b$ in $10^{-2}$ & 1.58 & $1.58 \pm 0.05$ & $-0.0190 \sigma$ \\
$y_s$ in $10^{-4}$ & 2.54 & $2.99 \pm 0.86$ & $-0.5206 \sigma$ \\[0.1pc]
$y_d$ in $10^{-6}$ & 15.8 & $15.9^{+6.8}_{-6.6}$ & $- 0.0081 \sigma$ \\  \midrule
$y_t$ & 0.938 & $0.936 \pm 0.016$  & $0.0827 \sigma$  \\
$y_c$ in $10^{-3}$ & 3.39 & $3.39 \pm 0.46$ & $0.0001 \sigma$   \\[0.1pc]
$y_u$ in $10^{-6}$ & 7.00 & $7.01^{+2.76}_{-2.30}$ & $-0.0063 \sigma$  \\ \midrule
$\theta_{12}^{\mathrm{CKM}}$ & 0.2257 & $0.2257^{+0.0009}_{-0.0010}$ & $-0.0302 \sigma$   \\[0.3pc]
$\theta_{23}^{\mathrm{CKM}}$ & 0.0412 & $0.0415^{+0.0011}_{-0.0012}$ & $-0.2811 \sigma$   \\[0.1pc]
$\theta_{13}^{\mathrm{CKM}}$ & 0.0037 & $0.0036 \pm 0.0002$ & $0.3309 \sigma$   \\[0.1pc]
$\delta_{\mathrm{CKM}}$ & 1.2850 & $1.2023^{+0.0786}_{-0.0431}$ & $1.0524 \sigma$   \\
\bottomrule
\end{tabular}
\caption[Fit Results for the $SO(3) \times SU(5)$ Model]{Fit results for the quark Yukawa couplings and mixing and the charged lepton Yukawa couplings at low energy compared to experimental data. A pictorial representation of the agreement between our predictions and experiment can also be found in Fig.~\ref{Fig:FitResultsPlot1}. \label{Tab:FitResults1}}
\end{center}
\end{table}

The Yukawa matrices for the quarks and charged leptons coupling to the light Higgs doublets are then given from Eqs.~\eqref{Eq:Yl}, \eqref{Eq:Yu}, \eqref{Eq:vacuumalignment} and \eqref{Eq:vacuumalignment2} by
\begin{align}
Y_u &= \begin{pmatrix} 2 a_{1} \epsilon^2_{23} & 0 &  0 \\ 0 & 3 a_{2} \epsilon^2_{123} +  (w^2-1) \tilde{a}_{2}^2 \tilde{\epsilon}^2_{23} & 0 \\ 0 &  0 & a_{3}  \end{pmatrix} \equiv \begin{pmatrix} \hat{y}_u & 0 &  0 \\ 0 & \hat{y}_c& 0 \\ 0 &  0 & \hat{y}_t  \end{pmatrix}, \\
Y_d &= \begin{pmatrix} 0 & \epsilon_{23} & - \epsilon_{23} \\ \epsilon_{123} & \epsilon_{123} + \ci \, \tilde{\epsilon}_{23} & \epsilon_{123} + w \tilde{\epsilon}_{23} \\ 0 & 0 & \epsilon_{3} \end{pmatrix}, \\
Y_e^T &= \begin{pmatrix} 0 & c_{23} \epsilon_{23} & - c_{23} \epsilon_{23} \\ c_{123} \epsilon_{123} & c_{123} \epsilon_{123} + \ci \, \tilde{c}_{23} \tilde{\epsilon}_{23} & c_{123} \epsilon_{123} + w \tilde{c}_{23} \tilde{\epsilon}_{23} \\ 0 & 0 & c_3 \epsilon_{3} \end{pmatrix} ,
\end{align}
where $c_{3}$, $c_{23}$, $\tilde c_{23}$ and $c_{123}$ are the CG factors arising from GUT symmetry breaking and we have used the notation described above for the flavon vevs with $v = - \ci$. We note that in the definition for the Yukawa matrices we have introduced a complex conjugation and therefore a phase factor of $+\mathrm{i}$ appears in the 2-2 elements of $Y_d$ and $Y_e$.

With the given representations of the flavon and Higgs fields we obtain
\begin{equation}
\quad c_{123} = -3/2 \;,  \quad c_{23} = -3/2 \;, \quad c_3 = -3/2 \;, \quad \tilde{c}_{23} = 6\;.
\end{equation}
Since we consider large values of $\tan \beta$, the 1-loop SUSY threshold corrections are important and, taking the actual  experimental values of the fermion masses into account, the GUT scale value of $y_\mu / y_s$ prefers $\tilde{c}_{23} = 6$ and the GUT scale value of $y_\tau / y_b$ prefers $c_3 = -3/2$, as argued in Chs.~\ref{Ch:AFirstGlance} and \ref{Ch:Pheno}.

\begin{figure}[t]
\centering
\includegraphics[scale=0.65]{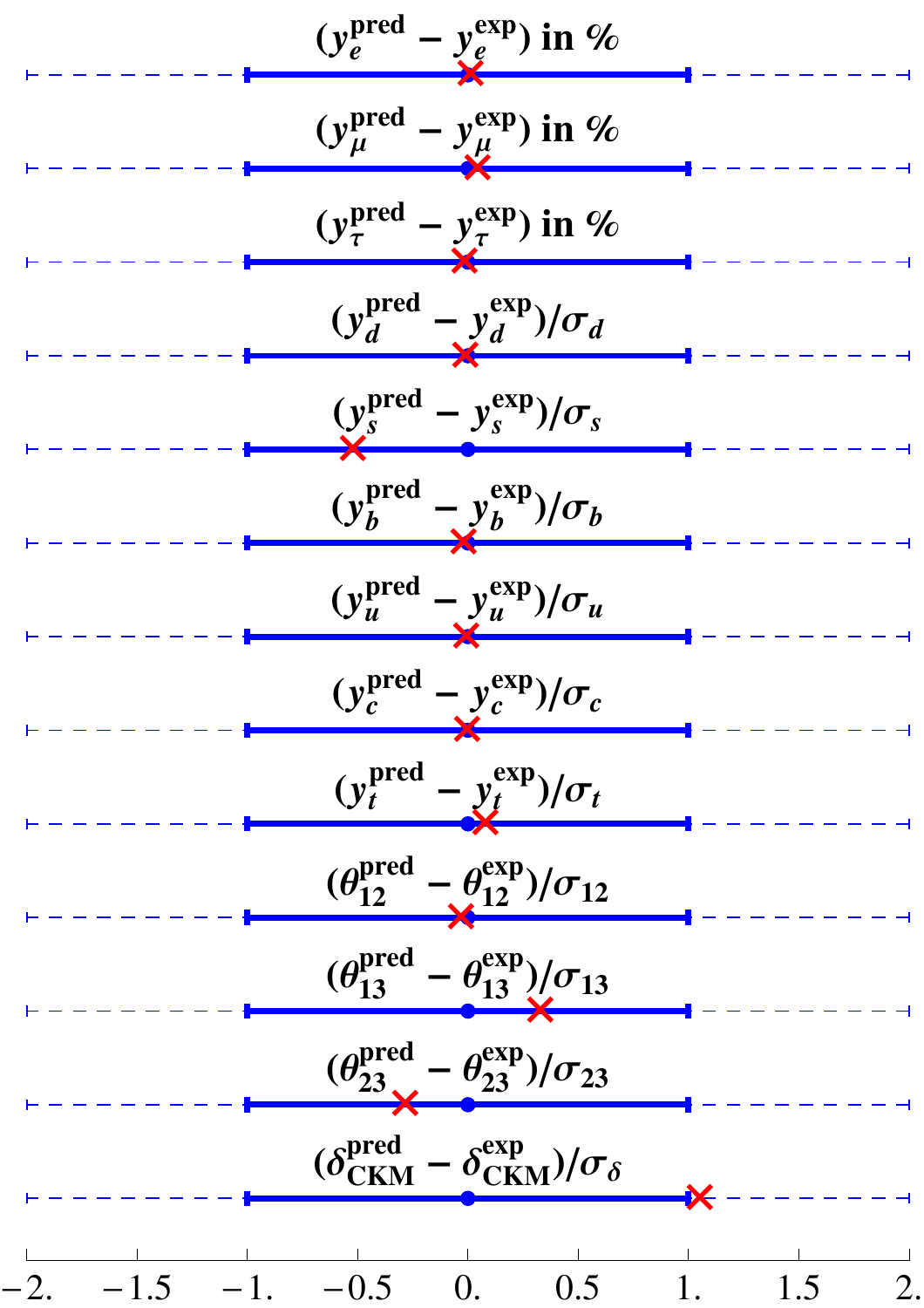}
\caption[Fit Results for the $SO(3) \times SU(5)$ Model]{Pictorial representation of the deviation of our predictions from low energy experimental data for the charged lepton Yukawa couplings and quark Yukawa couplings and mixing parameters.  The deviations of the charged lepton masses are given in~\% while all other deviations are given in units of standard deviations $\sigma$. The thick blue line gives the 1\% (1$\sigma$)  bound while the dashed line gives the 2\% (2$\sigma$) bound. The red crosses denote our predictions. \label{Fig:FitResultsPlot1} }
\end{figure}

For the detailed fit of the model to the data we have applied the following procedure:
\begin{itemize}
\item In a first step,  we took the fermion masses at the scale $m_t(m_t)$ as provided by \cite{Xing:2007fb} and the low energy values for the CKM parameters and calculated  their RG evolution numerically with the REAP software package \cite{Antusch:2005gp} without SUSY threshold corrections up to the GUT scale $M_{\mathrm{GUT}} = 2 \times 10^{16}$~GeV within the MSSM with $\tan \beta = 30$ and a SUSY scale of $M_{\mathrm{SUSY}} = 500$~GeV. There, we determined a first rough estimate of the parameters. We should mention that it is not possible to determine $a_2$ and $\tilde{a}_2$ simultaneously since they are fixed by the charm quark mass only. Therefore, we can effectively determine the three parameters $\hat{y}_u$, $\hat{y}_c$ and $\hat{y}_t$ only.

\item In a second step, we have implemented the SUSY threshold corrections in the RGE evolution via the matching conditions
\begin{align}
y_{u,c,t}^{\mathrm{MSSM}} &= \frac{y_{u,c,t}^{\mathrm{SM}}}{\sin\beta} \label{Eq:LeptonMatching} \;,\\
y_{e,\mu,\tau}^{\mathrm{MSSM}} &= (1 - \eta_l \tan \beta) \frac{y_{e,\mu,\tau}^{\mathrm{SM}}}{\cos\beta} \label{Eq:UpMatching} \;,\\
y_{d,s}^{\mathrm{MSSM}} &= (1 - \eta_q \tan \beta) \frac{y_{d,s}^{\mathrm{SM}}}{\cos\beta} \label{Eq:12DownMatching} \;,\\
y_{b}^{\mathrm{MSSM}} &= (1 - (\eta_q + \eta_u) \tan \beta ) \frac{y_{b}^{\mathrm{SM}}}{\cos\beta} \label{Eq:3DownMatching} \;,\\
\theta_{ij}^{\mathrm{MSSM}} &= \frac{1 - \eta_q \tan \beta}{1 - (\eta_q + \delta_{3j} \eta_u) \tan \beta} \theta_{ij}^{\mathrm{SM}} \label{Eq:MixingMatching} \;,\\
\delta_{\mathrm{CKM}}^{\mathrm{MSSM}} &=  \delta_{\mathrm{CKM}}^{\mathrm{SM}} \;, \label{Eq:deltaMixingMatching} 
\end{align}
at the SUSY scale $M_{\mathrm{SUSY}} = 500$~GeV, cf.\ Ch.~\ref{Ch:SUSYThresholdCorrections} and \cite{Blazek:1995nv, Antusch:2008tf}. In Ch.~\ref{Ch:SUSYThresholdCorrections} we have also given ranges for the values of $\eta_i$, which are consistent with the fitted values, where
\begin{align}
\eta_l &\approx \eta_\tau^B + \eta_\tau^W \;,\\
\eta_q &\approx \eta_b^G + \eta_b^B + \eta_b^W \;,\\
\eta_u &\approx \eta^y \;.
\end{align}
At the low energy scale $m_t(m_t)$ we have performed a $\chi^2$ fit with the GUT scale parameters as input. The fit gave a total $\chi^2$ of 1.6 where we have assumed a relative error of 1\% for the charged lepton masses. Since we have 11 parameters and 13 observables this corresponds to a $\chi^2/\mathrm{dof}$ of 0.8 which is very good.
\end{itemize}

The results for the GUT scale parameters are listed in Tab.~\ref{Tab:Parameters1}. In Tab.~\ref{Tab:FitResults1} the low energy results are shown and compared to experimental data. A graphical illustration of the deviations is given in Fig.~\ref{Fig:FitResultsPlot1}. They illustrate that our minimal example model, with the assumed vacuum alignment of Eqs.~\eqref{Eq:vacuumalignment} and \eqref{Eq:vacuumalignment2} with the $v=-\ci$, can fit the data well and leads to testable predictions.

We would like to remark that the parameters depend on $\tan \beta$ and $M_{\mathrm{SUSY}}$ and are also subject to several theoretical uncertainties. For example, we have not included the full flavour structure of the soft parameters. In a more sophisticated approach, there can be deviations from the matching formulas \eqref{Eq:LeptonMatching}-\eqref{Eq:deltaMixingMatching}. Due to such additional theoretical uncertainties, we do not explicitly give the errors on the high energy parameters or low energy predictions. The important input parameters for us are the charged lepton masses and quark mixing angles which have an experimental error much smaller than these uncertainties.

\subsection{The Neutrino Sector}

We now turn to the neutrino sector. We start with the derivation of the mass matrix for the light neutrino states.
The neutrino Yukawa matrix is obtained from Eq.~\eqref{Eq:Ynu} as
\begin{align}
Y_{\nu} &= \begin{pmatrix} 0 &  a_{\nu_2} \epsilon_{123} \\  a_{\nu_1} \epsilon_{23} &  a_{\nu_2} \epsilon_{123} \\ -  a_{\nu_1} \epsilon_{23} &  a_{\nu_2} \epsilon_{123}   \end{pmatrix} .
\end{align}
Additionally, we have a diagonal mass matrix for the two right-handed neutrinos from Eq.~\eqref{Eq:MR},
\begin{equation}
M_R = \begin{pmatrix} 2  a_{R_1} \epsilon_{23}^2 & 0 \\ 0 & 3  a_{R_2} \epsilon_{123}^2 + 3 (w^2-1) \tilde{a}_{R_2} \tilde{\epsilon}_{23}^2  \end{pmatrix} \;.
\end{equation}
Using the seesaw relation
\begin{equation}
m_\nu = - v_{u}^2 Y_\nu M_R^{-1} Y_\nu^T \;,
\end{equation}
we obtain for the neutrino mass matrix
\begin{equation}
m_\nu =
\frac{m_3}{2} \begin{pmatrix} 0 & 0 &0 \\ 0 & 1 & -1 \\ 0 & -1 & 1 \end{pmatrix}
+
\frac{m_2}{3} \begin{pmatrix} 1 & 1 &1 \\ 1 & 1 & 1 \\ 1 & 1 & 1 \end{pmatrix},
\end{equation}
with
\begin{equation}
m_2 = -   v_u^2 \frac{a_{\nu_2}^2 \epsilon_{123}^2}{a_{R_2} \epsilon_{123}^2 + (w^2-1) \tilde{a}_{R_2} \tilde{\epsilon}_{23}^2 }   \quad \text{and} \quad
m_3 = -   v_u^2 \frac{a_{\nu_1}^2 }{a_{R_1}}  \;.
\end{equation}
These two parameters can have either sign.

From the structure of $m_\nu$, we obtain TB mixing in the neutrino sector,
\begin{equation}
\theta_{13}^\nu = 0^\circ \;,\quad \theta_{23}^\nu = 45^\circ \;, \quad \theta_{12}^\nu = \arcsin \frac{1}{\sqrt{3}} \approx 35.3^\circ  \:.
\end{equation}
From the lepton sector we get the additional mixing contributions
\begin{equation}
\theta_{13}^e = 0^\circ \;,\quad \theta_{23}^e = 0^\circ \;, \quad \vert \theta_{12}^e \vert  = \left\vert \frac{c_{123} \epsilon_{123}}{c_{123} \epsilon_{123} - \ci \, \tilde{c}_{23} \tilde{\epsilon}_{23} } \right\vert \approx 4.2^\circ \;.
\end{equation}
There is also a complex phase introduced by the charged lepton Yukawa matrix which can be calculated to
\begin{equation}
 \delta_{12}^e =  \arctan \left( \frac{\tilde{c}_{23} \tilde{\epsilon}_{23} }{c_{123} \epsilon_{123}} \right)\approx  85.8^\circ \label{Eq:deltae} \;.
\end{equation}

For the approximate calculation of the MNS mixing parameters at the GUT scale we can use \cite{Antusch:2008yc,
Boudjemaa:2008jf, *Antusch:2007vw, *Antusch:2007ib, *King:2007pr}:
\begin{equation}
\begin{split}
s^{\mathrm{MNS}}_{23} \mathrm{e}^{- \mathrm{i} \delta^{\mathrm{MNS}}_{23} } &\approx s_{23}^\nu \mathrm{e}^{- \mathrm{i} \delta^{\nu}_{23} } - \theta_{23}^e \;, \\
s^{\mathrm{MNS}}_{13} \mathrm{e}^{- \mathrm{i} \delta^{\mathrm{MNS}}_{13} } &\approx \theta_{13}^\nu  \mathrm{e}^{- \mathrm{i} \delta^{\nu}_{13} } - s_{23}^\nu \theta_{12}^e \mathrm{e}^{- \mathrm{i} (\delta^\nu_{23} + \delta_{12}^e)} \;, \\
s^{\mathrm{MNS}}_{12} \mathrm{e}^{- \mathrm{i} \delta^{\mathrm{MNS}}_{12} } &\approx s_{12}^\nu \mathrm{e}^{- \mathrm{i} \delta^{\nu}_{12} }  - c_{23}^\nu c_{12}^\nu  \theta_{12}^e \mathrm{e}^{- \mathrm{i} \delta_{12}^e} \;,
\end{split}
\end{equation}
where we have already discarded RG corrections which are small for the case of hierarchical neutrino masses \cite{Antusch:2003kp,Antusch:2005gp}. For simplicity, we first want to assume that $m_2$ and $m_3$ have the same sign. In this case, the phases $\delta^\nu_{ij}$ are trivial. For the total leptonic  mixing angles we then obtain
\begin{equation} \label{Eq:prediction}
\begin{split}
\theta_{12}^{\mathrm{MNS}}  & \approx 35.2^\circ \;, \\
\theta_{13}^{\mathrm{MNS}}  & \approx  3.0^\circ \;, \\
\theta_{23}^{\mathrm{MNS}}  & \approx 45^\circ\;.
\end{split}
\end{equation}
For the phases we have $\delta^{\mathrm{MNS}}_{13} = \delta_{12}^e - \pi \approx  -94.2^\circ$, $\delta^{\mathrm{MNS}}_{12} = -4.2^\circ$ and $\delta^{\mathrm{MNS}}_{23} = 0^\circ$ from which the final MNS phases can be calculated according to \cite{Antusch:2008yc,
Boudjemaa:2008jf, *Antusch:2007vw, *Antusch:2007ib, *King:2007pr}
\begin{equation}\label{Eq:prediction_phases}
\begin{split}
\delta_{\mathrm{MNS}} &= \delta_{13}^{\mathrm{MNS}} - \delta^{\mathrm{MNS}}_{12} \approx  -90^\circ \;,\\
\alpha_1 &= 2 (\delta_{12}^{\mathrm{MNS}} + \delta_{23}^{\mathrm{MNS}}) = 2 \delta_{12}^{\mathrm{MNS}} \approx -8.4^\circ \;,\\
\alpha_2 &= 2 \delta_{23}^{\mathrm{MNS}} \approx 0^\circ \;,
\end{split}
\end{equation}
where $\alpha_1$ and $\alpha_2$ are the Majorana phases as in the PDG parameterisation where they are contained in a diagonal matrix $\mathrm{diag}(\mathrm{e}^{\mathrm{i} \alpha_1/2}, \mathrm{e}^{\mathrm{i} \alpha_2/2}, 1)$. This is not the whole story yet, since we made the assumption that $m_2$ and $m_3$ have the same sign. If they have opposite signs, we get
$\alpha_2 = 180^\circ$,
while the other phases remain the same.

Thus, in summary, the predictions of our model for the leptonic mixing parameters are compatible with the experimental 1$\sigma$ ranges at low energy which are: $\theta_{12}^{\mathrm{MNS}} = (34.5\pm 1.0)^\circ$, $\theta_{13}^{\mathrm{MNS}} =  (5.7^{+3.0}_{-3.9})^\circ$ and $\theta_{23}^{\mathrm{MNS}} = (42.3^{+5.3}_{-2.8})^\circ$, taken from \cite{GonzalezGarcia:2010er}.

We note that with the mixing pattern of our model, i.e.\ TB mixing produced in the neutrino sector, and charged lepton mixing corrections from $\theta_{12}^e$ only,   the leptonic mixing angles and the Dirac CP phase $\delta_{\mathrm{MNS}}$ satisfy the lepton mixing sum rule \cite{King:2005bj, Masina:2005hf, Antusch:2007rk, Antusch:2005kw}
\begin{equation}
\theta_{12}^{\mathrm{MNS}} - \theta_{13}^{\mathrm{MNS}} \cos (\delta_{\mathrm{MNS}}) \approx \arcsin (1/\sqrt{3}) \:.
\end{equation}
The approximately maximal CP violation, i.e.\ $\delta_{\mathrm{MNS}} \approx -90^\circ$, only leads to small deviations of the solar mixing angle from its TB value of $\arcsin (1/\sqrt{3})$, although the charged lepton corrections generate $\theta_{13}^{\mathrm{MNS}} \approx   3.0^\circ$.

The predictions of our model for the leptonic mixing angles and Dirac CP phase $\delta_{\mathrm{MNS}}$ stated in Eqs.~\eqref{Eq:prediction} and \eqref{Eq:prediction_phases} can be tested accurately by ongoing and future precision neutrino oscillation experiments \cite{Bandyopadhyay:2007kx}.

The kinematic mass accessible in the single beta decay end-point experiment KATRIN is
\begin{equation}
m^2_\beta \equiv m_1^2 c_{12}^2 c_{13}^2 + m_2^2 s_{12}^2 c_{13}^2  + m_3^2 s_{13}^2 \;.
\end{equation}
In our case, neutrino masses are strictly hierarchical $m_3 \gg m_2 > m_1 = 0$ such that we can determine the masses $m_2$ and $m_3$ from the mass squared differences and obtain
\begin{equation}
m^2_\beta = \left(3.2^{+ 0.2}_{-0.3} \right) \times 10^{-5} \; \text{eV}^2 \;,
\end{equation} 
which is beyond the reach of the upcoming experiments. 

The effective mass relevant for neutrinoless double beta decay reads
\begin{equation} 
m_{ee}  = \vert m_1 c_{12}^2 c_{13}^2 e^{\ci \, \alpha_1} + m_2 s_{12}^2 c_{13}^2 e^{\ci \, \alpha_2} + m_3 s_{13}^2 e^{2 \, \ci \, \delta_{\mathrm{MNS}}} \vert \;,
\end{equation}
which is calculated to
\begin{equation}
m_{ee} = (2.8 \pm 0.1) \times 10^{-3} \; \text{eV} \quad \text{or} \quad  m_{ee} = (3.0 \pm 0.1) \times 10^{-3} \; \text{eV} \;,
\end{equation} 
depending on the Majorana phase $\alpha_2$. This is also beyond the reach of upcoming experiments. However, we note that neutrinoless double beta decay is an unavoidable consequence in the this model.

%% file: kap_10_Textures.tex
\chapter[Quark Mixing Sum Rules and the Right Unitarity Triangle]{Quark Mixing Sum Rules and the\\ Right Unitarity Triangle} \label{Ch:Textures}

In the last part we have discussed Yukawa couplings in SUSY GUTs for medium and large $\tan \beta$. In this part we turn our attention to the case of small $\tan \beta$. We start our discussion of that particular case in this chapter with a discussion of quark mixing sum rules and their relation to the right unitarity triangle in the quark sector based on \cite{Antusch:2009hq}.

We assume here that the Yukawa matrices are generated by the vacuum alignment of some family symmetry breaking flavon fields. This point of view defines a preferred basis, which we shall refer to as the \emph{flavour basis}. We adopt this point of view since in such frameworks, the resulting low energy effective Yukawa matrices are expected to have a correspondingly simple form in the flavour basis associated with the high energy simple flavon vacuum alignment. This suggests that it may be useful to look for simple Yukawa matrix structures in a particular basis, since such patterns may provide a bottom-up route towards a theory of flavour based on a spontaneously broken family symmetry.

Unfortunately, experiment does not tell us directly the structure of the Yukawa matrices and the complexity of the problem, in particular the basis ambiguity from the bottom-up perspective, generally hinders the prospects of deducing even the basic features of the underlying flavour theory from the experimental data. We are left with little alternative but to follow an {\it ad hoc} approach pioneered some time ago by Fritzsch \cite{Fritzsch:1979zq, Fritzsch:1999ee} and currently represented by the myriads of proposed effective Yukawa textures, see, e.g.\  \cite{Fritzsch:1979zq, Fritzsch:1999ee, Leontaris:2009pi, *Dev:2009he, *Adhikary:2009kz, *Goswami:2009bd, *Goswami:2008uv, *Choubey:2008tb, *Branco:2007nb, *Alhendi:2007iu, *Kaneko:2007ea, *Branco:2006wv, *Lam:2006wm, *Kaneko:2006wi, *Fuki:2006xw, *Haba:2005ds, *Kim:2004ki, *Jack:2003pb, *Jack:2003qg, *Caravaglios:2002br, *Everett:2000up, *Berezhiani:2000cg, *Kuo:1999dt, *Falcone:1998us, Roberts:2001zy, *Ramond:1993kv, Chiu:2000gw,Fritzsch:1999rb}, whose starting assumption is that in some basis the Yukawa matrices exhibit certain nice features such as symmetries or zeros in specific elements which have become known as \emph{texture zeros}. For example, in his classic paper, Fritzsch pioneered the idea of having six texture zeros in the 1-1, 2-2, 1-3 entries of the Hermitian up and down quark Yukawa (or mass) matrices \cite{Fritzsch:1979zq}.

Unfortunately, these six-zero textures are no longer consistent with experimental data, since they imply the bad prediction $|V_{cb}|\sim \sqrt{m_s /m_b}$, so texture zerologists have been forced to retreat to the (at most) four-zero schemes discussed, for example, in \cite{Roberts:2001zy, *Ramond:1993kv, Chiu:2000gw, Fritzsch:1999rb} which give up on the 2-2 texture zeros allowing the good prediction $|V_{cb}|\sim m_s /m_b$.

However, it turns out that four-zero textures featuring zeros in the 1-1 and 1-3 entries of both up and down Hermitian mass matrices may also lead to the bad prediction $|V_{ub}|/|V_{cb}|\sim \sqrt{m_u /m_c}$ unless $|V_{cb}|$ results from the cancellation of quite sizeable up- and down-type quark 2-3 mixing angles, leading to non-negligible induced 1-3 up- and down-type quark mixing \cite{Fritzsch:1999rb}. Another possibility is to give up on the 1-3 texture zeros, as well as the 2-2 texture zeros, retaining only two texture zeros in the 1-1 entries of the up and down quark matrices \cite{Roberts:2001zy, *Ramond:1993kv}. We reject here both of these options, and instead choose to maintain up to four texture zeros, without invoking cancellations, for example by making the 1-1 element of the up (but not down) quark mass matrix nonzero, while retaining 1-3 texture zeros in both the up and down quark Hermitian matrices, as suggested in \cite{Chiu:2000gw}.

In this chapter we discuss phenomenologically viable textures for hierarchical quark mass matrices  
which have both 1-3 texture zeros and negligible 1-3 mixing in both the up and down quark mass matrices.
We derive quark mixing sum rules applicable to textures of this type, in which $V_{ub}$ is generated from $V_{cb}$ as a result of 1-2 up-type mixing in Sec.~\ref{Sec:SumRules}, in direct analogy to the lepton sum rules derived in \cite{King:2005bj, Masina:2005hf, Antusch:2005kw, Antusch:2008yc, Boudjemaa:2008jf, *Antusch:2007vw, *Antusch:2007ib, *King:2007pr, Antusch:2007rk}, and especially discuss how  to use the sum rules to show how the right-angled unitarity triangle, i.e., $\alpha \approx 90^\circ$, relates to the phases in the up and down quark mass matrices.

In Sec.~\ref{Sec:Textures} we show how this phase structure can be accounted for by a remarkably simple scheme involving real mass matrices apart from a single element of either the up or down quark mass matrix being purely imaginary. Fritzsch and Xing have previously emphasised how their four-zero scheme with 1-1 and 1-3 texture zeros in the Hermitian up and down quark mass matrices can be used to accommodate right unitarity triangles \cite{Fritzsch:1999rb}, but since their scheme involves large 2-3 and non-negligible 1-3 up and down quark mixing, our sum rules are not applicable to their case. Therefore, the textures in \cite{Roberts:2001zy, *Ramond:1993kv} and \cite{Fritzsch:1999rb} do not allow us to explain $\alpha \approx 90^\circ$ by simple structures 
with a combination of purely real and purely imaginary matrix elements.  Recently, it has become increasingly clear that current data is indeed consistent with the hypothesis of a right unitarity triangle, with the best fits giving $\left( \alpha = 90.7^{+4.5}_{-2.9} \right)^\circ$ \cite{Charles:2004jd, *Bona:2007qta, *Sordini:2009gu}, and this provides additional impetus for our scheme. The phenomenological observation that $\alpha \approx \pi /2$ has also motivated other approaches, 
see, e.g.\  \cite{Fritzsch:1997fw, *Fritzsch:1997st, Harrison:2007yn, *Harrison:2008ff, *Harrison:2009bb, *Harrison:2009bz, *Harrison:2009pw, Couture:2009it}, which are complementary to the approach developed in this chapter. In Sec.~\ref{Sec:Textures} we discuss also textures with nonzero 1-3 elements in the up sector which, however, turn out to be disfavoured.

We conclude this chapter in Sec.~\ref{Sec:QuarkLeptonMixingRelations} with a discussion of the implications of zero 1-3 mixing for the charged lepton and neutrino sectors in the framework of GUTs and show how the quark mixing sum rules may be used to yield an accurate prediction for the reactor mixing angle.

\section{Quark Mixing Sum Rules} \label{Sec:SumRules}

We start our derivation of quark mixing sum rules with the derivation of sum rules for the mixing angles and afterwards we derive a sum rule for the phases. In this whole discussion we always suppose that $\theta_{13}^d = \theta_{13}^u =0$. This can be understood as a direct result from a flavon vacuum alignment,  see, e.g.\  \cite{Barr:1990td, *Sogami:1992av, *Ibanez:1994ig, *Jain:1994hd, *Pomarol:1995xc, *Mondragon:1998gy, *Barbieri:1999pe, *Barbieri:1998em, *Babu:2004tn, *Haba:2005ds, *Masina:2006pe, *Bazzocchi:2007na, *Babu:2009nn, *Ishimori:2009ns, *Morisi:2010rk}. Therefore, although from the SM point of view this corresponds only to a convenient choice, it becomes a nontrivial assumption at the level of a specific flavour model.
 
\subsection{Mixing Angle Sum Rules}

We use the conventions for the CKM matrix as defined in App.~\ref{App:CKM}. For $\theta_{13}^d = \theta_{13}^u =0$, Eq.~\eqref{Eq:Param3} simplifies to
\begin{equation}
\label{Eq:Param5}
V'_{\mathrm{CKM}} = {U^{u_L}_{12}}^\dagger {U^{u_L}_{23}}^\dagger U^{d_L}_{23} U^{d_L}_{12}
 \;.
\end{equation}
Then, by equating the right-hand sides of Eqs.~\eqref{Eq:Param4} and \eqref{Eq:Param5} and expanding to leading order in the small mixing angles, we obtain the following relations (up to cubic terms in the physical quark mixing angles):
\begin{align} \label{Eq:F1}
{\theta_{23}}e^{-i\delta_{23}}&=
{\theta_{23}^{d}}e^{-i\delta_{23}^{d}}
-{\theta_{23}^{u}}e^{-i\delta_{23}^{u}}\;,
\\
\label{Eq:F2} {\theta_{13}}e^{-i\delta_{13}}&=
-{\theta_{12}^{u}}e^{-i\delta_{12}^{u}}
({\theta_{23}^{d}}e^{-i\delta_{23}^{d}} - {\theta_{23}^{u}}e^{-i\delta_{23}^{u}})
\;,\\
\label{Eq:F3} {\theta_{12}}e^{-i\delta_{12}}&=
{\theta_{12}^{d}}e^{-i\delta_{12}^{d}}
-{\theta_{12}^{u}}e^{-i\delta_{12}^{u}} \;.
\end{align}
For convenience we drop in this chapter the superscript CKM in the quark mixing angles $\theta_{ij}^{\mathrm{CKM}}$ . Let us first consider Eq.~\eqref{Eq:F2}, which can be transformed into
\begin{equation}\label{Eq:theta13withphases}
{\theta_{13}}e^{-i\delta_{13}} = -{\theta_{12}^{u}} {\theta_{23}}
e^{-i(\delta_{12}^{u}+\delta_{23})} \;,
\end{equation}
where $\theta_{13}$ and $\theta_{23}$ stand for the measurable 1-3 and 2-3 mixing angles in the quark sector, respectively.  Taking the modulus of  Eq.~\eqref{Eq:theta13withphases}, the 1-2 angle entering the up-sector rotation ($V_{u_L}$) in the flavour basis obeys
\begin{equation}\label{Eq:QuarkRelation}
{\theta_{12}^{u}} =
\frac{\theta_{13}}{\theta_{23}} = \left( 4.96 \pm 0.30
\right)^\circ \;.
\end{equation}
where the 1$\sigma$ errors are displayed \cite{Amsler:2008zzb}.

Similarly, combining Eq.~\eqref{Eq:F3} with Eq.~\eqref{Eq:theta13withphases} one receives
\begin{equation}
{\theta_{12}} - \frac{\theta_{13}}{\theta_{23}} e^{-i(\delta_{13}
- \delta_{23} - \delta_{12})} =
{\theta_{12}^{d}}e^{-i(\delta_{12}^{d} -  \delta_{12})}\;.
\end{equation}
This, together with the identification Eq.~\eqref{Eq:deltafromparam1} gives rise to the quark sector sum rule\footnote{We would like to remark that for $\theta_{12}^{u} \ll \theta_{12}^{d}$, the sum rule may be further simplified to $\theta_{12} - \frac{\theta_{13}}{\theta_{23}} \cos \delta = \theta_{12}^d$. For similar considerations in the lepton sector, see, e.g.\  \cite{King:2005bj, Masina:2005hf, Antusch:2005kw}.}
\begin{equation}\label{Eq:QuarkSumRule}
\theta_{12}^d = \left|\theta_{12} - \frac{\theta_{13}}{\theta_{23}} e^{- i\delta_\mathrm{CKM}} \right| = \left( 12.0^{+0.39}_{-0.22} \right)^\circ  
\end{equation}
which is valid up to higher order corrections. The present best-fit value and the 1$\sigma$ errors are also displayed.

Needless to say, the relations \eqref{Eq:QuarkRelation} and \eqref{Eq:QuarkSumRule} apply at the scale at which the flavour structure emerges, often close to the scale of Grand Unification. Thus, in principle, the renormalisation group (RG) effects should be taken into account. However, due to the smallness of the mixing angles in the quark sector and the hierarchy of the quark masses, the RG corrections to the above relations are small and can be neglected here to a good approximation.

\subsection{Phase Sum Rule}

It is interesting that, with the 1-2 mixing angles in the up and down sector derived from the physical parameters, the 1-2 phase difference in the up and down sectors can also be determined. Indeed, combining all three equations
\eqref{Eq:F1}, \eqref{Eq:F2} and \eqref{Eq:F3}, one obtains
\begin{equation} \label{Eq:deltadu}
\frac{\theta_{13} \theta_{12}}{\theta_{23}} e^{i \delta_{\rm CKM}}
= - \theta_{12}^{u}
({\theta_{12}^{d}}e^{-i(\delta_{12}^{d}-\delta_{12}^{u})}-{\theta_{12}^{u}}) \;.
\end{equation}
Using Eqs.~\eqref{Eq:QuarkRelation} and \eqref{Eq:QuarkSumRule} we can solve Eq.~\eqref{Eq:deltadu} for $\delta_{12}^{d} - \delta_{12}^{u}$ and obtain (at 1$\sigma$ level)
\begin{equation}
\delta_{12}^{d} - \delta_{12}^{u} = (91.5^{+5.5}_{-4.0})^\circ\;,
\label{Eq:phasediff}
\end{equation}
which is remarkably close to $\pi/2$. We emphasise that this is a consequence of zero 1-3 mixing in the up and down sectors, $\theta_{13}^d = \theta_{13}^u = 0$. Otherwise the dependence on $\theta_{23}$ and $\delta_{23}$ would not cancel and the phase $\delta_{13}$ could give a contribution.

We show now that, assuming quark textures with negligible 1-3 up and down quark mixing, corresponding to 1-3 texture zeros for hierarchical quark mass matrices,  $\delta_{12}^{d} - \delta_{12}^{u}$ is approximately equal to the unitarity triangle angle $\alpha$. This comes from its definition:
\begin{equation}
\begin{split}
 \alpha &= \arg \left( - \frac{V_{td} V_{tb}^*}{V_{ud} V_{ub}^*} \right)  \\
& =  \arg \left( - \frac{(s_{12} s_{23} - c_{12} c_{23} s_{13} e^{i \delta_{\mathrm{CKM}} })
c_{23} c_{13}}{ c_{12} c_{13} s_{13} e^{i\delta_{\mathrm{CKM}}}} \right)  \\
& \approx  \arg \left( 1 -
\frac{\theta_{12} \theta_{23} }{\theta_{13}} e^{-i\delta_{\mathrm{CKM}}} \right) \;.
\end{split}
\end{equation}
For the second term in the argument, we can use Eqs.~\eqref{Eq:F1}, \eqref{Eq:F2} and \eqref{Eq:F3}:
\begin{equation}
\begin{split}
 \alpha &\approx \arg \left( 1 +\frac{\theta_{23} e^{i\delta_{23} }
 (\theta_{12}^d e^{i\delta_{12}^d} - \theta_{12}^u e^{i \delta_{12}^u})}{\theta_{12}^u e^{i \delta_{12}^u} \theta_{23}
 e^{i\delta_{23} } } \right) \\
&= \arg \left( \frac{\theta_{12}^d}{\theta_{12}^u} e^{i
(\delta_{12}^d - \delta_{12}^u)} \right) = \delta_{12}^d -
\delta_{12}^u \;.
\end{split}
\end{equation}
Thus, one can see that the angle $\alpha$ is nothing but the phase difference $\delta_{12}^d - \delta_{12}^u$, corresponding to a very simple phase sum rule
\begin{equation}
\label{Eq:phase}
 \alpha \approx \delta_{12}^d - \delta_{12}^u \;.
\end{equation}

\section{Quark Mass Matrices with 1-3 Texture Zeros} \label{Sec:Textures}

Since we have now the quark mixing sum rules at hand we apply them to various types of textures. We end this section with a discussion of textures with nonzero 1-3 elements to which our sum rules are not directly applicable. However, we find in this case a severe deviation between our prediction and experimental data as long as we stick to our proposed simple phase structure of the mass matrices.

\subsection[Real/Imaginary Matrix Elements for $\alpha = 90^\circ$]{Real/Imaginary Matrix Elements for $\boldsymbol{\alpha = 90^\circ}$} \label{Sec:deltaCKM}

According to the phase sum rule in Eq.~\eqref{Eq:phase}, the experimental observation that $\alpha \approx 90^\circ$, or the equivalent determination in Eq.~\eqref{Eq:phasediff}, suggests looking at quark mass matrices with 1-3 texture zeros and with $\delta_{12}^{d}$ or $\delta_{12}^{u}$ at the special values $\pm \pi/2$. This would correspond to a set of rather specific textures of the quark mass matrices with, for example, purely imaginary 1-2 elements in either $M_u$ or $M_d$ while the 2-2 elements remain real. For a discussion of the relation between the phases of the mixing angles and the phases of the matrix elements see, e.g.\  \cite{King:2002nf}.  For instance, the following patterns naturally emerge:
\begin{equation}
M_u =
\begin{pmatrix}
a_u   & -i b_u & 0 \\
* & c_u & d_u \\
*   & *   & e_u
\end{pmatrix}
\; , \quad
M_d =
\begin{pmatrix}
a_u   &  b_d & 0 \\
* & c_d & d_d \\
*   & *   & e_d
\end{pmatrix} ,
\end{equation}
or
\begin{equation}
M_u =
\begin{pmatrix}
a_u   &  b_u & 0 \\
* & c_u & d_u \\
*   & *   & e_u
\end{pmatrix}
\; , \quad
M_d =
\begin{pmatrix}
a_u   & i b_d & 0 \\
* & c_d & d_d \\
*   & *   & e_d
\end{pmatrix} ,
\end{equation}
where $a_u$, $b_u$, $c_u$, $d_u$, $e_u$ and $a_d$, $b_d$, $c_d$, $d_d$, $e_d$ are real parameters, and the elements marked by ``*'' are irrelevant as long as the hierarchy of the mass matrix is large enough, or, equivalently, as long as the mixing angles in $V_{u_R}$ and $V_{d_R}$ are small. These textures are all phenomenologically viable, and consistent with $\alpha= 90^\circ$, and their simple phase structure provides a post justification of our assumption of 1-3 texture zeros and negligible 1-3 up- and down-type quark mixing. However, the above textures are clearly not the most predictive ones and, for example, do not relate the up and down quark 1-2 mixing angles to masses. This requires additional assumptions, such as additional texture zeros and Hermitian or symmetric matrices, as we  discuss now.

\subsection{Four-Zero Textures Confront the Sum Rules} \label{Sec:SumrulePlusGST}

Under the additional assumptions of symmetric or Hermitian mass matrices in the 1-2 block and zero textures in the 1-1 positions of the quark mass matrices, i.e.,
\begin{equation}\label{Eq:texture0}
M_u =
\begin{pmatrix}
0   & b_u & 0 \\
b_u & c_u & d_u \\
*   & *   & e_u
\end{pmatrix}
\quad \text{and} \quad M_d =
\begin{pmatrix}
0   & i b_d & 0 \\
\pm i b_d & c_d & d_d \\
*   & *   & e_d
\end{pmatrix},
\end{equation}
we obtain as additional predictions the GST relations \cite{Gatto:1968ss} with 1$\sigma$ errors displayed,
\begin{align}
\label{Eq:m_u_over_m_c}\theta_{12}^{u} &= \sqrt{\frac{m_u}{m_c}} = \left( 2.61^{+0.54}_{-0.46} \right)^\circ \;, \\
\label{Eq:m_d_over_m_s}\theta_{12}^{d} &= \sqrt{\frac{m_d}{m_s}} = \left( 13.2^{+3.4}_{-3.3} \right)^\circ\;.
\end{align}
Here we already see a conflict in the up sector. The prediction for $\theta_{12}^u$ from the sum rule in Eq.~\eqref{Eq:QuarkRelation} is quite different (several $\sigma$ away) from the GST relation above. That suggests that the texture in the up sector should be modified to be in good agreement with experiment. By contrast the prediction from the sum rule in Eq.~\eqref{Eq:QuarkSumRule} for $\theta_{12}^d$ is in very good agreement within the errors with the GST result in Eq.~\eqref{Eq:m_d_over_m_s}, and therefore it is quite plausible to keep the simple texture ansatz for the down sector.

Combining Eqs.~\eqref{Eq:m_u_over_m_c} and \eqref{Eq:m_d_over_m_s} with the sum rules in
Eqs.~\eqref{Eq:QuarkRelation} and \eqref{Eq:QuarkSumRule}, the two relations
\begin{equation}
\label{Eq:QSplusGST1}
\left|\theta_{12} -
\frac{\theta_{13}}{\theta_{23}} e^{- i\delta_\mathrm{CKM}} \right|
= \sqrt{\frac{m_d}{m_s}}
\end{equation}
and
\begin{equation}
\label{Eq:QSplusGST2}
\frac{\theta_{13}}{\theta_{23}}=\sqrt{\frac{m_u}{m_c}}
\end{equation}
emerge. We emphasise that these results do not hold for the textures in \cite{Fritzsch:1999rb} where the 2-3 up and down quark mixings are large and the 1-3 up and down quark mixings are non-negligible.

\begin{figure}
 \centering
 \includegraphics[scale=0.5]{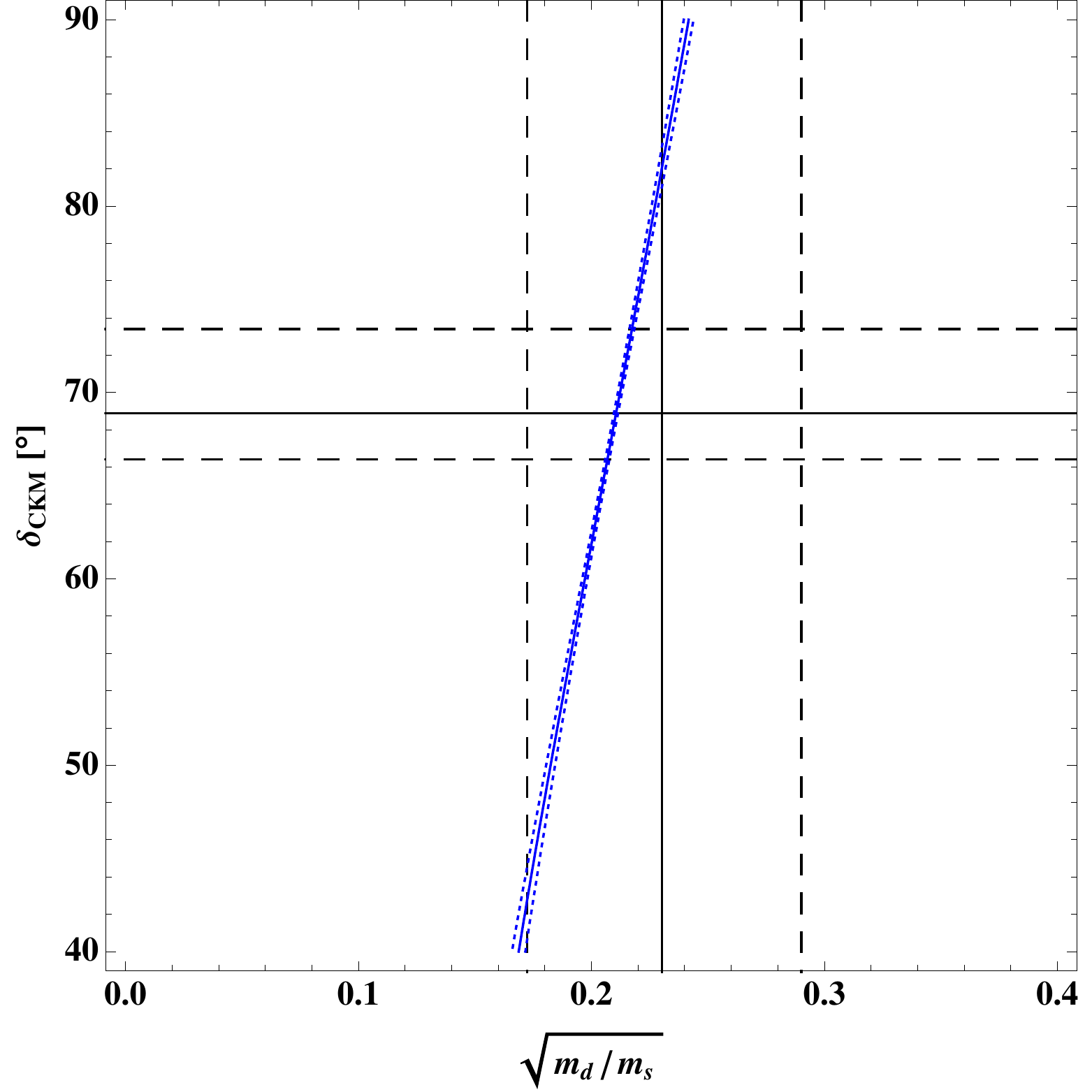}
 \caption[Relation between CKM Phase and Quark Masses]{Graphical illustration of the relation of Eq.~\eqref{Eq:QSplusGST1}.  The (dashed) blue lines indicate the predicted best-fit (1$\sigma$) values of  $\delta_\mathrm{CKM}$ for given $\sqrt{m_d/m_s}$  under the assumptions of Sec.~\ref{Sec:SumrulePlusGST}, and the dashed horizontal and vertical black lines (and solid black lines) show the $1\sigma$ errors (and best-fit values) for  $\delta_\mathrm{CKM}$ and $\sqrt{m_d/m_s}$, respectively. \label{Fig:dCKMvsMasses}}
\end{figure}

The compatibility of Eq.~\eqref{Eq:QSplusGST1} with the experimental results for the down-type quark masses and mixing parameters \cite{Amsler:2008zzb} is illustrated in Fig.~\ref{Fig:dCKMvsMasses}. We note that RG running for the quark masses, as well as their potential SUSY threshold corrections, are very similar for the first two generations and thus cancel out in their ratio. For our estimates, we have considered the running quark masses at the top mass scale $m_t (m_t)$ \cite{Xing:2007fb}. The CKM phase $\delta_\mathrm{CKM}$ is extracted for given $\sqrt{m_d/m_s}$. The solid blue line shows the relation for best-fit values of the parameters while the dashed blue lines indicate the range with $1\sigma$ errors included. The dashed horizontal and vertical black lines (and solid black lines) show the $1\sigma$ errors (and best-fit values) for $\delta_\mathrm{CKM}$ and $\sqrt{m_d/m_s}$, respectively. The relation of Eq.~\eqref{Eq:QSplusGST1} is well compatible with the present data. Future more precise experimental measurements (for instance at LHCb or $B$ factories) and, in particular, an improved knowledge on $m_d$, e.g.\  from lattice QCD, are required to test it more accurately.

In the following, we consider some examples of possible modifications to the textures in the up sector which are phenomenologically acceptable, while leaving the successful down sector texture unchanged, and retaining the successful real and imaginary scheme which leads to the right unitarity triangle. Then we discuss in Sec.~\ref{Sec:Nonzero13}, the idea of relaxing the up quark 1-3 texture zero which turns out to be disfavoured. Therefore we first restrict ourselves to either relaxing the up quark 1-1 texture zero, or relaxing symmetry in the 1-2 up quark sector.

\subsection{Relaxing the Up Quark 1-1 Texture Zero}

One possible modification to the four zero textures is to introduce a nonzero element in
the 1-1 position of the up quark mass matrix, i.e.
\begin{equation}
M_u =
\begin{pmatrix}
a_u   & b_u & 0 \\
b_u & c_u & d_u \\
*   & *   & e_u
\end{pmatrix}
\quad \text{and} \quad M_d =
\begin{pmatrix}
0   & i b_d & 0 \\
\pm i b_d & c_d & d_d \\
*   & *   & e_d
\end{pmatrix} \;.
\end{equation}
As a result, we obtain the up sector relation
\begin{equation}
m_u \approx a_u - \frac{b^2_u}{c_u}
\end{equation}
which allows to adjust $a_u$, which is of the order of the up quark mass, while $b_u/c_u \approx \theta_{12}^u$ has to be equal to the value obtained in Eq.~\eqref{Eq:QuarkRelation} using the sum rule. For the down sector, there is still the successful prediction from Eq.~\eqref{Eq:m_d_over_m_s} leading to the successful sum rule relation of Eq.~\eqref{Eq:QSplusGST1}. Furthermore, as discussed in Sec.~\ref{Sec:deltaCKM}, the Dirac phase of the CKM matrix is correct. We note that there exist several variants of the texture. For example, we can choose the 1-2 element of $M_d$ real, the 1-2 element of $M_u$ purely imaginary and all the other elements also real. These variants are valid as long as $b_u/c_u$ is real and $b_d/c_d$ is purely imaginary, or {\em vice versa}.

We emphasise that the elements marked by ``*'' are irrelevant as long as the hierarchy of the mass matrix is large enough, so they may be replaced by zeros or, if the matrices are Hermitian, the 3-1 elements may be zero while the 3-2 elements are determined by Hermiticity. Since the sum rule in Eq.~\eqref{Eq:F1} shows that $V_{cb}$ is determined only by the difference in 2-3 mixing angles in the
up and down sectors, it is also possible to set either $d_u$ or $d_d$ equal to zero without changing the physical predictions. In this way it is possible to arrive at some of the four-zero textures discussed, for example, in \cite{Chiu:2000gw}. However, we emphasise that here we are additionally assuming the real and imaginary structures consistent with the right unitarity triangle and this was not discussed in \cite{Chiu:2000gw}.

\subsection{Relaxing the Up Quark 1-2 Symmetry}

A second option for a texture consistent with experimental data consists in relaxing the symmetry of the 1-2 block in the up sector, while keeping the texture zero in the 1-1 position:
\begin{equation}
M_u =
\begin{pmatrix}
0   & b_u & 0 \\
b'_u & c_u & d_u \\
*   & *   & e_u
\end{pmatrix}
\quad \text{and} \quad
M_d =
\begin{pmatrix}
0   & i b_d & 0 \\
\pm i b_d & c_d & d_d \\
*   & *   & e_d
\end{pmatrix} \;.
\end{equation}
The two up-sector relations
\begin{equation}
m_u \approx b_u b'_u/c_u\;,\quad
m_c \approx c_u\;,\quad
\theta_{12}^u = \frac{b_u}{c_u}
\end{equation}
can be simultaneously fulfilled by choosing $b_u$, $b'_u$ and $c_u$ appropriately. The prediction from Eq.~\eqref{Eq:m_d_over_m_s} 
and the prediction for $\delta_{\rm CKM}$ do not change and remain compatible with data. We note that there exist several variants of the texture. As before, it is sufficient to have $b_u/c_u$ real and $b_d/c_d$  purely imaginary, or {\em vice versa}.

 \subsection{Textures with Nonzero 1-3 Elements} \label{Sec:Nonzero13}
 
With nonzero 1-3 elements, $\delta_{\mathrm{CKM}}$ depends not only on $\delta_{12}^{d} - \delta_{12}^{u}$ but also on other parameters (in particular $\delta_{13}^{u,d}$ and $\delta_{23}^{u,d}$) and the simple quark mixing sum rules in Eqs.~\eqref{Eq:QuarkRelation} and \eqref{Eq:QuarkSumRule} are no longer valid. Examples of this type of texture include (with real parameter $f_u$)
\begin{equation} \label{Eq:fuTexture1}
M_u =
\begin{pmatrix}
0   & b_u & f_u \\
b_u & c_u & d_u \\
*   & *   & e_u
\end{pmatrix}
\; , \quad
M_d =
\begin{pmatrix}
0   & i b_d & 0 \\
i b_d & c_d & d_d \\
*   & *   & e_d
\end{pmatrix},
\end{equation}
and
\begin{equation}
M_u =
\begin{pmatrix}
0   &  b_u & i f_u \\
 b_u & c_u & d_u \\
*   & *   & e_u
\end{pmatrix}
\; , \quad
M_d =
\begin{pmatrix}
0   &  i b_d & 0 \\
i b_d & c_d & d_d \\
*   & *   & e_d
\end{pmatrix},
\end{equation}
but also variations with different elements chosen either purely imaginary or real.

We demonstrate our approach for this case by means of the texture in Eq.~\eqref{Eq:fuTexture1}. The starting point is here (similar to Eqs.~\eqref{Eq:F1}-\eqref{Eq:F3}):
\begin{align} \label{Eq:G1}
{\theta_{23}}e^{-i\delta_{23}}&=
{\theta_{23}^{d}}e^{-i\delta_{23}^{d}}
-{\theta_{23}^{u}}e^{-i\delta_{23}^{u}}\;,
\\
\label{Eq:G2} {\theta_{13}}e^{-i\delta_{13}}&=
-{\theta_{13}^{u}}e^{-i\delta_{13}^{u}}
-{\theta_{12}^{u} \theta_{23}}e^{-i(\delta_{12}^{u} + \delta_{23})} 
\;,\\
\label{Eq:G3} {\theta_{12}}e^{-i\delta_{12}}&=
{\theta_{12}^{d}}e^{-i\delta_{12}^{d}}
-{\theta_{12}^{u}}e^{-i\delta_{12}^{u}} \;,
\end{align}
where we have also neglected terms of order ${\cal O}(\theta_{13} \theta_{ij})$. From our texture ansatz, we know the phases $\delta_{12}^{u/d}$, $\delta_{23}^{u/d}$ and $\delta_{13}^{u}$. For the values of $\theta_{12}^{u/d}$, we take the values from the GST relations, i.e.\ Eq.~\eqref{Eq:m_u_over_m_c}, which hold here because of the zeros in the 1-1 position and the symmetric structure for the first two generations. Then we can calculate $\theta_{13}^{u}$ and $\delta_{\rm CKM}$ in terms of the known quantities and obtain $\delta_{\rm CKM} = (78.83^{+3.62}_{-3.35})^\circ$. This result is several standard deviations away from the measurements.

Beyond the particular example discussed above, we found that the inconsistency of the prediction for $\delta_{\rm CKM}$ also appears in all other cases with $\delta_{23}$, $\delta_{12}^{u/d}$ and $\delta_{13}^{u/d}$ being either zero or $\pm \pi$. Furthermore, the same happens for textures with $f_u = 0$ and $f_d \neq 0$, where $f_d$ denotes the 1-3 element of $M_d$. We conclude that under these conditions textures with nonzero 1-3 elements are disfavoured.

\section{Quark-Lepton Mixing Relations} \label{Sec:QuarkLeptonMixingRelations}

Extending the notion of zero 1-3 mixing to the lepton sector (i.e.\ under the assumption of $\theta^\nu_{13} = \theta^e_{13} = 0$), the presently unknown mixing angle $\theta^{\mathrm{MNS}}_{13}$ of the leptonic MNS mixing matrix satisfies the relation (analogous to Eq.~\eqref{Eq:QuarkRelation})
\begin{equation}
\theta^{\mathrm{MNS}}_{13} = \sin \theta^{\mathrm{MNS}}_{23} \theta^e_{12} \;,
\end{equation}
where $\theta^e_{12}$ is the 1-2 mixing in the charged lepton mass matrix $M_e$. This relation has emerged before, for example, in the context of lepton sum rules in \cite{King:2005bj, Masina:2005hf, Antusch:2005kw}.

In many classes of GUT models of flavour, the 1-2 mixing angles corresponding to $M_e$ and $M_d$ are related by a group theoretical Clebsch factor, for example $\theta^d_{12} = 3 \theta^e_{12}$ \cite{Georgi:1979df}. In general, it is usually assumed that $\theta^d_{12}$ is of the order of the Cabibbo angle, leading to a prediction $\theta^{\mathrm{MNS}}_{13}\sim 3^\circ$ \cite{King:2005bj, Masina:2005hf, Antusch:2005kw, Antusch:2010es, Antusch:2010xx}. However, in the context of Fritzsch-type textures, which are based on Hermitian matrices with 1-1, 2-2 and 1-3 texture zeros, this prediction can be made more precise by using the sum rule which relates $\theta^d_{12}$ to down-type quark masses. Thus, applying Eq.~\eqref{Eq:QuarkSumRule} at low energies and taking the present experimental data for the quark mixing angles, and for $\theta^{\mathrm{MNS}}_{23}$ (taken from \cite{Schwetz:2008er}), one can make the rather precise prediction
\begin{equation}
\label{Eq:MNS13mixing}
\theta^{\mathrm{MNS}}_{13} = \left( 2.84^{+0.22}_{-0.18} \right)^\circ
\end{equation}
which gives $\sin^2 \theta^{\mathrm{MNS}}_{13} = 0.0025^{+0.0004}_{-0.0003}$ and holds under the assumption of texture zeros in the 1-3 elements of the mass matrices (or more precisely $\theta^u_{13} =\theta^d_{13} =\theta^\nu_{13} = \theta^e_{13} = 0$) and $\theta^d_{12} = 3 \theta^e_{12}$. Of course, Eq.~\eqref{Eq:MNS13mixing} is only a single example out of a larger variety of predictions which may arise in unified flavour models, see the chapters before or \cite{Antusch:2009gu}. We emphasise that the main use of Eq.~\eqref{Eq:QuarkSumRule} in this context is that it allows to ``determine'' the down quark mixing $\theta^d_{12}$, which is generically involved in relations between quark and lepton mixing angles, from measurable quantities.

We note that in the lepton sector, the RG corrections, see, e.g.\  \cite{Antusch:2003kp, Antusch:2005gp}, can be significant, depending on the absolute neutrino mass scale (and on $\tan \beta$ in a SUSY framework) and other effects, such as canonical normalisation on the mixing angles, can also be sizeable \cite{Antusch:2005kw, Antusch:2008yc, Boudjemaa:2008jf, *Antusch:2007vw, *Antusch:2007ib, *King:2007pr, Antusch:2007rk}. Furthermore, relaxing the 1-1 texture zero in the up quark sector may switch on a nonzero 1-3 mixing angle in the neutrino sector via partially constrained sequential dominance  \cite{King:1998jw, King:1999cm, *King:1999mb, *Antusch:2004gf, *Antusch:2010tf, King:2002nf, King:2005bj}.

%% file: kap_11_Model.tex
\chapter{A GUT Flavour Model for Small $\boldsymbol{\tan \beta}$} \label{Ch:Model}

In this chapter, based on \cite{Antusch:2010es}, we propose a type II upgrade \cite{Antusch:2004xd} model based on an $A_4$ family symmetry, amended by some discrete $\mathbb{Z}_3^2\times \mathbb{Z}_2^3$ symmetries with $SU(5)$ grand unification. The model proposed here is similar to the first flavour model we proposed, but there are also some substantial differences, like, e.g.\  the type II seesaw contribution in the neutrino sector. We predict TB neutrino mixing via CSD  \cite{King:1998jw, King:1999cm, *King:1999mb, *Antusch:2004gf, *Antusch:2010tf, King:2002nf, King:2005bj} like in Ch.~\ref{Ch:GUTImplications}.

It is interesting to ask in what class of theories would we learn the most about the neutrino mass scale from the discovery of neutrinoless double beta decay? Clearly the answer would be those theories which predict the effective neutrino mass parameter $m_{ee}$ uniquely as a function of the neutrino masses without ambiguities from unknown phases, but the next question is whether such theories do exist? Perhaps surprisingly the answer is in the affirmative, and, even more surprisingly, the class of theories which have this property turn out to uniquely specify the way that the seesaw mechanism is implemented in terms of a particular interplay between
the type I and type II seesaw mechanisms.

The effect of a type II upgrade unit matrix structure in the neutrino sector implies that for quasi-degenerate neutrino masses the Majorana CP phases are small and thus $m_{ee}\approx m_{\mathrm{lightest}}$. In general having quasi-degenerate or hierarchical neutrino masses does not lead to a sharp prediction for the neutrinoless double beta decay observable $m_{ee}$ as a function of the neutrino masses due to the presence of unknown phases in the neutrino mass matrix \cite{Feruglio:2002af}. Allowing for arbitrary Majorana phases and considering a quasi-degenerate neutrino mass spectrum and TB mixing, $m_{ee}$ can still be in the approximate interval $[m_{\mathrm{lightest}}/3,m_{\mathrm{lightest}}]$.
For the quarks and charged leptons further flavons are needed, misaligned to the neutrino flavons, to fit the observable masses and mixing angles. For all flavons we assume here the same vacuum alignment as in Ch.~\ref{Ch:GUTImplications}.
As expected, the type II upgrade model predicts the neutrinoless double beta decay mass observable to be approximately equal to the neutrino mass scale, independently of phases. 

In order for radiative corrections not to modify too much the TB mixing predictions for quasi-degenerate neutrinos \cite{Varzielas:2008jm}, we shall restrict ourselves to low values of $\tan \beta <1.5$. For such low $\tan \beta$, a viable GUT scale ratio of $y_\mu/y_s$ is achieved within SUSY $SU(5)$ GUTs using a CG factor of $9/2$, as proposed in the previous chapters. For the third generation we use $b$-$\tau$-Yukawa coupling unification $y_\tau/y_b = 1$ at the GUT scale which is viable for low $\tan \beta$, see, e.g.\ \cite{Ross:2007az}.

The layout of the remainder of this chapter is as follows. In Sec.~\ref{Sec:Symmetries2} we present the symmetries and the field content of the model and in Sec.~\ref{Sec:Superpotential2} its superpotential. Afterwards we perform a numerical fit to the quark and charged lepton masses and quark mixing angles and CP phase in Sec.~\ref{Sec:SMFit2}. We conclude this chapter with a discussion of the neutrino sector in Sec.~\ref{Sec:Neutrinos2}.

\section{Symmetries and Field Content} \label{Sec:Symmetries2}

\begin{table}
\centering
\begin{tabular}{cccccccc}
\toprule
& $SU(5)$ & $A_4$ & $\mathbb{Z}_2$ & $\mathbb{Z}'_3$ & $\mathbb{Z}'_2$ & $\mathbb{Z}_3$ & $\mathbb{Z}_2^{\mathrm{MP}}$\\
\midrule
\multicolumn{8}{l}{Chiral Matter}  \\
\midrule
$F$ &  $\mathbf{\overline{5}}$ & $\mathbf{3}$ & + & 0 & + & 0 & - \\
$T_1$, $T_2$, $T_3$  &  $\mathbf{10}$, $\mathbf{10}$, $\mathbf{10}$ &  $\mathbf{1}$, $\mathbf{1}$, $\mathbf{1}$ & +, +, - &  0, 1, 0 &  +, +, + &  1, 0, 0 & -, -, -  \\
$N_1$, $N_2$ &  $\mathbf{1}$, $\mathbf{1}$ &  $\mathbf{1}$, $\mathbf{1}$ & +, + &  0, 1 &  +, + &  1, 0  & -, -  \\
\midrule
\multicolumn{8}{l}{Flavons \& Higgs Multiplets}  \\
\midrule
$\phi_{23}$, $\phi_{123}$, $\phi_{3}$ & $\mathbf{1}$ & $\mathbf{3}$  & +, +, - & 0, 2, 0 & +, + ,+ & 2, 0, 0 & +, + ,+ \\
$\tilde{\phi}_{23}$ & $\mathbf{24}$ & $\mathbf{3}$  & + & 2 & - & 0   & +\\
$H_5$, $\bar{H}_5$ &  $\mathbf{5}$, $\mathbf{\overline{5}}$ & $\mathbf{1}$, $\mathbf{1}$ & +, + & 0, 0  & +, + & 0, 0 & +, +\\
$H_{15}$, $\bar{H}_{15}$ &  $\mathbf{15}$, $\mathbf{\overline{15}}$ &  $\mathbf{1}$, $\mathbf{1}$  & +, + &  0, 0 &  +, + & 0, 0 & +, +\\
$H_{45}$, $\bar{H}_{45}$ &  $\mathbf{45}$, $\mathbf{\overline{45}}$ &  $\mathbf{1}$, $\mathbf{1}$  & +, + &  2, 1 &  -, - & 0, 0 & +, +\\
\midrule
\multicolumn{8}{l}{Matter-like Messengers } \\
\midrule
$A_5$, $\bar{A}_5$ &  $\mathbf{5}$, $\mathbf{\overline{5}}$ & $\mathbf{1}$, $\mathbf{1}$ & +, + &   1, 2 & -, - &  0, 0 & -, - \\
$A_{10}$, $\bar{A}_{10}$ &  $\mathbf{10}$, $\mathbf{\overline{10}}$ & $\mathbf{3}$, $\mathbf{3}$ & +, + & 0, 0 &  +, + &  0, 0   & -, - \\
$A_{1}$ &  $\mathbf{1}$ & $\mathbf{3}$ & + &   0 &  + &  0   & -\\
\midrule
\multicolumn{8}{l}{Higgs-like Messengers }  \\
\midrule
$B$, $\bar{B}$ & $\mathbf{5}$, $\mathbf{\overline{5}}$  & $\mathbf{1}$, $\mathbf{1}$ & +, + & 1, 2 & +, + &  0, 0  & +, +  \\
$C_1$, $\bar{C}_1$ & $\mathbf{1}$, $\mathbf{1}$  & $\mathbf{1}$, $\mathbf{1}$ & +, + &  0, 0  & +, + &  2, 1 & +, +  \\
$C_2$, $\bar{C}_2$ & $\mathbf{1}$, $\mathbf{1}$  & $\mathbf{1}$, $\mathbf{1}$ & +, + &  2, 1  & +, + &  0, 0 & +, +  \\
\bottomrule
\end{tabular}
\caption[Representations and Charges in the $A_4 \times SU(5)$ Flavour Model]{Representations and charges of the superfields. The subscript $i$ on the fields $T_i$, $N_i$ and $C_i$ is a family index. The flavon fields $\phi_i$, $\tilde{\phi}_{23}$ can be associated to a family via their charges under $\mathbb{Z}_3^2\times \mathbb{Z}_2^3$.  The subscripts on the Higgs fields $H$, $\bar{H}$ and extra vector-like matter fields $A$, $\bar{A}$ denote the transformation properties under $SU(5)$. MP stands for matter parity. \label{Tab:Symmetries2}}
\end{table}

The full model is specified in Tab.~\ref{Tab:Symmetries2}. Compared to the previous model specified in Tab.~\ref{Tab:Symmetries1} the first thing to notice is that we use here only the discrete subgroup $A_4$ instead of the full $SO(3)$ symmetry group (for a brief description of $A_4$ see App.~\ref{App:A4}). Nevertheless, there is no principle difference between these two possibilities in our models since in both cases only the singlet and the real triplet representations appear which have the same transformation properties under $SO(3)$ and $A_4$. Furthermore, the additional symmetry is here $\mathbb{Z}_3^2\times \mathbb{Z}_2^3$ with one $\mathbb{Z}_2$ factor less than before.

The main difference between the two models lies in the field content, especially in the messenger sector and we have here no $H_{24}$ but additional Higgs fields in the 15-dimensional representations of $SU(5)$. They contain  $SU(2)_L$-triplet Higgs fields that obtain an induced vev after electroweak symmetry breaking. This induces a type II seesaw contribution to the neutrino mass matrix which is, to leading order, proportional to the unit matrix and can increase the neutrino mass scale without modifying the prediction for the leptonic mixing angles. There is another difference regarding the Higgs sector. We consider here instead of two additional five-dimensional Higgs representations $H_5'$ and $\bar{H}_5'$ two additional 45-dimensional Higgs representations $H_{45}$ and $\bar{H}_{45}$ which have some influence on the CG factors as we discuss later.

The matter sector, namely the fields $F_i$, $T_i$ and $N_i$ are the same as before. 

$SU(5)$ is spontaneously broken by the vev of the $\tilde{\phi}_{23}$ field, electroweak symmetry is broken by the vevs of the Higgs fields $H_5$, $\bar H_5$, $H_{45}$, $\bar H_{45}$ and $A_4$ is spontaneously broken by the vevs of the flavon fields, i.e.\ the family symmetry breaking Higgs fields, $\phi_{123}$, $\phi_{23}$, $\phi_{3}$ and  $\tilde{\phi}_{23}$.

The heavy messenger fields are specified in Tab.~\ref{Tab:Symmetries2} and give rise to higher-dimensional operators generating the Yukawa coupling matrices as well as the mass matrix of the gauge singlet (right-handed) neutrinos $N_i$ after effectively integrating them out of the theory. 

We would like to remark that we do not explicitly consider the full flavour and GUT Higgs sector of the model and just assume that the $SU(5)$ and $A_4$ breaking vevs are aligned in the desired directions of field space. We assume that in these sectors issues like doublet-triplet splitting are resolved. Without specifying these sectors, a reliable calculation of the proton decay rate is not possible and beyond the scope of the present thesis. The focus of this chapter is to illustrate that quasi-degenerate light neutrino masses can be realised via a type II upgrade in a $SU(5)$ GUT framework.

There is an additional contribution to the neutrino mass matrix proportional to the unit matrix from the messenger field $A_1$ which is a singlet under $SU(5)$ and a triplet under $A_4$. When it is integrated out, it induces a contribution to the neutrino mass operator which is proportional to the unit matrix like the type II upgrade part.

\addtocounter{section}{1}
\sectionmark{The $\boldsymbol{A_4\times SU(5)}$ Symmetric Superpotential}
\addtocounter{section}{-1}
\section[The $A_4\times SU(5)$ Symmetric Superpotential]{The $\boldsymbol{A_4\times SU(5)}$ Symmetric Superpotential} \label{Sec:Superpotential2}
\sectionmark{The $\boldsymbol{A_4\times SU(5)}$ Symmetric Superpotential}

With the field content and symmetries specified in Tab.~\ref{Tab:Symmetries2} the superpotential contains the following renormalisable terms:
\begin{align}
W_{H} &= \mu_{5} H_5 \bar{H}_5 + \mu_{15} H_{15} \bar{H}_{15} + \mu_{45} H_{45} \bar{H}_{45}+ \bar{\lambda}_{15} \bar{H}_{15}  H_5 H_5 + \lambda_{15} H_{15} \bar{H}_5 \bar{H}_5 \;, \label{Eq:mu2} \\
W_A &= M_{A_{10}} A_{10} \bar{A}_{10} + M_{A_5} A_5 \bar{A}_5 + M_{A_1} A^2_1 + M_B B \bar{B}  + M_{C_1} \bar{C}_1 C_1 + M_{C_2} \bar{C}_2 C_2 \label{Eq:Xmass2} \\
W_{\mathrm{int}} &=  \kappa_{Fi} F \phi_i A_{10} + \tilde{\kappa}_{F2} F \tilde{\phi}_{23} A_5 + \kappa_{Ti} T_i \bar{H}_5 \bar{A}_{10} + \tilde{\kappa}_{T2} T_2 \bar{H}_{45} \bar{A}_5 + y_\Delta H_{15} F F \nonumber\\
& + \lambda H_5 A_{10} \bar{A}_{10} + \tilde{\lambda} H_5 \bar{C}_2 \bar{B} + \kappa'_T C_2 T_2^2 + \kappa'_{\phi} C_2 \phi_{23}^2 + \tilde{\kappa}'_{\phi} \bar{C}_2 \tilde{\phi}_{23}^2 + a_t H_5 T_3^2 \nonumber\\
& + \bar{\kappa}_{F} F \bar{H}_5 A_1 + \kappa_{N i} N_i \phi_i A_1 + \xi_{\phi i} C_i \phi_i^2 + \xi_{N i} \bar{C}_i N_i^2 \label{Eq:Wint2} \;.
\end{align}

\begin{figure}
\centering
\includegraphics[scale=0.8]{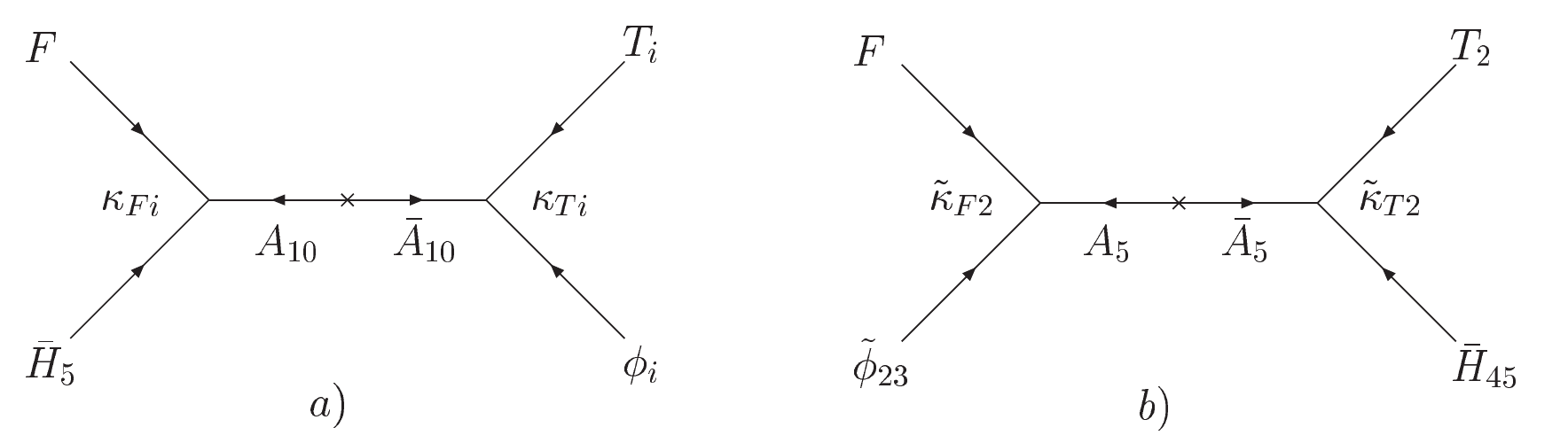}
\caption[Supergraphs for Down-type Fermion Masses in the $A_4 \times SU(5)$ Model]{Supergraph diagrams inducing effective superpotential operators for the down-type quarks and charged leptons. \label{Fig:messenger2_d}}
\end{figure}

\begin{figure}
\centering
\includegraphics[scale=0.7]{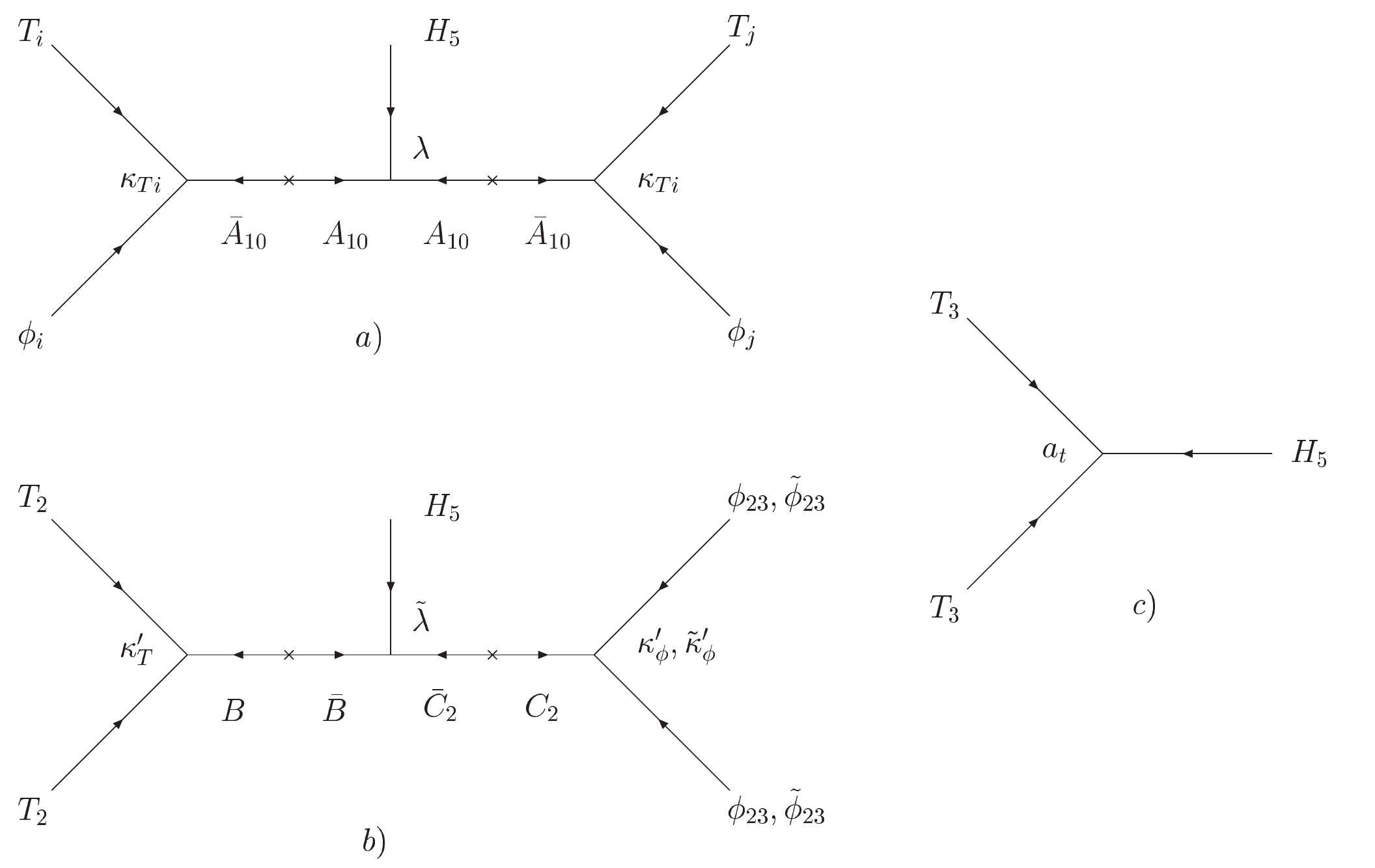}
\caption[Supergraphs for Up-type Quark Masses in the $A_4 \times SU(5)$ Model]{Supergraph diagrams inducing effective superpotential operators for the up-type quarks. \label{Fig:messenger2_u}}
\end{figure}

\begin{figure}
\centering
\includegraphics[scale=0.7]{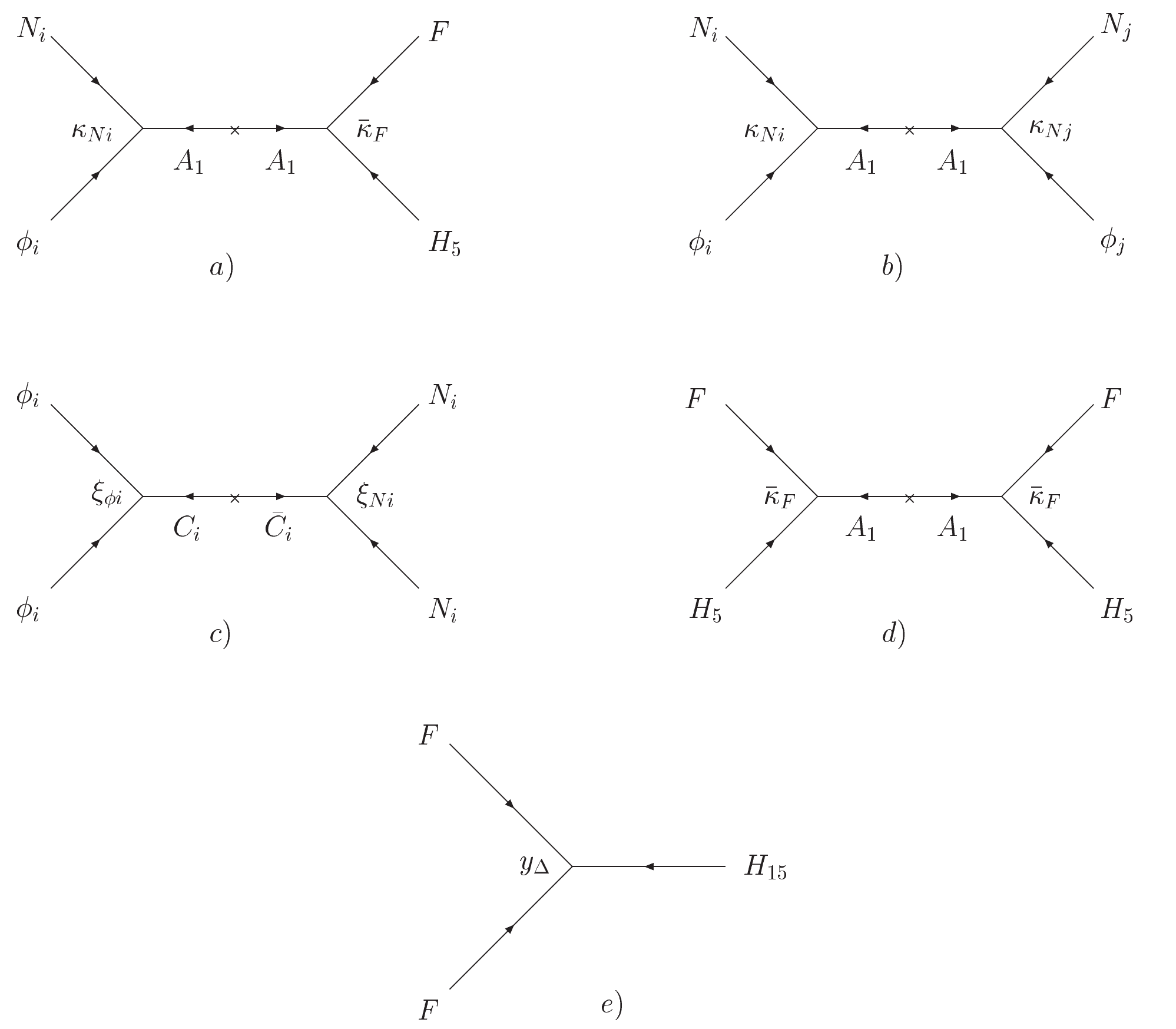}
\caption[Supergraphs for Neutrino Masses in the $A_4 \times SU(5)$ Model]{Supergraph diagrams inducing effective superpotential operators for the neutrino sector. \label{Fig:messenger2_n}}
\end{figure}

Integrating out the heavy messenger superfields denoted by $A_i$, $B$, $C_i$ and their conjugates, the Feynman diagrams in Figs.~\ref{Fig:messenger2_d}, \ref{Fig:messenger2_u} and \ref{Fig:messenger2_n} lead to the following effective non-renormalisable superpotential terms in the $SU(5)$ and $A_4$ unbroken phase:
\begin{align}
W_{Y_l} &= \frac{1}{M_{A_{10}}} F \left( b_1 \phi_{23} T_1 + b_2 \phi_{123} T_2 + b_3 \phi_{3} T_3 \right) \bar{H}_5 + \frac{\tilde{b}_2}{M_{A_5}} F \tilde{\phi}_{23} T_2 \bar{H}_{45} \:,\label{Eq:Yl2} \\
W_{Y_u} &= \left( \frac{a_{12}}{M_{A_{10}}^2} T_1 T_2 (\phi_{123} \cdot \phi_{23}) + \frac{a_{13}}{M_{A_{10}}^2} T_1 T_3 (\phi_3 \cdot \phi_{23}) + \frac{a_{23}}{M_{A_{10}}^2} T_2 T_3 (\phi_{123} \cdot \phi_3)    \right) H_5\:, \nonumber\\
& + \left( a_{33} T_3^2 + \frac{a_{22}}{M_{A_{10}}^2} T_2^2 \phi_{123}^2 + \frac{a_{11}}{M_{A_{10}}^2} T_1^2 \phi_{23}^2 + \frac{\tilde{a}_{22}}{M_{A_5}^2} T_2^2 \tilde{\phi}_{23}^2   \right) H_5 \label{Eq:Yu2} \:,\\
W_{Y_\nu} &= \frac{1}{M_{A_{10}}} F \left( a_{\nu_1} \phi_{23} N_1 + a_{\nu_2} \phi_{123} N_2 \right) H_5 \:, \label{Eq:Ynu2} \\
W_{\nu}^{\Delta} &= y_\Delta H_{15} F F\:, \label{Eq:Delta2} \\
W_{\nu}^{d=5} &= \frac{\bar{\kappa}_F^2}{M_{A_1}} F H_5 F H_5 \:, \label{Eq:d=5} \\
W_{\nu}^{M_R} &= \frac{a_{R_{11}}}{M_{A_{10}}^2} \phi_{23}^2 N_1^2 + \frac{a_{R_{22}}}{M_{A_{10}}^2} \phi_{123}^2 N_2^2 + \frac{a_{R_{12}}}{M_{A_{10}}^2} \left( \phi_{123} \cdot \phi_{23} \right) N_1 N_2 \:.\label{Eq:MR2}
\end{align}

After GUT symmetry breaking, the $SU(2)_L$ doublet components from $H_5$ and $H_{45}$ as well as $\bar{H}_5$ and $\bar{H}_{45}$ respectively mix and only the light states acquire the $SU(2)_L$ breaking vevs which give the fermion masses. We parameterise the Higgs mixing with the mixing angles $\gamma$ and  $\bar{\gamma}$ respectively
\begin{equation}
\begin{split}
\begin{pmatrix} H_5 \\ H_{45} \end{pmatrix} &= \begin{pmatrix} c_\gamma & - s_\gamma \\  s_\gamma & c_\gamma \end{pmatrix} \begin{pmatrix} H_l \\ H_{h} \end{pmatrix} \;, \\
\begin{pmatrix} \bar{H}_{5} \\ \bar{H}_{45} \end{pmatrix} &= \begin{pmatrix} c_{\bar{\gamma}} & - s_{\bar{\gamma}} \\  s_{\bar{\gamma}} & c_{\bar{\gamma}} \end{pmatrix} \begin{pmatrix} \bar{H}_l \\ \bar{H}_{h} \end{pmatrix}  \;,
\end{split}
\end{equation}
where we have used the common abbreviations $c_\gamma \equiv \cos \gamma$ and $s_\gamma \equiv \sin \gamma$ and similar for the other angle $\bar{\gamma}$. The light Higgs doublets are denoted with an index $l$ while the heavy Higgs doublets are denoted with an index $h$. We use here the same labels for the Higgs mixing angles and other quantities as in Ch.~\ref{Ch:GUTImplications}. Nevertheless from the context it should be clear which quantity belongs to which model.

The effective couplings $a$ and $b$ appearing in the effective superpotential can be expressed in terms of the fundamental couplings from Eq.~\eqref{Eq:Wint2}, the messenger masses from \eqref{Eq:Xmass2} and the Higgs mixing angles as
\begin{gather}
a_{11} = \frac{c_{\bar{\gamma}}^2 }{c_\gamma} \frac{\lambda}{\kappa_{F1}^2 } \;, \quad a_{22} = \frac{c_{\bar{\gamma}}^2 }{c_\gamma} \frac{\lambda}{\kappa_{F2}^2 } + \frac{c_{\bar{\gamma}}^2 }{c_\gamma} \frac{M_{A_5}^2}{M_{B} M_{C_2}} \frac{ \kappa'_T \tilde{\lambda} \kappa'_{\phi} }{\kappa_{F2}^2 \kappa_{T2}^2 } \;, \quad a_{33} = \frac{a_t}{c_{\gamma}} + \frac{c_{\bar{\gamma}}^2 }{c_\gamma} \frac{\lambda}{\kappa_{F3} \kappa_{F3} } \;,\\
\tilde{a}_{22} = \frac{s_{\bar{\gamma}}^2 }{c_\gamma} \frac{\lambda }{\tilde{\kappa}_{F2}^2 } + \frac{s_{\bar{\gamma}}^2 }{c_\gamma} \frac{M_{A_5}^2}{M_{B} M_{C_2}} \frac{\kappa'_T \tilde{\lambda} \tilde{\kappa}'_{\phi} }{\tilde{\kappa}_{F2}^2 \tilde{\kappa}_{T2}^2 } \;, \quad  \quad a_{ij} = \frac{c_{\bar{\gamma}}^2 }{c_\gamma} \frac{\lambda}{\kappa_{Fi} \kappa_{Fj} } \; \; \text{for $i \neq j$} \;,
\end{gather}
\begin{equation}
\label{Eq:bis2}
b_i = \frac{\kappa_{Fi} \kappa_{Ti}}{c_{\bar{\gamma}}} \;, \quad \tilde{b}_2 = \frac{\tilde{\kappa}_{Fi} \tilde{\kappa}_{Ti}}{s_{\bar{\gamma}}} \;,
\end{equation}
\begin{equation}
a_{\nu_i} = \frac{c_{\bar{\gamma}}}{c_\gamma} \frac{M_{A_{10}}}{M_{A_1}} \frac{\kappa_{Ni} \bar{\kappa}_F}{\kappa_{Fi} \kappa_{Ti}} \;,
\end{equation}
\begin{equation}
\label{Eq:aRs2}
a_{R_i}= c_{\bar{\gamma}}^2 \frac{M_{A_{10}}^2}{M_{A_1}} \frac{\kappa_{Ni}^2}{\kappa_{Fi}^2 \kappa_{Ti}^2} + c_{\bar{\gamma}}^2 \frac{M_{A_{10}}^2}{M_{C_i}} \frac{\xi_{\phi i} \xi_{N i} }{\kappa_{Fi}^2 \kappa_{Ti}^2}   \;,\quad a_{R_{12}} = c_{\bar{\gamma}}^2 \frac{M_{A_{10}}^2}{M_{A_1}} \frac{\kappa_{N1} \kappa_{N2}}{\kappa_{F1} \kappa_{T1} \kappa_{F2} \kappa_{T2}} \;.
\end{equation}

We assume a vacuum alignment for the $A_4$ breaking flavons very similar to Eqs.~\eqref{Eq:vacuumalignment} and \eqref{Eq:vacuumalignment2}. There is only one difference due to the different messenger sector, resulting in different coefficients. Explicitly we make the ansatz
\begin{equation} \label{Eq:VacAlign}
b_1 \frac{\langle \phi_{23} \rangle }{M_{A_{10}}}  = \begin{pmatrix}  0 \\  1\\  - 1  \end{pmatrix} \epsilon_{23} \;, \quad
b_2 \frac{\langle \phi_{123} \rangle }{M_{A_{10}}}  = \begin{pmatrix} 1 \\  1 \\ 1  \end{pmatrix} \epsilon_{123} \;, \quad
b_3 \frac{\langle \phi_{3} \rangle }{M_{A_{10}}}  = \begin{pmatrix}  0 \\  0\\  1  \end{pmatrix} \epsilon_{3} \;,
\end{equation}
which also defines the quantities $\epsilon_{123}$, $\epsilon_{23}$ and $\epsilon_{3}$. The breaking of $A_4$ along the field directions of $\phi_{123}$ and $\phi_{23}$  allows us to realise TB neutrino mixing via CSD  \cite{King:1998jw, King:1999cm, *King:1999mb, *Antusch:2004gf, *Antusch:2010tf, King:2002nf, King:2005bj}. It is also worth noting that the flavon vevs $\langle \phi_{123} \rangle$ and $\langle \phi_{23} \rangle$
are orthogonal, causing some of the terms in the superpotential to give a vanishing contribution to the mass matrices.
In the following, we assume that CP is only broken spontaneously by the vev of the flavon $\tilde{\phi}_{23}$.  For the flavon $\tilde{\phi}_{23}$ we use the ansatz
\begin{equation} \label{Eq:VacAlign2}
\tilde{b}_2 \frac{\langle \tilde\phi_{23} \rangle }{M_{A_5}}  = \begin{pmatrix}  0 \\  - \ci \\  w  \end{pmatrix} \tilde{\epsilon}_{23} \;,
\end{equation}
which is again motivated empirically.

\section{Numerical Fit to the SM Fermion Masses and Mixings} \label{Sec:SMFit2}

The convention for the Yukawa matrices used here is fixed in \eqref{Eq:GutYukConvention}.
From Eqs.~\eqref{Eq:Yl2}, \eqref{Eq:Yu2}, \eqref{Eq:VacAlign} and \eqref{Eq:VacAlign2} the Yukawa matrix coupling the up-type quarks to the light up-type Higgs doublet is given as
\begin{align}
Y_u &= \begin{pmatrix} 2 a_{11} \epsilon^2_{23} & 0 &  a_{13} \epsilon_{23} \epsilon_3 \\ 0 & 3 a_{22} \epsilon^2_{123} +  (w^2-1) \tilde{a}_{22}^2 \tilde{\epsilon}^2_{23} & a_{23} \epsilon_{123} \epsilon_3 \\  a_{13} \epsilon_{23} \epsilon_3 &  a_{23} \epsilon_{123} \epsilon_{3} & a_{33}  \end{pmatrix}, \label{Eq:GUTYu2}
\end{align}
whereas the Yukawa matrices coupling the down-type quarks and charged leptons to the light down-type Higgs doublet are given as
\begin{align}
Y_d &= \begin{pmatrix} 0 & \epsilon_{23} & - \epsilon_{23} \\ \epsilon_{123} & \epsilon_{123} + \ci \, \tilde{\epsilon}_{23} & \epsilon_{123} + w \tilde{\epsilon}_{23} \\ 0 & 0 & \epsilon_{3} \end{pmatrix}, \label{Eq:GUTYd2}\\
Y_e^T &= \begin{pmatrix} 0 & c_{23} \epsilon_{23} & - c_{23} \epsilon_{23} \\ c_{123} \epsilon_{123} & c_{123} \epsilon_{123} + \ci \, \tilde{c}_{23} \tilde{\epsilon}_{23} & c_{123} \epsilon_{123} + w \tilde{c}_{23} \tilde{\epsilon}_{23} \\ 0 & 0 & c_3 \epsilon_{3} \end{pmatrix} , \label{Eq:GUTYe2}
\end{align}
where $c_{3}$, $c_{23}$, $\tilde c_{23}$ and $c_{123}$ are the CG factors arising from GUT symmetry breaking, see Ch.~\ref{Ch:GUTYukawaRelations}. In $Y_u$ we have used the orthogonality of the flavon vevs $\langle \phi_{23} \rangle$ and $\langle \phi_{123} \rangle$. We note that in the definition for the Yukawa matrices we have introduced a complex conjugation and thus a phase factor of $+\mathrm{i}$ appears in the 2-2 elements of $Y_d$ and $Y_e$.

With the given representations of the flavon and Higgs fields, see Tab.~\ref{Tab:Symmetries2}, we obtain the following CG coefficients
\begin{equation}
\quad c_{123} = 1 \;,  \quad c_{23} = 1 \;, \quad c_3 = 1 \;, \quad \tilde{c}_{23} = 9/2 \;.
\end{equation}
For such small values of $\tan \beta$ as we consider, the 1-loop SUSY threshold corrections are small and, taking the actual  experimental values of the strange quark and muon masses into account, the GUT scale value of $y_\mu / y_s$ prefers $\tilde{c}_{23} = 9/2$, as argued in Chs.~\ref{Ch:AFirstGlance} and \ref{Ch:Pheno}.

The texture in Eq.~\eqref{Eq:GUTYd2} is pretty similar to the ones suggested in the last chapter. We have here also all matrix elements purely real apart from the 2-2 element which is purely imaginary. Nevertheless, we cannot apply here the sum rules derived in the last chapter since the 1-3 mixing in the up- and down-sector does not vanish. Therefore, although this choice of phases is inspired by the previous considerations it is not clear beforehand that the textures appearing in this model are successful in describing the quark CP phase. However,  we show now that this choice is meaningful.

\begin{table}
\begin{center}
\begin{tabular}{cc}
\toprule
Parameter & Value \\ \midrule
$2 a_{11} \epsilon_{23}^2 $ in $10^{-6}$ & $9.62$ \\
$3 a_{22} \epsilon_{123}^2$ in $10^{-4}$ & $-1.10$ \\
$(w^2 - 1) \tilde{a}_{22} \tilde{\epsilon}_{23}^2$ in $10^{-3}$ & $-1.10$ \\
$a_{13} \epsilon_{23} \epsilon_3$ in $10^{-3}$  & $-2.92$ \\
$a_{23} \epsilon_{123} \epsilon_3$ in $10^{-2}$ & $3.21$ \\
$a_{33}$ & $2.44$ \\ \midrule
$\epsilon_{123}$ in $10^{-5}$ & $5.88$ \\
$\epsilon_{23}$ in $10^{-5}$ & $4.30$ \\
$\tilde{\epsilon}_{23}$ in $10^{-4}$ & $-1.61$ \\
$\epsilon_{3}$ in $10^{-2}$ & $1.12$ \\
$w$ & $1.44$ \\
\bottomrule
\end{tabular}
\caption[Fitted Parameters for the $A_4 \times SU(5)$ Model]{The model parameters for $\tan \beta = 1.4$ and $M_{\mathrm{SUSY}} = 500$ GeV from a fit to the experimental data.
\label{Tab:Parameters2}}
\end{center}
\end{table}

\begin{table}
\begin{center}
\begin{tabular}{cccc}
\toprule
Quantity (at $m_t(m_t)$) & Model & Experiment & Deviation  \\ \midrule
$y_\tau$ in $10^{-2}$ & 1.00 & 1.00 & $-0.027$\% \\
$y_\mu$ in $10^{-4}$  & 5.89 & 5.89 & $-0.029$\% \\
$y_e$ in $10^{-6}$ & 2.79 & 2.79 & $-0.130$\% \\  \midrule
$y_b$ in $10^{-2}$ & 1.58 & $1.58 \pm 0.05$ & $0.086 \sigma $  \\
$y_s$ in $10^{-4}$ & 2.83 & $2.99 \pm 0.86$ & $- 0.184 \sigma $ \\[0.1pc]
$y_d$ in $10^{-6}$ & 27.6 & $15.9^{+6.8}_{-6.6}$ & $1.723 \sigma$   \\  \midrule
$y_t$ & 0.938 & $0.936 \pm 0.016$ & $0.084 \sigma $   \\
$y_c$ in $10^{-3}$ & 3.54 & $3.39 \pm 0.46$ & $0.318 \sigma$   \\[0.1pc]
$y_u$ in $10^{-6}$ & 6.70 & $7.01^{+2.76}_{-2.30}$ & $- 0.134 \sigma$  \\ \midrule
$\theta_{12}^{\mathrm{CKM}}$ & 0.2257 & $0.2257^{+0.0009}_{-0.0010}$ & $- 0.022 \sigma $  \\[0.3pc]
$\theta_{23}^{\mathrm{CKM}}$ & 0.0413 & $0.0415^{+0.0011}_{-0.0012}$ & $0.004 \sigma $ \\[0.1pc]
$\theta_{13}^{\mathrm{CKM}}$ & 0.0036 & $0.0036 \pm 0.0002$  & $-0.157 \sigma $  \\[0.1pc]
$\delta_{\mathrm{CKM}}$ & 1.1782 & $1.2023^{+0.0786}_{-0.0431}$  & $-0.560 \sigma $  \\
\bottomrule
\end{tabular}
\caption[Fit Results for the $A_4 \times SU(5)$ Model]{Fit results for the quark Yukawa couplings and mixing and the charged lepton Yukawa couplings at low energy compared to experimental data. A pictorial representation of the agreement between our predictions and experiment can also be found in Fig.~\ref{Fig:FitResultsPlot2}. \label{Tab:FitResults2}}
\end{center}
\end{table}

From the charged lepton Yukawa matrix we can derive the following approximate relations for the eigenvalues
\begin{equation}
y_\tau \approx c_3 \epsilon_3  \;, \quad y_\mu \approx \vert c_{123} \epsilon_{123} + \ci \, \tilde{c}_{23} \tilde{\epsilon}_{23} \vert \;, \quad y_e \approx \frac{c_{23} \epsilon_{23} c_{123} \epsilon_{123} }{y_\mu } \;.
\end{equation}
Furthermore, since there is no 1-2 mixing from the up-sector, the mixing angle $\theta_{12}$ is approximately given as
\begin{equation}
\theta^{\mathrm{CKM}}_{12} \approx \left\vert \frac{\epsilon_{23} }{\epsilon_{123} + \ci \, \tilde{\epsilon}_{23}} \right\vert \;.
\end{equation}
From these four equations the four $\epsilon$'s can be calculated and the relation for the CKM phase gives at the GUT scale
\begin{equation}
\vert \tan \delta_{\mathrm{CKM}} \vert \approx \left\vert \frac{\tilde{\epsilon}_{23}}{\epsilon_{123}} \right\vert \approx 1.22 \;. \label{Eq:deltaCKM2}
\end{equation}
The RG evolution of the measured value for $\delta_{\mathrm{CKM}}$ gives a GUT scale value of $1.20$. So the prediction for the CKM phase is already remarkably good if we only take the lepton masses and the value for $\theta_{12}$ into account which are measured to a high accuracy.\footnote{We would like to remark that with the assumed spontaneous CP violation, real $\det Y_u$ and $\det Y_d$ and with the small $|\tilde{\epsilon}_{23}| = {\cal O}(10^{-4})$, the model might also provide a solution to the strong CP problem, along the lines discussed in \cite{Nelson:1983zb, *Barr:1984fh}.}

\begin{figure}
\centering
\includegraphics[scale=0.65]{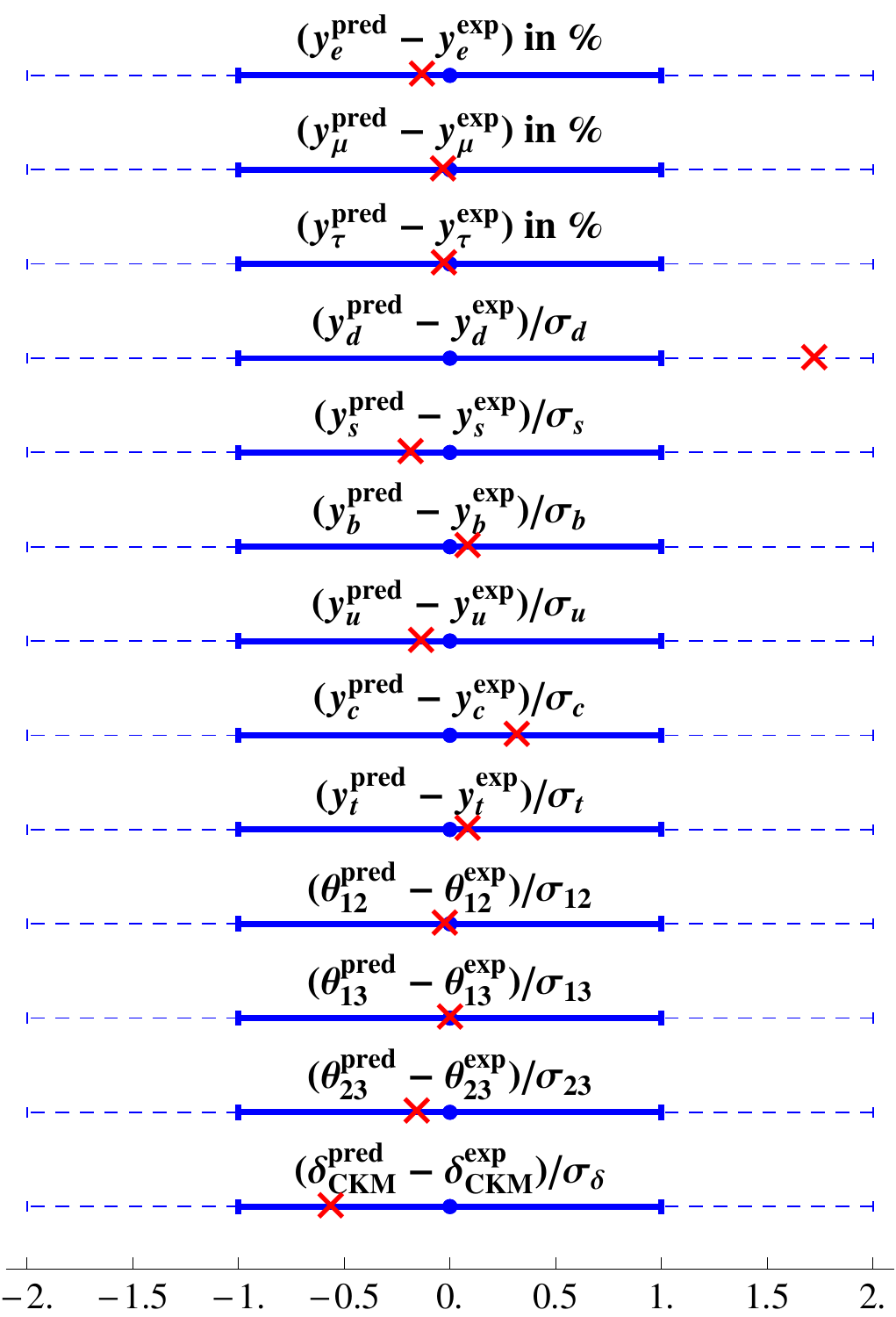}
\caption[Fit Results for the $A_4 \times SU(5)$ Model]{Pictorial representation of the deviation of our predictions from low energy experimental data for the charged lepton Yukawa couplings and quark Yukawa couplings and mixing parameters.  The deviations of the charged lepton masses are given in~\% while all other deviations are given in units of standard deviations $\sigma$. The thick blue line gives the 1\% (1$\sigma$)  bound while the dashed line gives the 2\% (2$\sigma$) bound. The red crosses denote our predictions. \label{Fig:FitResultsPlot2} }
\end{figure}

For the detailed fit of the model to the data we applied the following procedure: We have taken the GUT scale Yukawa matrices from Eqs.~\eqref{Eq:GUTYu2}, \eqref{Eq:GUTYd2} and \eqref{Eq:GUTYe2} and calculated their RG evolution down to the scale $m_t(m_t)$ for $\tan \beta = 1.4$ \footnote{We note that in the MSSM small values of $\tan \beta$ are somewhat constrained due to bounds on the Higgs mass. However, we emphasise that our model may well be formulated in the context of the NMSSM or other non-minimal SUSY models where $\tan \beta$ of order one can readily be realised without these constraints.} and $M_{\mathrm{SUSY}} = 500$~GeV with the REAP software package \cite{Antusch:2005gp}. At the low scale we performed a $\chi^2$ fit to the quark masses and mixing and charged lepton masses depending on the parameters of the GUT scale Yukawa matrices. The fit gave a total $\chi^2$ of about 3.5 where we have assumed a relative error of 1\% for the charged lepton masses and for the other observables we have taken the experimental errors.  Since we have 11 parameters and 13 observables this corresponds to a $\chi^2/\mathrm{dof}$ of about 1.6. This is a good fit since we have neglected theoretical uncertainties like, e.g.\  threshold corrections which could be treated as additional errors on the data lowering the total $\chi^2$.

The results for the GUT scale parameters are listed in Tab.~\ref{Tab:Parameters2}. We remark that these parameters depend on $\tan \beta$ and $M_{\mathrm{SUSY}}$ and are also subject to several theoretical uncertainties. For example, we note that the Higgs fields $H_{15}$ and $\bar{H}_{15}$ containing the Higgs triplets of the type II seesaw mechanism have masses at an intermediate energy scale between $M_{\mathrm{GUT}}$ and $M_{\mathrm{EW}}$. Their effects are not included in the RG analysis. The effects are small and may be neglected if $y_{\Delta}$ is small, but they could be sizeable if $y_{\Delta}$ is large.\footnote{Since the coupling $y_{\Delta}$ gives a contribution proportional to the unit matrix, it affects only  the RG evolution of the mass eigenvalues, but not of the mixing angles. Nevertheless, the possibility of additional RG effects from $y_{\Delta}$ provides a theoretical uncertainty in our setup.} Due to the additional theoretical uncertainties, we do not explicitly give the errors on the high energy parameters or low energy predictions. The important input parameters for us are the charged lepton masses and quark mixing angles which have a experimental error much smaller than these uncertainties.

In Tab.~\ref{Tab:FitResults2} the low energy results are shown and compared to experimental data. A graphical illustration is given in Fig.~\ref{Fig:FitResultsPlot2}. They illustrate that our minimal example model, with the assumed vacuum alignment of Eqs.~\eqref{Eq:VacAlign} and \eqref{Eq:VacAlign2}, can fit the data well and leads to testable predictions. We turn now to the results for the neutrino sector.

\section{The Neutrino Sector} \label{Sec:Neutrinos2}

The neutrino Yukawa matrix is obtained from Eq.~\eqref{Eq:Ynu2} as
\begin{align}
Y_{\nu} &= \begin{pmatrix} 0 &  a_{\nu_2} \epsilon_{123} \\  a_{\nu_1} \epsilon_{23} &  a_{\nu_2} \epsilon_{123} \\ -  a_{\nu_1} \epsilon_{23} &  a_{\nu_2} \epsilon_{123}   \end{pmatrix} \;.
\end{align}
Additionally we have a diagonal mass matrix for the two right-handed neutrinos from Eq.~\eqref{Eq:MR2},
\begin{equation}
M_R = \begin{pmatrix} 2 a_{R_1} \epsilon_{23}^2 & 0 \\ 0 & 3 a_{R_2} \epsilon_{123}^2  \end{pmatrix} \;,
\end{equation}
and a diagonal type II seesaw contribution coming from Eqs.~\eqref{Eq:Delta2} and \eqref{Eq:d=5},
\begin{equation}
M_L = \begin{pmatrix} m_0 & 0 &0 \\ 0 & m_0 & 0 \\ 0 & 0 & m_0 \end{pmatrix} \;.
\end{equation}

Using the seesaw relation,
\begin{equation}
m_\nu = M_L - v_{u}^2 Y_\nu M_R^{-1} Y_\nu^T \;,
\end{equation}
we obtain for the neutrino mass matrix
\begin{equation}
m_\nu =
m_0 \begin{pmatrix} 1 & 0 &0 \\ 0 & 1 & 0 \\ 0 & 0 & 1 \end{pmatrix}
+
\frac{m^{I}_3}{2} \begin{pmatrix} 0 & 0 &0 \\ 0 & 1 & -1 \\ 0 & -1 & 1 \end{pmatrix}
+
\frac{m^{I}_2}{3} \begin{pmatrix} 1 & 1 &1 \\ 1 & 1 & 1 \\ 1 & 1 & 1 \end{pmatrix} \;,
\end{equation}
with
\begin{equation}
m_0 = c_\gamma^2 v_u^2 \frac{y_{\Delta} \bar{\lambda}_{15}}{\mu_{15}} + c_\gamma^2 v_u^2 \frac{\bar{\kappa}_F^2}{M_{A_1}} \;, \quad
m^{I}_2 = -  v_u^2 \frac{a_{\nu_2}^2}{a_{R_2}}  \quad \text{and} \quad
m^{I}_3 = -  v_u^2 \frac{a_{\nu_1}^2}{a_{R_1}}  \;.
\end{equation}
In our model we therefore identify the neutrino masses as 
$m_1=m_0$, $m_2=m_0+m_2^I$ and $m_3=m_0+m_3^I$, where, without loss of generality,
we can take $m_0$ to be positive and real while $m_2^I,m_3^I$ are real but can take 
either sign.
With $\vert m_0 \vert \gg \vert m_3^I \vert$,~$\vert m_2^I \vert$ a quasi-degenerate mass spectrum of the light neutrinos can be explained in a natural way. Below, we will mainly restrict ourselves to this case.

From the structure of $m_\nu$ we obtain TB mixing in the neutrino sector,
\begin{equation}
\theta_{13}^\nu = 0 \;,\quad \theta_{23}^\nu = 45^\circ \;, \quad \theta_{12}^\nu = \arcsin \frac{1}{\sqrt{3}} \approx 35.3^\circ  \:.
\end{equation}
From the lepton sector we get the additional mixing contributions
\begin{equation}
\theta_{13}^e = 0 \;,\quad \theta_{23}^e = 0 \;, \quad \vert \theta_{12}^e \vert  = \left\vert \frac{c_{123} \epsilon_{123}}{c_{123} \epsilon_{123} - \ci \, \tilde{c}_{23} \tilde{\epsilon}_{23} } \right\vert \approx 4.6^\circ \;.
\end{equation}
There is also a complex phase introduced by the charged lepton Yukawa matrix which can be calculated in the same way as in the quark sector
\begin{equation}
\delta_{12}^e =  \arctan \left( \frac{\tilde{c}_{23} \tilde{\epsilon}_{23} }{c_{123} \epsilon_{123}}  \right) \approx  - 85.4^\circ \label{Eq:deltae2} \;.
\end{equation}

For the approximate calculation of the MNS mixing parameters at the GUT scale we can use \cite{Antusch:2008yc,
Boudjemaa:2008jf, *Antusch:2007vw, *Antusch:2007ib, *King:2007pr}:
\begin{equation}
\begin{split}
s^{\mathrm{MNS}}_{23} &\approx s_{23}^\nu - \theta_{23}^e \;, \\
s^{\mathrm{MNS}}_{13} \mathrm{e}^{- \mathrm{i} \delta^{\mathrm{MNS}}_{13} } &\approx \theta_{13}^\nu - s_{23}^\nu \theta_{12}^e \mathrm{e}^{- \mathrm{i} \delta_{12}^e} \;, \\
s^{\mathrm{MNS}}_{12} \mathrm{e}^{- \mathrm{i} \delta^{\mathrm{MNS}}_{12} } &\approx s_{12}^\nu - c_{23}^\nu c_{12}^\nu  \theta_{12}^e \mathrm{e}^{- \mathrm{i} \delta_{12}^e} \;,
\end{split}
\end{equation}
where we have already discarded all trivial phases and RG corrections which we discuss later. For the total leptonic mixing angles we obtain
\begin{equation}\label{eq:prediction}
\begin{split}
\theta_{12}^{\mathrm{MNS}}  & \approx 35.1^\circ \;, \\
\theta_{13}^{\mathrm{MNS}}  & \approx   3.3^\circ \;, \\
\theta_{23}^{\mathrm{MNS}}  & \approx 45.0^\circ\;.
\end{split}
\end{equation}
For the phases we have $\delta^{\mathrm{MNS}}_{13} = \pi - \delta_{12}^e \approx  94.6^\circ$, $\delta^{\mathrm{MNS}}_{12} = 4.6^\circ$ and $\delta^{\mathrm{MNS}}_{23} = 0^\circ$ from which the final MNS phases can be calculated according to \cite{Antusch:2008yc,
Boudjemaa:2008jf, *Antusch:2007vw, *Antusch:2007ib, *King:2007pr}
\begin{equation}\label{eq:prediction_phases}
\begin{split}
\delta_{\mathrm{MNS}} &= \delta_{13}^{\mathrm{MNS}} - \delta^{\mathrm{MNS}}_{12} \approx  90.0^\circ \;,\\
\alpha_1 &= 2 (\delta_{12}^{\mathrm{MNS}} + \delta_{23}^{\mathrm{MNS}}) = 2 \delta_{12}^{\mathrm{MNS}} \approx 9.3^\circ \;,\\
\alpha_2 &= 2 \delta_{23}^{\mathrm{MNS}} \approx 0^\circ \;,
\end{split}
\end{equation}
where $\alpha_1$ and $\alpha_2$ are the Majorana phases as in the PDG parameterisation where they are contained in a diagonal matrix $\mathrm{diag}(\mathrm{e}^{\mathrm{i} \alpha_1/2}, \mathrm{e}^{\mathrm{i} \alpha_2/2}, 1)$.

Similar to the model proposed in Ch.~\ref{Ch:GUTImplications} the leptonic mixing angles and the Dirac CP phase $\delta_{\mathrm{MNS}}$ satisfy the lepton mixing sum rule \cite{King:2005bj, Masina:2005hf, Antusch:2007rk, Antusch:2005kw}
\begin{equation}
\theta_{12}^{\mathrm{MNS}} - \theta_{13}^{\mathrm{MNS}} \cos (\delta_{\mathrm{MNS}}) \approx \arcsin (1/\sqrt{3}) \;,
\end{equation}
where the approximately maximal CP violation, i.e.\ $\delta_{\mathrm{MNS}} \approx 90^\circ$, leads only to small deviations of the solar mixing angle from its TB value of $\arcsin (1/\sqrt{3})$ although the charged lepton corrections generate $\theta_{13}^{\mathrm{MNS}} \approx   3.3^\circ$.

So far, we have discussed the neutrino mixing parameters at the GUT scale. To calculate the predictions at low energies we have to take RG running of the parameters into account.

\subsection{Renormalisation Group Corrections}

For a quasi-degenerate neutrino mass spectrum, RG corrections to the neutrino parameters can in principle change the high scale predictions dramatically. However, as has been discussed for type II upgraded seesaw models in \cite{Antusch:2004xd} and more generally in \cite{Antusch:2003kp,Antusch:2005gp}, for small $\tan \beta$ and small neutrino Yukawa couplings (in our example model they are much smaller than $y_\tau$) the corrections to the mixing angles and CP phases are under control. Setting the small Majorana phases to zero and with $\delta_{\mathrm{MNS}} = 90^\circ$ we can estimate in leading order \cite{Antusch:2003kp,Antusch:2005gp}
\begin{align}
\frac{\mathrm{d} \, \theta^{\mathrm{MNS}}_{12}}{\mathrm{d} \ln (\mu/\mu_0)}  &\approx
- \frac{y_\tau^2}{32 \pi^2} \sin (2 \theta^{\mathrm{MNS}}_{12}) (s_{23}^{\mathrm{MNS}})^2  \frac{|m_1 + m_2|^2}{\Delta m^2_{\mathrm{sol}}}\:,\\
\frac{\mathrm{d} \, \theta^{\mathrm{MNS}}_{13}}{\mathrm{d} \ln (\mu/\mu_0)}  &\approx 0 \:,\\
\frac{\mathrm{d} \, \theta^{\mathrm{MNS}}_{23}}{\mathrm{d} \ln (\mu/\mu_0)}  &\approx
- \frac{y_\tau^2}{32 \pi^2} \sin (2 \theta^{\mathrm{MNS}}_{23})  \frac{(c_{12}^{\mathrm{MNS}})^2  |m_2+ m_3|^2 + (s_{12}^{\mathrm{MNS}})^2  |m_1+ m_3|^2 }{\Delta m^2_{\mathrm{atm}}} \:,
\end{align}
where $\mu$ is the renormalisation scale. In the case of quasi-degenerate neutrino masses, we can further use the approximation $m_3 \approx m_2 \approx m_1 = m_0 $. Integrating these equations approximately with the parameters on the right side taken constant and equal to their GUT scale values,  we obtain the following estimated low energy values of the mixing angles
\begin{align}
\theta^{\mathrm{MNS}}_{12}|_{m_t(m_t)} &\approx \theta^{\mathrm{MNS}}_{12}|_{M_{\mathrm{GUT}}} + 0.15^\circ \frac{m_0^2}{(0.1 \:\mathrm{eV})^2} \:,\\
\theta^{\mathrm{MNS}}_{13}|_{m_t(m_t)} &\approx \theta^{\mathrm{MNS}}_{13}|_{M_{\mathrm{GUT}}}  \:,\\
\theta^{\mathrm{MNS}}_{23}|_{m_t(m_t)} &\approx \theta^{\mathrm{MNS}}_{23}|_{M_{\mathrm{GUT}}} \pm 0.01^\circ \frac{m_0^2}{(0.1  \:\mathrm{eV})^2} \:.
\end{align}
In the last equation, the ``+'' applies for a normal neutrino mass ordering, whereas the ``$-$'' applies for an inverted mass ordering, i.e. the case $\Delta m^2_{\mathrm{atm}} < 0$. It is important to note that both mass orderings can be realised in our model. The strong suppression for the RG running of $\theta^{\mathrm{MNS}}_{13}$ is caused by the particular values of the CP violating phases in our model. For similar reasons the running of the CP phases themselves is also suppressed, as can be seen using the analytic results in \cite{Antusch:2003kp,Antusch:2005gp}. In summary, RG correction are under control in our setup and only cause comparatively small corrections to the mixing parameters in the lepton sector.

In summary, the predictions of our model for the leptonic mixing parameters are compatible with the experimental 1$\sigma$ ranges at low energy which are: $\theta_{12}^{\mathrm{MNS}} = (34.5\pm 1.0)^\circ$, $\theta_{13}^{\mathrm{MNS}} =  (5.7^{+3.0}_{-3.9})^\circ$ and $\theta_{23}^{\mathrm{MNS}} = (42.3^{+5.3}_{-2.8})^\circ$, taken from \cite{GonzalezGarcia:2010er}, as long as $m_0$ is smaller than the cosmological bounds suggest, $m_0 \lesssim 0.2$~eV \cite{Jarosik:2010iu}. The predictions of our model for the leptonic mixing angles and Dirac CP phase $\delta_{\mathrm{MNS}}$ stated in Eqs.~\eqref{eq:prediction} and \eqref{eq:prediction_phases} can be tested accurately by ongoing and future precision neutrino oscillation experiments \cite{Bandyopadhyay:2007kx}.

\subsection{Predictions for Beta Decay Experiments}

\begin{figure}[ht]
\centering
\includegraphics[scale=0.55]{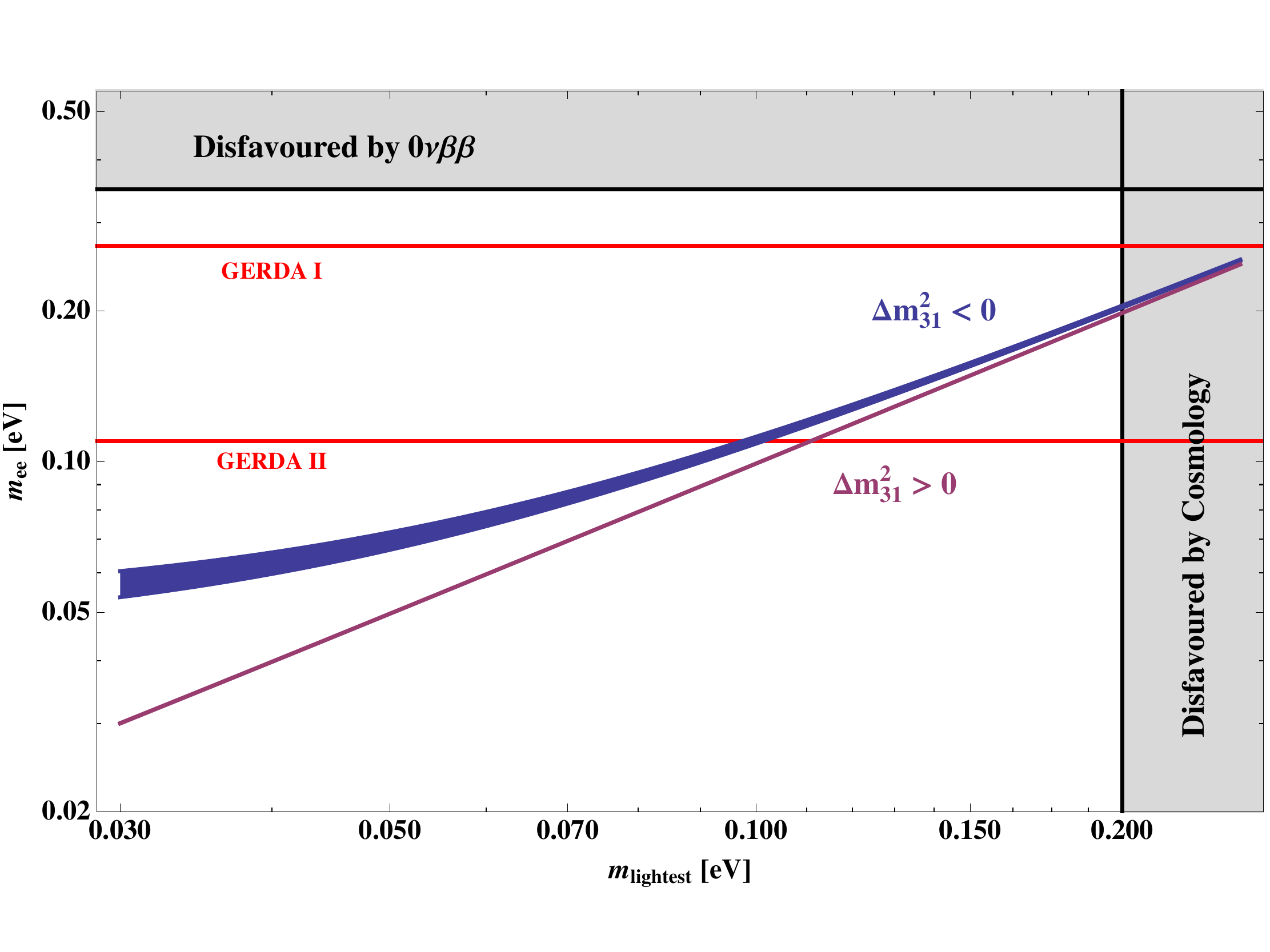}
\caption[Predictions for $m_{ee}$ in the $A_4 \times SU(5)$ Model]{Prediction for effective mass $m_{ee}$ relevant for neutrinoless double beta decay as a function of the mass of the lightest neutrino $m_\mathrm{lightest}$, for an inverted neutrino mass ordering ($\Delta m_{31}^2 < 0$, upper line) and for a normal mass ordering ($\Delta m_{31}^2 > 0$, lower line). The bands represent the experimental uncertainties of the mass squared differences. The mass bounds from cosmology \protect\cite{Jarosik:2010iu} and from the Heidelberg-Moscow experiment \protect\cite{KlapdorKleingrothaus:2000sn} are displayed as grey shaded regions. The red lines show the expected sensitivities of the  GERDA experiment in phase I and II \protect\cite{Smolnikov:2008fu}.  \label{Fig:meePlot2}}
\end{figure}

The effective mass relevant for neutrinoless double beta decay is
\begin{equation}\label{eq:mee_formula}
m_{ee}  = \vert m_1 c_{12}^2 c_{13}^2 e^{\ci \, \alpha_1} + m_2 s_{12}^2 c_{13}^2 e^{\ci \, \alpha_2} + m_3 s_{13}^2 e^{2 \, \ci \, \delta_{\mathrm{MNS}}} \vert \;,
\end{equation}
while the kinematic mass accessible in the single beta decay end-point experiment KATRIN is
\begin{equation}
m^2_\beta \equiv m_1^2 c_{12}^2 c_{13}^2 + m_2^2 s_{12}^2 c_{13}^2  + m_3^2 s_{13}^2 \;.
\end{equation}
For a quasi-degenerate neutrino mass spectrum ($m_0 = m_1 \simeq m_2 \simeq m_3$) we obtain that
\begin{equation}
m_\beta \approx m_0
\end{equation}
directly gives information about the absolute neutrino mass scale.

On the other hand, due to the phases appearing in Eq.~\eqref{eq:mee_formula} there is typically a sizeable ambiguity in the relation between $m_{ee}$ and $m_0$, as long as the Majorana CP phases are not predicted.
Allowing, for instance, for arbitrary Majorana phases and considering a quasi-degenerate neutrino mass spectrum ($m_0 = m_1 \simeq m_2 \simeq m_3$) with small $\theta^{\mathrm{MNS}}_{13}$, $m_{ee}$ can still be in the approximate interval $m_{ee} \in[m_1/3,m_1]$.

This ambiguity is resolved in our model, and in type II upgraded seesaw models in general, since the type II contribution to the neutrino mass matrix being proportional to the unit matrix (with $\vert m_0 \vert \gg \vert m_3^I \vert$,~$\vert m_2^I \vert$) results in small Majorana CP phases and thus we predict:
\begin{equation}
m_{ee} \approx m_0  \quad \text{(type II upgraded models).}
\end{equation}
The assumed dominance of $m_0$ in type II upgraded models allows to realise a quasi-degenerate neutrino spectrum (with normal or inverted mass ordering) in a natural way, without any tuning of parameters.
Our prediction for $m_{ee}$ as a function of the lightest neutrino mass $m_{\mathrm{lightest}}$ is shown in Fig.~\ref{Fig:meePlot2}.

We would like to remark that for smaller $m_0$ one can also naturally extend the model to hierarchical or inverted hierarchical neutrino masses without changing the predictions for the leptonic mixing angles. With $\vert m_0 \vert \approx \vert m_2^I \vert$ or $\vert m_0 \vert \approx \vert m_3^I \vert$ or both, we also encounter cases where the Majorana phases are close to $\pi$. For a quasi-degenerate spectrum, the model disfavours these unnatural cases since they would correspond to heavily fine-tuned parameters of the model. Similarly, an inverted strongly hierarchical spectrum would require unnatural tuning between $m_0$ and $m_3^I$ to make $m_3 = |m_0 - m_3^{I}|$ very small. By contrast, for a typical parameter choice of the model, a normally ordered hierarchical spectrum simply corresponds to $\vert m_0 \vert \ll \vert m_2^I \vert$,~$\vert m_3^I \vert$ and does not require any tuning at all.
It is also interesting to note that with the predicted phases there is {\em no} possibility to have cancellations in Eq.~\eqref{eq:mee_formula} that could make $m_{ee}$ vanish exactly.\footnote{In fact we find numerically that $m_{ee} \gtrsim 0.007$ eV.} Neutrinoless double beta decay is thus, also for smaller $m_0$, an unavoidable consequence in the considered type II upgraded seesaw model.

%% file: kap_12_SummaryConclusions.tex
\chapter{Summary and Conclusions} \label{Ch:SummaryConclusions}

In this thesis we have derived possible GUT predictions for the ratios $y_e/y_d$, $y_\mu/y_s$, $y_\tau/y_b$ and $y_t/y_b$ arising after GUT symmetry breaking at the unification scale. Such relations are a characteristic property of unified flavour models. We have checked their phenomenological viability and found interesting new options for model building. We have also discussed sum rules for quark mixing parameters which are valid for a special class of mass matrix textures. Finally, we have used our improved understanding of the GUT scale Yukawa matrices to construct GUT flavour models in the $SU(5)$ context with interesting phenomenological consequences in the neutrino sector.

After deriving possible Yukawa coupling ratios from higher-dimensional operators in $SU(5)$ and PS we have investigated $\tan \beta$-enhanced SUSY threshold corrections to the down-type quark and charged lepton Yukawa couplings in the electroweak unbroken phase which turn out to give significant corrections. Thus, the inclusion of SUSY threshold corrections in the renormalisation group evolution of the Yukawa couplings is necessary in the large $\tan \beta$ regime. The threshold corrections are very sensitive to the soft SUSY breaking parameters which are (partially) defined by the three parameter sets $g_\pm$, inspired by CMSSM with both signs of the $\mu$ parameter, and $a$, inspired by AMSB with positive $\mu$ parameter.

Remarkably, the sign of the corrections is determined by the signs of the soft SUSY parameters which either results in an increase or decrease of the corrected Yukawa couplings compared to their uncorrected values. The resulting ranges for the GUT scale Yukawa couplings and ratios point towards alternative relations aside from the commonly assumed (partial) third family Yukawa coupling unification and the Georgi--Jarlskog relations. However, they can also help to recover these relations which are somewhat challenged without the inclusion of SUSY threshold corrections in the large $\tan \beta$ regime.

Nevertheless, these findings ask for a more sophisticated analysis including full expressions for the SUSY threshold corrections which take into account effects from electroweak symmetry breaking. For this analysis, the whole SUSY spectrum was calculated for the three common SUSY breaking schemes mAMSB, mGMSB and CMSSM. Thus we are also able to calculate various observables, namely the sparticle masses, the electroweak precision observables $M_W$ and $\sin^2 \theta_{\mathrm{eff}}$,  $\mathrm{BR}(b \to s \gamma)$, $\mathrm{BR}(B_s \to \mu^+ \mu^-)$, the anomalous magnetic moment of the muon and the relic density of the lightest neutralino. We have used these variables as phenomenological constraints for the soft SUSY parameters and thus also for the resulting GUT scale Yukawa coupling ratios.

We have found new viable GUT relations for the Yukawa couplings besides the (partial) third family Yukawa coupling unification and the Georgi--Jarlskog relations. For example, in $SU(5)$, we find the relations $y_\tau/y_b = \pm 3/2$ and $y_\mu/y_s = 6$ in the mGMSB and CMSSM scenarios as well as $y_\mu/y_s = 9/2$ in all three scenarios compatible with the phenomenological constraints. For the PS model, we find the relations $2 y_t = 2 y_b = y_\tau$ in the mGMSB and CMSSM scenario and $y_t = 2 y_b = 2 y _\tau$ in the mAMSB scenario viable.

These results have some implications for GUT model building. First of all, it influences the field content of the model if the newly proposed GUT scale ratios shall be used. Furthermore, it is quite common in GUT models that the charged lepton Yukawa mixing matrices are non-diagonal which leads to a nonzero 1-3 mixing angle in the leptonic mixing matrix even if the 1-3 neutrino mixing vanishes as it happens to be the case, for example, in models incorporating tri-bimaximal mixing in the neutrino sector.

For large $\tan \beta$ we have constructed a flavour model based on the family symmetry $SO(3)$, amended by some discrete symmetries within a $SU(5)$ GUT which incorporates our newly proposed GUT relations $y_\tau / y_b = -3/2$ and $y_\mu / y_s = 6$ for $\tan \beta = 30$. With a simple ansatz for the flavon vevs we are able to fit all fermion masses and mixing angles accurately.

The proposed model has interesting features in the neutrino sector. First of all, we predict the Dirac CP phase to be $\delta_{\mathrm{MNS}} \approx - 90^\circ$. Since we predict $\theta^{\mathrm{MNS}}_{13} \approx 3.0^\circ$, future neutrino oscillation experiments will be able to check the prediction for the yet unknown reactor angle and the Dirac CP phase. The predictions for the remaining leptonic mixing angles are close to TB mixing which will also be tested to higher accuracy in the near future. Furthermore, we also predict the neutrinoless double beta decay mass observable $m_{ee}$ to be greater than zero but too small to be detected by ongoing or proposed experiments due to the strongly hierarchical neutrino masses. Nevertheless, neutrinoless double beta decay is an unavoidable consequence in our model.

After the large $\tan \beta$ case, we focus our attention on the case of small $\tan \beta$. We have started our discussion of this case with quark mass matrix textures with vanishing 1-3 elements for which we derived quark mixing sum rules. Interestingly, for such textures there is a simple relation between the unitarity triangle angle $\alpha$ and the phases of the 1-2 and 2-2 elements. This relation suggests a simple phase pattern for the mass matrices where one element is purely imaginary while all others are purely real. We discussed variants of such textures and their phenomenological viability. Also, we have found that the experimental result $\alpha \approx 90^\circ$ provides an impetus for having hierarchical textures compatible with negligible 1-3 mixing in both the up and down quark mass matrices. This can be related to the leptonic mixing angles in the context of unified theories. In particular, it could  influence the reactor mixing angle, although such considerations are strongly model dependent.

In addition, we have also constructed a flavour model for small $\tan \beta$ within $SU(5)$ and an $A_4$ family symmetry amended by additional discrete symmetries. We note that both models can be formulated with an $A_4$ or a $SO(3)$ family symmetry since we only need real triplet representations and therefore all other groups accommodating real triplets also work. We have used $b$-$\tau$ Yukawa coupling unification and our newly proposed relation for the second family $y_\mu / y_s = 9 / 2$, both of which are viable for this $\tan \beta$ regime. We have used the same ansatz for the flavon vevs as in our previous model with which we can describe all known fermion masses and mixing angles correctly. Furthermore, this model implements a type II seesaw upgrade of a type I seesaw model and can thus easily describe quasi-degenerate neutrino masses. In the quasi-degenerate regime, the prediction for the effective neutrino mass parameter $m_{ee}$ is in the reach of forthcoming neutrinoless double beta decay experiments. Neutrino oscillation experiments can check the validity of this model as well, since we predict TB mixing with small deviations. Noteworthy are the predictions $\theta^{\mathrm{MNS}}_{13} \approx 3.3^\circ$ and $\delta_{\mathrm{MNS}} \approx 90^\circ$.

We began our discussion of flavour model building in supersymmetric unification with group theoretical considerations which depend on the underlying gauge structure only and could therefore be used in non-supersymmetric models as well. Nevertheless, this analysis was suggested by our investigation of GUT scale Yukawa couplings and ratios for which a careful inclusion of SUSY threshold corrections is important, especially for the case of large $\tan \beta$. These two parts fit together and open up new possibilities in GUT model building by choosing operators which generate the Yukawa couplings, different from the usual choices. We have also considered mixing in the form of quark mixing sum rules emerging in a special class of textures for small $\tan \beta$, which suggest a simple phase pattern in the quark mass matrices. All these insights are collected in the two predictive GUT models of flavour presented in this thesis.

The flavour models can be falsified, for example, by neutrinoless double beta decay experiments since in both models the effective neutrino mass does not vanish. Quasi-degenerate neutrinos which can be described by the second model will be tested soon by the next generation of experiments. Indeed, if neutrinoless double beta decay were observed in the near future this would herald another neutrino revolution in which the interest in models with quasi-degenerate neutrinos would explode. Among the many possible models of quasi-degenerate neutrinos, the type II upgrade models are distinguished by their prediction that the neutrinoless double beta decay mass observable is approximately equal to the neutrino mass scale.

Neutrino oscillation experiments can also shed more light on the fundamental flavour structure. In this context, it is interesting to test deviations from TB mixing and, especially, the size of the last missing leptonic mixing angle in order to to distinguish between different models. Also, the leptonic Dirac CP phase is very interesting to measure since, as we have shown in GUT models, this could be related to the Dirac CP Phase in the quark sector.

However, this is not the only possibility to gain additional insight into the structure of fundamental physics. The LHC has very recently started operation with a centre-of-mass energy of 7~TeV, an energy region never tested before in a laboratory. The LHC will probably test the mechanism of electroweak symmetry breaking and if low-energy supersymmetry is realised in nature it is also likely to be found. Knowledge of the sparticle spectrum may point towards a certain supersymmetry breaking scheme which in turn points towards certain GUT scale Yukawa coupling ratios and thus eventually towards the correct underlying gauge group.

From this point of view, we live in exciting times since in the next few years we can expect interesting data which will give us a much deeper insight not only into the flavour structure of the Standard Model of particle physics but also into the flavour structure of new physics which could even give a hint at the correct fundamental description of nature.

Nevertheless, there are still a lot of issues to examine. For the first two generations, a better understanding of the quark masses is essential to constrain (GUT) flavour models, as we have seen. The errors on the light quark masses are still pretty large and need further improvement on experimental and theoretical side. Especially lattice calculations are expected to improve our knowledge about the light quark masses.

Furthermore, many other open theoretical issues are worth further studies. First of all, there is the flavon alignment for which we only employed a phenomenological ansatz. It would be interesting to explicitly construct a mechanism which can generate such a texture or flavon alignment. In this regard, it is especially interesting to reproduce the simple phase structure in the context of spontaneous CP violation.

Also, the flavour models have some impact on the structure of the soft supersymmetry breaking parameters which needs to be analysed. There are severe experimental constraints from flavour physics and electric dipole moments which could further constrain the parameter space of the flavour models.

Nevertheless, many ideas of how to extend the Standard Model of particle physics are on the market and only experiments will tell which one is correct. One single observation will certainly not be enough to distinguish between the vast variety of possible extensions. Instead, the interplay and the correlations between different signals will point towards classes of new physics. Flavour physics is the perfect playground for such measurements due to the vast amount of observables related to the flavour sector some of which are already measured with high precision, providing a perfect touchstone of new physics.

%% file: app_01_Conventions.tex
\chapter{Notation and Conventions} \label{App:Conventions}

In this appendix we briefly summarise the notations and conventions used in this thesis. We work in natural units where $\hbar = c =1$ and use the Minkowski metric with signature $\{+,-,-,- \}$. 

\section{Conventions for the CKM Matrix} \label{App:CKM}

In this section we fix the conventions for the CKM matrix $V_{\mathrm{CKM}}$ used throughout this thesis. The CKM matrix is defined as the unitary matrix occurring in the charged current part of the SM interaction Lagrangian expressed in terms of the quark mass eigenstates. These mass eigenstates can be determined from the mass matrices in the Yukawa sector, namely
\begin{equation}
\mathcal{L}_{Y}=-\bar{u}^i_L (M_u)_{ij} u^j_R - \bar{d}^i_L (M_d)_{ij} d^j_R + \mathrm{h.c.} \;,
\end{equation}
where $M_u$ and $M_d$ are the mass matrices of the up-type and down-type quarks, respectively. The change from the flavour to the mass basis is achieved via bi-unitary transformations
\begin{equation}
\begin{split}
V_{u_L} M_u V_{u_R}^\dagger &= \mathrm{diag}(m_u, m_c, m_t) \;, \\
V_{d_L} M_d V_{d_R}^\dagger &= \mathrm{diag}(m_d, m_s, m_b) \;,
\end{split}
\end{equation}
where $V_{u_{L}}$, $V_{u_{R}}$, $V_{d_{L}}$ and $V_{d_{R}}$ are unitary $3\times 3$ matrices. The CKM matrix $V'_{\mathrm{CKM}}$ (in the raw form, i.e.\ before the unphysical phases are absorbed into redefinitions of the quark mass eigenstate field operators) is then given by
\begin{equation} \label{Eq:UCKM_VuVd}
V'_{\mathrm{CKM}} =V_{u_L} V_{d_L}^\dagger \;.
\end{equation}

In this thesis we shall use the standard, or so-called PDG \cite{Amsler:2008zzb}, parameterisation for the CKM matrix after eliminating the unphysical phases with the structure
\begin{equation}
\label{Eq:UPDG} V_{\mathrm{CKM}}=R_{23}U_{13}R_{12} \;,
\end{equation}
where $R_{23}$, $R_{12}$ denote real, i.e.\ orthogonal, matrices, and the unitary matrix $U_{13}$ contains the observable phase $\delta_{\mathrm{CKM}}$. Alternative parameterisations, motivated by the observation that the unitarity triangle angle $\alpha$ is approximately $90^\circ$, see, e.g.\  \cite{Xing:2009eg}, have been suggested, but we prefer to stick to the standard one.

However, in order to construct the \emph{physical} CKM matrix $V_{\mathrm{CKM}}$ in any given theory of flavour one should begin with the raw CKM matrix $V'_{\mathrm{CKM}}$ defined in Eq.~\eqref{Eq:UCKM_VuVd}, where $V_{u_L}$ and $V_{d_L}$ are general unitary matrices. Recall that a generic 3$\times$3 unitary matrix $V^{\dagger}$ can always be written in terms of three angles $\theta_{ij}$, three phases $\delta_{ij}$ ($i<j$) and three phases $\gamma_{i}$ in the form \cite{King:2002nf}
\begin{equation}
\label{Eq:Param1}
V^{\dagger}=U_{23} U_{13} U_{12} \,{\rm diag}(e^{i\gamma_1},e^{i\gamma_2},e^{i \gamma_3})\;,
\end{equation}
where the three unitary transformations $U_{23}$, $U_{13}$ and $U_{12}$ are defined as
\begin{equation}\label{Eq:U12}
U_{12}= \begin{pmatrix}
  c_{12} & s_{12}e^{-i\delta_{12}} & 0\\
  -s_{12}e^{i\delta_{12}}&c_{12} & 0\\
  0 & 0 & 1 \end{pmatrix} \;,
\end{equation}
and analogously for $U_{13}$ and $U_{23}$. As usual, $s_{ij}$ and $c_{ij}$ are abbreviations for $\sin \theta^{\mathrm{CKM}}_{ij}$ and $\cos \theta^{\mathrm{CKM}}_{ij}$, and the $\theta^{\mathrm{CKM}}_{ij}$ angles can always be made positive by a suitable choice of the $\delta_{ij}$'s. It is convenient to use this parameterisation for both $V^\dagger_{u_L}$ and $V^\dagger_{d_L}$, where the phases $\gamma_i$ can immediately be absorbed into the quark mass eigenstates. Thus, they can be dropped and one is effectively left with
\begin{equation}
\label{Eq:Param2}
V^{\dagger}_{u_L} = U^{u_L}_{23} U^{u_L}_{13} U^{u_L}_{12}\quad\text{and}\quad
V^{\dagger}_{d_L} = U^{d_L}_{23} U^{d_L}_{13} U^{d_L}_{12} \;,
\end{equation}
where $V^{\dagger}_{u_L}$ involves the angles $\theta^u_{ij}$ and phases $\delta^u_{ij}$, while $V^{\dagger}_{d_L}$ involves the angles $\theta^d_{ij}$ and phases $\delta^d_{ij}$. Using Eqs.~\eqref{Eq:UCKM_VuVd} and \eqref{Eq:Param2}, $V'_{\mathrm{CKM}}$ can be written as
\begin{equation}
\label{Eq:Param3}
V'_{\mathrm{CKM}} = {U^{u_L}_{12}}^\dagger {U^{u_L}_{13}}^\dagger {U^{u_L}_{23}}^\dagger
U^{d_L}_{23}U^{d_L}_{13} U^{d_L}_{12}  \;.
\end{equation}
On the other hand, $V'_{\mathrm{CKM}}$ can also be parameterised along the lines of Eq.~\eqref{Eq:Param1},
\begin{equation}
\label{Eq:Param4}
V'_{\mathrm{CKM}}= U_{23} U_{13} U_{12} \, {\mathrm{diag}}(e^{i\gamma_1},e^{i\gamma_2},e^{i \gamma_3})\;.
\end{equation}
By comparing Eq.~\eqref{Eq:Param4} to Eq.~\eqref{Eq:UPDG}, we see that the angles $\theta_{ij}$ are the standard PDG ones in $V_{\mathrm{CKM}}$, and five of the six phases of  $V'_{\mathrm{CKM}}$ in Eq.~\eqref{Eq:Param4} may be removed leaving the standard PDG phase in $V_{\mathrm{CKM}}$ identified as \cite{King:2002nf}
\begin{equation} \label{Eq:deltafromparam1}
\delta_{\mathrm{CKM}} = \delta_{13}-\delta_{23}-\delta_{12} \;.
\end{equation}

\section{Pauli and Dirac Matrices} \label{App:PauliDirac}

For the Pauli matrices we use the convention
\begin{equation}
  \sigma^1 = \begin{pmatrix} 0 & 1 \\ 1 & 0 \end{pmatrix} \;, \quad  \sigma^2 = \begin{pmatrix} 0 & - \ci \\ \ci & 0 \end{pmatrix} \;, \quad  \sigma^3 = \begin{pmatrix} 1 & 0 \\ 0 & -1 \end{pmatrix} \; ,
\end{equation}
which obey the commutation relation
\begin{equation}
 [\sigma^i, \sigma^j] = 2 \, \ci \, \epsilon^{ijk} \sigma^k \;,
\end{equation}
where $\epsilon^{ijk}$ is the totally antisymmetric tensor with three indices and $\epsilon^{123} = +1$.

The Dirac matrices in four dimensions can be defined in terms of the Pauli matrices to be
\begin{equation}
  \gamma^0 = \begin{pmatrix} 0 & \mathds{1} \\ \mathds{1} & 0 \end{pmatrix} \;, \quad \gamma^i = \begin{pmatrix} 0 & \sigma^i \\ -\sigma^i & 0 \end{pmatrix} \;, \quad \gamma_5 = \ci \gamma^0 \gamma^1 \gamma^2 \gamma^3 = \begin{pmatrix} -\mathds{1} & 0 \\ 0 & \mathds{1} \end{pmatrix} \; ,
\end{equation}
where all entries have to be understood as $2 \times 2$~matrices. The Dirac matrices obey the Clifford algebra
\begin{equation}
 \{ \gamma_\mu, \gamma_\nu \} = 2 g_{\mu \nu} \;,
\end{equation}
and the $\gamma_\mu$ matrices anticommute with the $\gamma_5$ matrix,
\begin{equation}
 \{ \gamma_\mu, \gamma_5 \} = 0 \;.
\end{equation}

\section{Weyl Spinors} \label{App:Weyl}

The left-chiral Weyl spinors $\psi_\alpha$ and their right-chiral conjugate $\bar{\psi}^{\dot{\alpha}}$ are two-component spinors transforming under Lorentz transformations as
\begin{equation}
 \psi_\alpha \rightarrow {M_\alpha}^\beta \psi_\beta \;,\quad \bar{\psi}^{\dot{\alpha}} \rightarrow {\left( \left(M^{-1}\right)^\dagger \right)^{\dot{\alpha}}}_{\dot{\beta}} \, \bar{\psi}^{\dot{\beta}} \;,
\end{equation}
where $M=M(\Lambda)$ is the two-dimensional spinor representation of the Lorentz transformation $\Lambda$. The spinor indices $\alpha, \beta =1,2$ and $\dot{\alpha}, \dot{\beta} = 1,2$ can be raised and lowered with the totally antisymmetric $\epsilon$ tensor where $\epsilon_{\alpha \beta} = \epsilon_{\dot{\alpha} \dot{\beta} } = - \epsilon^{\alpha \beta} = - \epsilon^{\dot{\alpha} \dot{\beta}}$ and $\epsilon_{12} = -1$.

The two spinors $\psi$ and $\bar{\psi}$ are related to each other via Hermitian conjugation
\begin{equation}
 \bar{\psi}_{\dot{\alpha}} \equiv (\psi_\alpha)^\dagger = (\psi^\dagger)_{\dot{\alpha}} \;,\quad \psi^\alpha \;, =(\bar{\psi}^{\dot{\alpha}})^\dagger \;.
\end{equation}
and the scalar product of two Weyl spinors $\xi$ and $\chi$ is defined to be
\begin{equation}
 \begin{split}
 \xi \chi &= \xi^\alpha \chi_\alpha = \chi^\alpha \xi_\alpha = \chi \xi \;, \\
 \bar{\xi} \bar{\chi} &= \bar{\xi}_{\dot{\alpha}} \bar{\chi}^{\dot{\alpha}} = \bar{\chi}_{\dot{\alpha}} \bar{\xi}^{\dot{\alpha}} = \bar{\chi} \bar{\xi} = (\xi \chi)^\dagger = (\chi \xi)^\dagger \;.
 \end{split}
\end{equation}
Keep in mind that $\xi^\alpha \chi_\alpha = - \xi_\alpha \chi^\alpha$ and $\bar{\xi}_{\dot{\alpha}} \bar{\chi}^{\dot{\alpha}} = - \bar{\xi}^{\dot{\alpha}} \bar{\chi}_{\dot{\alpha}}$.

We define two four-vectors of Pauli matrices via
\begin{equation}
 (\sigma^\mu)_{\alpha \dot{\beta}} \equiv (\mathds{1}, \sigma^i)_{\alpha \dot{\beta}} \quad \mathrm{and} \quad (\bar{\sigma}^\mu)^{\dot{\alpha} \beta} \equiv ( \mathds{1}, - \sigma^i)^{\dot{\alpha} \beta} \;. \label{Eq:sigmamu}
\end{equation}
There are also antisymmetrised products of these four-vectors
\begin{equation}
 {(\sigma^{\mu \nu})_\alpha}^\beta \equiv \frac{i}{2} {(\sigma^\mu \bar{\sigma}^\nu - \sigma^\nu \bar{\sigma}^\mu)_\alpha}^\beta \quad \mathrm{and} \quad {(\bar{\sigma}^{\mu \nu})^{\dot{\alpha}}}_{\dot{\beta}} \equiv \frac{i}{2} {(\bar{\sigma}^\mu \sigma^\nu - \bar{\sigma}^\nu \sigma^\mu)^{\dot{\alpha}}}_{\dot{\beta}} \;. \label{Eq:sigmamunu}
\end{equation}

\section{Dirac and Majorana Spinors} \label{App:DiracMajorana}

A Dirac spinor $\Psi$ has four components and consists of two Weyl spinors $\xi_\alpha$ and $\chi_\beta$
\begin{equation}
 \Psi = \begin{pmatrix} \xi_\alpha \\ \bar{\chi}^{\dot{\beta}} \end{pmatrix} \quad \mathrm{and} \quad \bar{\Psi} = \Psi^\dagger \gamma^0 = \begin{pmatrix} \chi^\beta & \bar{\xi}_{\dot{\alpha}} \end{pmatrix} \;.
\end{equation}
The charge conjugation of a Dirac spinor is defined to be
\begin{equation}
 \Psi^c \equiv C \bar{\Psi}^T = \begin{pmatrix} \chi_\beta \\ \bar{\xi}^{\dot{\alpha}} \end{pmatrix} \;,
\end{equation}
where $C = \ci \gamma^0 \gamma^2$ is the charge conjugation matrix which satisfies $C^{-1} \gamma^\mu C = - (\gamma^\mu)^T$.

For Dirac spinors we can also define the chiral projection operators $P_{L/R} = \frac{1}{2}(1 \mp \gamma_5)$ which project out the left- and right-chiral states of a Dirac field
\begin{equation}
 \Psi_L \equiv P_L \Psi = \begin{pmatrix} \xi_\alpha \\ 0 \end{pmatrix} \quad \mathrm{and} \quad \Psi_R \equiv P_R \Psi = \begin{pmatrix} 0 \\ \bar{\chi}^{\dot{\alpha}} \end{pmatrix} \;.
\end{equation}
Therefore $\xi_\alpha$ is called a left-chiral Weyl spinor and $\bar{\chi}_{\dot{\alpha}}$ a right-chiral Weyl spinor.

If a spinor $\Psi$ fulfils the Majorana condition
\begin{equation}
 \Psi^c = \Psi \;,
\end{equation}
it consists of two identical Weyl spinors,
\begin{equation}
 \Psi =  \begin{pmatrix} \xi_\alpha \\ \bar{\xi}^{\dot{\beta}} \end{pmatrix} \quad \mathrm{and} \quad \bar{\Psi} = \Psi^\dagger \gamma^0 = \begin{pmatrix} \xi^\beta & \bar{\xi}_{\dot{\alpha}} \end{pmatrix} \;.
\end{equation}
This spinor is called a Majorana spinor. The corresponding field is its own antiparticle.

\section{Grassmann Numbers} \label{App:Grassmann}

Grassmann numbers $\theta_\alpha$ are anticommuting like fermions,
\begin{equation}
 \{ \theta_\alpha , \theta_\beta \} = \{ \bar{\theta}_{\dot{\alpha}} , \bar{\theta}_{\dot{\beta}} \} = \{ \theta_\alpha, \bar{\theta}_{\dot{\beta}} \} = 0 \;. 
\end{equation}
From these relations, it follows that the square of a Grassmann number $\theta_\alpha$ has to vanish, $\theta_\alpha \theta_\alpha = 0$, and the product of more than two Grassmann numbers is zero, $\theta^\alpha \theta^\beta \theta^\gamma \cdots = 0$, since in our case $\alpha, \beta, \gamma, \ldots \in \{ 1,2 \}$. The (Lorentz invariant) product of two Grassmann spinors is defined via
\begin{equation}
 \theta \theta \equiv \theta^\alpha \theta_\alpha = \epsilon_{\alpha \beta} \theta^\alpha \theta^\beta = \epsilon^{\alpha \beta} \theta_\beta \theta_\alpha \;,
\end{equation}
which gives the relations
\begin{equation}
\begin{split}
 \theta^\alpha \theta^\beta = - \frac{1}{2} \epsilon^{\alpha \beta} (\theta \theta) \;,\quad  \theta_\alpha \theta_\beta = + \frac{1}{2} \epsilon_{\alpha \beta}(\theta \theta) \;, \\
\bar{\theta}^{\dot{\alpha}} \bar{\theta}^{\dot{\beta}} = + \frac{1}{2} \epsilon^{\dot{\alpha} \dot{\beta}} (\bar{\theta} \bar{\theta}) \;,\quad \bar{\theta}_{\dot{\alpha}} \bar{\theta}_{\dot{\beta}} = - \frac{1}{2} \epsilon_{\dot{\alpha}  \dot{\beta} }  (\bar{\theta} \bar{\theta}) \;.  \\
\end{split}
\end{equation}

The derivatives of the Grassmann variables are defined as
\begin{equation}
 \frac{\partial \theta^\beta}{\partial \theta^\alpha} \equiv \partial_\alpha \theta^\beta = {\delta_\alpha}^\beta \;, \quad \frac{\partial \bar{\theta}_{\dot{\beta}} }{\partial \bar{\theta}_{\dot{\alpha}} } \equiv \bar{\partial}^{\dot{\alpha}} \bar{\theta}_{\dot{\beta}} = {\delta^{\dot{\alpha}} }_{\dot{\beta}} \;, \quad \frac{\partial \bar{\theta}_{\dot{\beta}} }{\partial \theta^\alpha} = \frac{\partial \theta^\beta}{\partial \bar{\theta}_{\dot{\alpha}} } = 0 \;,
\end{equation}
where we use the conventions
\begin{equation}
 \partial_\alpha = - \epsilon_{\alpha \beta} \partial^\beta \;, \quad \partial^\alpha = - \epsilon^{\alpha \beta} \partial_\beta \;, \quad \bar{\partial}_{\dot{\alpha}} = - \epsilon_{\dot{\alpha} \dot{\beta}} \bar{\partial}^{\dot{\beta}} \;, \quad \bar{\partial}^{\dot{\alpha}} = - \epsilon^{\dot{\alpha} \dot{\beta}} \bar{\partial}_{\dot{\beta}} \;,
\end{equation}
to ensure, for example,  ${\delta_\alpha}^\beta = \partial_\alpha \theta^\beta = \partial^\beta \theta_\alpha$.

The integration over Grassmann variables is defined as
\begin{gather}
\begin{split}
 \int \dd \theta_\alpha = 0 \;&, \quad \int \dd \theta_\alpha \, \theta_\beta = \delta_{\alpha \beta} \;, \\
 \int \dd^2 \theta = 0 \;, \quad \int \dd^2 \theta \, \theta^\alpha =0 \;&, \quad \int \dd^2 \theta \, \theta^\alpha \theta^\beta = - \frac{1}{2} \epsilon^{\alpha \beta} \;, \quad \int \dd^2 \theta (\theta \theta) = 1 \;.
\end{split}
\end{gather}

If $\theta$ and $\bar{\theta}$ are independent we have the following integral identities
\begin{equation}
 \int \dd \theta \dd \bar{\theta} \, \bar{\theta} \theta = 1 \quad \mathrm{and} \quad \int \dd^4 \theta (\theta \theta) (\bar{\theta} \bar{\theta}) = 1 \;.
\end{equation}

%% file: app_02_SUNReps.tex
\chapter{$\boldsymbol{SU(N)}$ Representations} \label{App:SUNReps}

In this chapter we want to discuss some formal aspects of $SU(N)$ representations based on \cite{Cheng:1985bj, *Ross:1985ai, *Georgi:1982jb} since they play a crucial role in this thesis. We start with the transformation law of $SU(N)$ tensors in Sec.~\ref{App:SUNtensors}. Afterwards we discuss the irreducible representations of $SU(N)$ and their diagrammatic representation as Young tableaux in Sec.~\ref{App:Young}. Then we show in Sec.~\ref{App:RepReduction} how to use those tableaux to calculate the products of irreducible representations. We conclude this chapter with Sec.~\ref{App:RepList} where a list of representations is given which are important for this work.

\section{Transformation Law of Tensors} \label{App:SUNtensors}

The $SU(N)$ group can be written in the form of $N \times N$ unitary matrices with unit determinant. Matrices act as linear transformations on a vector space. In the case of $SU(N)$, the matrices give rise to linear invertible transformations on the complex $N$-dimensional vector space, $\mathds{C}_N$. Let us denote an element of $\mathds{C}_N$ by $\psi = (\psi_1, \psi_2, \ldots,\psi_N)^T$ and an element of $SU(N)$ by $U$ with $U U^\dagger = U^\dagger U = \mathds{1}$ and $\det U = 1$. Then the transformation of the vector in component notations reads as
\begin{equation}
 \psi_i \rightarrow \psi'_i = U_{ij} \psi_j \;, \label{Eq:SUNtransform}
\end{equation}
where $\psi'$ is the transformed vector.

We can define a scalar product
\begin{equation}
 (\psi,\phi) \equiv \psi^\dagger \phi =  \psi_i^* \phi_i \label{Eq:SUNscalarproduct}
\end{equation}
on the vector space $\mathds{C}_N$ which is invariant under $SU(N)$ transformations. The transformation law for the conjugate vector is given by
\begin{equation}
 \psi_i^* \rightarrow \psi'^*_i = U_{ij}^* \psi_j^* = \psi_j^* U_{ji}^\dagger \;. \label{Eq:SUNctransform}
\end{equation}
It is common to distinguish a vector from its conjugate by the position of its index. Conjugated vectors have upper indices, i.e.
\begin{equation}
 \psi^i \equiv \psi_i^* \,, \; {U_i}^j \equiv U_{ij} \; \; \text{and} \;\; {U^i}_j \equiv U_{ij}^* \;.
\end{equation}
In other words complex conjugation raises lower indices and vice versa. In this notation the transformation laws for vectors become
\begin{equation}
\begin{split}
 \psi_i &\rightarrow \psi'_i = {U_i}^j \psi_j \;, \\
 \psi^i &\rightarrow \psi'^i = {U^i}_j \psi^j \;.
\end{split}
\end{equation}
The scalar product, see Eq.~\eqref{Eq:SUNscalarproduct}, simplifies to
\begin{equation}
 (\psi, \phi) = \psi^i \phi_i \;,
\end{equation}
and the unitarity condition becomes
\begin{equation}
 {U^k}_i {U_k}^j = {\delta_i}^j \;.
\end{equation}
One of the advantages of this notation is that a summation always involves upper and lower indices so that we can contract indices as it is known for Lorentz indices. The $\psi_i$ vectors are the basis of the \emph{fundamental} representation, denoted by $\mathbf{N}$ of $SU(N)$ and the $\psi^i$ form the conjugate representation $\mathbf{\overline{N}}$. Throughout this thesis we label representations with a bold typed $\mathbf{d}$ where $d$ is the dimension of the given representation. By definition $\psi^i$ and $\psi_i$ are elements of a $N$-dimensional vector space and hence for the fundamental representation this rule is already fulfilled.

Higher-rank tensors are defined to have the same transformation properties as the corresponding direct product of vectors. Therefore they have in general upper and lower indices. Their transformation law is accordingly
\begin{equation}
 {\psi'}^{i_1 i_2 \ldots i_p}_{j_1 j_2 \ldots j_q} = ({U^{i_1}}_{k_1} {U^{i_2}}_{k_2} \ldots {U^{i_p}}_{k_p}) ({U_{j_1}}^{l_1} {U_{j_2}}^{l_2} \ldots {U_{j_q}}^{l_q}) \psi^{k_1 k_2 \ldots k_p}_{l_1 l_2 \ldots l_q} \;. \label{Eq:HighRankTensors}
\end{equation}
These tensors give the bases for higher-dimensional representations of $SU(N)$.

There are two important tensors which are invariant under $SU(N)$ transformations. First of all there is the Kronecker $\delta$
\begin{equation}
 {\delta'^i}_j = {U^i}_k {U_j}^l {\delta^k}_l = {U^i}_k {U_j}^k = {\delta^i}_j \;,
\end{equation}
where we have used the unitarity condition. The second invariant tensor is the totally antisymmetric tensor in $N$ dimensions, the Levi--Civita or $\epsilon$ tensor,
\begin{equation}
 \epsilon'_{i_1 i_2 \ldots i_N} = {U_{i_1}}^{j_1} {U_{i_2}}^{j_2} \ldots {U_{i_N}}^{j_N} \epsilon_{j_1 j_2 \ldots j_N} = (\det U) \epsilon_{i_1 i_2 \ldots i_N} = \epsilon_{i_1 i_2 \ldots i_N} \;,
\end{equation}
where we have used the definition of the determinant and the fact that the elements of $SU(N)$ have unit determinant. We stress here that the $\epsilon$ tensor can be used to lower and raise $SU(N)$ indices
\begin{equation}
 \epsilon^{i_1 i_2 \ldots i_N} \psi_{i_2 \ldots i_N} = \psi^{i_1} \;.
\end{equation}
Hence, with the help of the $\epsilon$ tensor new representations can be constructed and every totally antisymmetric $SU(N)$ tensor is proportional to the $\epsilon$ tensor and transforms as a scalar like the $\epsilon$ tensor.

\section{Irreducible Representations and Young Tableaux}\label{App:Young}

The irreducible representations play a crucial role in the discussion of representations. The higher-rank tensors as in Eq.~\eqref{Eq:HighRankTensors} are usually the basis for reducible representations. In order to decompose the reducible representations we make a symmetry decomposition regarding the tensor indices.

The group transformations and an index permutation commute with each other since a group transformation consists of identical matrices. As an illustration we discuss the example of the second-rank tensor $\psi_{ij}$. We define the permutation operator $P_{12}$ which simply interchanges the two indices, $P_{12} \psi_{ij} = \psi_{ji}$, and commutes with the group transformation
\begin{equation}
 P_{12} \psi'_{ij} = \psi'_{ji} = {U_j}^l {U_i}^k \psi_{lk} = {U_i}^k {U_j}^l  P_{12} \psi_{kl} \;.
\end{equation}
The eigenstates of $P_{12}$ are the combinations
\begin{equation}
 S_{ij} = \frac{1}{2} (\psi_{ij} + \psi_{ji}) \quad \text{and} \quad A_{ij} = \frac{1}{2} (\psi_{ij}-\psi_{ji}) \;,
\end{equation}
which are symmetric and antisymmetric respectively under $P_{12}$. A second rank tensor of $SU(N)$ can always be decomposed in terms of those two tensors which cannot be decomposed any further. The second rank tensors $S_{ij}$ and $A_{ij}$ form two irreducible representations of $SU(N)$.

This statement can be generalised to a tensor of arbitrary rank. It is sufficient to find all symmetry decompositions to have all irreducible representations of a given higher-rank tensor.

This problem can be easily solved in terms of Young tableaux. A general Young tableau is an arrangement of $f$ boxes in rows and columns such that the length of rows does not increase from top to bottom: $f_1 \geq f_2 \geq \ldots$ and $\sum_i f_i = f$. For our needs $f$ is the rank of the tensor and we write the tensor always in a form with only lower indices (upper indices can be lowered with the $\epsilon$ tensor). An example Young tableau corresponding to a rank eight tensor $\psi_{i_1 i_2 i_3 i_4 ; i_5 i_6 i_7; i_8}$ can take the form
\begin{equation}
 \yng(4,3,1) \;.
\end{equation}
The shape of the Young tableau also defines the symmetry of the indices by the following two rules:
\begin{enumerate}
 \item Indices appearing in the same row of the tableau are first subject to symmetrisation.
 \item Subsequently, indices appearing in the same column are subject to antisymmetrisation.
\end{enumerate}
To illustrate these rules we give a simple example. The Young diagram
\begin{equation}
 \young(ij,k) 
\end{equation}
defines the tensor $\psi_{ij;k} = \psi_{ijk} + \psi_{jik} - \psi_{kji} - \psi_{jki}$. First the indices $i$ and $j$ are symmetrised and afterwards the indices $i$ and $k$ are antisymmetrised.

There is a \emph{fundamental theorem} relating Young tableaux and $SU(N)$ representations \cite{Hamermesh:1962gt}: A tensor corresponding to the Young tableau of a given pattern forms the basis of an irreducible representation of $SU(N)$. Moreover if we enumerate all possible Young tableaux under the restriction that there should be no more than $N-1$ rows, the corresponding tensors form a complete set, in the sense that all finite-dimensional irreducible representations of the group are counted only once. This theorem is the basis for our above statements but we do not give a proof here.

Another advantage of Young tableaux is that it is easy to calculate the dimension of the corresponding representation from them. The dimension $d$ of an irreducible $SU(N)$ representation is given by
\begin{equation}
 d = \prod_i \frac{N+D_i}{h_i} \;, \label{Eq:RepDimension}
\end{equation}
where $D_i$ is the distance to the first box and $h_i$ is the hook length associated to the $i$th box. The product is taken over all boxes in the tableau. The distance to the first box is defined by counting the number of steps going from the upper left corner of the tableau, the first box, to the $i$th box. For every step to the right $D_i$ increases by one unit, whereas for every step downwards $D_i$ decreases by one unit. As an example in the following tableau $D_i$ is written in every box
\begin{equation}
 \young(012,\mone0,\mtwo) \;. \label{Eq:Distance}
\end{equation}

The hook length is defined by putting in the $i$th box a hook and then for every box, the hook passes, one unit is added to the hook length. The hook has two arms. One arm goes horizontally to the right and the other one vertically downwards. For example the hook length of the first box in the tableau
\begin{equation}
 \young(\bullet \bullet \bullet,\bullet \hfil )
\end{equation}
is four (the hook is denoted by the bullets) whereas the hook length of the fifth box in the tableau
\begin{equation}
 \young(\hfil \hfil \hfil,\hfil \bullet \bullet)
\end{equation}
is equal to two. The hook always passes at least the starting box and hence the minimal hook length is one.

To illuminate the dimension formula \eqref{Eq:RepDimension} we give an example in $SU(6)$. The Young tableau is
\begin{equation}
 \young(12,34) \;, \label{Eq:YoungExample}
\end{equation}
where we have labelled the boxes from 1 to 4. The distances to the first box are $D_1 = 0$, $D_2 = 1$, $D_3 = -1$ and $D_4 = 0$, cf.\  \eqref{Eq:Distance}. The hook lengths are $h_1 = 3$, $h_2 = 2$, $h_3 = 2$ and $h_1 = 1$. Therefore the dimension of the corresponding representation is
\begin{equation}
 d = \prod_i \frac{6+D_i}{h_i} = 2 \cdot \frac{7}{2} \cdot \frac{5}{2} \cdot 6 = 105 \;.
\end{equation}
So the Young tableau in Eq.~\eqref{Eq:YoungExample} depicts in $SU(6)$ the 105-dimensional representation.

\section{Reduction of Product Representations} \label{App:RepReduction}

We now turn to a further application of Young tableaux. With their help it is easy to calculate the irreducible representations of the product of two irreducible representations. By induction every product of representations can be evaluated. The algorithm how to do this is:
\begin{enumerate}
 \item Assign the same symbol, say $a$, to all boxes in the first row in the tableau of the first factor. Then assign $b$ to the second row, $c$ to the third row and so on.
 \item Attach boxes labelled with $a$ to the tableau of the second factor in all possible ways, such that that the resulting tableau is still a Young tableau and does not have more than $N$ rows in any column. It is also not allowed to have more than one $a$ appearing in any column. Repeat this with the boxes labelled with a $b$ and so on.
 \item After all boxes of the first factor have been added to the tableau of the second factor, the added symbols have to be read from \emph{right} to \emph{left} in the first row, then the second row and so on. The resulting sequence of symbols must form a lattice permutation, that is a permutation where there are no fewer $a$'s than $b$'s to the left of any symbol, no fewer $b$'s than $c$'s, and so on. If the sequence of symbols does not form a lattice permutation the corresponding tableau is discarded from the product. For example the permutation $baa$ is not a lattice permutation while the permutation $aba$ is a lattice permutation.
\end{enumerate}

We give now two examples in $SU(3)$ to illustrate this algorithm. The first example is
\begin{equation}
 \mathbf{3} \times \mathbf{3} = \young(a) \times \yng(1) = \young(\hfil,a) + \young(\hfil a) = \mathbf{\overline{3}} + \mathbf{6} \;.
\end{equation}
Physically it can be interpreted as the combination of two quarks. Since two quarks can be combined to form an anti-triplet, three quarks can be combined to form a singlet of $SU(3)_C$ which is a baryon.

The second example, corresponding to the combination of an anti-quark and a gluon, is a little bit more complicated
\begin{equation}
\begin{split}
 \mathbf{\overline{3}} \times \mathbf{8} &= \young(a,b) \times \yng(2,1)  \\
&= \young(b) \times \left( \young(\hfil \hfil a,\hfil) + \young(\hfil \hfil,\hfil a) + \young(\hfil \hfil,\hfil,a)  \right) \\
&= \young(\hfil \hfil a,\hfil b) + \young(\hfil \hfil a,\hfil ,b) + \young(\hfil \hfil,\hfil a,b) \\
&= \mathbf{\overline{15}} + \mathbf{6} + \mathbf{\overline{3}} \;,
\end{split}
\end{equation}
where we have already discarded the tableaux in the third step which do not give rise to a lattice permutation. A good cross-check is that the product of the dimensions on the left hand side has to be equal to the sum of the dimensions on the right hand side. There is a comprehensive list of product representations in the appendix of \cite{Slansky:1981yr}. Full columns ($N$ rows) in a tableau can be crossed out since they give only a trivial factor of one in the calculation of the dimension.

The fact that full columns give only trivial contributions to the dimension of the representation offers a simple possibility for determining the Young tableau of the conjugate of a given representation. Simply fill out the given Young tableau to form a rectangle with $N$ rows where you may not add any columns. A full rectangle gives the trivial singlet representation under $SU(N)$. The added boxes, rotated by 180 degrees, give the Young tableau of the conjugate representation. We give an example in $SU(4)$. The Young tableau,
\begin{equation}
 \yng(2,1) = \mathbf{20} \;,
\end{equation}
has the conjugate representation
\begin{equation}
 \yng(2,2,1) = \mathbf{\overline{20}} \;.
\end{equation}

\section{List of Used Representations} \label{App:RepList}

In this section we list the $SU(N)$ representations and their Young tableaux which are important for this work. We only give either the representation or the conjugated one depending on which Young tableau is more compact. The other representation can easily be derived as described above.

\newpage

\begin{table}
 \begin{center}
 \begin{tabular}{cc} \toprule
 Representation & Young Tableau \\  \midrule
 $\mathbf{2}$ &  \tiny{$\yng(1)$}  \\[0.5pc]
 $\mathbf{3}$ &  \tiny{$\yng(2)$}  \\[0.5pc]
 \bottomrule
 \end{tabular}
 \caption[Relevant $SU(2)$ Representations]{$SU(2)$ representations and their Young tableaux relevant for this work.}
 \end{center}
\end{table}

\begin{table}
 \begin{center}
 \begin{tabular}{cc} \toprule
 Representation & Young Tableau \\ \midrule
 $\mathbf{4}$ &  \tiny{$\yng(1)$}  \\[0.5pc]
 $\mathbf{15}$ &  \tiny{$\yng(2,1,1)$}  \\[0.5pc]
 \bottomrule
 \end{tabular}
 \caption[Relevant $SU(4)$ Representations]{$SU(4)$ representations and their Young tableaux relevant for this work.}
 \end{center}
\end{table}

\begin{table}
 \begin{center}
 \begin{tabular}{cc} \toprule
 Representation & Young Tableau \\ \midrule
 $\mathbf{5}$ & \tiny{$\yng(1)$} \\[0.5pc]
 $\mathbf{10}$ &  \tiny{$\yng(1,1)$} \\[0.5pc]
 $\mathbf{15}$ &  \tiny{$\yng(2)$} \\[0.5pc]
 $\mathbf{24}$ &  \tiny{$\yng(2,1,1,1)$} \\[1.0pc]
 $\mathbf{\overline{40}}$ &  \tiny{$\yng(2,1)$} \\[1.0pc]
 $\mathbf{\overline{45}}$ &  \tiny{$\yng(2,1,1)$}  \\[1.0pc]
 $\mathbf{\overline{50}}$ &  \tiny{$\yng(2,2)$}  \\[1.0pc]
 $\mathbf{75}$ &  \tiny{$\yng(2,2,1)$} \\[0.5pc]
 \bottomrule
 \end{tabular}
 \caption[Relevant $SU(5)$ Representations]{$SU(5)$ representations and their Young tableaux relevant for this work.}
 \end{center}
\end{table}

%% file: app_03_Plots.tex
\chapter[Detailed Plots for the Phenomenological Scans]{Detailed Plots for the\\ Phenomenological Scans} \label{App:Plots}

In this appendix, we show detailed plots from our parameter scan described in Ch.~\ref{Ch:Pheno}. We show here the impact of eaech experimental constraint, see Sec.~\ref{Sec:ExperimentalConstraints}, on the allowed GUT scale Yukawa coupling ratios for the three SUSY breaking schemes mAMSB, mGMSB and CMSSM. The scan ranges are given in Tab~\ref{Tab:SUSYParameterRanges}.

For each of these scans, we plot the results for every experimental constraint defined in Sec.~\ref{Sec:ExperimentalConstraints}, for the Yukawa coupling ratios $y_e/y_d$, $y_\mu/y_s$, $y_\tau/y_b$ and $y_t/y_b$. Black points in the plots are allowed by the respective experimental constraint, while red points are excluded. There is no one-to-one correspondence between GUT scale Yukawa coupling ratios and the SUSY parameters, so that black and red points may overlap when allowed and disfavoured parameter points give the same ratios.

For the mGMSB scenario we do not show the plots for the CDM constraint, since in this scenario the LSP is the gravitino which is not included within our calculations. Hence the corresponding plot has no significance for the rest of our considerations.

We have not included quark mass errors in the plots here, however, we show them in the final in Fig.~\ref{Fig:FinalResults}.

\begin{figure}
 \centering
 \includegraphics[scale=0.43]{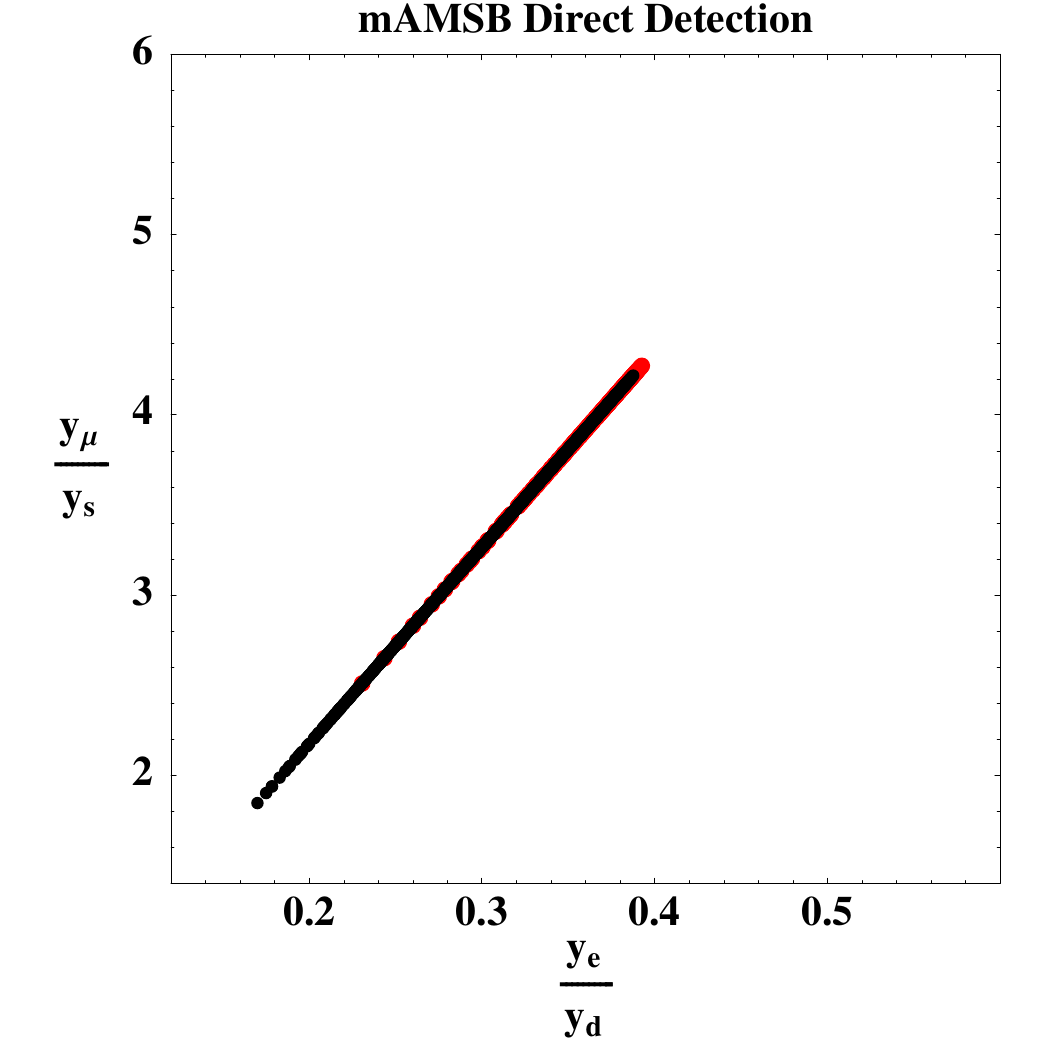}
 \includegraphics[scale=0.43]{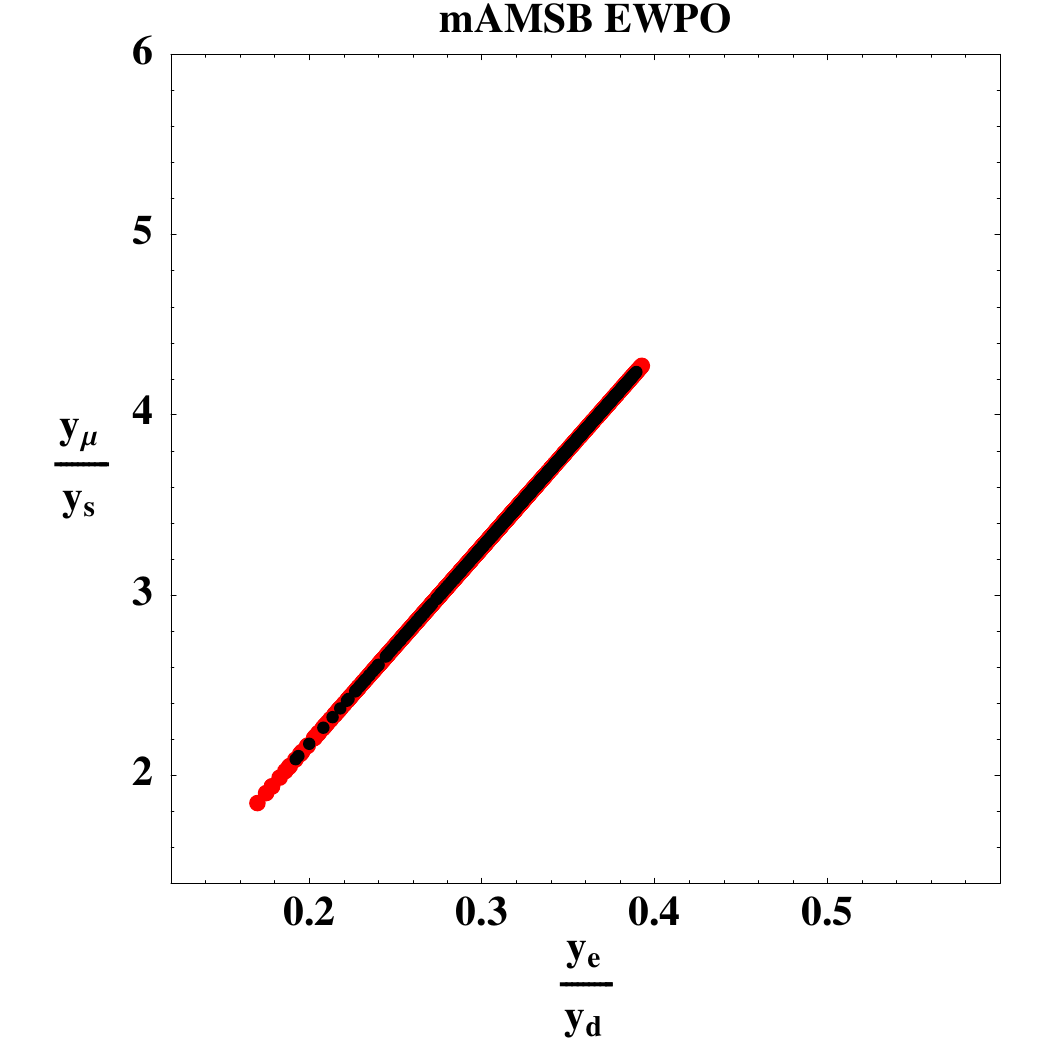}
 \includegraphics[scale=0.43]{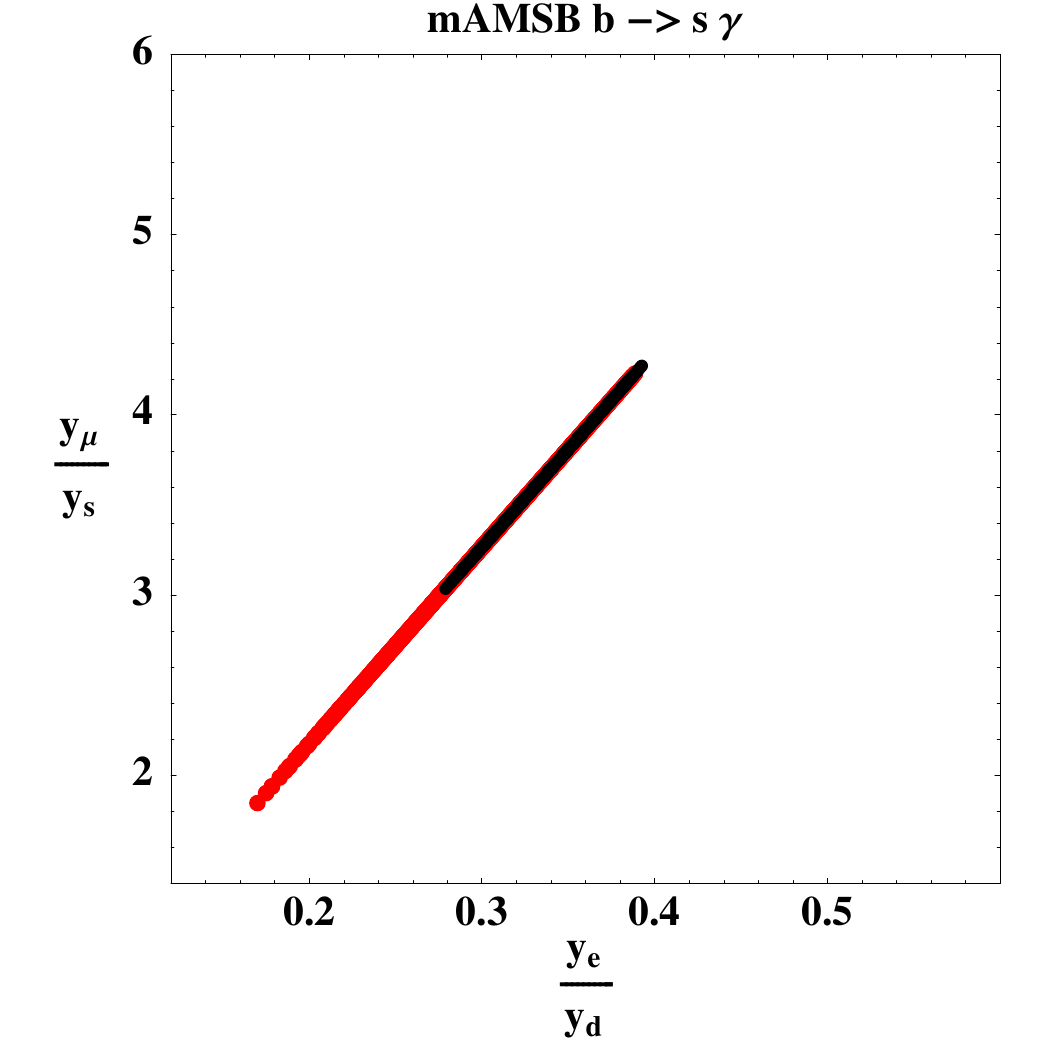}
 \includegraphics[scale=0.43]{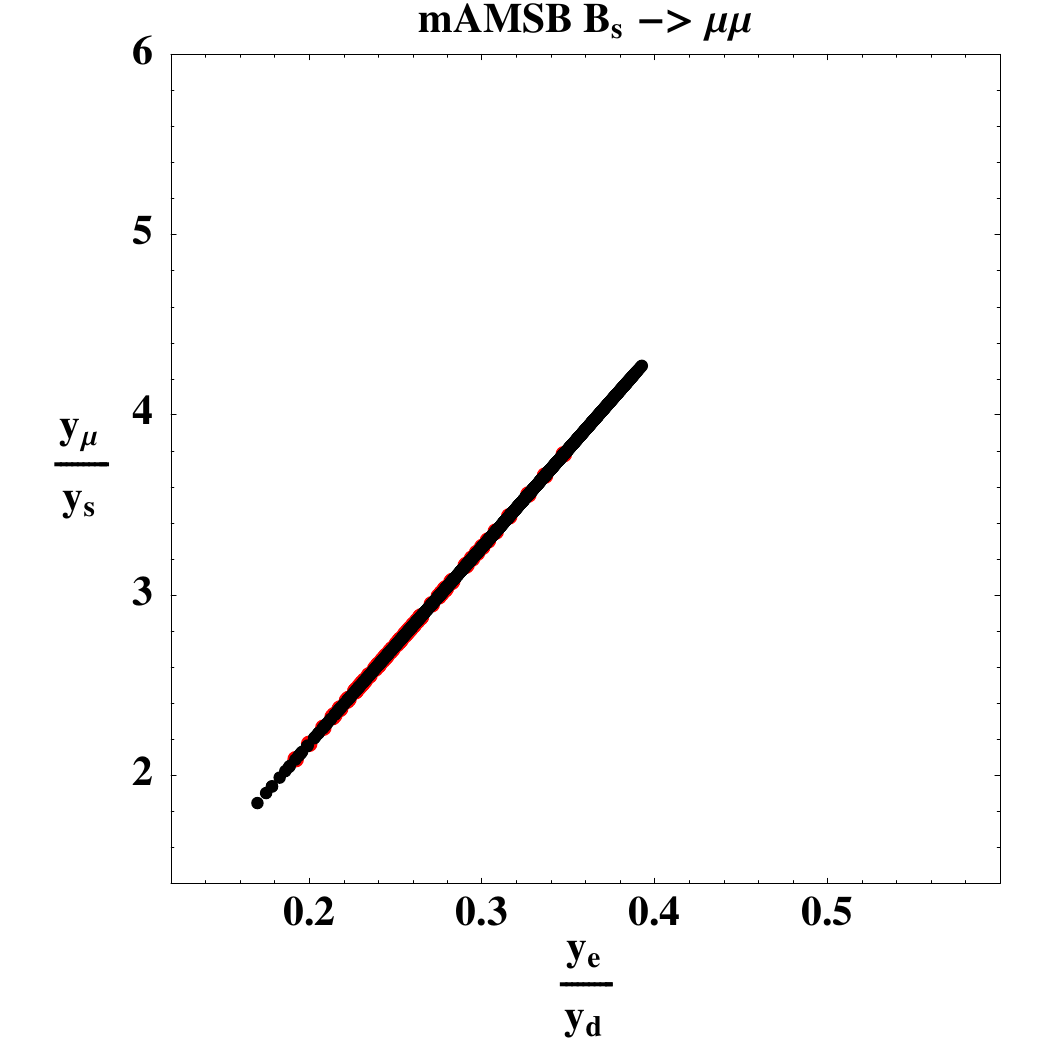}
 \includegraphics[scale=0.43]{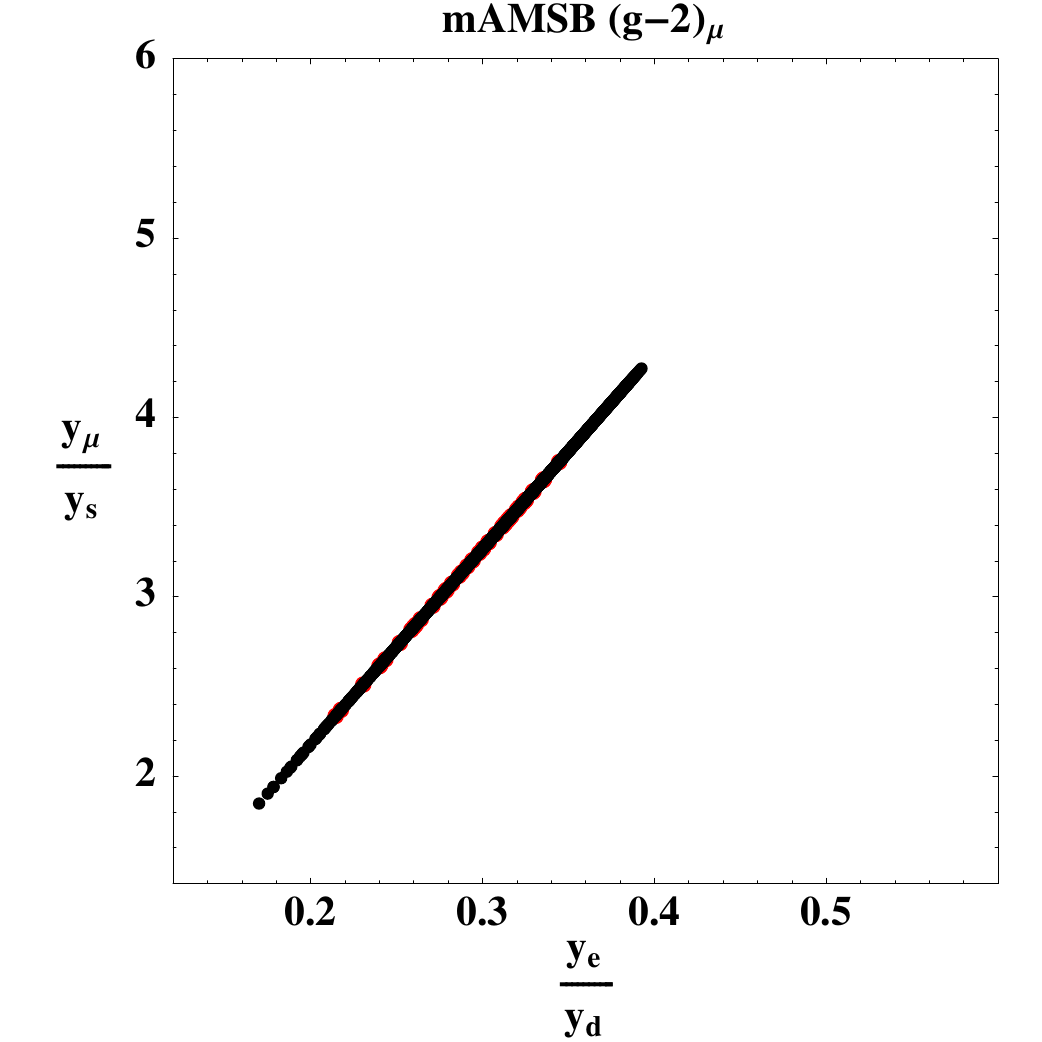}
 \includegraphics[scale=0.43]{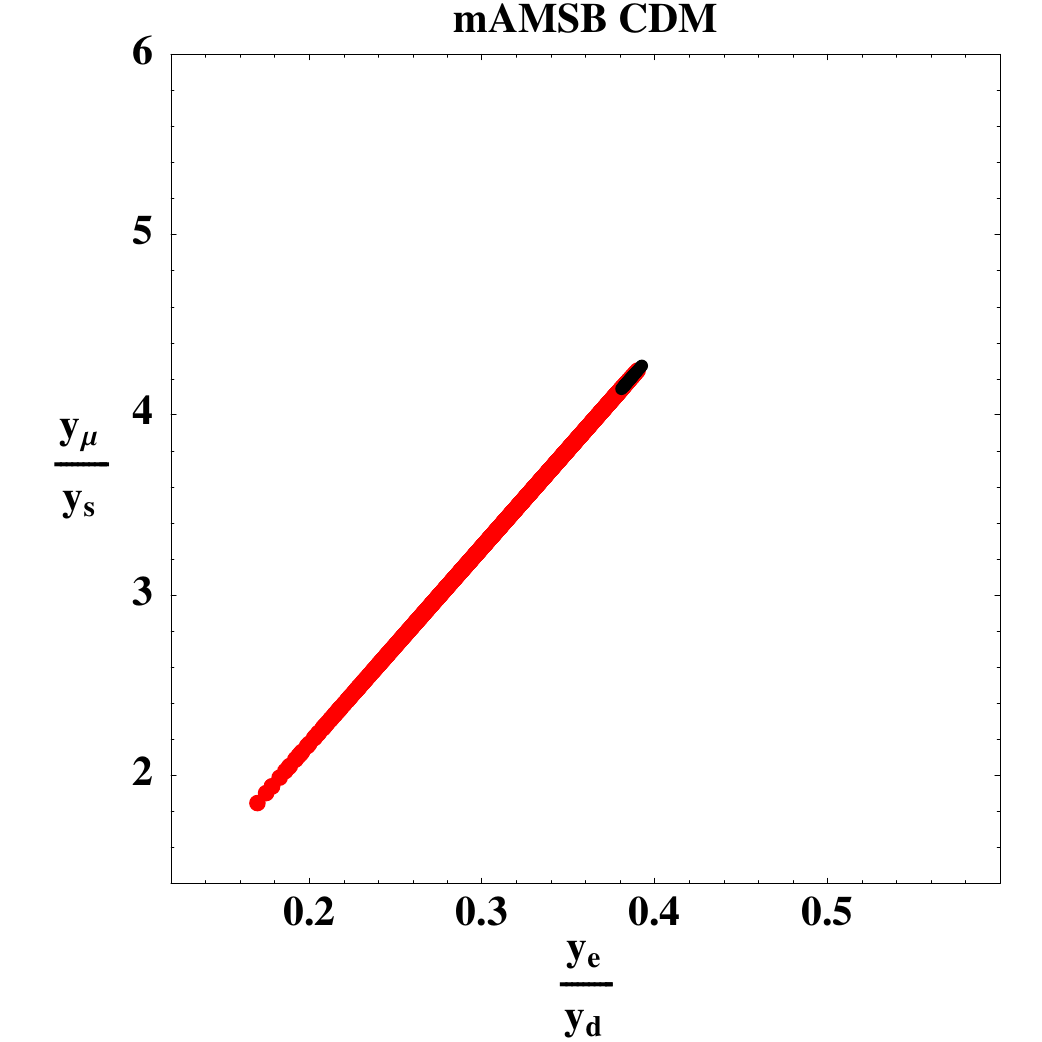}

 \includegraphics[scale=0.43]{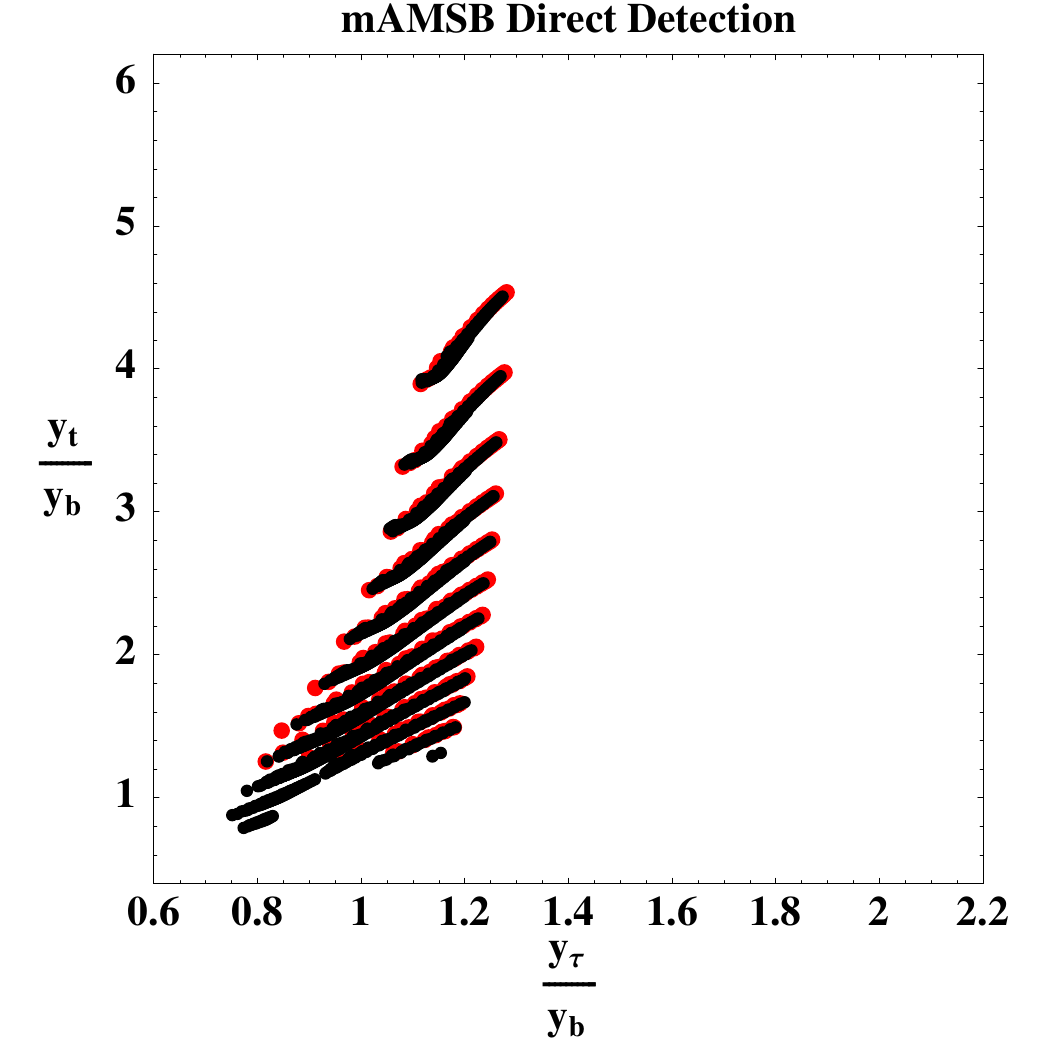} 
 \includegraphics[scale=0.43]{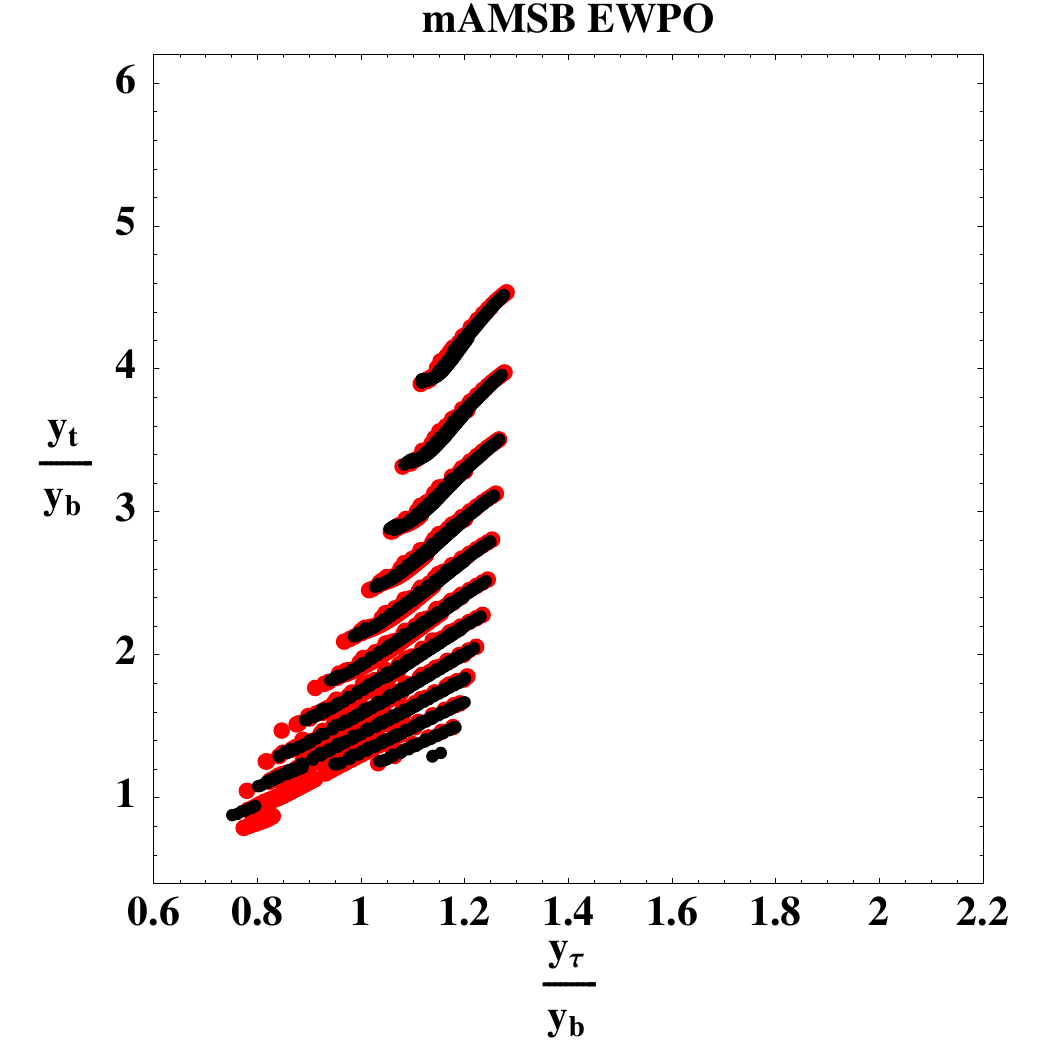}
 \includegraphics[scale=0.43]{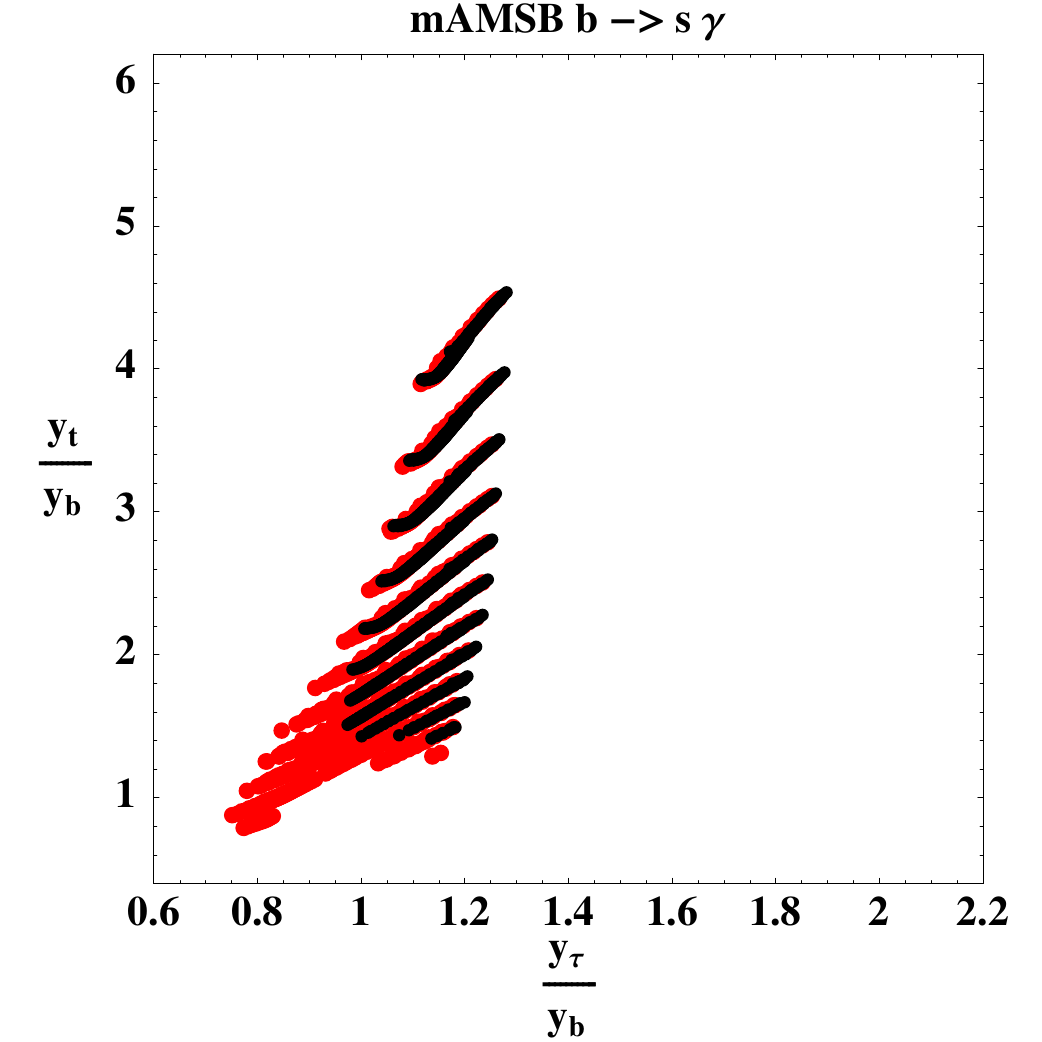}
 \includegraphics[scale=0.43]{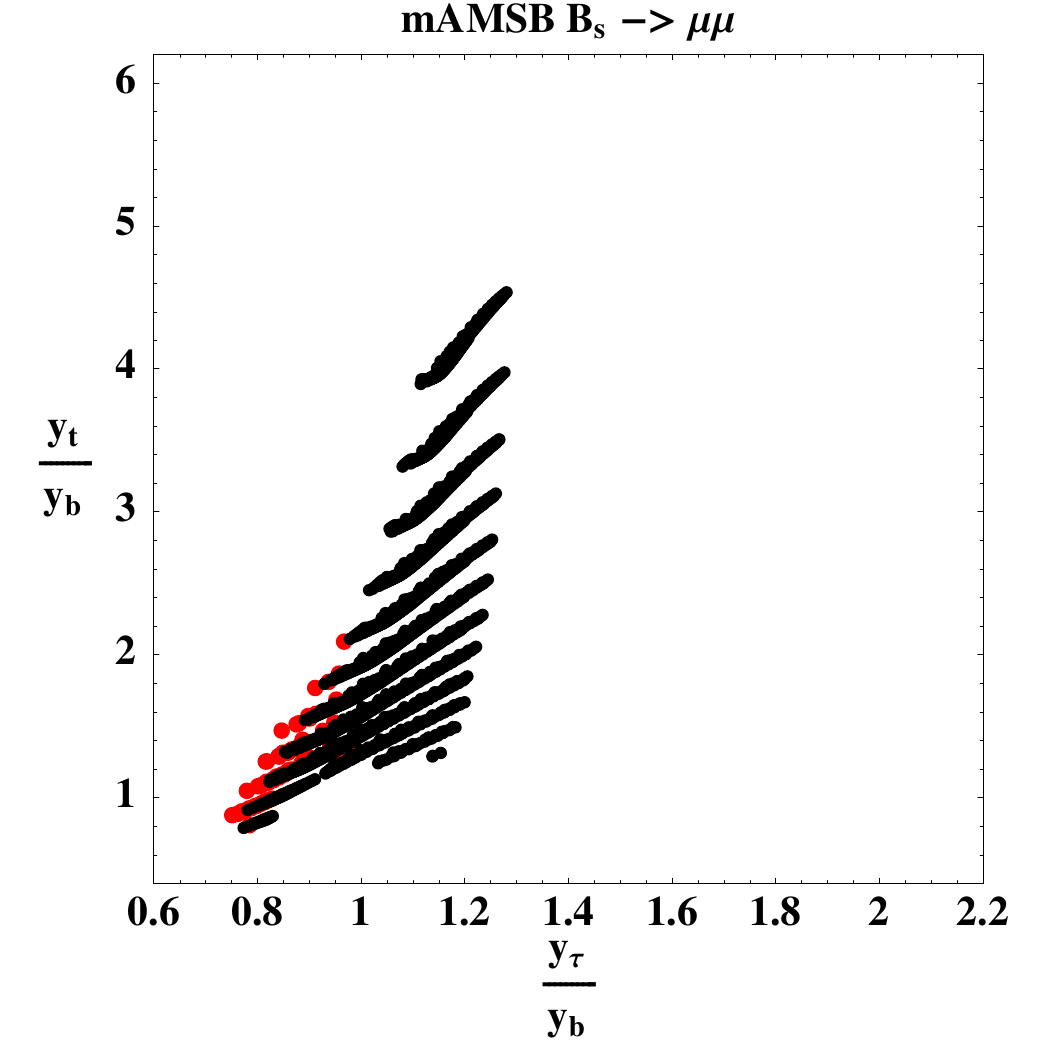}
 \includegraphics[scale=0.43]{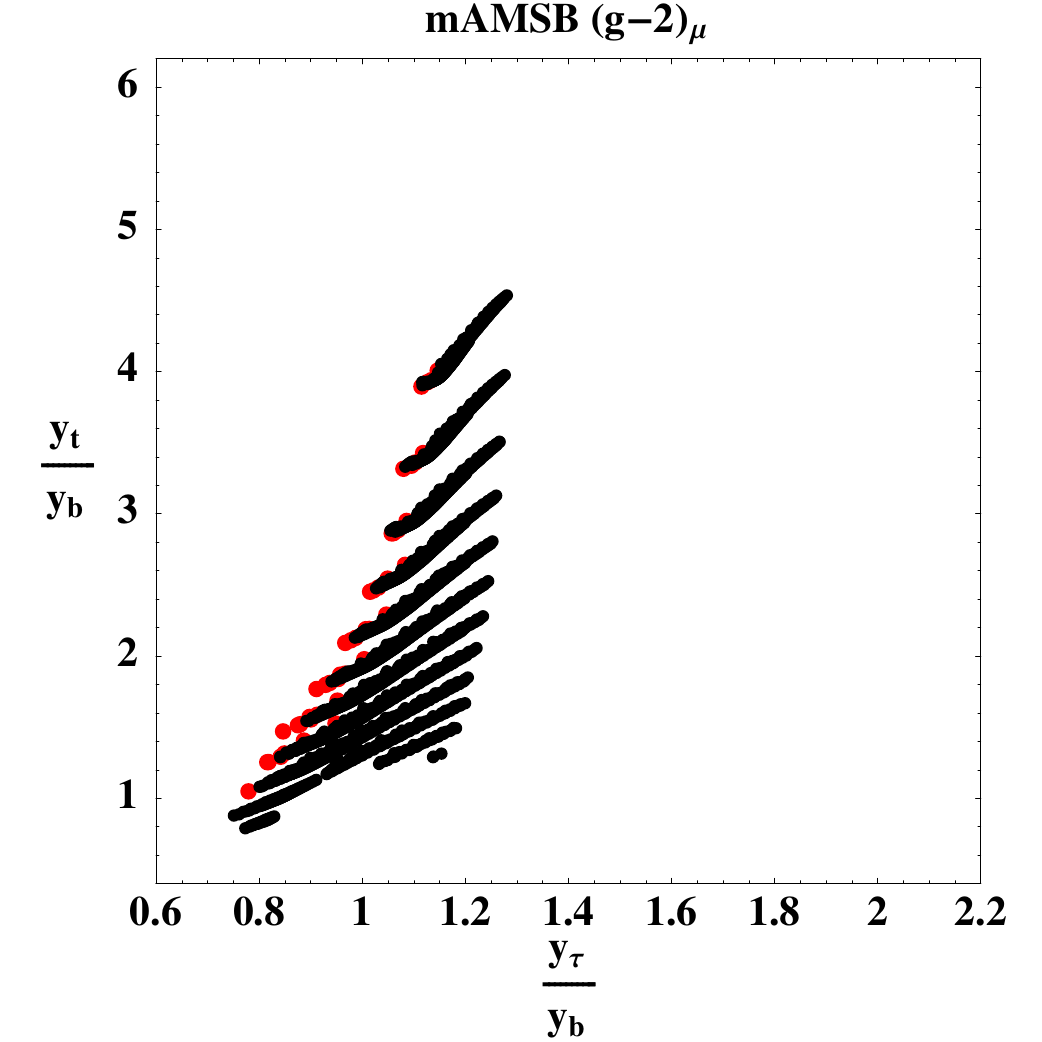}
 \includegraphics[scale=0.43]{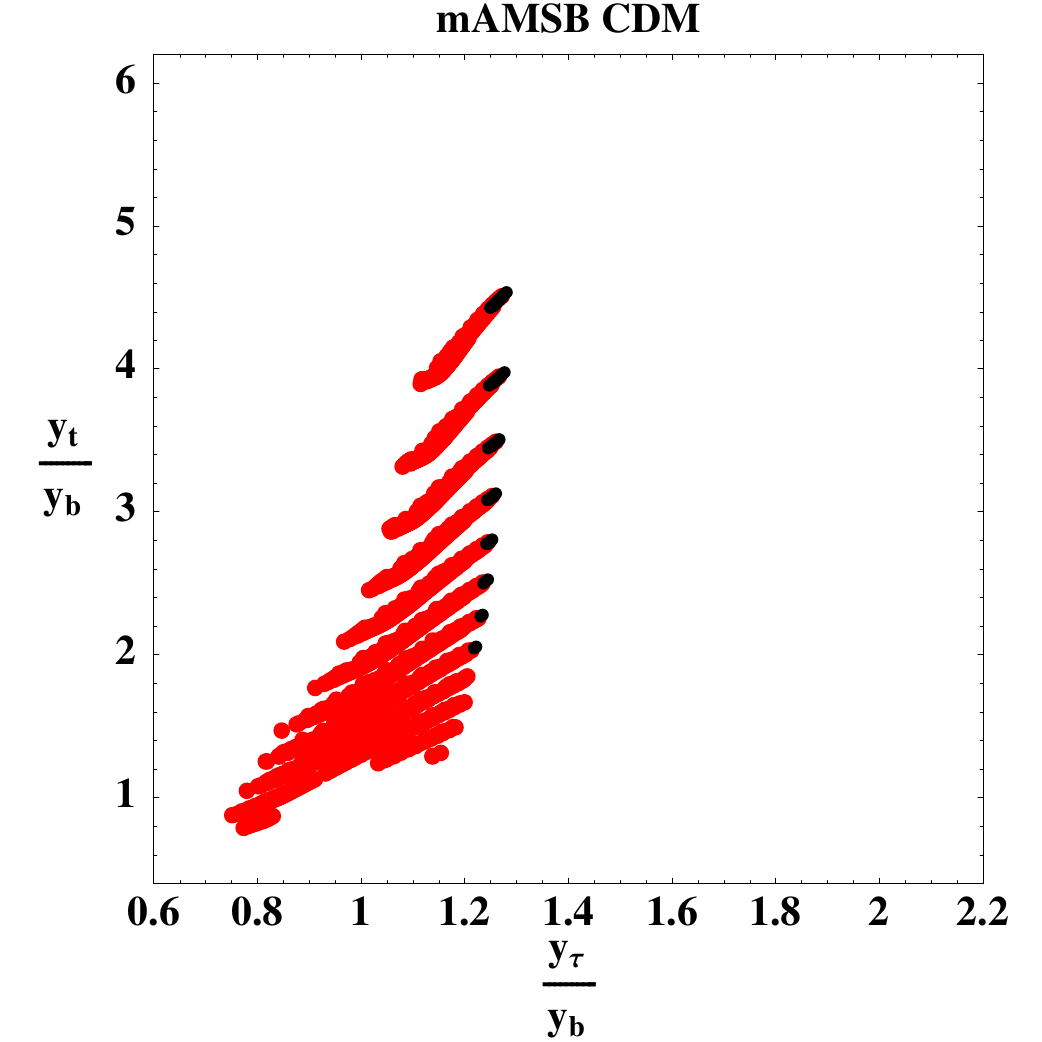}
 \caption[Plots for Each Experimental Constraint in mAMSB]{Impact of experimental constraints from direct detection, EWPO,  $b \rightarrow s \gamma$, $B_s \to \mu^+ \mu^-$, $(g-2)_\mu$ and CDM on $y_e/y_d$ and $y_\mu/y_s$ (upper plots) and $y_\tau/y_b$ and $y_t/y_b$ (lower plots) in mAMSB, cf.\ Ch.~\ref{Ch:Pheno}. Red dots denote parameter points which are excluded by the constraint, while black dots indicate parameter points which are allowed. In the lower plots for the third generation, the different lines of points correspond to different values of $\tan \beta$, increasing from top to bottom. \label{Fig:Plots_mAMSB} }
\end{figure}

\begin{figure}
 \centering
 \includegraphics[scale=0.43]{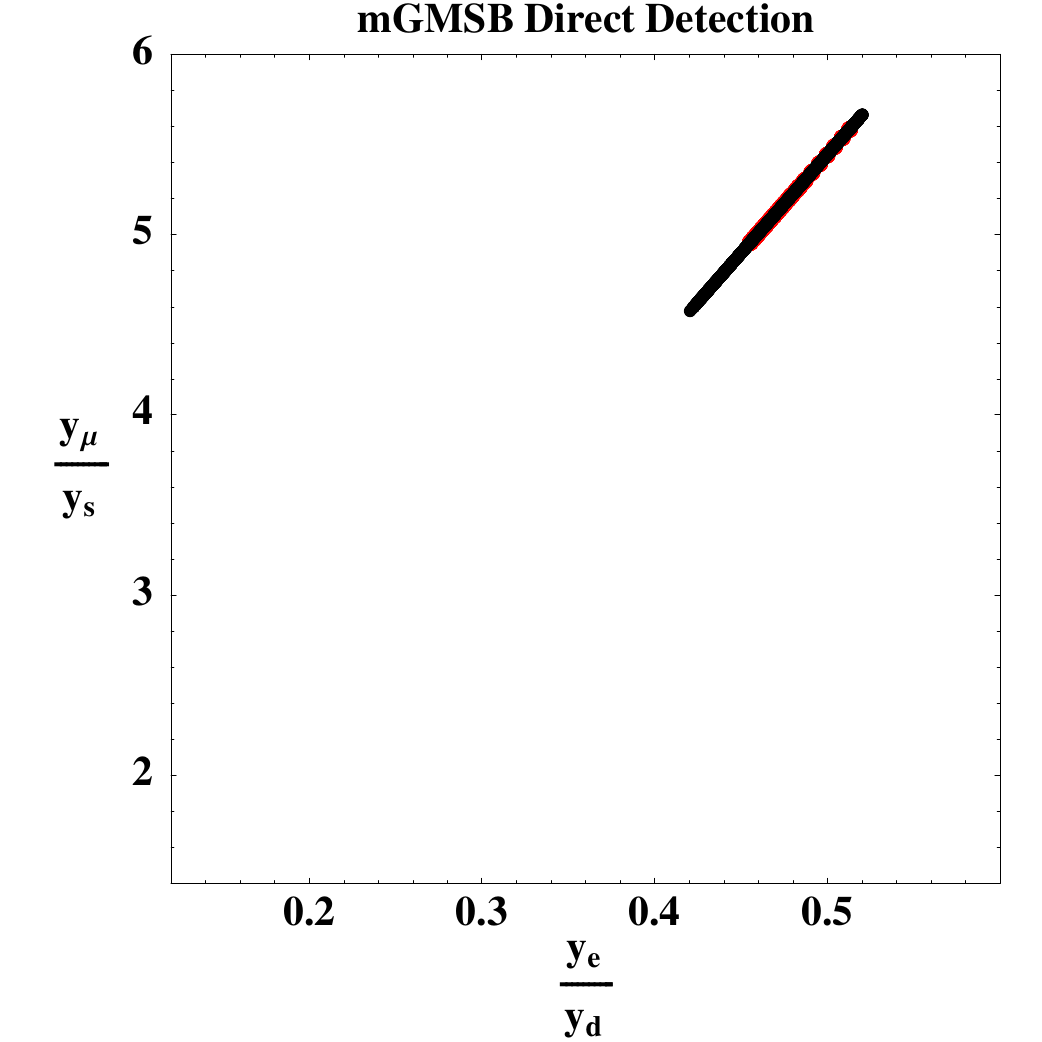}
 \includegraphics[scale=0.43]{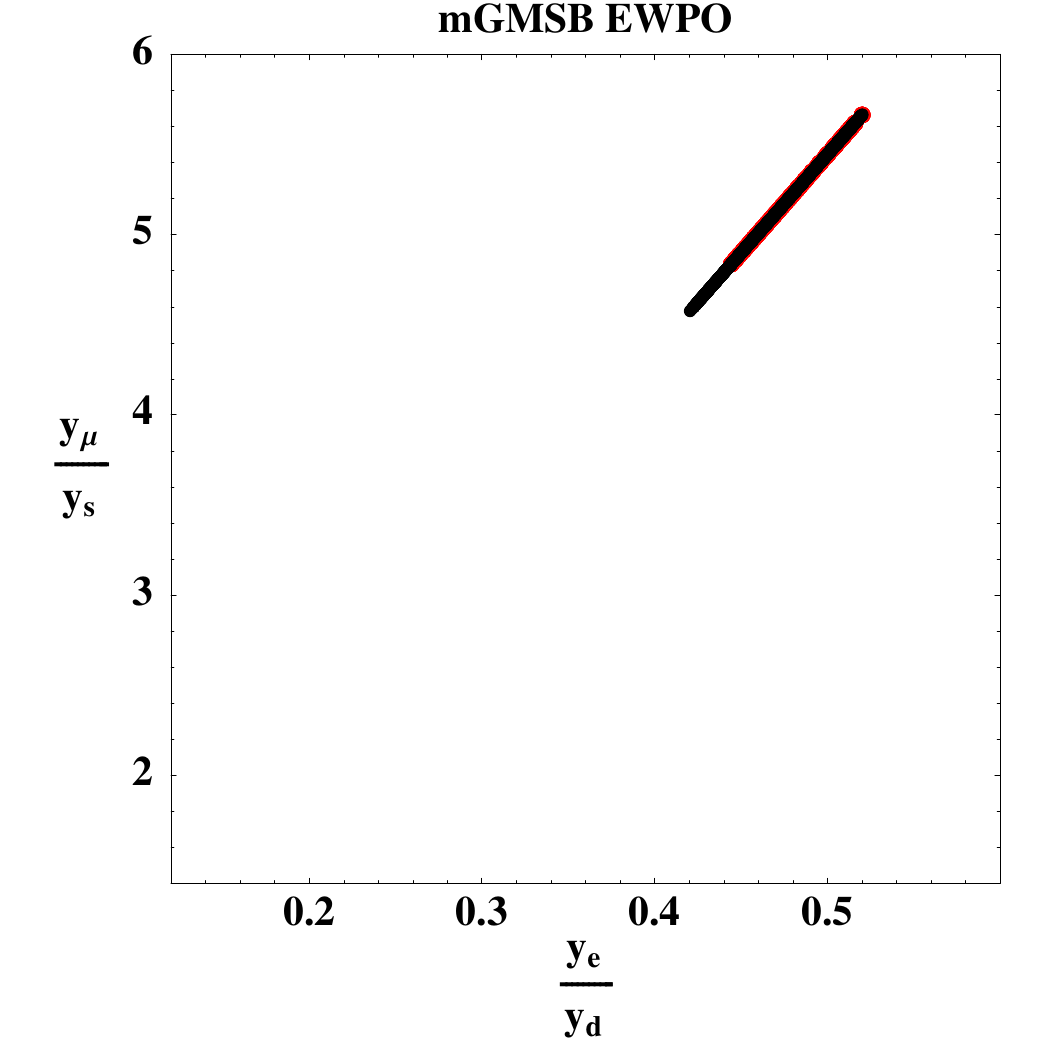}
 \includegraphics[scale=0.43]{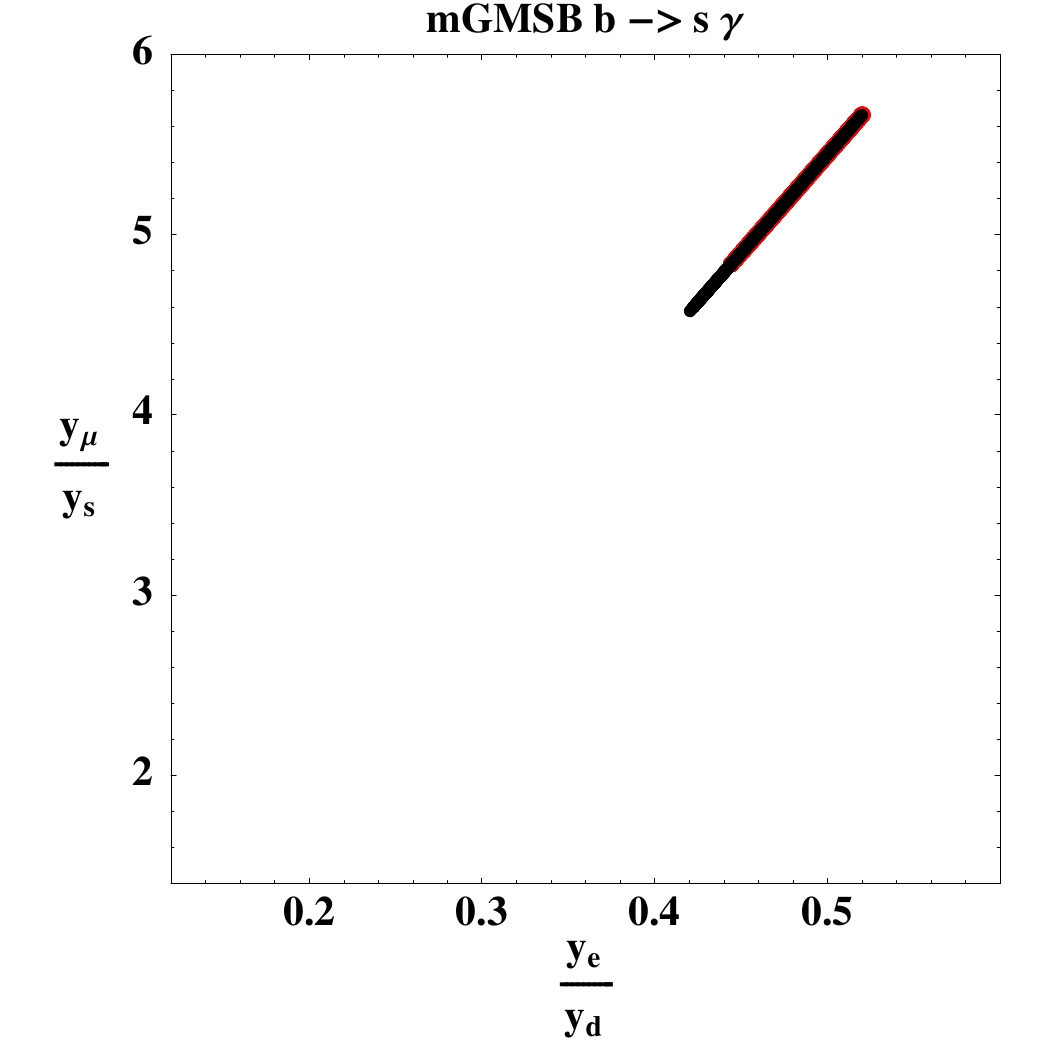}
 \includegraphics[scale=0.43]{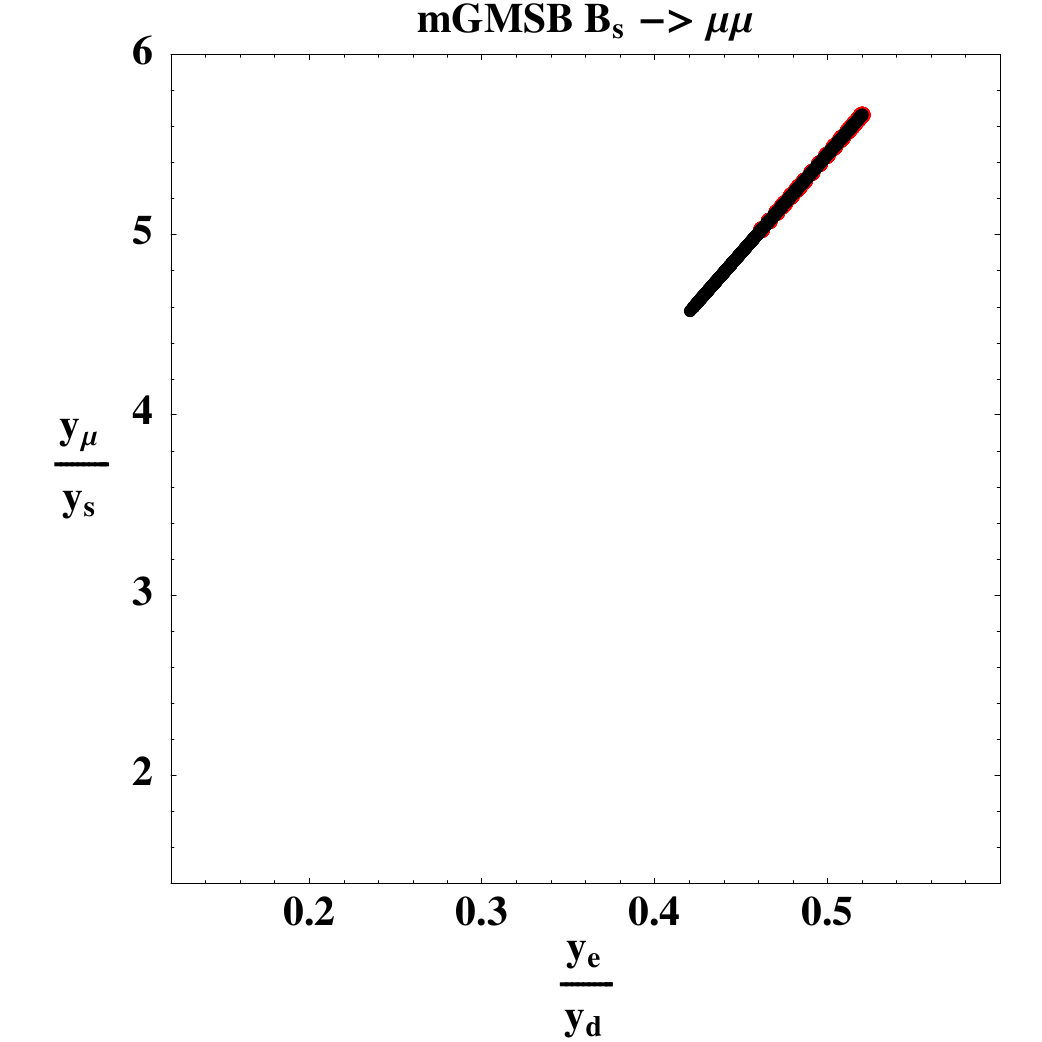}
 \includegraphics[scale=0.43]{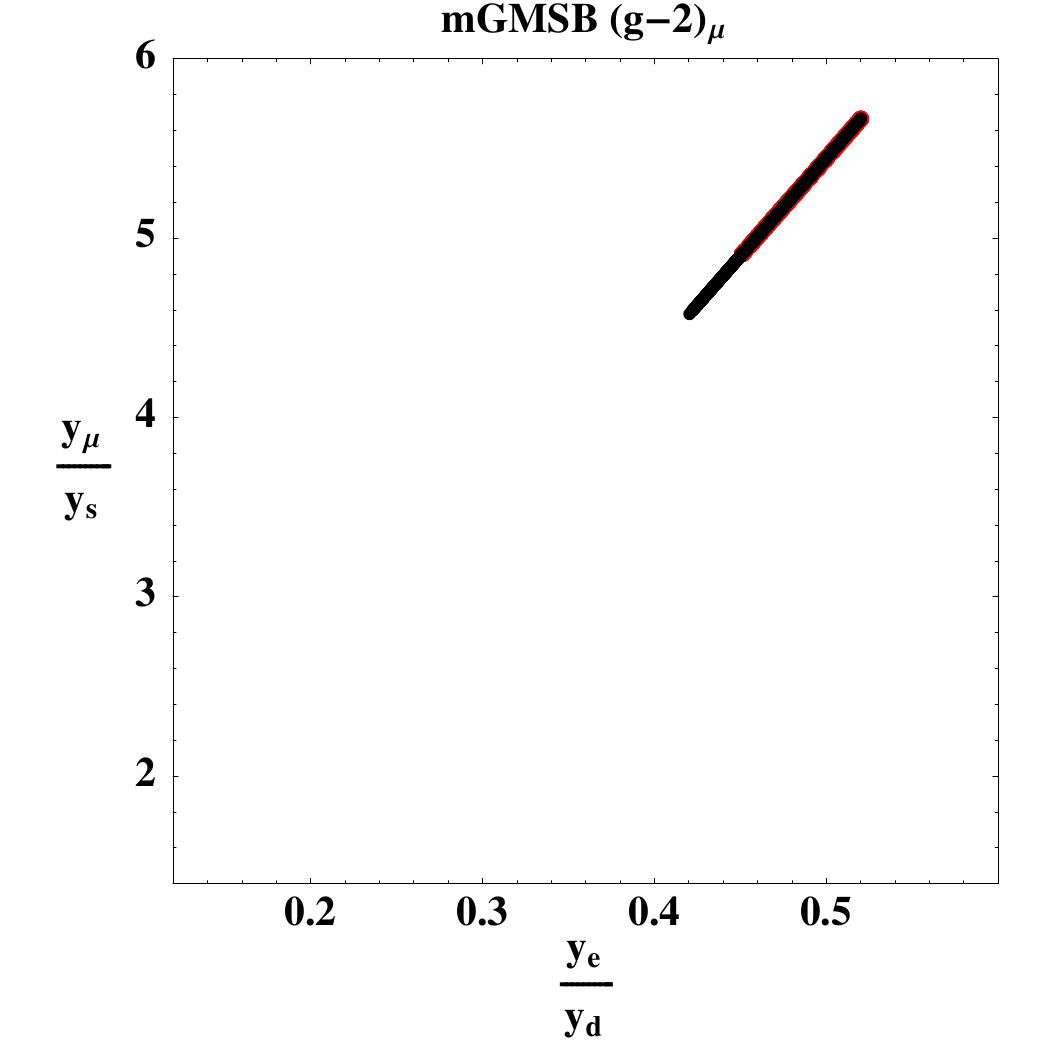}

 \includegraphics[scale=0.43]{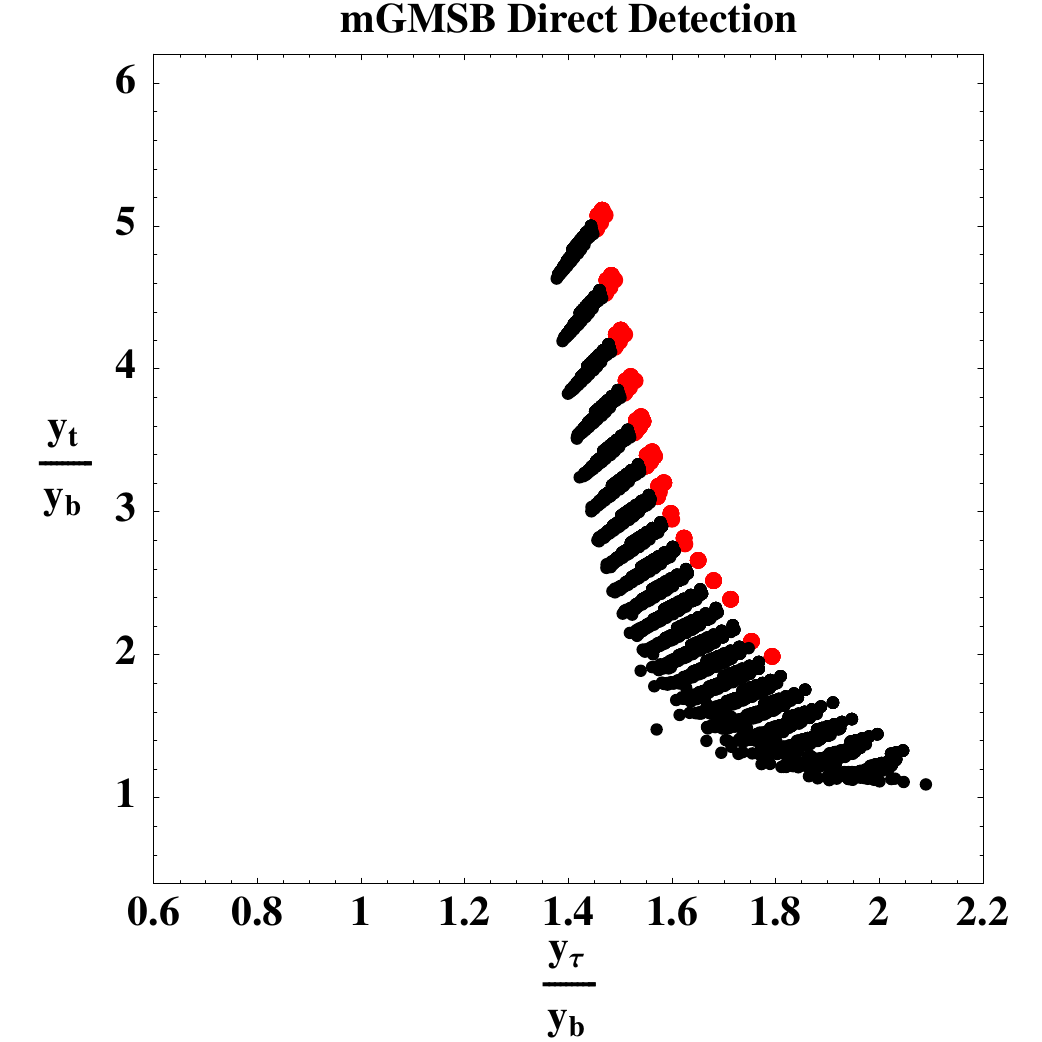} 
 \includegraphics[scale=0.43]{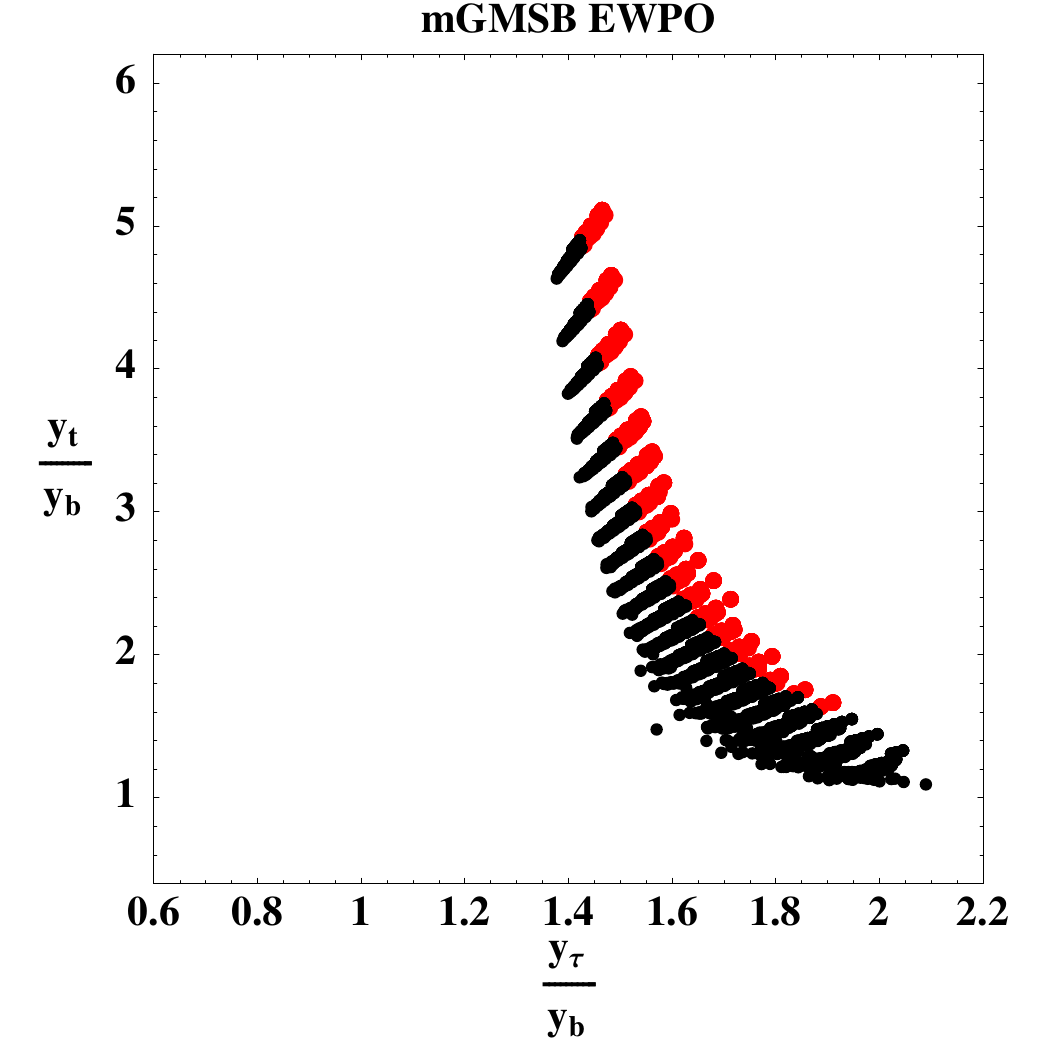}
 \includegraphics[scale=0.43]{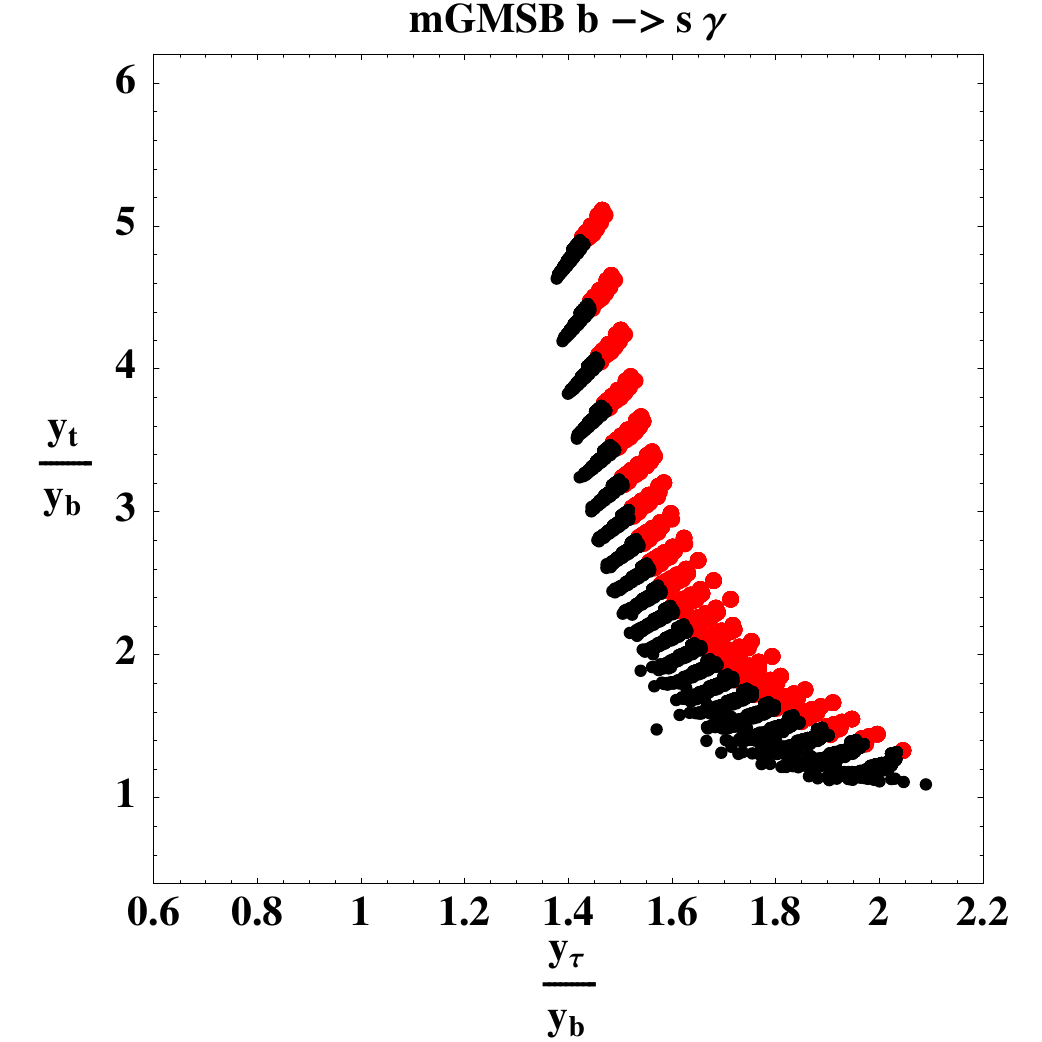}
 \includegraphics[scale=0.43]{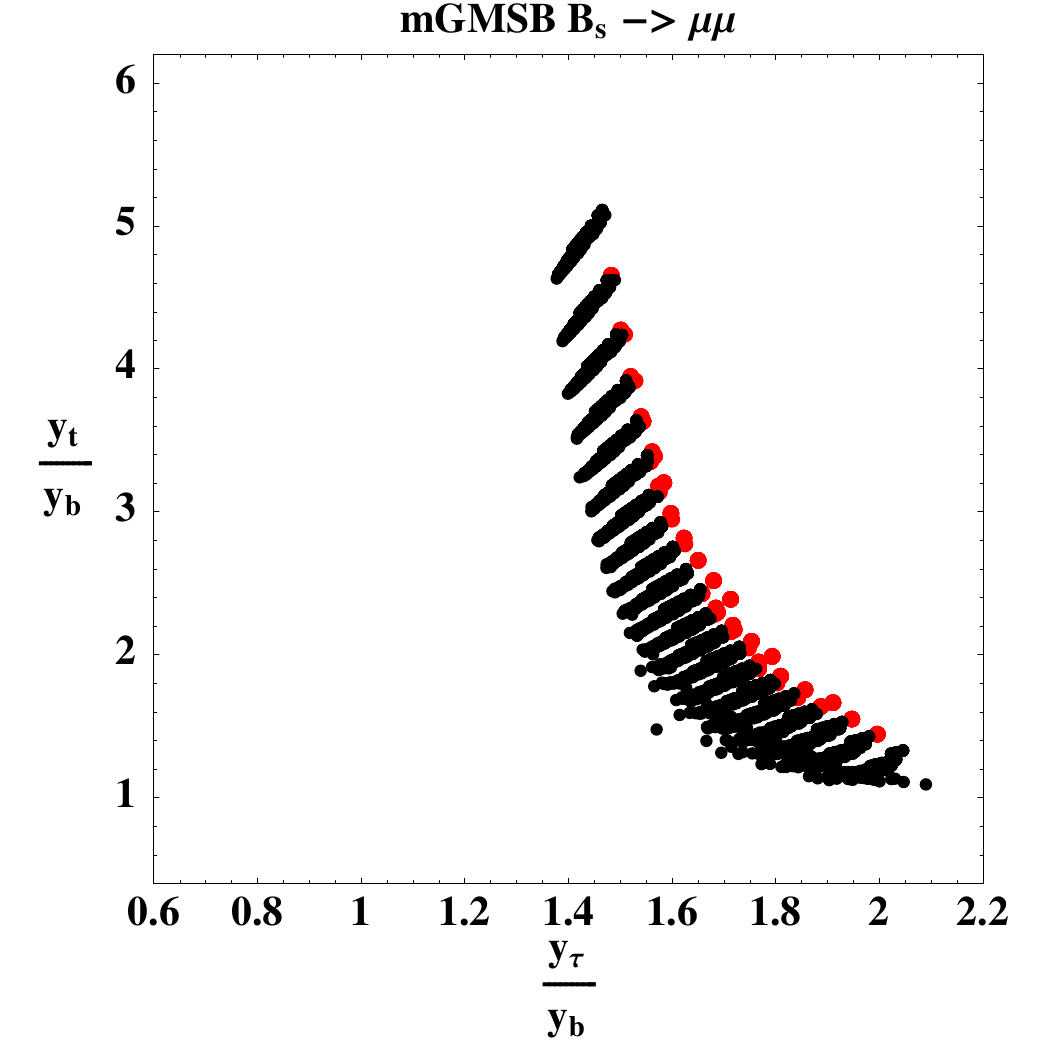}
 \includegraphics[scale=0.43]{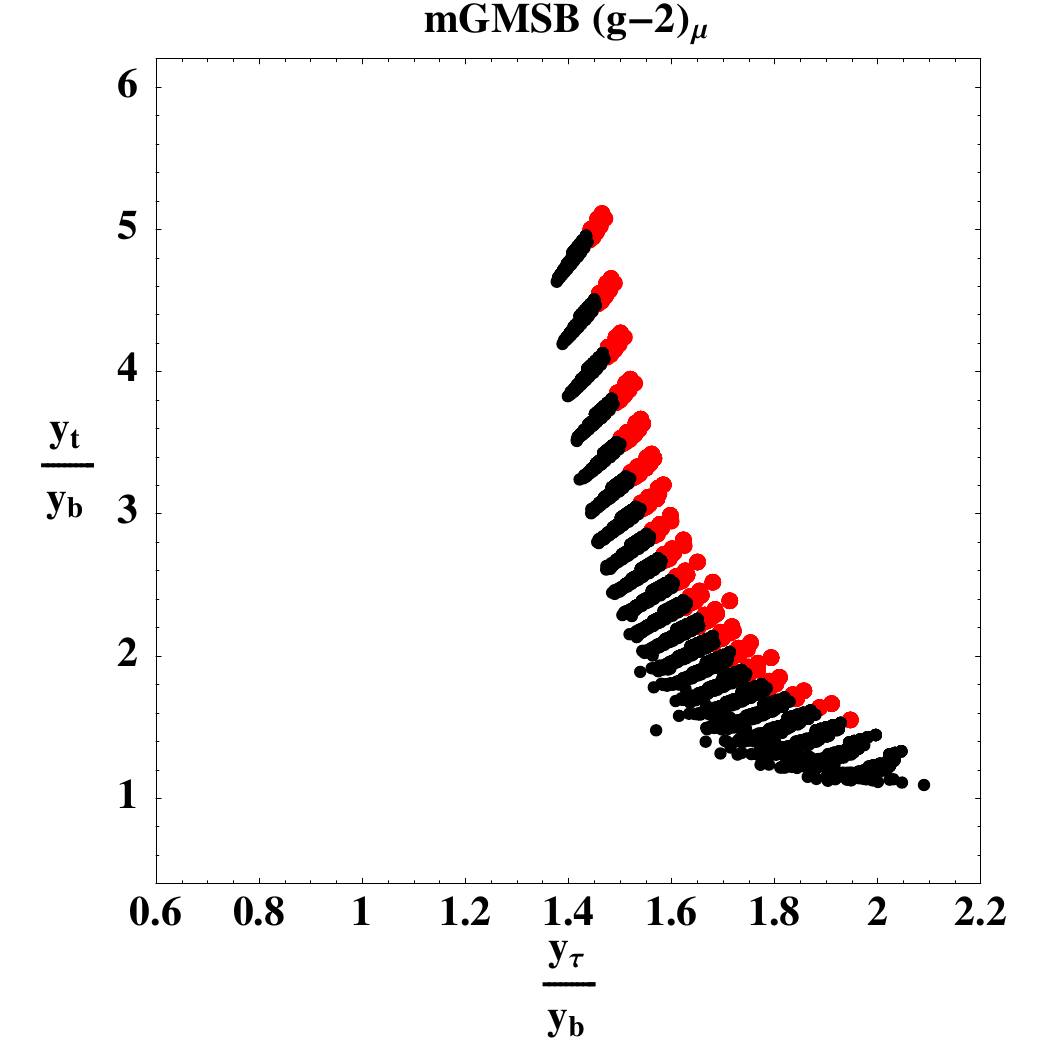}
 \caption[Plots for Each Experimental Constraint in mGMSB]{Impact of experimental constraints from direct detection, EWPO,  $b \rightarrow s \gamma$, $B_s \to \mu^+ \mu^-$ and $(g-2)_\mu$ on $y_e/y_d$ and $y_\mu/y_s$ (upper plots) and $y_\tau/y_b$ and $y_t/y_b$ (lower plots) in mGMSB, cf.\ Ch.~\ref{Ch:Pheno}. Red dots denote parameter points which are excluded by the constraint, while black dots indicate parameter points which are allowed. In the lower plots for the third generation, the different lines of points correspond to different values of $\tan \beta$, increasing from top to bottom. \label{Fig:Plots_mGMSB} }
\end{figure}

\begin{figure}
 \centering
 \includegraphics[scale=0.43]{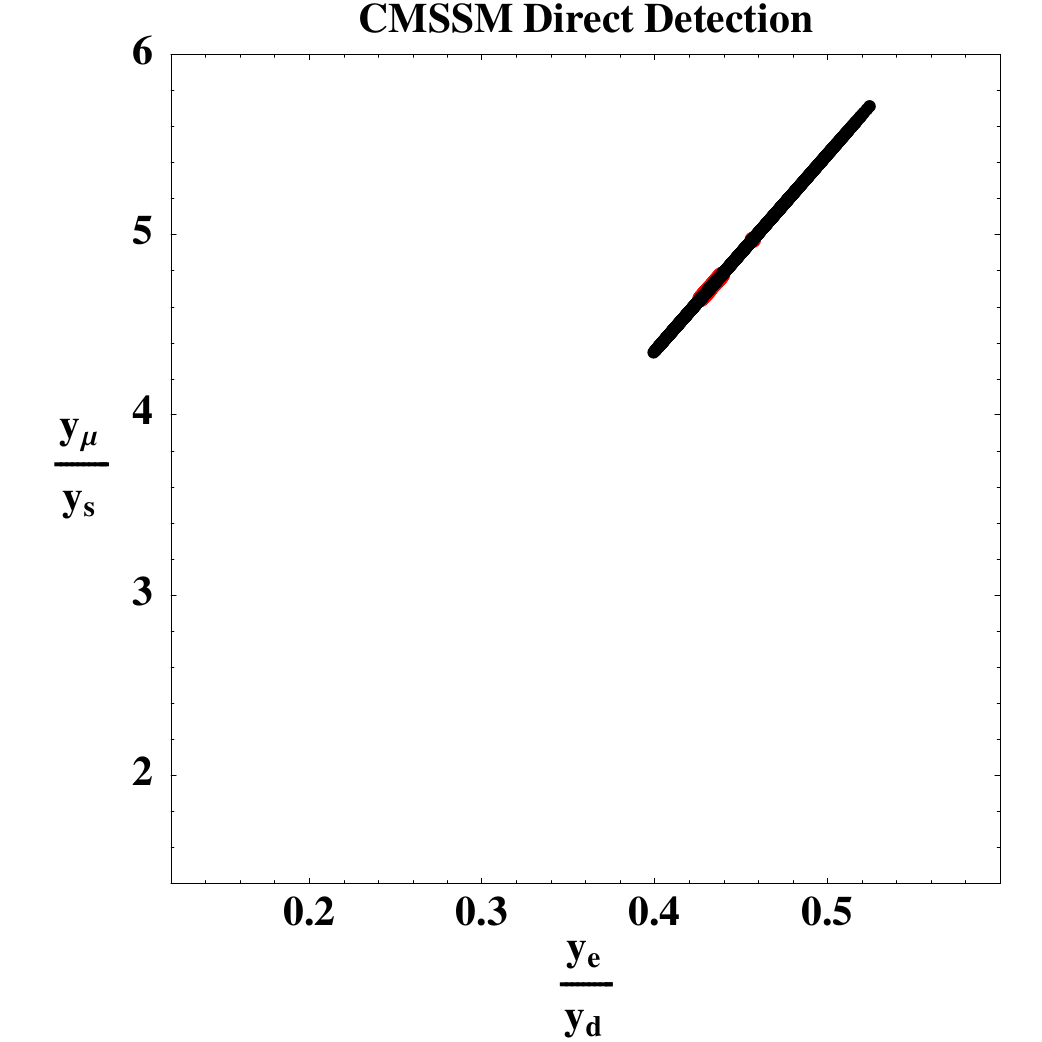}
 \includegraphics[scale=0.43]{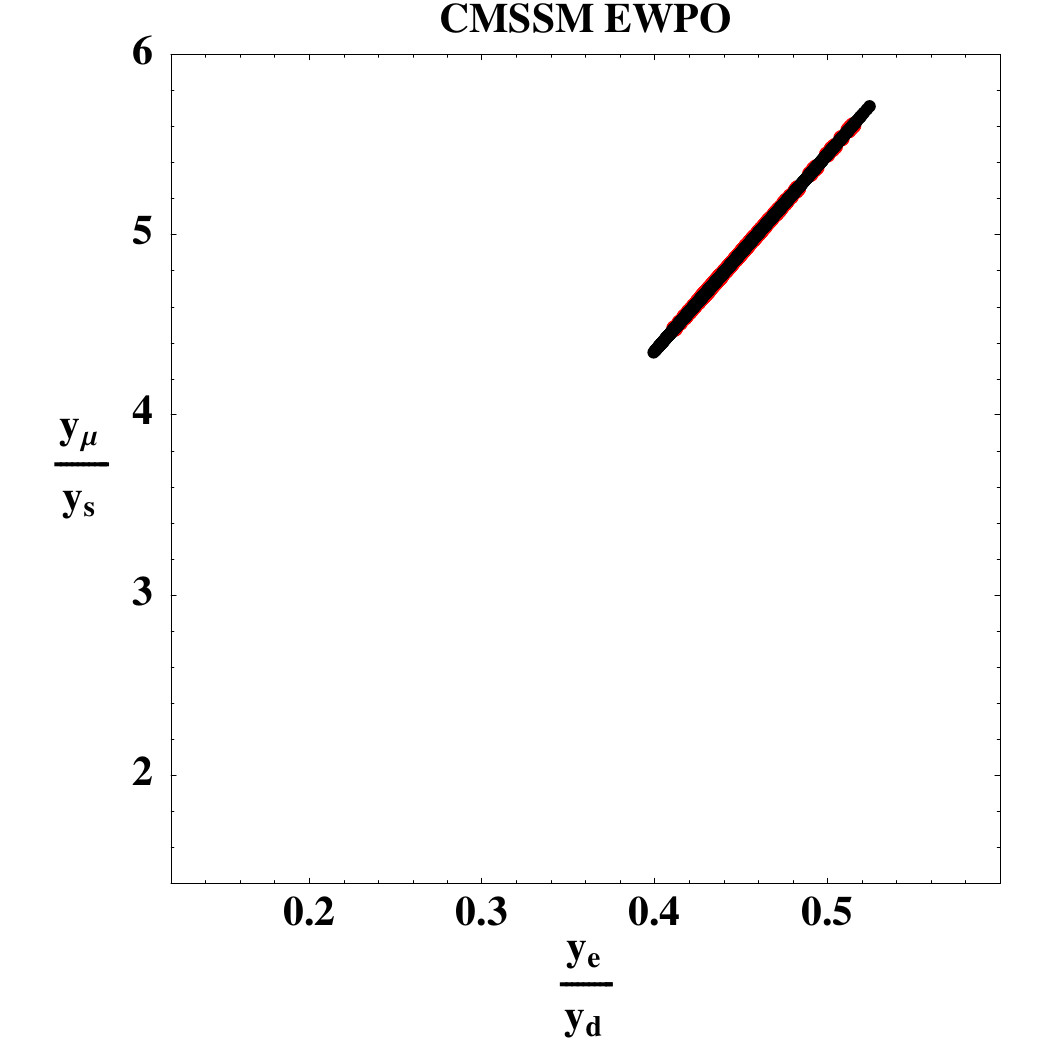}
 \includegraphics[scale=0.43]{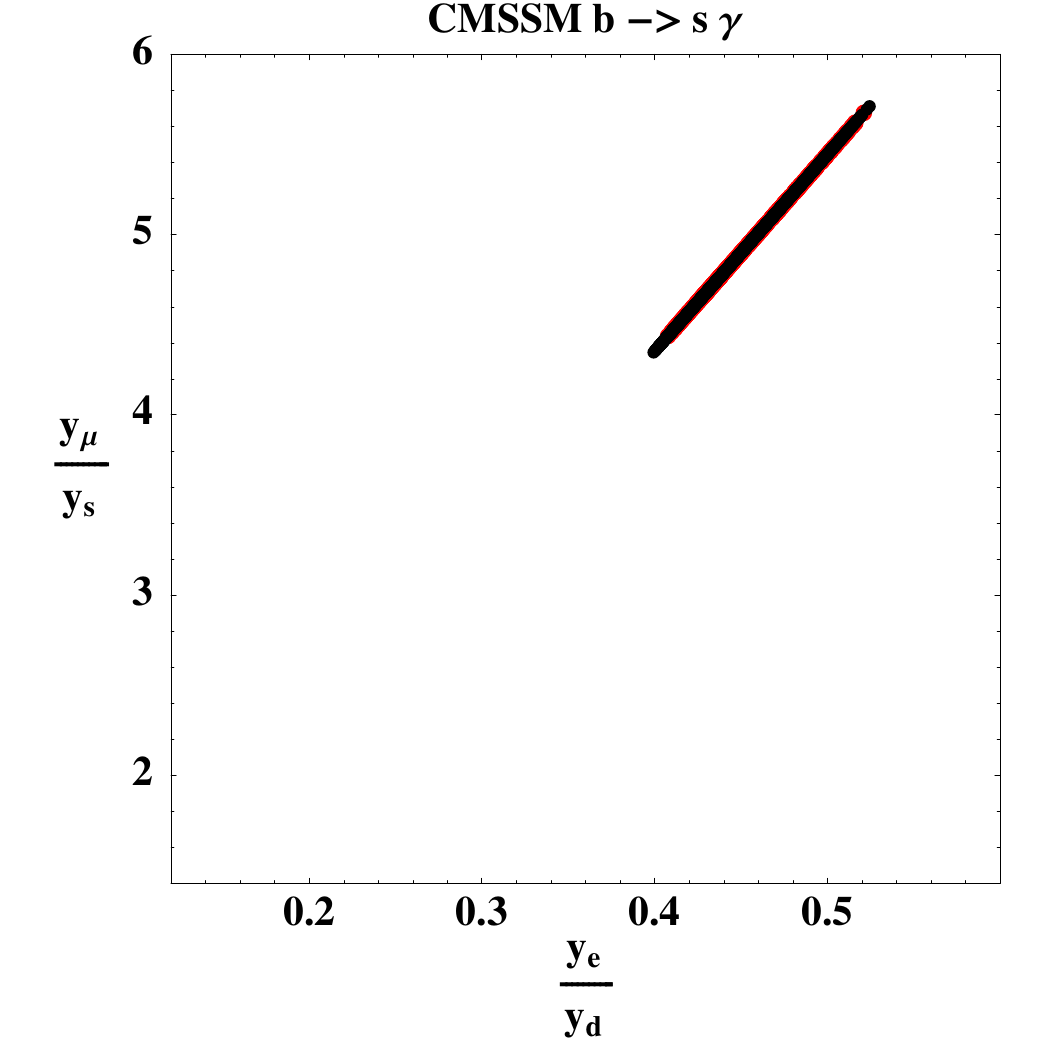}
 \includegraphics[scale=0.43]{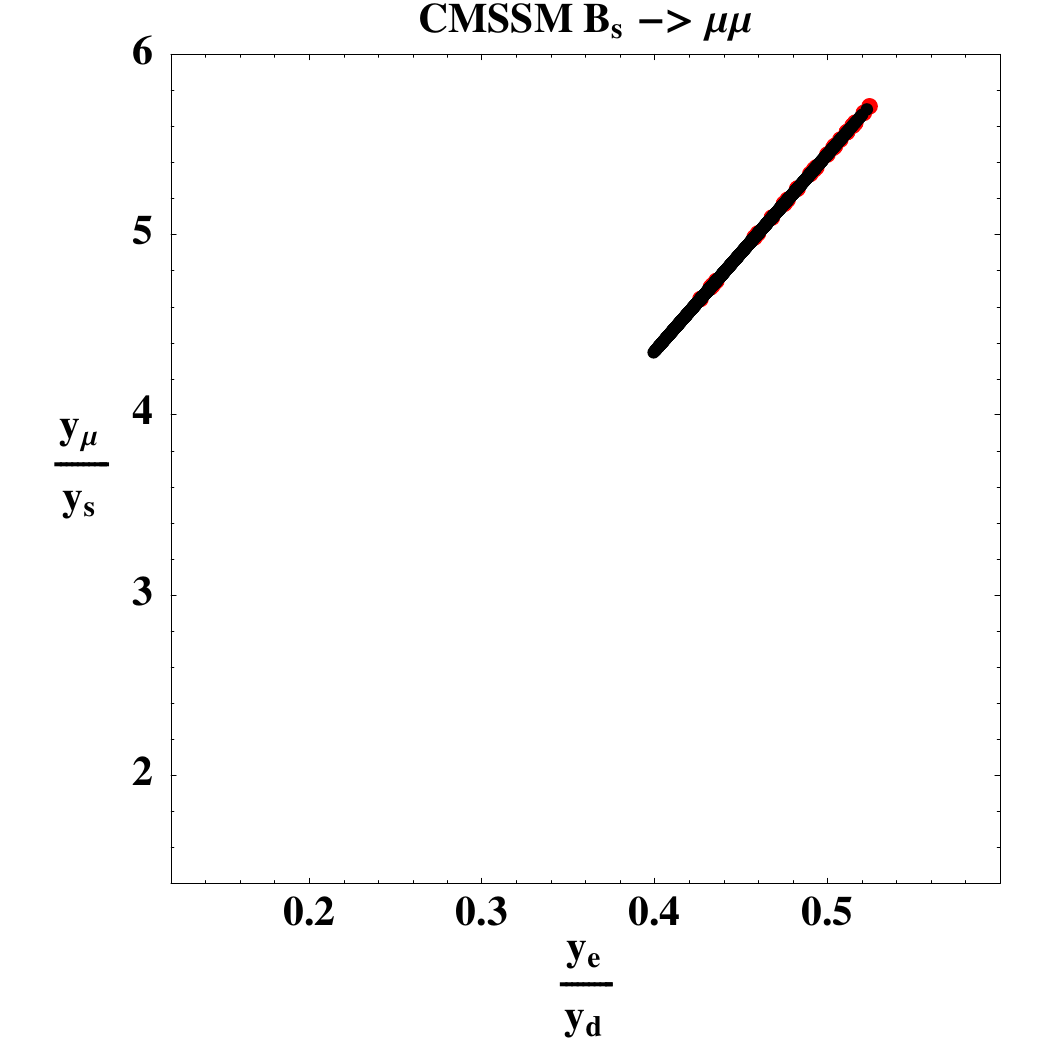}
 \includegraphics[scale=0.43]{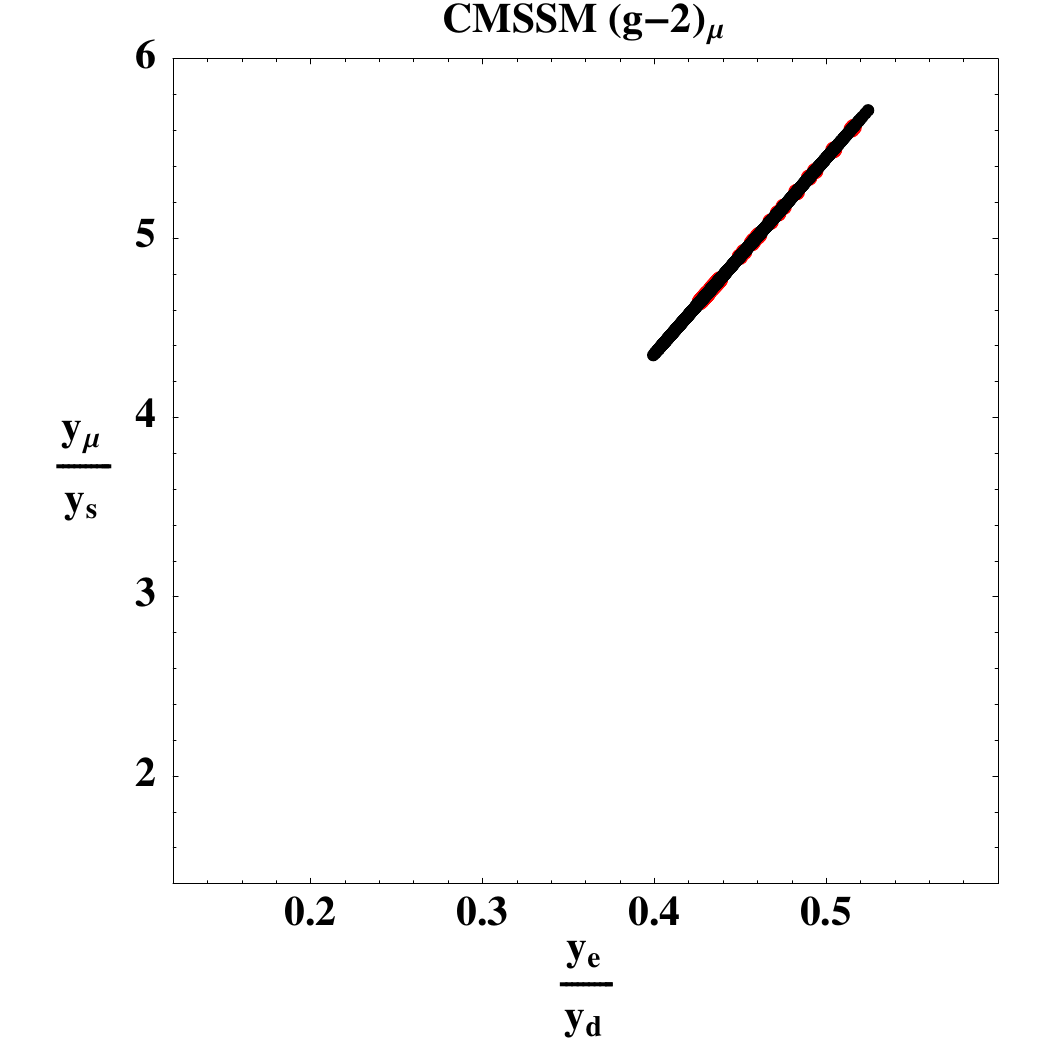}
 \includegraphics[scale=0.43]{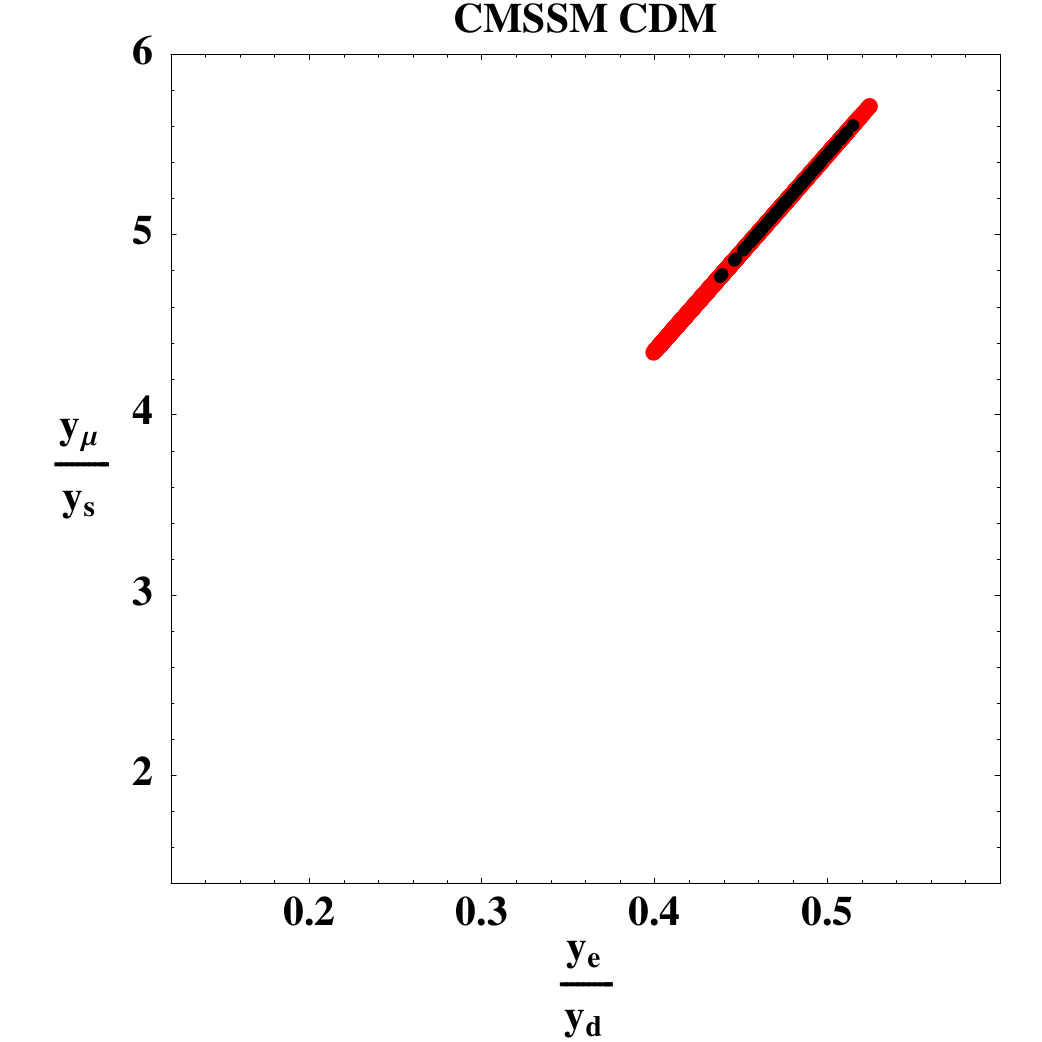}

 \includegraphics[scale=0.43]{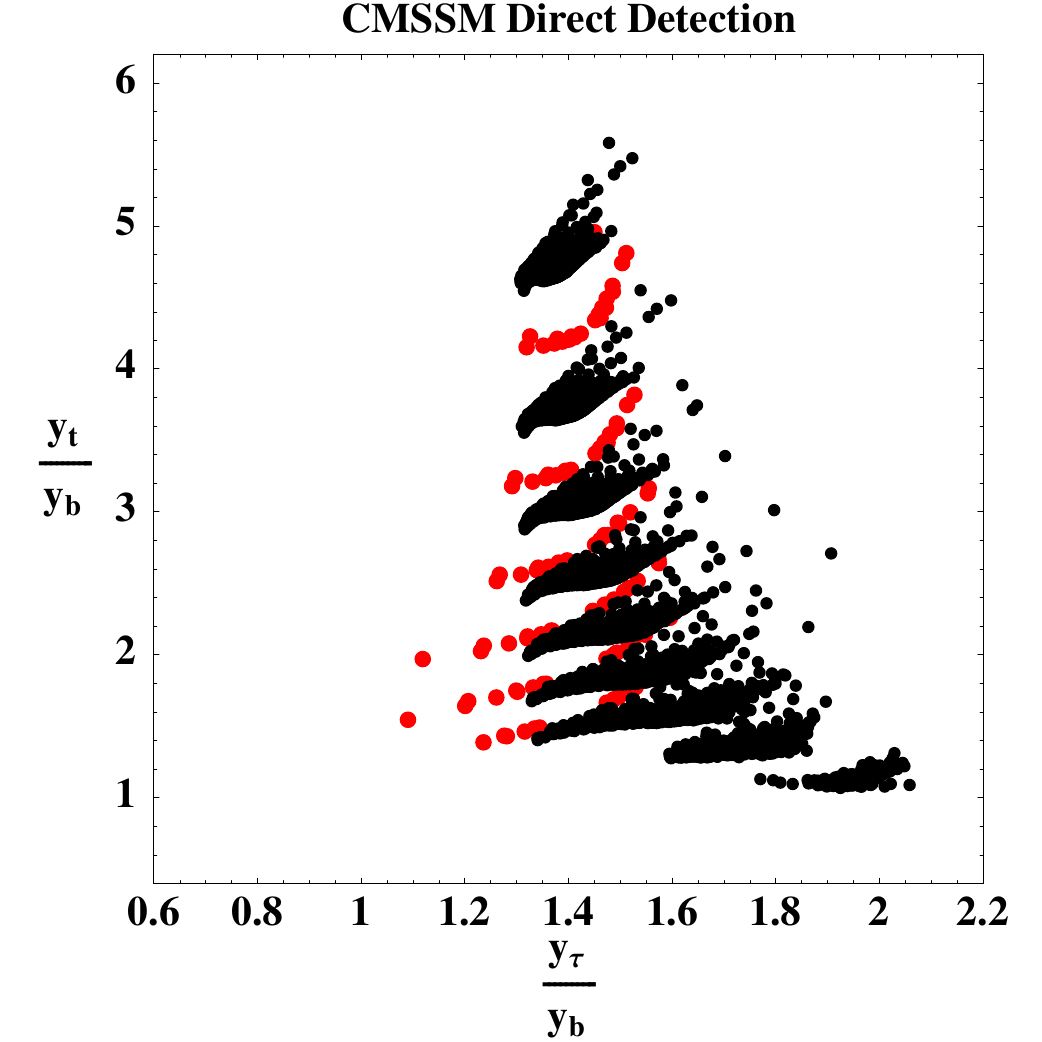} 
 \includegraphics[scale=0.43]{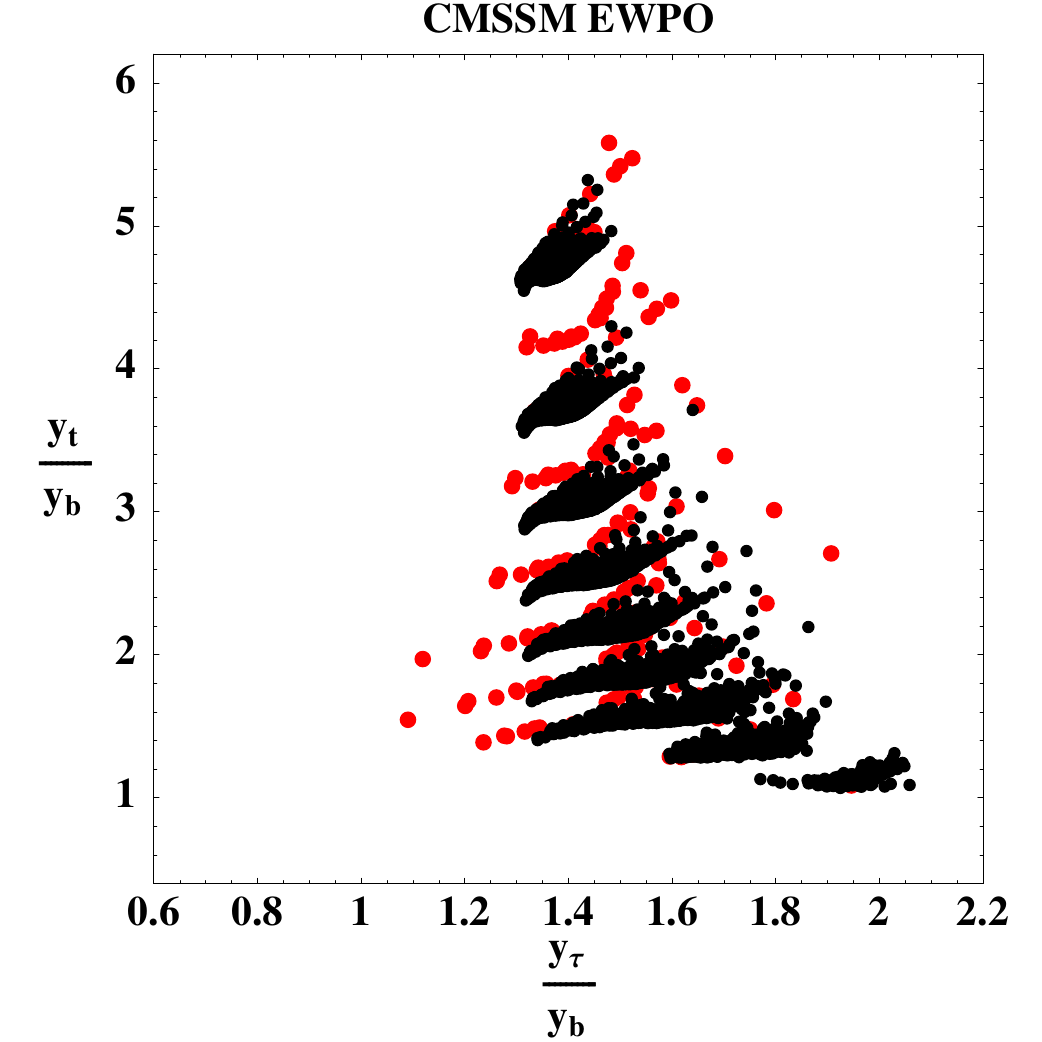}
 \includegraphics[scale=0.43]{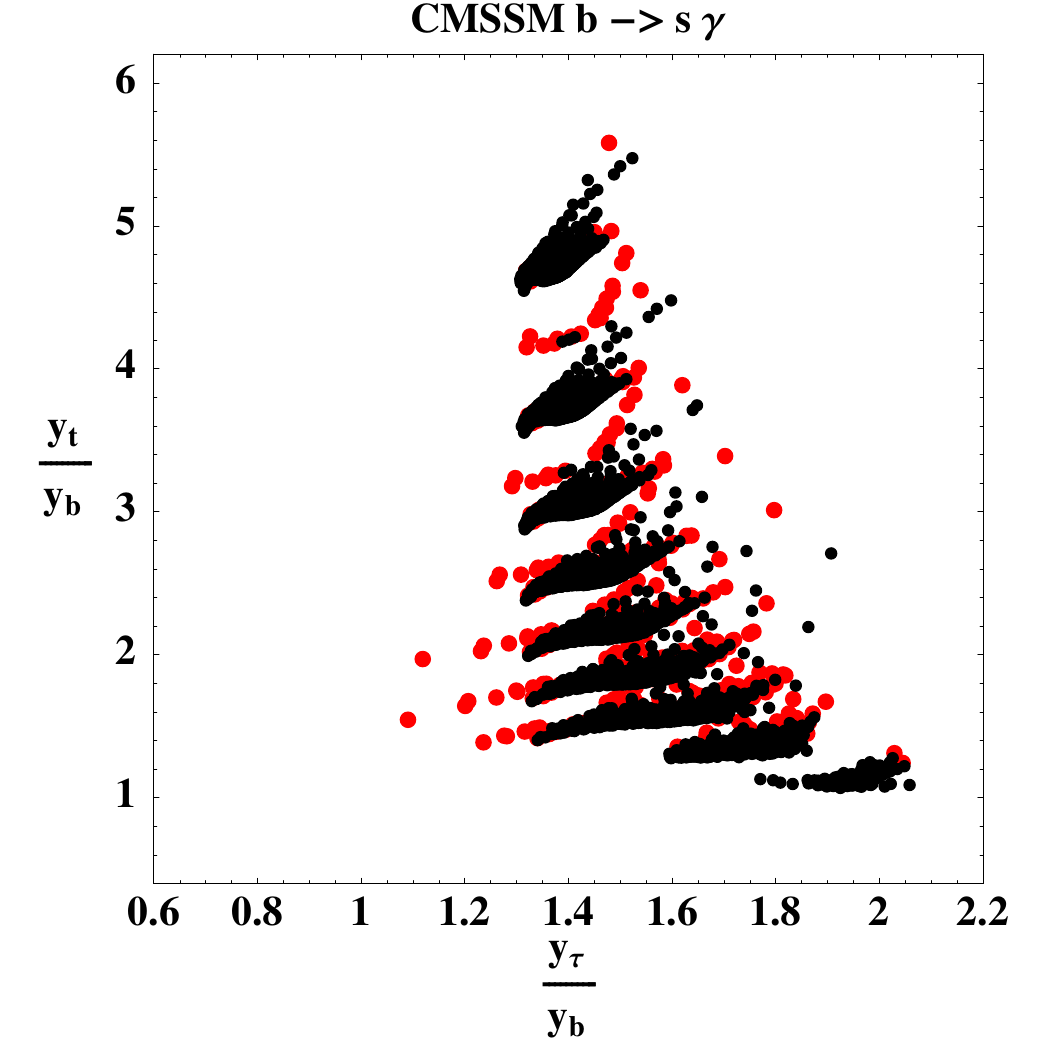}
 \includegraphics[scale=0.43]{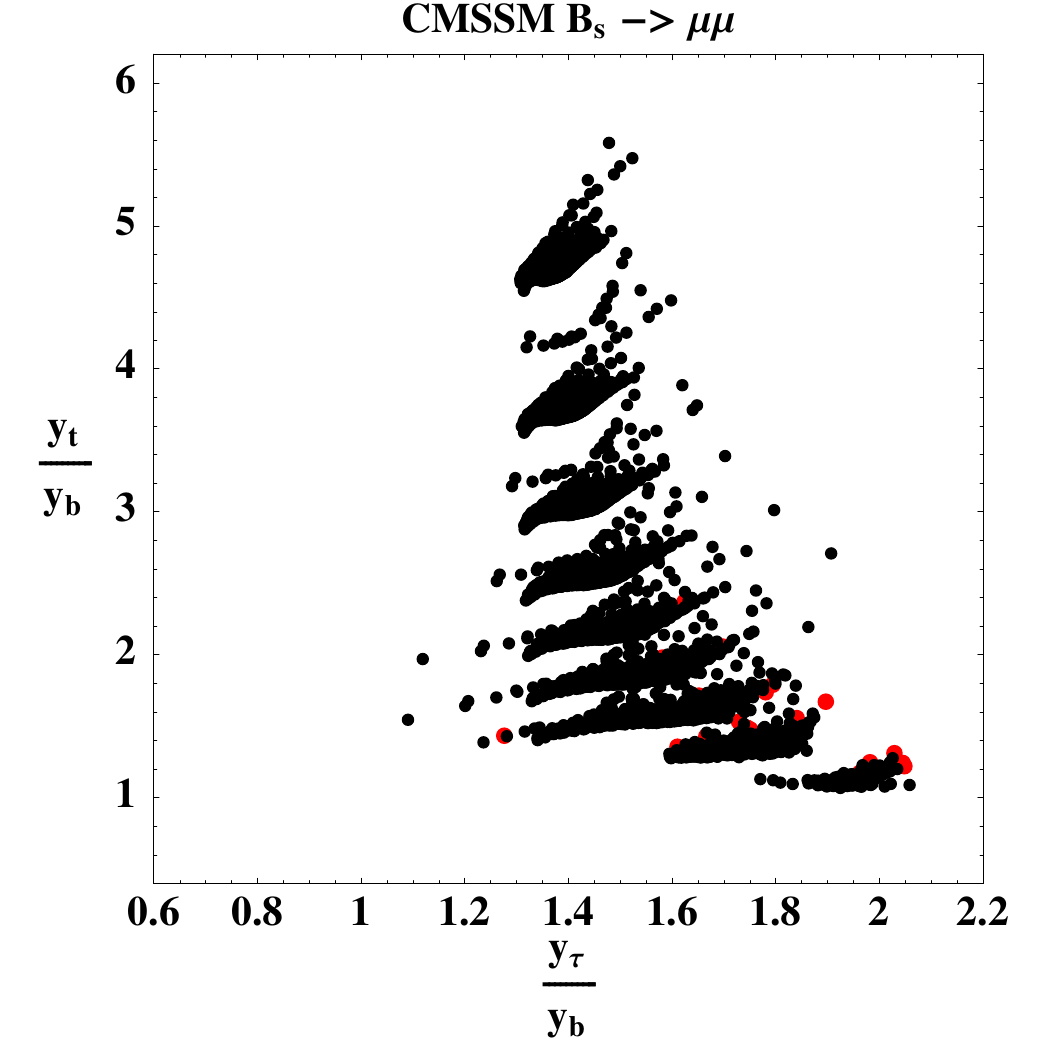}
 \includegraphics[scale=0.43]{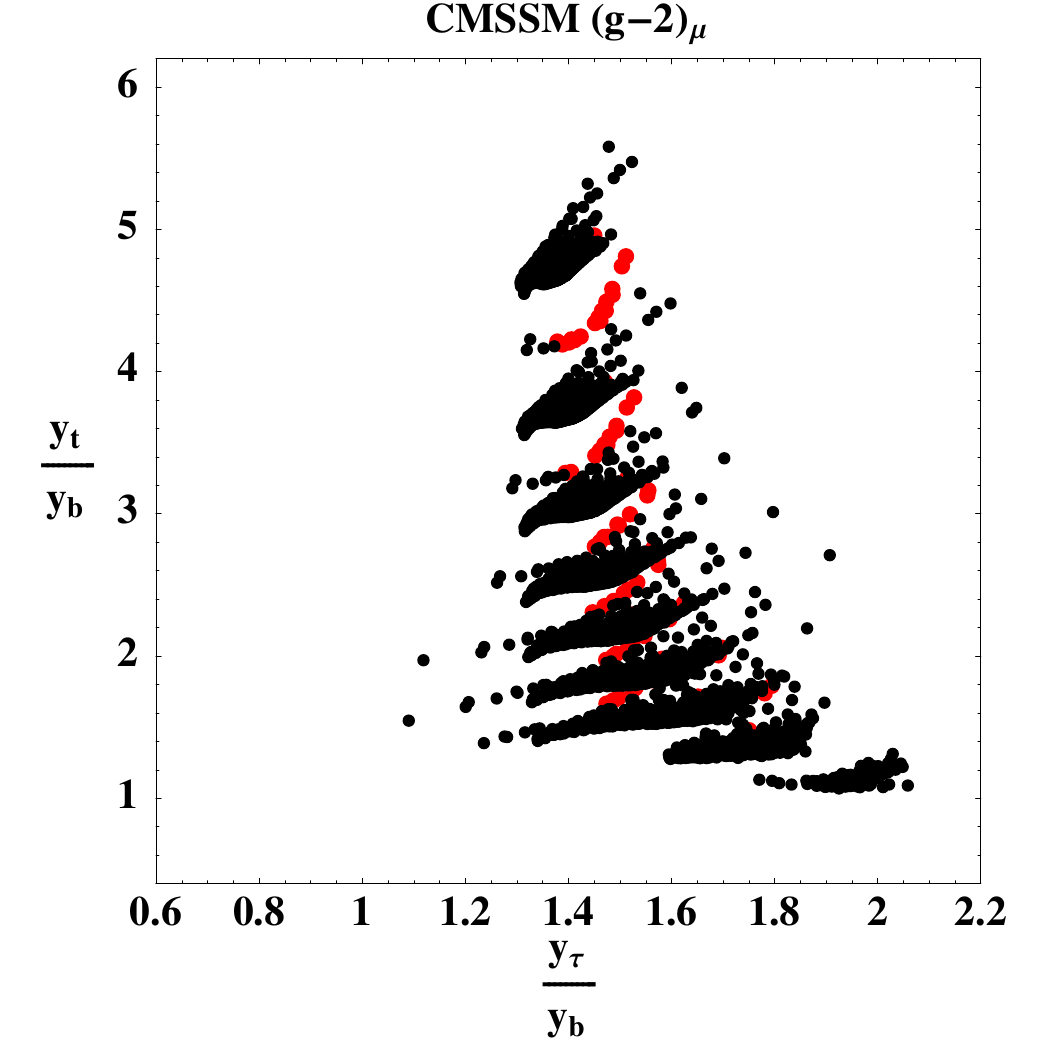}
 \includegraphics[scale=0.43]{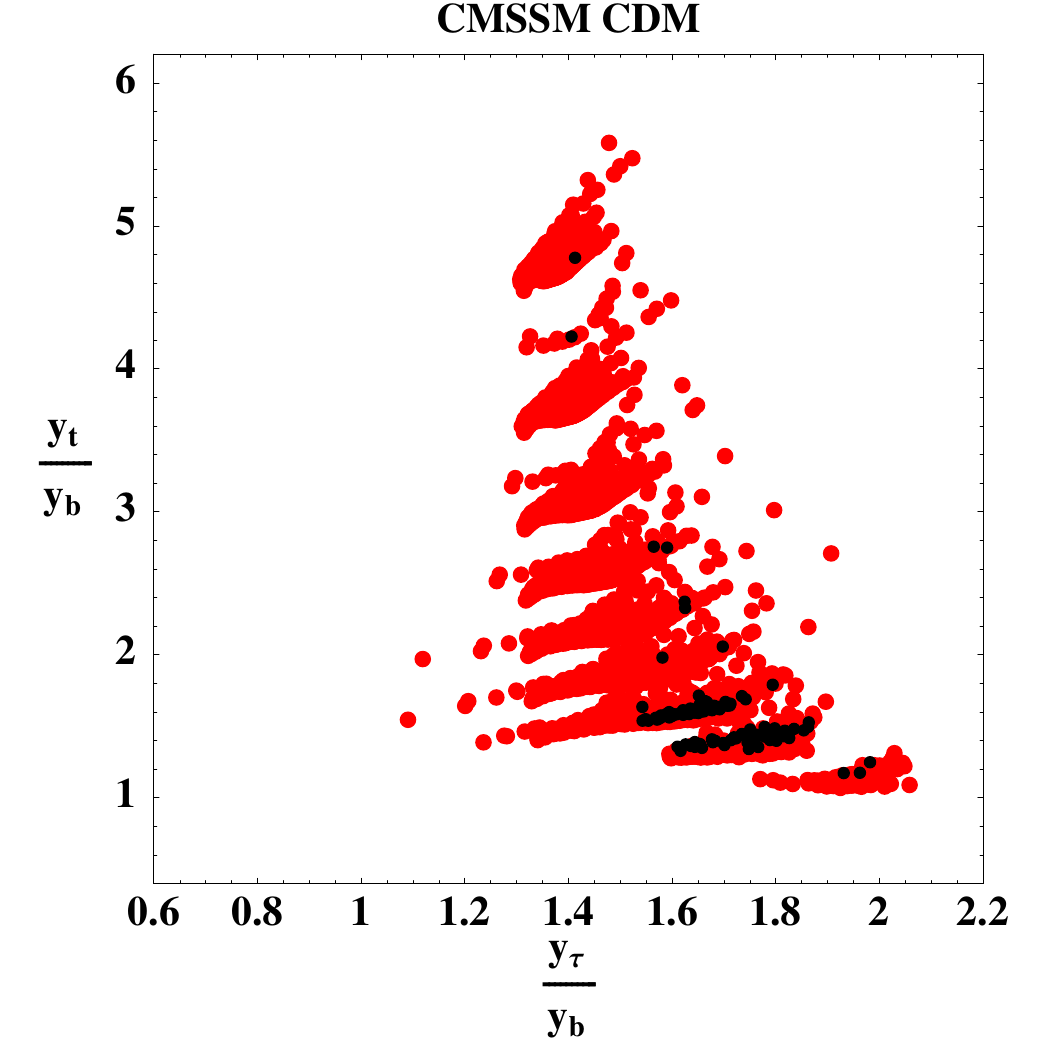}
 \caption[Plots for Each Experimental Constraint in CMSSM]{Impact of experimental constraints from direct detection, EWPO,  $b \rightarrow s \gamma$, $B_s \to \mu^+ \mu^-$, $(g-2)_\mu$ and CDM on $y_e/y_d$ and $y_\mu/y_s$ (upper plots) and $y_\tau/y_b$ and $y_t/y_b$ (lower plots) in CMSSM, cf.\ Ch.~\ref{Ch:Pheno}. Red dots denote parameter points which are excluded by the constraint, while black dots indicate parameter points which are allowed. In the lower plots for the third generation, the different lines of points correspond to different values of $\tan \beta$, increasing from top to bottom. \label{Fig:Plots_CMSSM} }
\end{figure}

%% file: app_04_Discrete.tex
\chapter{Discrete Symmetries} \label{App:Discrete}

In this chapter we briefly discuss some properties of the discrete symmetries we have used as flavour symmetries. We start with a discussion of $\mathbb{Z}_n$ symmetries and conclude with a brief introduction to $A_4$ based on \cite{Altarelli:2010gt}.

\section{$\mathbb{Z}_n$ Symmetries} \label{App:Zn}

One of the simplest classes of discrete symmetries are the $\mathbb{Z}_n$ symmetries with $n \geq 2$. The $\mathbb{Z}_n$ groups form subgroups of $U(1)$ and are Abelian as well. The simplest example for this class of symmetries is $\mathbb{Z}_2$ under which a field can have either positive or negative unit charge. An operator is neutral under such a $\mathbb{Z}_2$ if and only if the product of all field charges is positive.

For $n \geq 3$ the charge assignments are a little bit more complicated and there are in principle two different ways how to label a charge. The charge $q_i$ of a field $\phi_i$ under $\mathbb{Z}_n$ can be written as
\begin{equation}
q_i = \e^{\frac{2 \pi \ci \alpha_i}{n}} \;,
\end{equation}
where $\alpha_i$ is an integer. In this thesis, especially in Chs.~\ref{Ch:GUTImplications} and \ref{Ch:Model}, we label the charge under $\mathbb{Z}_n$ with $n \geq 3$ with $\alpha_i$ instead of the $q_i$. A field combination  is neutral under $\mathbb{Z}_n$ when the condition
\begin{equation}
\prod_i \, q_i = 1 \;,
\end{equation}
is fulfilled, which can be rewritten in terms of $\alpha_i$,
\begin{equation}
\sum_i \alpha_i \;\; \text{mod} \; n = 0 \;.
\end{equation}
Here, we see why this class of symmetries is denoted with $\mathbb{Z}_n$, namely the charges $\alpha_i$ can be interpreted as elements of the residue class ring modulo $n$ where multiples of $n$ are identical to the zero element.

From the point of view of the $q_i$ charges we would prefer to call these groups cyclic since all group elements of $\mathbb{Z}_n$ are powers of one single element $a$ with $a^n = 1$ where we can identify $a$ with a $q_i$ where $\alpha_i$ equals one.

\section[The $A_4$ Symmetry]{The $\boldsymbol{A_4}$ Symmetry} \label{App:A4}

The alternating group $A_4$ is the group of even permutations on four objects. It is the symmetry group of a regular tetrahedron as well. The group $A_4$ is a non-Abelian discrete group with twelve elements and is generated by two elements $S$ and $T$ obeying
\begin{equation}
 S^2 = (S T)^3 = T^3 = 1 \;.
\end{equation}
$A_4$ has three one-dimensional representations given by
\begin{alignat}{3}
&\mathbf{1}:   &\quad& S = 1 \;,&\quad& T=1 \;, \\
&\mathbf{1}':  &\quad& S = 1 \;,&\quad& T=\e^{2 \pi \ci /3} \equiv \omega \;, \\
&\mathbf{1}'': &\quad& S = 1 \;,&\quad& T=\e^{4 \pi \ci /3} \equiv \omega^2 \;.
\end{alignat}
The Lagrangian has to form the singlet $\mathbf{1}$ since the two remaining one-dimensional representations transform non-trivially under $A_4$.

Besides the one-dimensional representations there is only one three-dimensional representation left given by
\begin{equation}
 T' = \begin{pmatrix} 0 & 1 & 0 \\ 0 & 0 & 1 \\ 1 & 0 & 0 \end{pmatrix} \quad \text{and} \quad S' =  \begin{pmatrix} 1 & 0 & 0 \\ 0 & -1 & 0 \\ 0 & 0 & -1 \end{pmatrix} \;,
\end{equation}
where we have used a basis in which $S'$ is diagonal. In this basis the singlet $\mathbf{1}$ combination of two triplets $a=(a_1,a_2,a_3)^T$ and $b=(b_1,b_2,b_3)^T$ is given through the combination $a_1 b_1 + a_2 b_2 + a_3 b_3$. Throughout this thesis we only use the singlet representation $\mathbf{1}$ and the triplet representation $\mathbf{3}$. The relevant tensor products are
\begin{equation}
\begin{split}
 \mathbf{1} \times \mathbf{1} &= \mathbf{1} \;, \\
 \mathbf{1} \times \mathbf{3} &= \mathbf{3} \;, \\
 \mathbf{3} \times \mathbf{3} &= \mathbf{1} + \mathbf{1}' + \mathbf{1}'' + \mathbf{3}_s + \mathbf{3}_a  \;.
\end{split}
\end{equation}
In the last product we have labelled the symmetric combination with an index $s$ and the antisymmetric one with an index $a$.

%% file: danksagung.tex
\addcontentsline{toc}{chapter}{\protect Acknowledgements}

\chapter*{Acknowledgements}

First of all, I thank my supervisor Dr.\ Stefan Antusch for his guidance during my doctoral studies and continuous flow of ideas. Secondly, I thank my other collaborators Prof.\ Dr.\ Stephen F.\ King, Dr.\ Michal Malinsk\'{y}, Dr.\ Lorenzo Calibbi and Vinzenz Maurer. Especially, I would like to thank Stephen F. King for his British sereneness. I also thank my thesis referees PD Dr.\ Georg Raffelt and Prof.\ Dr.\ Gerhard Buchalla for agreeing on reading this thesis.

Many thanks to Dr.\ Frank Daniel Steffen (FDS) for making the IMPRS program of our MPI such a valuable experience. I will really miss Ringberg.

Furthermore, I want to thank my colleagues who proofread (parts of) this thesis, namely Jochen Baumann, Daniel H\"artl, Clemens Kie\ss{}ig and Philipp Kostka. Moreover, I would like to thank all my (former) colleagues at the MPI for such a good working atmosphere and their patience during lunch. A special warm thank goes to my office mates Clemens Kie\ss{}ig and Philipp Kostka. Our office is the real ``Ministry of Fun''.

Special thanks to Stephan Schulte, Julia G\"ortz and Annika W\"ohner as well. You know why.

And last but not least I would like to thank my family for giving me all the support they could give me.

%% file: diss.bbl
\providecommand{\href}[2]{#2}\begingroup\raggedright\begin{thebibliography}{10%
0}

\bibitem{Antusch:2008tf}
S.~Antusch and M.~Spinrath, {\it {Quark and lepton masses at the GUT scale
  including SUSY threshold corrections}},  Phys. Rev. {\bf D78} (2008) 075020
  [\href{http://xxx.lanl.gov/abs/0804.0717}{{\tt arXiv:0804.0717}}].

\bibitem{Antusch:2009gu}
S.~Antusch and M.~Spinrath, {\it {New GUT predictions for quark and lepton mass
  ratios confronted with phenomenology}},  Phys. Rev. {\bf D79} (2009) 095004
  [\href{http://xxx.lanl.gov/abs/0902.4644}{{\tt arXiv:0902.4644}}].

\bibitem{Antusch:2009hq}
S.~Antusch, S.~F. King, M.~Malinsky and M.~Spinrath, {\it {Quark mixing sum
  rules and the right unitarity triangle}},  Phys. Rev. {\bf D81} (2010) 033008
  [\href{http://xxx.lanl.gov/abs/0910.5127}{{\tt arXiv:0910.5127}}].

\bibitem{Antusch:2010es}
S.~Antusch, S.~F. King and M.~Spinrath, {\it {Measurable Neutrino Mass Scale in
  $A_4 \times SU(5)$}},  [\href{http://xxx.lanl.gov/abs/1005.0708}{{\tt
  arXiv:1005.0708}}].

\bibitem{Glashow:1961tr}
S.~L. Glashow, {\it {Partial Symmetries of Weak Interactions}},  Nucl. Phys.
  {\bf 22} (1961) 579--588.

\bibitem{Weinberg:1967tq}
S.~Weinberg, {\it {A Model of Leptons}},  Phys. Rev. Lett. {\bf 19} (1967)
  1264--1266.

\bibitem{Salam:1968rm}
A.~Salam, {\it {Weak and Electromagnetic Interactions}},  Originally printed in
  *Svartholm: Elementary Particle Theory, Proceedings Of The Nobel Symposium
  Held 1968 At Lerum, Sweden*, Stockholm 1968, 367-377 (1968).

\bibitem{Glashow:1970gm}
S.~L. Glashow, J.~Iliopoulos and L.~Maiani, {\it {Weak Interactions with
  Lepton-Hadron Symmetry}},  Phys. Rev. {\bf D2} (1970) 1285--1292.

\bibitem{Fritzsch:1973pi}
H.~Fritzsch, M.~Gell-Mann and H.~Leutwyler, {\it {Advantages of the Color Octet
  Gluon Picture}},  Phys. Lett. {\bf B47} (1973) 365--368.

\bibitem{Gross:1973ju}
D.~J. Gross and F.~Wilczek, {\it {Asymptotically Free Gauge Theories. 1}},
  Phys. Rev. {\bf D8} (1973) 3633--3652.

\bibitem{Gross:1974cs}
D.~J. Gross and F.~Wilczek, {\it {Asymptotically Free Gauge Theories. 2}},
  Phys. Rev. {\bf D9} (1974) 980--993.

\bibitem{Politzer:1973fx}
H.~D. Politzer, {\it {Reliable Perturbative Results for Strong Interactions?}},
   Phys. Rev. Lett. {\bf 30} (1973) 1346--1349.

\bibitem{Amsler:2008zzb}
{\bf Particle Data Group} Collaboration, C.~Amsler {\em et.~al.}, {\it {Review
  of particle physics}},  Phys. Lett. {\bf B667} (2008) 1.

\bibitem{Wess:1973kz}
J.~Wess and B.~Zumino, {\it {A Lagrangian Model Invariant Under Supergauge
  Transformations}},  Phys. Lett. {\bf B49} (1974) 52.

\bibitem{Volkov:1973ix}
D.~V. Volkov and V.~P. Akulov, {\it {Is the Neutrino a Goldstone Particle?}},
  Phys. Lett. {\bf B46} (1973) 109--110.

\bibitem{Wess:1974tw}
J.~Wess and B.~Zumino, {\it {Supergauge Transformations in Four-Dimensions}},
  Nucl. Phys. {\bf B70} (1974) 39--50.

\bibitem{Witten:1981nf}
E.~Witten, {\it {Dynamical Breaking of Supersymmetry}},  Nucl. Phys. {\bf B188}
  (1981) 513.

\bibitem{Kaul:1981hi}
R.~K. Kaul and P.~Majumdar, {\it {Cancellation of Quadratically Divergent Mass
  Corrections in Globally Supersymmetric Spontaneously Broken Gauge Theories}},
   Nucl. Phys. {\bf B199} (1982) 36.

\bibitem{Ibanez:1981yh}
L.~E. Ibanez and G.~G. Ross, {\it {Low-Energy Predictions in Supersymmetric
  Grand Unified Theories}},  Phys. Lett. {\bf B105} (1981) 439.

\bibitem{Dimopoulos:1981yj}
S.~Dimopoulos, S.~Raby and F.~Wilczek, {\it {Supersymmetry and the Scale of
  Unification}},  Phys. Rev. {\bf D24} (1981) 1681--1683.

\bibitem{Ellis:1983ew}
J.~R. Ellis, J.~S. Hagelin, D.~V. Nanopoulos, K.~A. Olive and M.~Srednicki,
  {\it {Supersymmetric relics from the big bang}},  Nucl. Phys. {\bf B238}
  (1984) 453--476.

\bibitem{Hinshaw:2008kr}
{\bf WMAP} Collaboration, G.~Hinshaw {\em et.~al.}, {\it {Five-Year Wilkinson
  Microwave Anisotropy Probe (WMAP) Observations: Data Processing, Sky Maps \&
  Basic Results}},  Astrophys. J. Suppl. {\bf 180} (2009) 225--245
  [\href{http://xxx.lanl.gov/abs/0803.0732}{{\tt arXiv:0803.0732}}].

\bibitem{Georgi:1974sy}
H.~Georgi and S.~L. Glashow, {\it {Unity of All Elementary Particle Forces}},
  Phys. Rev. Lett. {\bf 32} (1974) 438--441.

\bibitem{Georgi:1974my}
H.~Georgi, {\it {The State of the Art - Gauge Theories. (Talk)}},  AIP Conf.
  Proc. {\bf 23} (1975) 575--582.

\bibitem{Fritzsch:1974nn}
H.~Fritzsch and P.~Minkowski, {\it {Unified Interactions of Leptons and
  Hadrons}},  Ann. Phys. {\bf 93} (1975) 193--266.

\bibitem{Georgi:1979df}
H.~Georgi and C.~Jarlskog, {\it {A New Lepton - Quark Mass Relation in a
  Unified Theory}},  Phys. Lett. {\bf B86} (1979) 297--300.

\bibitem{Hall:1993gn}
L.~J. Hall, R.~Rattazzi and U.~Sarid, {\it {The Top quark mass in
  supersymmetric $SO(10)$ unification}},  Phys. Rev. {\bf D50} (1994)
  7048--7065 [\href{http://xxx.lanl.gov/abs/hep-ph/9306309}{{\tt
  hep-ph/9306309}}].

\bibitem{Blazek:1995nv}
T.~Blazek, S.~Raby and S.~Pokorski, {\it {Finite supersymmetric threshold
  corrections to CKM matrix elements in the large $\tan \beta$ regime}},  Phys.
  Rev. {\bf D52} (1995) 4151--4158
  [\href{http://xxx.lanl.gov/abs/hep-ph/9504364}{{\tt hep-ph/9504364}}].

\bibitem{Carena:1994bv}
M.~S. Carena, M.~Olechowski, S.~Pokorski and C.~E.~M. Wagner, {\it {Electroweak
  symmetry breaking and bottom - top Yukawa unification}},  Nucl. Phys. {\bf
  B426} (1994) 269--300 [\href{http://xxx.lanl.gov/abs/hep-ph/9402253}{{\tt
  hep-ph/9402253}}].

\bibitem{Hempfling:1993kv}
R.~Hempfling, {\it {Yukawa coupling unification with supersymmetric threshold
  corrections}},  Phys. Rev. {\bf D49} (1994) 6168--6172.

\bibitem{Minkowski:1977sc}
P.~Minkowski, {\it {$\mu \to e \gamma$ at a Rate of One Out of 1-Billion Muon
  Decays?}},  Phys. Lett. {\bf B67} (1977) 421.

\bibitem{GellMann:1980vs}
M.~Gell-Mann, P.~Ramond and R.~Slansky, {\it {Complex Spinors and Unified
  Theories}},  Print-80-0576 (CERN).

\bibitem{Yanagida:1979as}
T.~Yanagida, {\it {Horizontal gauge symmetry and masses of neutrinos}},  In
  Proceedings of the Workshop on the Baryon Number of the Universe and Unified
  Theories, Tsukuba, Japan, 13-14 Feb 1979.

\bibitem{Glashow:1979nm}
S.~L. Glashow, {\it {The Future of Elementary Particle Physics}},  NATO Adv.
  Study Inst. Ser. B Phys. {\bf 59} (1980) 687.

\bibitem{Mohapatra:1980yp}
R.~N. Mohapatra and G.~Senjanovic, {\it {Neutrino Masses and Mixings in Gauge
  Models with Spontaneous Parity Violation}},  Phys. Rev. {\bf D23} (1981) 165.

\bibitem{Schechter:1981cv}
J.~Schechter and J.~W.~F. Valle, {\it {Neutrino Decay and Spontaneous Violation
  of Lepton Number}},  Phys. Rev. {\bf D25} (1982) 774.

\bibitem{Fusaoka:1998vc}
H.~Fusaoka and Y.~Koide, {\it {Updated estimate of running quark masses}},
  Phys. Rev. {\bf D57} (1998) 3986--4001
  [\href{http://xxx.lanl.gov/abs/hep-ph/9712201}{{\tt hep-ph/9712201}}].

\bibitem{Xing:2007fb}
Z.-z. Xing, H.~Zhang and S.~Zhou, {\it {Updated Values of Running Quark and
  Lepton Masses}},  Phys. Rev. {\bf D77} (2008) 113016
  [\href{http://xxx.lanl.gov/abs/0712.1419}{{\tt arXiv:0712.1419}}].

\bibitem{Fritzsch:1979zq}
H.~Fritzsch, {\it {Quark Masses and Flavor Mixing}},  Nucl. Phys. {\bf B155}
  (1979) 189.

\bibitem{Fritzsch:1999ee}
H.~Fritzsch and Z.-z. Xing, {\it {Mass and flavor mixing schemes of quarks and
  leptons}},  Prog. Part. Nucl. Phys. {\bf 45} (2000) 1--81
  [\href{http://xxx.lanl.gov/abs/hep-ph/9912358}{{\tt hep-ph/9912358}}].

\bibitem{Leontaris:2009pi}
G.~K. Leontaris and N.~D. Vlachos, {\it {D-brane Inspired Fermion Mass
  Textures}},  JHEP {\bf 01} (2010) 016
  [\href{http://xxx.lanl.gov/abs/0909.4701}{{\tt arXiv:0909.4701}}].

\bibitem{Dev:2009he}
S.~Dev, S.~Verma and S.~Gupta, {\it {Phenomenological Analysis of Hybrid
  Textures of Neutrinos}},  Phys. Lett. {\bf B687} (2010) 53--60
  [\href{http://xxx.lanl.gov/abs/0909.3182}{{\tt arXiv:0909.3182}}].

\bibitem{Adhikary:2009kz}
B.~Adhikary, A.~Ghosal and P.~Roy, {\it {'Mu-Tau' symmetry, tribimaximal mixing
  and four zero neutrino Yukawa textures}},  JHEP {\bf 10} (2009) 040
  [\href{http://xxx.lanl.gov/abs/0908.2686}{{\tt arXiv:0908.2686}}].

\bibitem{Goswami:2009bd}
S.~Goswami, S.~Khan and W.~Rodejohann, {\it {Minimal Textures in Seesaw Mass
  Matrices and their low and high Energy Phenomenology}},  Phys. Lett. {\bf
  B680} (2009) 255--262 [\href{http://xxx.lanl.gov/abs/0905.2739}{{\tt
  arXiv:0905.2739}}].

\bibitem{Goswami:2008uv}
S.~Goswami, S.~Khan and A.~Watanabe, {\it {Hybrid textures in minimal seesaw
  mass matrices}},  [\href{http://xxx.lanl.gov/abs/0811.4744}{{\tt
  arXiv:0811.4744}}].

\bibitem{Choubey:2008tb}
S.~Choubey, W.~Rodejohann and P.~Roy, {\it {Phenomenological consequences of
  four zero neutrino Yukawa textures}},  Nucl. Phys. {\bf B808} (2009) 272--291
  [\href{http://xxx.lanl.gov/abs/0807.4289}{{\tt arXiv:0807.4289}}].

\bibitem{Branco:2007nb}
G.~C. Branco, D.~Emmanuel-Costa, M.~N. Rebelo and P.~Roy, {\it {Four Zero
  Neutrino Yukawa Textures in the Minimal Seesaw Framework}},  Phys. Rev. {\bf
  D77} (2008) 053011 [\href{http://xxx.lanl.gov/abs/0712.0774}{{\tt
  arXiv:0712.0774}}].

\bibitem{Alhendi:2007iu}
H.~A. Alhendi, E.~I. Lashin and A.~A. Mudlej, {\it {Textures with two traceless
  submatrices of the neutrino mass matrix}},  Phys. Rev. {\bf D77} (2008)
  013009 [\href{http://xxx.lanl.gov/abs/0708.2007}{{\tt arXiv:0708.2007}}].

\bibitem{Kaneko:2007ea}
S.~Kaneko, H.~Sawanaka, T.~Shingai, M.~Tanimoto and K.~Yoshioka, {\it {New
  approach to texture-zeros with S(3) symmetry: Flavor symmetry and vacuum
  aligned mass textures}},  [\href{http://xxx.lanl.gov/abs/hep-ph/0703250}{{\tt
  hep-ph/0703250}}].

\bibitem{Branco:2006wv}
G.~C. Branco, M.~N. Rebelo and J.~I. Silva-Marcos, {\it {Yukawa Textures, New
  Physics and Nondecoupling}},  Phys. Rev. {\bf D76} (2007) 033008
  [\href{http://xxx.lanl.gov/abs/hep-ph/0612252}{{\tt hep-ph/0612252}}].

\bibitem{Lam:2006wm}
C.~S. Lam, {\it {Mass Independent Textures and Symmetry}},  Phys. Rev. {\bf
  D74} (2006) 113004 [\href{http://xxx.lanl.gov/abs/hep-ph/0611017}{{\tt
  hep-ph/0611017}}].

\bibitem{Kaneko:2006wi}
S.~Kaneko, H.~Sawanaka, T.~Shingai, M.~Tanimoto and K.~Yoshioka, {\it {Flavor
  Symmetry and Vacuum Aligned Mass Textures}},  Prog. Theor. Phys. {\bf 117}
  (2007) 161--181 [\href{http://xxx.lanl.gov/abs/hep-ph/0609220}{{\tt
  hep-ph/0609220}}].

\bibitem{Fuki:2006xw}
K.~Fuki and M.~Yasue, {\it {Two categories of approximately mu - tau symmetric
  neutrino mass textures}},  Nucl. Phys. {\bf B783} (2007) 31--56
  [\href{http://xxx.lanl.gov/abs/hep-ph/0608042}{{\tt hep-ph/0608042}}].

\bibitem{Haba:2005ds}
N.~Haba and K.~Yoshioka, {\it {Discrete flavor symmetry, dynamical mass
  textures, and grand unification}},  Nucl. Phys. {\bf B739} (2006) 254--284
  [\href{http://xxx.lanl.gov/abs/hep-ph/0511108}{{\tt hep-ph/0511108}}].

\bibitem{Kim:2004ki}
H.~D. Kim, S.~Raby and L.~Schradin, {\it {Quark mass textures and $\sin (2
  \beta)$}},  Phys. Rev. {\bf D69} (2004) 092002
  [\href{http://xxx.lanl.gov/abs/hep-ph/0401169}{{\tt hep-ph/0401169}}].

\bibitem{Jack:2003pb}
I.~Jack, D.~R.~T. Jones and R.~Wild, {\it {Yukawa textures and the mu-term}},
  Phys. Lett. {\bf B580} (2004) 72--78
  [\href{http://xxx.lanl.gov/abs/hep-ph/0309165}{{\tt hep-ph/0309165}}].

\bibitem{Jack:2003qg}
I.~Jack and D.~R.~T. Jones, {\it {Yukawa textures and anomaly mediated
  supersymmetry breaking}},  Nucl. Phys. {\bf B662} (2003) 63--88
  [\href{http://xxx.lanl.gov/abs/hep-ph/0301163}{{\tt hep-ph/0301163}}].

\bibitem{Caravaglios:2002br}
F.~Caravaglios, P.~Roudeau and A.~Stocchi, {\it {Precision test of quark mass
  textures: A Model independent approach}},  Nucl. Phys. {\bf B633} (2002)
  193--211 [\href{http://xxx.lanl.gov/abs/hep-ph/0202055}{{\tt
  hep-ph/0202055}}].

\bibitem{Everett:2000up}
L.~L. Everett, G.~L. Kane and S.~F. King, {\it {D branes and textures}},  JHEP
  {\bf 08} (2000) 012 [\href{http://xxx.lanl.gov/abs/hep-ph/0005204}{{\tt
  hep-ph/0005204}}].

\bibitem{Berezhiani:2000cg}
Z.~Berezhiani and A.~Rossi, {\it {Predictive grand unified textures for quark
  and neutrino masses and mixings}},  Nucl. Phys. {\bf B594} (2001) 113--168
  [\href{http://xxx.lanl.gov/abs/hep-ph/0003084}{{\tt hep-ph/0003084}}].

\bibitem{Kuo:1999dt}
T.-K. Kuo, S.~W. Mansour and G.-H. Wu, {\it {Triangular textures for quark mass
  matrices}},  Phys. Rev. {\bf D60} (1999) 093004
  [\href{http://xxx.lanl.gov/abs/hep-ph/9907314}{{\tt hep-ph/9907314}}].

\bibitem{Falcone:1998us}
D.~Falcone and F.~Tramontano, {\it {Relation between quark masses and weak
  mixings}},  Phys. Rev. {\bf D59} (1999) 017302
  [\href{http://xxx.lanl.gov/abs/hep-ph/9806496}{{\tt hep-ph/9806496}}].

\bibitem{Roberts:2001zy}
R.~G. Roberts, A.~Romanino, G.~G. Ross and L.~Velasco-Sevilla, {\it {Precision
  test of a Fermion mass texture}},  Nucl. Phys. {\bf B615} (2001) 358--384
  [\href{http://xxx.lanl.gov/abs/hep-ph/0104088}{{\tt hep-ph/0104088}}].

\bibitem{Ramond:1993kv}
P.~Ramond, R.~G. Roberts and G.~G. Ross, {\it {Stitching the Yukawa quilt}},
  Nucl. Phys. {\bf B406} (1993) 19--42
  [\href{http://xxx.lanl.gov/abs/hep-ph/9303320}{{\tt hep-ph/9303320}}].

\bibitem{Chiu:2000gw}
S.-H. Chiu, T.-K. Kuo and G.-H. Wu, {\it {Hermitian quark mass matrices with
  four texture zeros}},  Phys. Rev. {\bf D62} (2000) 053014
  [\href{http://xxx.lanl.gov/abs/hep-ph/0003224}{{\tt hep-ph/0003224}}].

\bibitem{Fritzsch:1999rb}
H.~Fritzsch and Z.-z. Xing, {\it {The light quark sector, CP violation, and the
  unitarity triangle}},  Nucl. Phys. {\bf B556} (1999) 49--75
  [\href{http://xxx.lanl.gov/abs/hep-ph/9904286}{{\tt hep-ph/9904286}}].

\bibitem{Jarosik:2010iu}
N.~Jarosik {\em et.~al.}, {\it {Seven-Year Wilkinson Microwave Anisotropy Probe
  (WMAP) Observations: Sky Maps, Systematic Errors, and Basic Results}},
  [\href{http://xxx.lanl.gov/abs/1001.4744}{{\tt arXiv:1001.4744}}].

\bibitem{Harrison:2002er}
P.~F. Harrison, D.~H. Perkins and W.~G. Scott, {\it {Tri-bimaximal mixing and
  the neutrino oscillation data}},  Phys. Lett. {\bf B530} (2002) 167
  [\href{http://xxx.lanl.gov/abs/hep-ph/0202074}{{\tt hep-ph/0202074}}].

\bibitem{Harrison:2003aw}
P.~F. Harrison and W.~G. Scott, {\it {Permutation symmetry, tri-bimaximal
  neutrino mixing and the S3 group characters}},  Phys. Lett. {\bf B557} (2003)
  76 [\href{http://xxx.lanl.gov/abs/hep-ph/0302025}{{\tt hep-ph/0302025}}].

\bibitem{King:2009ap}
S.~F. King and C.~Luhn, {\it {On the origin of neutrino flavour symmetry}},
  JHEP {\bf 10} (2009) 093 [\href{http://xxx.lanl.gov/abs/0908.1897}{{\tt
  arXiv:0908.1897}}].

\bibitem{Altarelli:2010gt}
G.~Altarelli and F.~Feruglio, {\it {Discrete Flavor Symmetries and Models of
  Neutrino Mixing}},  [\href{http://xxx.lanl.gov/abs/1002.0211}{{\tt
  arXiv:1002.0211}}].

\bibitem{King:2009tj}
S.~F. King and C.~Luhn, {\it {A Supersymmetric Grand Unified Theory of Flavour
  with $PSL(2,7) \times SO(10)$}},
  [\href{http://xxx.lanl.gov/abs/0912.1344}{{\tt arXiv:0912.1344}}].

\bibitem{Altarelli:2005yx}
G.~Altarelli and F.~Feruglio, {\it {Tri-Bimaximal Neutrino Mixing, $A_4$ and
  the Modular Symmetry}},  Nucl. Phys. {\bf B741} (2006) 215--235
  [\href{http://xxx.lanl.gov/abs/hep-ph/0512103}{{\tt hep-ph/0512103}}].

\bibitem{Ma:2001dn}
E.~Ma and G.~Rajasekaran, {\it {Softly broken $A_4$ symmetry for nearly
  degenerate neutrino masses}},  Phys. Rev. {\bf D64} (2001) 113012
  [\href{http://xxx.lanl.gov/abs/hep-ph/0106291}{{\tt hep-ph/0106291}}].

\bibitem{Ma:2004zv}
E.~Ma, {\it {$A_4$ origin of the neutrino mass matrix}},  Phys. Rev. {\bf D70}
  (2004) 031901 [\href{http://xxx.lanl.gov/abs/hep-ph/0404199}{{\tt
  hep-ph/0404199}}].

\bibitem{Altarelli:2005yp}
G.~Altarelli and F.~Feruglio, {\it {Tri-bimaximal neutrino mixing from discrete
  symmetry in extra dimensions}},  Nucl. Phys. {\bf B720} (2005) 64--88
  [\href{http://xxx.lanl.gov/abs/hep-ph/0504165}{{\tt hep-ph/0504165}}].

\bibitem{Ma:2005qf}
E.~Ma, {\it {Tribimaximal neutrino mixing from a supersymmetric model with
  $A_4$ family symmetry}},  Phys. Rev. {\bf D73} (2006) 057304
  [\href{http://xxx.lanl.gov/abs/hep-ph/0511133}{{\tt hep-ph/0511133}}].

\bibitem{Ma:2005mw}
E.~Ma, {\it {Tetrahedral family symmetry and the neutrino mixing matrix}},
  Mod. Phys. Lett. {\bf A20} (2005) 2601--2606
  [\href{http://xxx.lanl.gov/abs/hep-ph/0508099}{{\tt hep-ph/0508099}}].

\bibitem{Ma:2005sha}
E.~Ma, {\it {Aspects of the tetrahedral neutrino mass matrix}},  Phys. Rev.
  {\bf D72} (2005) 037301 [\href{http://xxx.lanl.gov/abs/hep-ph/0505209}{{\tt
  hep-ph/0505209}}].

\bibitem{Chen:2005jm}
S.-L. Chen, M.~Frigerio and E.~Ma, {\it {Hybrid seesaw neutrino masses with
  $A_4$ family symmetry}},  Nucl. Phys. {\bf B724} (2005) 423--431
  [\href{http://xxx.lanl.gov/abs/hep-ph/0504181}{{\tt hep-ph/0504181}}].

\bibitem{Altarelli:2006kg}
G.~Altarelli, F.~Feruglio and Y.~Lin, {\it {Tri-bimaximal neutrino mixing from
  orbifolding}},  Nucl. Phys. {\bf B775} (2007) 31--44
  [\href{http://xxx.lanl.gov/abs/hep-ph/0610165}{{\tt hep-ph/0610165}}].

\bibitem{Ma:2006vq}
E.~Ma, {\it {Supersymmetric $A_4 \times Z_3$ and $A_4$ Realizations of Neutrino
  Tribimaximal Mixing Without and With Corrections}},  Mod. Phys. Lett. {\bf
  A22} (2007) 101--106 [\href{http://xxx.lanl.gov/abs/hep-ph/0610342}{{\tt
  hep-ph/0610342}}].

\bibitem{Ma:2006wm}
E.~Ma, {\it {Suitability of $A_4$ as a Family Symmetry in Grand Unification}},
  Mod. Phys. Lett. {\bf A21} (2006) 2931--2936
  [\href{http://xxx.lanl.gov/abs/hep-ph/0607190}{{\tt hep-ph/0607190}}].

\bibitem{Ma:2006sk}
E.~Ma, H.~Sawanaka and M.~Tanimoto, {\it {Quark masses and mixing with $A_4$
  family symmetry}},  Phys. Lett. {\bf B641} (2006) 301--304
  [\href{http://xxx.lanl.gov/abs/hep-ph/0606103}{{\tt hep-ph/0606103}}].

\bibitem{Adhikary:2006wi}
B.~Adhikary, B.~Brahmachari, A.~Ghosal, E.~Ma and M.~K. Parida, {\it {$A_4$
  symmetry and prediction of $U_{e3}$ in a modified Altarelli-Feruglio model}},
   Phys. Lett. {\bf B638} (2006) 345--349
  [\href{http://xxx.lanl.gov/abs/hep-ph/0603059}{{\tt hep-ph/0603059}}].

\bibitem{He:2006et}
X.-G. He, {\it {$A_4$ group and tri-bimaximal neutrino mixing: A renormalizable
  model}},  Nucl. Phys. Proc. Suppl. {\bf 168} (2007) 350--352
  [\href{http://xxx.lanl.gov/abs/hep-ph/0612080}{{\tt hep-ph/0612080}}].

\bibitem{King:2006np}
S.~F. King and M.~Malinsky, {\it {$A_4$ family symmetry and quark-lepton
  unification}},  Phys. Lett. {\bf B645} (2007) 351--357
  [\href{http://xxx.lanl.gov/abs/hep-ph/0610250}{{\tt hep-ph/0610250}}].

\bibitem{Ma:2007ku}
E.~Ma, {\it {New lepton family symmetry and neutrino tribimaximal mixing}},
  Europhys. Lett. {\bf 79} (2007) 61001
  [\href{http://xxx.lanl.gov/abs/hep-ph/0701016}{{\tt hep-ph/0701016}}].

\bibitem{Feruglio:2007uu}
F.~Feruglio, C.~Hagedorn, Y.~Lin and L.~Merlo, {\it {Tri-bimaximal Neutrino
  Mixing and Quark Masses from a Discrete Flavour Symmetry}},  Nucl. Phys. {\bf
  B775} (2007) 120--142 [\href{http://xxx.lanl.gov/abs/hep-ph/0702194}{{\tt
  hep-ph/0702194}}].

\bibitem{Chen:2009um}
M.-C. Chen and S.~F. King, {\it {$A_4$ See-Saw Models and Form Dominance}},
  JHEP {\bf 06} (2009) 072 [\href{http://xxx.lanl.gov/abs/0903.0125}{{\tt
  arXiv:0903.0125}}].

\bibitem{Altarelli:2008bg}
G.~Altarelli, F.~Feruglio and C.~Hagedorn, {\it {A SUSY $SU(5)$ Grand Unified
  Model of Tri-Bimaximal Mixing from $A_4$}},  JHEP {\bf 03} (2008) 052--052
  [\href{http://xxx.lanl.gov/abs/0802.0090}{{\tt arXiv:0802.0090}}].

\bibitem{Burrows:2009pi}
T.~J. Burrows and S.~F. King, {\it {$A_4$ Family Symmetry from $SU(5)$ SUSY
  GUTs in 6d}},  [\href{http://xxx.lanl.gov/abs/0909.1433}{{\tt
  arXiv:0909.1433}}].

\bibitem{Ciafaloni:2009ub}
P.~Ciafaloni, M.~Picariello, E.~Torrente-Lujan and A.~Urbano, {\it {Neutrino
  masses and tribimaximal mixing in Minimal renormalizable SUSY $SU(5)$ Grand
  Unified Model with $A_4$ Flavor symmetry}},  Phys. Rev. {\bf D79} (2009)
  116010 [\href{http://xxx.lanl.gov/abs/0901.2236}{{\tt arXiv:0901.2236}}].

\bibitem{Ciafaloni:2009qs}
P.~Ciafaloni, M.~Picariello, A.~Urbano and E.~Torrente-Lujan, {\it {Toward
  minimal renormalizable SUSY $SU(5)$ Grand Unified Model with tribimaximal
  mixing from $A_4$ Flavor symmetry}},  Phys. Rev. {\bf D81} (2010) 016004
  [\href{http://xxx.lanl.gov/abs/0909.2553}{{\tt arXiv:0909.2553}}].

\bibitem{Barbieri:1999km}
R.~Barbieri, L.~J. Hall, G.~L. Kane and G.~G. Ross, {\it {Nearly degenerate
  neutrinos and broken flavour symmetry}},
  [\href{http://xxx.lanl.gov/abs/hep-ph/9901228}{{\tt hep-ph/9901228}}].

\bibitem{Xing:2002sw}
Z.-z. Xing, {\it {Nearly tri-bimaximal neutrino mixing and CP violation}},
  Phys. Lett. {\bf B533} (2002) 85--93
  [\href{http://xxx.lanl.gov/abs/hep-ph/0204049}{{\tt hep-ph/0204049}}].

\bibitem{King:2003rf}
S.~F. King and G.~G. Ross, {\it {Fermion masses and mixing angles from $SU(3)$
  family symmetry and unification}},  Phys. Lett. {\bf B574} (2003) 239--252
  [\href{http://xxx.lanl.gov/abs/hep-ph/0307190}{{\tt hep-ph/0307190}}].

\bibitem{deMedeirosVarzielas:2005qg}
I.~de~Medeiros~Varzielas, S.~F. King and G.~G. Ross, {\it {Tri-bimaximal
  neutrino mixing from discrete subgroups of $SU(3)$ and $SO(3)$ family
  symmetry}},  Phys. Lett. {\bf B644} (2007) 153--157
  [\href{http://xxx.lanl.gov/abs/hep-ph/0512313}{{\tt hep-ph/0512313}}].

\bibitem{Kang:2005bg}
S.~K. Kang, Z.-z. Xing and S.~Zhou, {\it {Possible deviation from the
  tri-bimaximal neutrino mixing in a seesaw model}},  Phys. Rev. {\bf D73}
  (2006) 013001 [\href{http://xxx.lanl.gov/abs/hep-ph/0511157}{{\tt
  hep-ph/0511157}}].

\bibitem{Luo:2005fc}
S.~Luo and Z.-z. Xing, {\it {Generalized tri-bimaximal neutrino mixing and its
  sensitivity to radiative corrections}},  Phys. Lett. {\bf B632} (2006)
  341--348 [\href{http://xxx.lanl.gov/abs/hep-ph/0509065}{{\tt
  hep-ph/0509065}}].

\bibitem{Xing:2005ur}
Z.-z. Xing, {\it {Nontrivial correlation between the CKM and MNS matrices}},
  Phys. Lett. {\bf B618} (2005) 141--149
  [\href{http://xxx.lanl.gov/abs/hep-ph/0503200}{{\tt hep-ph/0503200}}].

\bibitem{Xing:2006xa}
Z.-z. Xing, H.~Zhang and S.~Zhou, {\it {Nearly tri-bimaximal neutrino mixing
  and CP violation from mu - tau symmetry breaking}},  Phys. Lett. {\bf B641}
  (2006) 189--197 [\href{http://xxx.lanl.gov/abs/hep-ph/0607091}{{\tt
  hep-ph/0607091}}].

\bibitem{Haba:2006dz}
N.~Haba, A.~Watanabe and K.~Yoshioka, {\it {Twisted flavors and tri/bi-maximal
  neutrino mixing}},  Phys. Rev. Lett. {\bf 97} (2006) 041601
  [\href{http://xxx.lanl.gov/abs/hep-ph/0603116}{{\tt hep-ph/0603116}}].

\bibitem{He:2006qd}
X.-G. He and A.~Zee, {\it {Minimal Modification To The Tri-bimaximal Neutrino
  Mixing}},  Phys. Lett. {\bf B645} (2007) 427--431
  [\href{http://xxx.lanl.gov/abs/hep-ph/0607163}{{\tt hep-ph/0607163}}].

\bibitem{Hirsch:2006je}
M.~Hirsch, E.~Ma, J.~C. Romao, J.~W.~F. Valle and A.~Villanova~del Moral, {\it
  {Minimal supergravity radiative effects on the tri- bimaximal neutrino mixing
  pattern}},  Phys. Rev. {\bf D75} (2007) 053006
  [\href{http://xxx.lanl.gov/abs/hep-ph/0606082}{{\tt hep-ph/0606082}}].

\bibitem{King:2006me}
S.~F. King and M.~Malinsky, {\it {Towards a complete theory of fermion masses
  and mixings with $SO(3)$ family symmetry and 5d $SO(10)$ unification}},  JHEP
  {\bf 11} (2006) 071 [\href{http://xxx.lanl.gov/abs/hep-ph/0608021}{{\tt
  hep-ph/0608021}}].

\bibitem{Mohapatra:2006pu}
R.~N. Mohapatra, S.~Nasri and H.-B. Yu, {\it {$S_3$ symmetry and tri-bimaximal
  mixing}},  Phys. Lett. {\bf B639} (2006) 318--321
  [\href{http://xxx.lanl.gov/abs/hep-ph/0605020}{{\tt hep-ph/0605020}}].

\bibitem{Mohapatra:2006se}
R.~N. Mohapatra and H.-B. Yu, {\it {Connecting Leptogenesis to CP Violation in
  Neutrino Mixings in a Tri-bimaximal Mixing model}},  Phys. Lett. {\bf B644}
  (2007) 346--351 [\href{http://xxx.lanl.gov/abs/hep-ph/0610023}{{\tt
  hep-ph/0610023}}].

\bibitem{Singh:2006dr}
N.~N. Singh, M.~Rajkhowa and A.~Borah, {\it {Deviation from tri-bimaximal
  mixings in two types of inverted hierarchical neutrino mass models}},
  Pramana {\bf 69} (2007) 533--550
  [\href{http://xxx.lanl.gov/abs/hep-ph/0603189}{{\tt hep-ph/0603189}}].

\bibitem{deMedeirosVarzielas:2006fc}
I.~de~Medeiros~Varzielas, S.~F. King and G.~G. Ross, {\it {Neutrino
  tri-bi-maximal mixing from a non-Abelian discrete family symmetry}},  Phys.
  Lett. {\bf B648} (2007) 201--206
  [\href{http://xxx.lanl.gov/abs/hep-ph/0607045}{{\tt hep-ph/0607045}}].

\bibitem{Ma:2006ip}
E.~Ma, {\it {Neutrino mass matrix from $\Delta(27)$ symmetry}},  Mod. Phys.
  Lett. {\bf A21} (2006) 1917
  [\href{http://xxx.lanl.gov/abs/hep-ph/0607056}{{\tt hep-ph/0607056}}].

\bibitem{Aranda:2007dp}
A.~Aranda, {\it {Neutrino mixing from the double tetrahedral group
  $T^{\prime}$}},  Phys. Rev. {\bf D76} (2007) 111301
  [\href{http://xxx.lanl.gov/abs/0707.3661}{{\tt arXiv:0707.3661}}].

\bibitem{Luhn:2007sy}
C.~Luhn, S.~Nasri and P.~Ramond, {\it {Tri-Bimaximal Neutrino Mixing and the
  Family Symmetry $Z_7 \rtimes Z_3$}},  Phys. Lett. {\bf B652} (2007) 27--33
  [\href{http://xxx.lanl.gov/abs/0706.2341}{{\tt arXiv:0706.2341}}].

\bibitem{Ma:2007wu}
E.~Ma, {\it {Near Tribimaximal Neutrino Mixing with $\Delta(27)$ Symmetry}},
  Phys. Lett. {\bf B660} (2008) 505--507
  [\href{http://xxx.lanl.gov/abs/0709.0507}{{\tt arXiv:0709.0507}}].

\bibitem{Chan:2007ng}
A.~H. Chan, H.~Fritzsch, S.~Luo and Z.-z. Xing, {\it {Deviations from
  Tri-bimaximal Neutrino Mixing in Type-II Seesaw and Leptogenesis}},  Phys.
  Rev. {\bf D76} (2007) 073009 [\href{http://xxx.lanl.gov/abs/0704.3153}{{\tt
  arXiv:0704.3153}}].

\bibitem{Chen:2007afa}
M.-C. Chen and K.~T. Mahanthappa, {\it {CKM and Tri-bimaximal MNS Matrices in a
  $SU(5) \times {}^{(d)}T$ Model}},  Phys. Lett. {\bf B652} (2007) 34--39
  [\href{http://xxx.lanl.gov/abs/0705.0714}{{\tt arXiv:0705.0714}}].

\bibitem{Chen:2007gp}
M.-C. Chen and K.~T. Mahanthappa, {\it {Tri-bimaximal Neutrino Mixing and CKM
  Matrix in a $SU(5) \times {}^{(d)}T$ Model}},
  [\href{http://xxx.lanl.gov/abs/0710.2118}{{\tt arXiv:0710.2118}}].

\bibitem{Chen:2008eq}
M.-C. Chen and K.~T. Mahanthappa, {\it {Neutrino Mass Models: circa 2008}},
  Nucl. Phys. Proc. Suppl. {\bf 188} (2009) 315--320
  [\href{http://xxx.lanl.gov/abs/0812.4981}{{\tt arXiv:0812.4981}}].

\bibitem{Plentinger:2008nv}
F.~Plentinger and G.~Seidl, {\it {Mapping out $SU(5)$ GUTs with non-Abelian
  discrete flavor symmetries}},  Phys. Rev. {\bf D78} (2008) 045004
  [\href{http://xxx.lanl.gov/abs/0803.2889}{{\tt arXiv:0803.2889}}].

\bibitem{Csaki:2008qq}
C.~Csaki, C.~Delaunay, C.~Grojean and Y.~Grossman, {\it {A Model of Lepton
  Masses from a Warped Extra Dimension}},  JHEP {\bf 10} (2008) 055
  [\href{http://xxx.lanl.gov/abs/0806.0356}{{\tt arXiv:0806.0356}}].

\bibitem{Lin:2008aj}
Y.~Lin, {\it {A predictive $A_4$ model, Charged Lepton Hierarchy and Tri-
  bimaximal Sum Rule}},  Nucl. Phys. {\bf B813} (2009) 91--105
  [\href{http://xxx.lanl.gov/abs/0804.2867}{{\tt arXiv:0804.2867}}].

\bibitem{Lam:2009hn}
C.~S. Lam, {\it {A bottom-up analysis of horizontal symmetry}},
  [\href{http://xxx.lanl.gov/abs/0907.2206}{{\tt arXiv:0907.2206}}].

\bibitem{Lin:2009ic}
Y.~Lin, {\it {A dynamical approach to link low energy phases with
  leptogenesis}},  Phys. Rev. {\bf D80} (2009) 076011
  [\href{http://xxx.lanl.gov/abs/0903.0831}{{\tt arXiv:0903.0831}}].

\bibitem{King:2009mk}
S.~F. King and C.~Luhn, {\it {A new family symmetry for $SO(10)$ GUTs}},  Nucl.
  Phys. {\bf B820} (2009) 269--289
  [\href{http://xxx.lanl.gov/abs/0905.1686}{{\tt arXiv:0905.1686}}].

\bibitem{Hagedorn:2010th}
C.~Hagedorn, S.~F. King and C.~Luhn, {\it {A SUSY GUT of Flavour with $S_4
  \times SU(5)$ to NLO}},  [\href{http://xxx.lanl.gov/abs/1003.4249}{{\tt
  arXiv:1003.4249}}].

\bibitem{Cooper:2010ik}
I.~K. Cooper, S.~F. King and C.~Luhn, {\it {SUSY $SU(5)$ with singlet plus
  adjoint matter and $A_4$ family symmetry}},
  [\href{http://xxx.lanl.gov/abs/1004.3243}{{\tt arXiv:1004.3243}}].

\bibitem{deMedeirosVarzielas:2005ax}
I.~de~Medeiros~Varzielas and G.~G. Ross, {\it {$SU(3)$ family symmetry and
  neutrino bi-tri-maximal mixing}},  Nucl. Phys. {\bf B733} (2006) 31--47
  [\href{http://xxx.lanl.gov/abs/hep-ph/0507176}{{\tt hep-ph/0507176}}].

\bibitem{Antusch:2004xd}
S.~Antusch and S.~F. King, {\it {From hierarchical to partially degenerate
  neutrinos via type II upgrade of type I see-saw models}},  Nucl. Phys. {\bf
  B705} (2005) 239--268 [\href{http://xxx.lanl.gov/abs/hep-ph/0402121}{{\tt
  hep-ph/0402121}}].

\bibitem{King:2005bj}
S.~F. King, {\it {Predicting neutrino parameters from $SO(3)$ family symmetry
  and quark-lepton unification}},  JHEP {\bf 08} (2005) 105
  [\href{http://xxx.lanl.gov/abs/hep-ph/0506297}{{\tt hep-ph/0506297}}].

\bibitem{Antusch:2010xx}
S.~Antusch, L.~Calibbi, V.~Maurer and M.~Spinrath, {\it {Work in progress}}.

\bibitem{Pati:1973uk}
J.~C. Pati and A.~Salam, {\it {Unified Lepton-Hadron Symmetry and a Gauge
  Theory of the Basic Interactions}},  Phys. Rev. {\bf D8} (1973) 1240--1251.

\bibitem{Aaltonen:2007ps}
{\bf CDF} Collaboration, T.~Aaltonen {\em et.~al.}, {\it {First Run II
  Measurement of the $W$ Boson Mass}},  Phys. Rev. {\bf D77} (2008) 112001
  [\href{http://xxx.lanl.gov/abs/0708.3642}{{\tt arXiv:0708.3642}}].

\bibitem{Camilleri:2008zz}
L.~Camilleri, E.~Lisi and J.~F. Wilkerson, {\it {Neutrino Masses and Mixings:
  Status and Prospects}},  Ann. Rev. Nucl. Part. Sci. {\bf 58} (2008) 343--369.

\bibitem{Dore:2008dp}
U.~Dore and D.~Orestano, {\it {Experimental results on neutrino oscillations}},
   Rept. Prog. Phys. {\bf 71} (2008) 106201
  [\href{http://xxx.lanl.gov/abs/0811.1194}{{\tt arXiv:0811.1194}}].

\bibitem{King:2007nw}
S.~F. King, {\it {Neutrino Physics}},
  [\href{http://xxx.lanl.gov/abs/0712.1750}{{\tt arXiv:0712.1750}}].

\bibitem{King:2003jb}
S.~F. King, {\it {Neutrino mass models}},  Rept. Prog. Phys. {\bf 67} (2004)
  107--158 [\href{http://xxx.lanl.gov/abs/hep-ph/0310204}{{\tt
  hep-ph/0310204}}].

\bibitem{Mohapatra:2005wg}
R.~N. Mohapatra {\em et.~al.}, {\it {Theory of neutrinos: A white paper}},
  Rept. Prog. Phys. {\bf 70} (2007) 1757--1867
  [\href{http://xxx.lanl.gov/abs/hep-ph/0510213}{{\tt hep-ph/0510213}}].

\bibitem{Mohapatra:2006gs}
R.~N. Mohapatra and A.~Y. Smirnov, {\it {Neutrino Mass and New Physics}},  Ann.
  Rev. Nucl. Part. Sci. {\bf 56} (2006) 569--628
  [\href{http://xxx.lanl.gov/abs/hep-ph/0603118}{{\tt hep-ph/0603118}}].

\bibitem{Albright:2009cn}
C.~H. Albright, {\it {Overview of Neutrino Mixing Models and Ways to
  Differentiate among Them}},  [\href{http://xxx.lanl.gov/abs/0905.0146}{{\tt
  arXiv:0905.0146}}].

\bibitem{Cabibbo:1963yz}
N.~Cabibbo, {\it {Unitary Symmetry and Leptonic Decays}},  Phys. Rev. Lett.
  {\bf 10} (1963) 531--533.

\bibitem{Kobayashi:1973fv}
M.~Kobayashi and T.~Maskawa, {\it {CP Violation in the Renormalizable Theory of
  Weak Interaction}},  Prog. Theor. Phys. {\bf 49} (1973) 652--657.

\bibitem{Pontecorvo:1957qd}
B.~Pontecorvo, {\it {Inverse beta processes and nonconservation of lepton
  charge}},  Sov. Phys. JETP {\bf 7} (1958) 172--173.

\bibitem{Maki:1962mu}
Z.~Maki, M.~Nakagawa and S.~Sakata, {\it {Remarks on the unified model of
  elementary particles}},  Prog. Theor. Phys. {\bf 28} (1962) 870--880.

\bibitem{Alcaraz:2009jr}
{\bf ALEPH} Collaboration, J.~Alcaraz, {\it {Precision Electroweak Measurements
  and Constraints on the Standard Model}},
  [\href{http://xxx.lanl.gov/abs/0911.2604}{{\tt arXiv:0911.2604}}].

\bibitem{Martin:1997ns}
S.~P. Martin, {\it {A Supersymmetry Primer}},
  [\href{http://xxx.lanl.gov/abs/hep-ph/9709356}{{\tt hep-ph/9709356}}].

\bibitem{Aitchison:2007fn}
I.~J.~R. Aitchison, {\it {Supersymmetry in particle physics: An elementary
  introduction}}, . SLAC-R-865.

\bibitem{Bailin:1994qt}
D.~Bailin and A.~Love, {\it {Supersymmetric gauge field theory and string
  theory}}, . Bristol, UK: IOP (1994) 322 p. (Graduate student series in
  physics).

\bibitem{Trenkel:2009phd}
M.~Trenkel, {Phenomenology of Supersymmetric Particle Production Processes at
  the LHC}.
\newblock PhD thesis, {Max-Planck-Institut f\"ur Physik/TU M\"unchen}, 2009.

\bibitem{Coleman:1967ad}
S.~R. Coleman and J.~Mandula, {\it All possible symmetries of the s matrix},
  Phys. Rev. {\bf 159} (1967) 1251--1256.

\bibitem{Haag:1974qh}
R.~Haag, J.~T. \L{}opusza\'nski and M.~Sohnius, {\it All possible generators of
  supersymmetries of the s matrix},  Nucl. Phys. {\bf B88} (1975) 257.

\bibitem{Cassel:2010px}
S.~Cassel, D.~M. Ghilencea and G.~G. Ross, {\it {Testing SUSY at the LHC:
  Electroweak and Dark matter fine tuning at two-loop order}},
  [\href{http://xxx.lanl.gov/abs/1001.3884}{{\tt arXiv:1001.3884}}].

\bibitem{Ellis:2008di}
J.~Ellis, {\it {Prospects for Discovering Supersymmetry at the LHC}},  Eur.
  Phys. J. {\bf C59} (2009) 335--343
  [\href{http://xxx.lanl.gov/abs/0810.1178}{{\tt arXiv:0810.1178}}].

\bibitem{Abdallah:2009zz}
{\bf ATLAS} Collaboration, J.~Abdallah {\em et.~al.}, {\it {Prospects for
  supersymmetry discovery based on inclusive searches}}, .
  ATL-PHYS-PUB-2009-066.

\bibitem{Bechtle:2009zz}
{\bf ATLAS} Collaboration, P.~Bechtle {\em et.~al.}, {\it {Measurements from
  supersymmetric events}}, . ATL-PHYS-PUB-2009-067.

\bibitem{of:2009qj}
{\bf CMS} Collaboration, M.~Pioppi, {\it {Search for SUSY at LHC in the first
  year of data-taking}},  [\href{http://xxx.lanl.gov/abs/0912.1189}{{\tt
  arXiv:0912.1189}}].

\bibitem{Lungu:2009nh}
{\bf CMS} Collaboration, G.~Lungu, {\it {Search for supersymmetry at the CMS in
  all-hadronic final state}},  [\href{http://xxx.lanl.gov/abs/0910.3310}{{\tt
  arXiv:0910.3310}}].

\bibitem{Germer:2010vn}
J.~Germer, W.~Hollik, E.~Mirabella and M.~K. Trenkel, {\it {Hadronic production
  of squark-squark pairs: The electroweak contributions}},
  [\href{http://xxx.lanl.gov/abs/1004.2621}{{\tt arXiv:1004.2621}}].

\bibitem{Ehrenfeld:2009rt}
W.~Ehrenfeld, A.~Freitas, A.~Landwehr and D.~Wyler, {\it {Distinguishing spins
  in decay chains with photons at the Large Hadron Collider}},  JHEP {\bf 07}
  (2009) 056 [\href{http://xxx.lanl.gov/abs/0904.1293}{{\tt arXiv:0904.1293}}].

\bibitem{Amaldi:1991cn}
U.~Amaldi, W.~de~Boer and H.~Furstenau, {\it {Comparison of grand unified
  theories with electroweak and strong coupling constants measured at LEP}},
  Phys. Lett. {\bf B260} (1991) 447--455.

\bibitem{deBoer:1994dg}
W.~de~Boer, {\it Grand unified theories and supersymmetry in particle physics
  and cosmology},  Prog. Part. Nucl. Phys. {\bf 33} (1994) 201--302
  [\href{http://xxx.lanl.gov/abs/hep-ph/9402266}{{\tt hep-ph/9402266}}].

\bibitem{Ellis:2010kf}
J.~Ellis and K.~A. Olive, {\it {Supersymmetric Dark Matter Candidates}},
  [\href{http://xxx.lanl.gov/abs/1001.3651}{{\tt arXiv:1001.3651}}].

\bibitem{Freedman:1976xh}
D.~Z. Freedman, P.~van Nieuwenhuizen and S.~Ferrara, {\it {Progress Toward a
  Theory of Supergravity}},  Phys. Rev. {\bf D13} (1976) 3214--3218.

\bibitem{Deser:1976eh}
S.~Deser and B.~Zumino, {\it {Consistent Supergravity}},  Phys. Lett. {\bf B62}
  (1976) 335.

\bibitem{Sohnius:1985qm}
M.~F. Sohnius, {\it {Introducing Supersymmetry}},  Phys. Rept. {\bf 128} (1985)
  39--204.

\bibitem{Nilles:1983ge}
H.~P. Nilles, {\it {Supersymmetry, Supergravity and Particle Physics}},  Phys.
  Rept. {\bf 110} (1984) 1--162.

\bibitem{Haber:1984rc}
H.~E. Haber and G.~L. Kane, {\it {The Search for Supersymmetry: Probing Physics
  Beyond the Standard Model}},  Phys. Rept. {\bf 117} (1985) 75--263.

\bibitem{Barbieri:1987xf}
R.~Barbieri, {\it {Looking Beyond the Standard Model: The Supersymmetric
  Option}},  Riv. Nuovo Cim. {\bf 11N4} (1988) 1--45.

\bibitem{Fayet:1976et}
P.~Fayet, {\it {Supersymmetry and Weak, Electromagnetic and Strong
  Interactions}},  Phys. Lett. {\bf B64} (1976) 159.

\bibitem{Sakai:1981pk}
N.~Sakai and T.~Yanagida, {\it {Proton Decay in a Class of Supersymmetric Grand
  Unified Models}},  Nucl. Phys. {\bf B197} (1982) 533.

\bibitem{Weinberg:1981wj}
S.~Weinberg, {\it {Supersymmetry at Ordinary Energies. 1. Masses and
  Conservation Laws}},  Phys. Rev. {\bf D26} (1982) 287.

\bibitem{Smirnov:1996bg}
A.~Y. Smirnov and F.~Vissani, {\it {Upper bound on all products of R-parity
  violating couplings $\lambda'$ and $\lambda''$ from proton decay}},  Phys.
  Lett. {\bf B380} (1996) 317--323
  [\href{http://xxx.lanl.gov/abs/hep-ph/9601387}{{\tt hep-ph/9601387}}].

\bibitem{Bhattacharyya:1998bx}
G.~Bhattacharyya and P.~B. Pal, {\it {Upper bounds on all R-parity-violating
  lambda lambda'' combinations from proton stability}},  Phys. Rev. {\bf D59}
  (1999) 097701 [\href{http://xxx.lanl.gov/abs/hep-ph/9809493}{{\tt
  hep-ph/9809493}}].

\bibitem{Barbier:2004ez}
R.~Barbier {\em et.~al.}, {\it {R-parity violating supersymmetry}},  Phys.
  Rept. {\bf 420} (2005) 1--202
  [\href{http://xxx.lanl.gov/abs/hep-ph/0406039}{{\tt hep-ph/0406039}}].

\bibitem{SuperK:2009gd}
{\bf Super-Kamiokande} Collaboration, H.~Nishino {\em et.~al.}, {\it Search for
  proton decay via $p \rightarrow e^+ \pi^0$ and $p \rightarrow \mu^+ \pi^0$ in
  a large water cherenkov detector},  Phys. Rev. Lett. {\bf 102} (2009) 141801
  [\href{http://xxx.lanl.gov/abs/0903.0676}{{\tt arXiv:0903.0676}}].

\bibitem{Raby:2008pd}
S.~Raby {\em et.~al.}, {\it {DUSEL Theory White Paper}},
  [\href{http://xxx.lanl.gov/abs/0810.4551}{{\tt arXiv:0810.4551}}].

\bibitem{Dreiner:2005rd}
H.~K. Dreiner, C.~Luhn and M.~Thormeier, {\it {What is the discrete gauge
  symmetry of the MSSM?}},  Phys. Rev. {\bf D73} (2006) 075007
  [\href{http://xxx.lanl.gov/abs/hep-ph/0512163}{{\tt hep-ph/0512163}}].

\bibitem{Ibanez:1991pr}
L.~E. Ibanez and G.~G. Ross, {\it {Discrete gauge symmetries and the origin of
  baryon and lepton number conservation in supersymmetric versions of the
  standard model}},  Nucl. Phys. {\bf B368} (1992) 3--37.

\bibitem{Ibanez:1991hv}
L.~E. Ibanez and G.~G. Ross, {\it {Discrete gauge symmetry anomalies}},  Phys.
  Lett. {\bf B260} (1991) 291--295.

\bibitem{Mohapatra:2007vd}
R.~N. Mohapatra and M.~Ratz, {\it {Gauged Discrete Symmetries and Proton
  Stability}},  Phys. Rev. {\bf D76} (2007) 095003
  [\href{http://xxx.lanl.gov/abs/0707.4070}{{\tt arXiv:0707.4070}}].

\bibitem{Farrar:1978xj}
G.~R. Farrar and P.~Fayet, {\it {Phenomenology of the Production, Decay, and
  Detection of New Hadronic States Associated with Supersymmetry}},  Phys.
  Lett. {\bf B76} (1978) 575--579.

\bibitem{Hall:1983iz}
L.~J. Hall, J.~D. Lykken and S.~Weinberg, {\it {Supergravity as the Messenger
  of Supersymmetry Breaking}},  Phys. Rev. {\bf D27} (1983) 2359--2378.

\bibitem{Kim:1983dt}
J.~E. Kim and H.~P. Nilles, {\it {The mu Problem and the Strong CP Problem}},
  Phys. Lett. {\bf B138} (1984) 150.

\bibitem{Inoue:1985cw}
K.~Inoue, A.~Kakuto and H.~Takano, {\it {Higgs as (Pseudo)Goldstone
  Particles}},  Prog. Theor. Phys. {\bf 75} (1986) 664.

\bibitem{Anselm:1986um}
A.~A. Anselm and A.~A. Johansen, {\it {SUSY GUT with Automatic Doublet -
  Triplet Hierarchy}},  Phys. Lett. {\bf B200} (1988) 331--334.

\bibitem{Giudice:1988yz}
G.~F. Giudice and A.~Masiero, {\it {A Natural Solution to the mu Problem in
  Supergravity Theories}},  Phys. Lett. {\bf B206} (1988) 480--484.

\bibitem{Fayet:1974pd}
P.~Fayet, {\it {Supergauge Invariant Extension of the Higgs Mechanism and a
  Model for the electron and Its Neutrino}},  Nucl. Phys. {\bf B90} (1975)
  104--124.

\bibitem{Nilles:1982dy}
H.~P. Nilles, M.~Srednicki and D.~Wyler, {\it {Weak Interaction Breakdown
  Induced by Supergravity}},  Phys. Lett. {\bf B120} (1983) 346.

\bibitem{Frere:1983ag}
J.~M. Frere, D.~R.~T. Jones and S.~Raby, {\it {Fermion Masses and Induction of
  the Weak Scale by Supergravity}},  Nucl. Phys. {\bf B222} (1983) 11.

\bibitem{Derendinger:1983bz}
J.~P. Derendinger and C.~A. Savoy, {\it {Quantum Effects and $SU(2) \times
  U(1)$ Breaking in Supergravity Gauge Theories}},  Nucl. Phys. {\bf B237}
  (1984) 307.

\bibitem{Greene:1986th}
B.~R. Greene and P.~J. Miron, {\it {Supersymmetric Cosmology with a Gauge
  Singlet}},  Phys. Lett. {\bf B168} (1986) 226.

\bibitem{Ellis:1986mq}
J.~R. Ellis {\em et.~al.}, {\it {Problems for (2,0) Compactifications}},  Phys.
  Lett. {\bf B176} (1986) 403.

\bibitem{Durand:1988rg}
L.~Durand and J.~L. Lopez, {\it {Upper Bounds on Higgs and Top Quark Masses in
  the Flipped $SU(5) \times U(1)$ Superstring Model}},  Phys. Lett. {\bf B217}
  (1989) 463.

\bibitem{Drees:1988fc}
M.~Drees, {\it {Supersymmetric Models with Extended Higgs Sector}},  Int. J.
  Mod. Phys. {\bf A4} (1989) 3635.

\bibitem{Ellis:1988er}
J.~R. Ellis, J.~F. Gunion, H.~E. Haber, L.~Roszkowski and F.~Zwirner, {\it
  {Higgs Bosons in a Nonminimal Supersymmetric Model}},  Phys. Rev. {\bf D39}
  (1989) 844.

\bibitem{Elliott:1994ht}
T.~Elliott, S.~F. King and P.~L. White, {\it {Unification constraints in the
  next-to-minimal supersymmetric standard model}},  Phys. Lett. {\bf B351}
  (1995) 213--219 [\href{http://xxx.lanl.gov/abs/hep-ph/9406303}{{\tt
  hep-ph/9406303}}].

\bibitem{King:1995vk}
S.~F. King and P.~L. White, {\it {Resolving the constrained minimal and
  next-to-minimal supersymmetric standard models}},  Phys. Rev. {\bf D52}
  (1995) 4183--4216 [\href{http://xxx.lanl.gov/abs/hep-ph/9505326}{{\tt
  hep-ph/9505326}}].

\bibitem{Chung:2003fi}
D.~J.~H. Chung {\em et.~al.}, {\it {The soft supersymmetry-breaking Lagrangian:
  Theory and applications}},  Phys. Rept. {\bf 407} (2005) 1--203
  [\href{http://xxx.lanl.gov/abs/hep-ph/0312378}{{\tt hep-ph/0312378}}].

\bibitem{O'Raifeartaigh:1975pr}
L.~O'Raifeartaigh, {\it {Spontaneous Symmetry Breaking for Chiral Scalar
  Superfields}},  Nucl. Phys. {\bf B96} (1975) 331.

\bibitem{Fayet:1974jb}
P.~Fayet and J.~Iliopoulos, {\it {Spontaneously Broken Supergauge Symmetries
  and Goldstone Spinors}},  Phys. Lett. {\bf B51} (1974) 461--464.

\bibitem{Randall:1998uk}
L.~Randall and R.~Sundrum, {\it {Out of this world supersymmetry breaking}},
  Nucl. Phys. {\bf B557} (1999) 79--118
  [\href{http://xxx.lanl.gov/abs/hep-th/9810155}{{\tt hep-th/9810155}}].

\bibitem{Giudice:1998xp}
G.~F. Giudice, M.~A. Luty, H.~Murayama and R.~Rattazzi, {\it {Gaugino Mass
  without Singlets}},  JHEP {\bf 12} (1998) 027
  [\href{http://xxx.lanl.gov/abs/hep-ph/9810442}{{\tt hep-ph/9810442}}].

\bibitem{Gherghetta:1999sw}
T.~Gherghetta, G.~F. Giudice and J.~D. Wells, {\it {Phenomenological
  consequences of supersymmetry with anomaly-induced masses}},  Nucl. Phys.
  {\bf B559} (1999) 27--47 [\href{http://xxx.lanl.gov/abs/hep-ph/9904378}{{\tt
  hep-ph/9904378}}].

\bibitem{Dine:1993yw}
M.~Dine and A.~E. Nelson, {\it {Dynamical supersymmetry breaking at
  low-energies}},  Phys. Rev. {\bf D48} (1993) 1277--1287
  [\href{http://xxx.lanl.gov/abs/hep-ph/9303230}{{\tt hep-ph/9303230}}].

\bibitem{Dine:1994vc}
M.~Dine, A.~E. Nelson and Y.~Shirman, {\it {Low-energy dynamical supersymmetry
  breaking simplified}},  Phys. Rev. {\bf D51} (1995) 1362--1370
  [\href{http://xxx.lanl.gov/abs/hep-ph/9408384}{{\tt hep-ph/9408384}}].

\bibitem{Dine:1995ag}
M.~Dine, A.~E. Nelson, Y.~Nir and Y.~Shirman, {\it {New tools for low-energy
  dynamical supersymmetry breaking}},  Phys. Rev. {\bf D53} (1996) 2658--2669
  [\href{http://xxx.lanl.gov/abs/hep-ph/9507378}{{\tt hep-ph/9507378}}].

\bibitem{Ambrosanio:1997rv}
S.~Ambrosanio, G.~D. Kribs and S.~P. Martin, {\it {Signals for gauge-mediated
  supersymmetry breaking models at the CERN LEP2 collider}},  Phys. Rev. {\bf
  D56} (1997) 1761--1777 [\href{http://xxx.lanl.gov/abs/hep-ph/9703211}{{\tt
  hep-ph/9703211}}].

\bibitem{Giudice:1998bp}
G.~F. Giudice and R.~Rattazzi, {\it {Theories with gauge-mediated supersymmetry
  breaking}},  Phys. Rept. {\bf 322} (1999) 419--499
  [\href{http://xxx.lanl.gov/abs/hep-ph/9801271}{{\tt hep-ph/9801271}}].

\bibitem{Nilles:1982ik}
H.~P. Nilles, {\it {Dynamically Broken Supergravity and the Hierarchy
  Problem}},  Phys. Lett. {\bf B115} (1982) 193.

\bibitem{Nilles:1982xx}
H.~P. Nilles, {\it {Supergravity Generates Hierarchies}},  Nucl. Phys. {\bf
  B217} (1983) 366.

\bibitem{Chamseddine:1982jx}
A.~H. Chamseddine, R.~L. Arnowitt and P.~Nath, {\it {Locally Supersymmetric
  Grand Unification}},  Phys. Rev. Lett. {\bf 49} (1982) 970.

\bibitem{Barbieri:1982eh}
R.~Barbieri, S.~Ferrara and C.~A. Savoy, {\it {Gauge Models with Spontaneously
  Broken Local Supersymmetry}},  Phys. Lett. {\bf B119} (1982) 343.

\bibitem{Antusch:2010va}
S.~Antusch {\em et.~al.}, {\it {Gauge Non-Singlet Inflation in SUSY GUTs}},
  [\href{http://xxx.lanl.gov/abs/1003.3233}{{\tt arXiv:1003.3233}}].

\bibitem{Gursey:1975ki}
F.~Gursey, P.~Ramond and P.~Sikivie, {\it {A Universal Gauge Theory Model Based
  on $E_6$}},  Phys. Lett. {\bf B60} (1976) 177.

\bibitem{Cheng:1985bj}
T.~P. Cheng and L.~F. Li, {Gauge Theory of Elementary Particle Physics}.
\newblock Oxford, Uk: Clarendon (1984) 536 P. (Oxford Science Publications).

\bibitem{Ross:1985ai}
G.~G. Ross, {Grand Unified Theories}.
\newblock Reading, USA: Benjamin/cummings (1984) 497 P. (Frontiers In Physics,
  60).

\bibitem{Georgi:1982jb}
H.~Georgi, {Lie Algebras in Particle Physics. From Isospin to Unified
  Theories}.
\newblock Reading, USA: Benjamin/cummings (1982) 255 P. (Frontiers in Physics,
  54).

\bibitem{Slansky:1981yr}
R.~Slansky, {\it {Group Theory for Unified Model Building}},  Phys. Rept. {\bf
  79} (1981) 1--128.

\bibitem{Babu:1989ex}
K.~S. Babu and R.~N. Mohapatra, {\it {Quantization of electric charge from
  anomaly constraints and a Majorana neutrino}},  Phys. Rev. {\bf D41} (1990)
  271.

\bibitem{Babu:1989tq}
K.~S. Babu and R.~N. Mohapatra, {\it {Is there a connection between
  quantization of electric charge and a Majorana neutrino?}},  Phys. Rev. Lett.
  {\bf 63} (1989) 938.

\bibitem{Senjanovic:2009kr}
G.~Senjanovic, {\it {Proton decay and grand unification}},
  [\href{http://xxx.lanl.gov/abs/0912.5375}{{\tt arXiv:0912.5375}}].

\bibitem{Babu:2010ej}
K.~S. Babu, J.~C. Pati and Z.~Tavartkiladze, {\it {Constraining Proton Lifetime
  in $SO(10)$ with Stabilized Doublet-Triplet Splitting}},
  [\href{http://xxx.lanl.gov/abs/1003.2625}{{\tt arXiv:1003.2625}}].

\bibitem{Gatto:1968ss}
R.~Gatto, G.~Sartori and M.~Tonin, {\it {Weak Selfmasses, Cabibbo Angle, and
  Broken $SU(2) \times SU(2)$}},  Phys. Lett. {\bf B28} (1968) 128--130.

\bibitem{Duque:2008ah}
L.~F. Duque, D.~A. Gutierrez, E.~Nardi and J.~Norena, {\it {Fermion mass
  hierarchy and non-hierarchical mass ratios in $SU(5) \times U(1)_F$}},  Phys.
  Rev. {\bf D78} (2008) 035003 [\href{http://xxx.lanl.gov/abs/0804.2865}{{\tt
  arXiv:0804.2865}}].

\bibitem{Ellis:1979fg}
J.~R. Ellis and M.~K. Gaillard, {\it {Fermion Masses and Higgs Representations
  in $SU(5)$}},  Phys. Lett. {\bf B88} (1979) 315.

\bibitem{Allanach:1996hz}
B.~C. Allanach, S.~F. King, G.~K. Leontaris and S.~Lola, {\it {A new approach
  to Yukawa textures in supersymmetric unified models with gauged family
  symmetries}},  Phys. Rev. {\bf D56} (1997) 2632--2655
  [\href{http://xxx.lanl.gov/abs/hep-ph/9610517}{{\tt hep-ph/9610517}}].

\bibitem{Bagger:1996ei}
J.~A. Bagger, K.~T. Matchev, D.~M. Pierce and R.-J. Zhang, {\it {Gauge and
  Yukawa unification in models with gauge-mediated supersymmetry breaking}},
  Phys. Rev. Lett. {\bf 78} (1997) 1002--1005
  [\href{http://xxx.lanl.gov/abs/hep-ph/9611229}{{\tt hep-ph/9611229}}].

\bibitem{King:2000vp}
S.~F. King and M.~Oliveira, {\it {Yukawa unification as a window into the soft
  supersymmetry breaking Lagrangian}},  Phys. Rev. {\bf D63} (2001) 015010
  [\href{http://xxx.lanl.gov/abs/hep-ph/0008183}{{\tt hep-ph/0008183}}].

\bibitem{Buras:2002vd}
A.~J. Buras, P.~H. Chankowski, J.~Rosiek and L.~Slawianowska, {\it {$\Delta
  M_{d,s}, B^0_{d,s} \to \mu^{+} \mu^{-}$ and $B \to X_{s} \gamma$ in
  supersymmetry at large $\tan\beta$}},  Nucl. Phys. {\bf B659} (2003) 3
  [\href{http://xxx.lanl.gov/abs/hep-ph/0210145}{{\tt hep-ph/0210145}}].

\bibitem{Carena:1999py}
M.~S. Carena, D.~Garcia, U.~Nierste and C.~E.~M. Wagner, {\it {Effective
  Lagrangian for the $\bar{t} b H^{+}$ interaction in the MSSM and charged
  Higgs phenomenology}},  Nucl. Phys. {\bf B577} (2000) 88--120
  [\href{http://xxx.lanl.gov/abs/hep-ph/9912516}{{\tt hep-ph/9912516}}].

\bibitem{Pierce:1996zz}
D.~M. Pierce, J.~A. Bagger, K.~T. Matchev and R.-j. Zhang, {\it {Precision
  corrections in the minimal supersymmetric standard model}},  Nucl. Phys. {\bf
  B491} (1997) 3--67 [\href{http://xxx.lanl.gov/abs/hep-ph/9606211}{{\tt
  hep-ph/9606211}}].

\bibitem{Freitas:2007dp}
A.~Freitas, E.~Gasser and U.~Haisch, {\it {Supersymmetric large tan(beta)
  corrections to $\Delta M_{d, s}$ and $B_{d, s} \to \mu^{+} \mu^{-}$
  revisited}},  Phys. Rev. {\bf D76} (2007) 014016
  [\href{http://xxx.lanl.gov/abs/hep-ph/0702267}{{\tt hep-ph/0702267}}].

\bibitem{Blazek:2001sb}
T.~Blazek, R.~Dermisek and S.~Raby, {\it {Predictions for Higgs and SUSY
  spectra from $SO(10)$ Yukawa unification with $\mu > 0$}},  Phys. Rev. Lett.
  {\bf 88} (2002) 111804 [\href{http://xxx.lanl.gov/abs/hep-ph/0107097}{{\tt
  hep-ph/0107097}}].

\bibitem{Altmannshofer:2008vr}
W.~Altmannshofer, D.~Guadagnoli, S.~Raby and D.~M. Straub, {\it {SUSY GUTs with
  Yukawa unification: A Go/no-go study using FCNC processes}},  Phys. Lett.
  {\bf B668} (2008) 385--391 [\href{http://xxx.lanl.gov/abs/0801.4363}{{\tt
  arXiv:0801.4363}}].

\bibitem{Ross:2007az}
G.~Ross and M.~Serna, {\it {Unification and Fermion Mass Structure}},  Phys.
  Lett. {\bf B664} (2008) 97--102
  [\href{http://xxx.lanl.gov/abs/0704.1248}{{\tt arXiv:0704.1248}}].

\bibitem{Antusch:2005gp}
S.~Antusch, J.~Kersten, M.~Lindner, M.~Ratz and M.~A. Schmidt, {\it {Running
  neutrino mass parameters in see-saw scenarios}},  JHEP {\bf 03} (2005) 024
  [\href{http://xxx.lanl.gov/abs/hep-ph/0501272}{{\tt hep-ph/0501272}}].

\bibitem{Antusch:2002rr}
S.~Antusch, J.~Kersten, M.~Lindner and M.~Ratz, {\it {Neutrino mass matrix
  running for non-degenerate see-saw scales}},  Phys. Lett. {\bf B538} (2002)
  87--95 [\href{http://xxx.lanl.gov/abs/hep-ph/0203233}{{\tt hep-ph/0203233}}].

\bibitem{King:1998jw}
S.~F. King, {\it {Atmospheric and solar neutrinos with a heavy singlet}},
  Phys. Lett. {\bf B439} (1998) 350--356
  [\href{http://xxx.lanl.gov/abs/hep-ph/9806440}{{\tt hep-ph/9806440}}].

\bibitem{King:1999cm}
S.~F. King, {\it {Atmospheric and solar neutrinos from single right-handed
  neutrino dominance and U(1) family symmetry}},  Nucl. Phys. {\bf B562} (1999)
  57--77 [\href{http://xxx.lanl.gov/abs/hep-ph/9904210}{{\tt hep-ph/9904210}}].

\bibitem{King:1999mb}
S.~F. King, {\it {Large mixing angle MSW and atmospheric neutrinos from single
  right-handed neutrino dominance and U(1) family symmetry}},  Nucl. Phys. {\bf
  B576} (2000) 85--105 [\href{http://xxx.lanl.gov/abs/hep-ph/9912492}{{\tt
  hep-ph/9912492}}].

\bibitem{Antusch:2004gf}
S.~Antusch and S.~F. King, {\it {Sequential dominance}},  New J. Phys. {\bf 6}
  (2004) 110 [\href{http://xxx.lanl.gov/abs/hep-ph/0405272}{{\tt
  hep-ph/0405272}}].

\bibitem{Antusch:2010tf}
S.~Antusch, S.~Boudjemaa and S.~F. King, {\it {Neutrino Mixing Angles in
  Sequential Dominance to NLO and NNLO}},
  [\href{http://xxx.lanl.gov/abs/1003.5498}{{\tt arXiv:1003.5498}}].

\bibitem{King:2002nf}
S.~F. King, {\it {Constructing the large mixing angle MNS matrix in see-saw
  models with right-handed neutrino dominance}},  JHEP {\bf 09} (2002) 011
  [\href{http://xxx.lanl.gov/abs/hep-ph/0204360}{{\tt hep-ph/0204360}}].

\bibitem{Antusch:2007dj}
S.~Antusch and S.~F. King, {\it {Lepton Flavour Violation in the Constrained
  MSSM with Constrained Sequential Dominance}},  Phys. Lett. {\bf B659} (2008)
  640--650 [\href{http://xxx.lanl.gov/abs/0709.0666}{{\tt arXiv:0709.0666}}].

\bibitem{Allanach:2001kg}
B.~C. Allanach, {\it {SOFTSUSY: A C++ program for calculating supersymmetric
  spectra}},  Comput. Phys. Commun. {\bf 143} (2002) 305--331
  [\href{http://xxx.lanl.gov/abs/hep-ph/0104145}{{\tt hep-ph/0104145}}].

\bibitem{Belanger:2001fz}
G.~Belanger, F.~Boudjema, A.~Pukhov and A.~Semenov, {\it {micrOMEGAs: A program
  for calculating the relic density in the MSSM}},  Comput. Phys. Commun. {\bf
  149} (2002) 103--120 [\href{http://xxx.lanl.gov/abs/hep-ph/0112278}{{\tt
  hep-ph/0112278}}].

\bibitem{Belanger:2004yn}
G.~Belanger, F.~Boudjema, A.~Pukhov and A.~Semenov, {\it {micrOMEGAs: Version
  1.3}},  Comput. Phys. Commun. {\bf 174} (2006) 577--604
  [\href{http://xxx.lanl.gov/abs/hep-ph/0405253}{{\tt hep-ph/0405253}}].

\bibitem{Belanger:2006is}
G.~Belanger, F.~Boudjema, A.~Pukhov and A.~Semenov, {\it {micrOMEGAs2.0: A
  program to calculate the relic density of dark matter in a generic model}},
  Comput. Phys. Commun. {\bf 176} (2007) 367--382
  [\href{http://xxx.lanl.gov/abs/hep-ph/0607059}{{\tt hep-ph/0607059}}].

\bibitem{Belanger:2008sj}
G.~Belanger, F.~Boudjema, A.~Pukhov and A.~Semenov, {\it {Dark matter direct
  detection rate in a generic model with micrOMEGAs2.2}},  Comput. Phys.
  Commun. {\bf 180} (2009) 747--767
  [\href{http://xxx.lanl.gov/abs/0803.2360}{{\tt arXiv:0803.2360}}].

\bibitem{Skands:2003cj}
P.~Z. Skands {\em et.~al.}, {\it {SUSY Les Houches Accord: Interfacing SUSY
  Spectrum Calculators, Decay Packages, and Event Generators}},  JHEP {\bf 07}
  (2004) 036 [\href{http://xxx.lanl.gov/abs/hep-ph/0311123}{{\tt
  hep-ph/0311123}}].

\bibitem{ALEPH:2005ema}
{\bf ALEPH} Collaboration, {\it {Precision electroweak measurements on the $Z$
  resonance}},  Phys. Rept. {\bf 427} (2006) 257
  [\href{http://xxx.lanl.gov/abs/hep-ex/0509008}{{\tt hep-ex/0509008}}].

\bibitem{Alcaraz:2006mx}
{\bf ALEPH} Collaboration, J.~Alcaraz {\em et.~al.}, {\it {A Combination of
  preliminary electroweak measurements and constraints on the standard model}},
   [\href{http://xxx.lanl.gov/abs/hep-ex/0612034}{{\tt hep-ex/0612034}}].

\bibitem{HFAG}
{Heavy Flavor Averaging Group}. {See: \tt{www.slac.stanford.edu/xorg/hfag}}.

\bibitem{Barate:1998vz}
{\bf ALEPH} Collaboration, R.~Barate {\em et.~al.}, {\it {A measurement of the
  inclusive $b \to s \gamma$ branching ratio}},  Phys. Lett. {\bf B429} (1998)
  169--187.

\bibitem{Chen:2001fja}
{\bf CLEO} Collaboration, S.~Chen {\em et.~al.}, {\it {Branching fraction and
  photon energy spectrum for $b \to s \gamma$}},  Phys. Rev. Lett. {\bf 87}
  (2001) 251807 [\href{http://xxx.lanl.gov/abs/hep-ex/0108032}{{\tt
  hep-ex/0108032}}].

\bibitem{Koppenburg:2004fz}
{\bf Belle} Collaboration, P.~Koppenburg {\em et.~al.}, {\it {An inclusive
  measurement of the photon energy spectrum in $b \to s \gamma$ decays}},
  Phys. Rev. Lett. {\bf 93} (2004) 061803
  [\href{http://xxx.lanl.gov/abs/hep-ex/0403004}{{\tt hep-ex/0403004}}].

\bibitem{Abe:2001hk}
{\bf Belle} Collaboration, K.~Abe {\em et.~al.}, {\it {A measurement of the
  branching fraction for the inclusive $B \to X_s \gamma $ decays with Belle}},
   Phys. Lett. {\bf B511} (2001) 151--158
  [\href{http://xxx.lanl.gov/abs/hep-ex/0103042}{{\tt hep-ex/0103042}}].

\bibitem{Aubert:2002pb}
{\bf BABAR} Collaboration, B.~Aubert {\em et.~al.}, {\it {$b \to s \gamma$
  using a sum of exclusive modes}},
  [\href{http://xxx.lanl.gov/abs/hep-ex/0207074}{{\tt hep-ex/0207074}}].

\bibitem{Aaltonen:2007kv}
{\bf CDF} Collaboration, T.~Aaltonen {\em et.~al.}, {\it {Search for $B^0_{s}
  \to \mu^{+} \mu^{-}$ and $B^0_{d} \to \mu^{+} \mu^{-}$ decays with $2fb^{-1}$
  of $p \bar{p}$ collisions}},  Phys. Rev. Lett. {\bf 100} (2008) 101802
  [\href{http://xxx.lanl.gov/abs/0712.1708}{{\tt arXiv:0712.1708}}].

\bibitem{Buchalla:1993bv}
G.~Buchalla and A.~J. Buras, {\it {QCD corrections to rare K and B decays for
  arbitrary top quark mass}},  Nucl. Phys. {\bf B400} (1993) 225--239.

\bibitem{Misiak:1999yg}
M.~Misiak and J.~Urban, {\it {QCD corrections to FCNC decays mediated by
  Z-penguins and W-boxes}},  Phys. Lett. {\bf B451} (1999) 161--169
  [\href{http://xxx.lanl.gov/abs/hep-ph/9901278}{{\tt hep-ph/9901278}}].

\bibitem{Buchalla:1998ba}
G.~Buchalla and A.~J. Buras, {\it {The rare decays $K \to \pi \nu \bar{\nu}$,
  $B \to X \nu \bar{\nu}$ and $B \to l^+ l^-$: An update}},  Nucl. Phys. {\bf
  B548} (1999) 309--327 [\href{http://xxx.lanl.gov/abs/hep-ph/9901288}{{\tt
  hep-ph/9901288}}].

\bibitem{Buras:2003td}
A.~J. Buras, {\it {Relations between $\Delta M_{s,d}$ and $B_{s, d} \to \mu
  \bar{\mu}$ in models with minimal flavor violation}},  Phys. Lett. {\bf B566}
  (2003) 115--119 [\href{http://xxx.lanl.gov/abs/hep-ph/0303060}{{\tt
  hep-ph/0303060}}].

\bibitem{Isidori:2001fv}
G.~Isidori and A.~Retico, {\it {Scalar flavor changing neutral currents in the
  large tan beta limit}},  JHEP {\bf 11} (2001) 001
  [\href{http://xxx.lanl.gov/abs/hep-ph/0110121}{{\tt hep-ph/0110121}}].

\bibitem{Hayashii:2005ih}
{\bf Belle} Collaboration, H.~Hayashii, {\it {Tau lepton physics at Belle}},
  PoS {\bf HEP2005} (2006) 291.

\bibitem{Davier:2007ua}
M.~Davier, {\it {The hadronic contribution to $(g-2)_\mu$}},  Nucl. Phys. Proc.
  Suppl. {\bf 169} (2007) 288--296
  [\href{http://xxx.lanl.gov/abs/hep-ph/0701163}{{\tt hep-ph/0701163}}].

\bibitem{Stockinger:2006zn}
D.~Stockinger, {\it {The muon magnetic moment and supersymmetry}},  J. Phys.
  {\bf G34} (2007) R45--R92 [\href{http://xxx.lanl.gov/abs/hep-ph/0609168}{{\tt
  hep-ph/0609168}}].

\bibitem{Stockinger:2007pe}
D.~Stockinger, {\it {$(g-2)_\mu$ and supersymmetry: status and prospects}},
  [\href{http://xxx.lanl.gov/abs/0710.2429}{{\tt arXiv:0710.2429}}].

\bibitem{Zhang:2008pka}
Z.~Zhang, {\it {Muon g-2: a mini review}},
  [\href{http://xxx.lanl.gov/abs/0801.4905}{{\tt arXiv:0801.4905}}].

\bibitem{Bennett:2004pv}
{\bf Muon g-2} Collaboration, G.~W. Bennett {\em et.~al.}, {\it {Measurement of
  the negative muon anomalous magnetic moment to 0.7-ppm}},  Phys. Rev. Lett.
  {\bf 92} (2004) 161802 [\href{http://xxx.lanl.gov/abs/hep-ex/0401008}{{\tt
  hep-ex/0401008}}].

\bibitem{Bennett:2006fi}
{\bf Muon G-2} Collaboration, G.~W. Bennett {\em et.~al.}, {\it {Final report
  of the muon E821 anomalous magnetic moment measurement at BNL}},  Phys. Rev.
  {\bf D73} (2006) 072003 [\href{http://xxx.lanl.gov/abs/hep-ex/0602035}{{\tt
  hep-ex/0602035}}].

\bibitem{Miller:2007kk}
J.~P. Miller, E.~de~Rafael and B.~L. Roberts, {\it {Muon g-2: Review of Theory
  and Experiment}},  Rept. Prog. Phys. {\bf 70} (2007) 795
  [\href{http://xxx.lanl.gov/abs/hep-ph/0703049}{{\tt hep-ph/0703049}}].

\bibitem{Jegerlehner:2007xe}
F.~Jegerlehner, {\it {Essentials of the Muon g-2}},  Acta Phys. Polon. {\bf
  B38} (2007) 3021 [\href{http://xxx.lanl.gov/abs/hep-ph/0703125}{{\tt
  hep-ph/0703125}}].

\bibitem{Hagiwara:2006jt}
K.~Hagiwara, A.~D. Martin, D.~Nomura and T.~Teubner, {\it {Improved predictions
  for g-2 of the muon and $\alpha_{\rm QED}(M_Z^2)$}},  Phys. Lett. {\bf B649}
  (2007) 173--179 [\href{http://xxx.lanl.gov/abs/hep-ph/0611102}{{\tt
  hep-ph/0611102}}].

\bibitem{Martin:2001st}
S.~P. Martin and J.~D. Wells, {\it {Muon anomalous magnetic dipole moment in
  supersymmetric theories}},  Phys. Rev. {\bf D64} (2001) 035003
  [\href{http://xxx.lanl.gov/abs/hep-ph/0103067}{{\tt hep-ph/0103067}}].

\bibitem{Feruglio:2002af}
F.~Feruglio, A.~Strumia and F.~Vissani, {\it {Neutrino oscillations and signals
  in $\beta$ and $0 \nu 2 \beta$ experiments}},  Nucl. Phys. {\bf B637} (2002)
  345--377 [\href{http://xxx.lanl.gov/abs/hep-ph/0201291}{{\tt
  hep-ph/0201291}}].

\bibitem{Blazek:2002ta}
T.~Blazek, R.~Dermisek and S.~Raby, {\it {Yukawa unification in $SO(10)$}},
  Phys. Rev. {\bf D65} (2002) 115004
  [\href{http://xxx.lanl.gov/abs/hep-ph/0201081}{{\tt hep-ph/0201081}}].

\bibitem{Antusch:2005ca}
S.~Antusch, S.~F. King and R.~N. Mohapatra, {\it {Quark lepton complementarity
  in unified theories}},  Phys. Lett. {\bf B618} (2005) 150--161
  [\href{http://xxx.lanl.gov/abs/hep-ph/0504007}{{\tt hep-ph/0504007}}].

\bibitem{Raidal:2004iw}
M.~Raidal, {\it {Relation between the neutrino and quark mixing angles and
  grand unification}},  Phys. Rev. Lett. {\bf 93} (2004) 161801
  [\href{http://xxx.lanl.gov/abs/hep-ph/0404046}{{\tt hep-ph/0404046}}].

\bibitem{Minakata:2004xt}
H.~Minakata and A.~Y. Smirnov, {\it {Neutrino Mixing and Quark-Lepton
  Complementarity}},  Phys. Rev. {\bf D70} (2004) 073009
  [\href{http://xxx.lanl.gov/abs/hep-ph/0405088}{{\tt hep-ph/0405088}}].

\bibitem{Antusch:2005kw}
S.~Antusch and S.~F. King, {\it {Charged lepton corrections to neutrino mixing
  angles and CP phases revisited}},  Phys. Lett. {\bf B631} (2005) 42--47
  [\href{http://xxx.lanl.gov/abs/hep-ph/0508044}{{\tt hep-ph/0508044}}].

\bibitem{Antusch:2008yc}
S.~Antusch, S.~F. King and M.~Malinsky, {\it {Perturbative Estimates of Lepton
  Mixing Angles in Unified Models}},  Nucl. Phys. {\bf B820} (2009) 32--46
  [\href{http://xxx.lanl.gov/abs/0810.3863}{{\tt arXiv:0810.3863}}].

\bibitem{Boudjemaa:2008jf}
S.~Boudjemaa and S.~F. King, {\it {Deviations from Tri-bimaximal Mixing:
  Charged Lepton Corrections and Renormalization Group Running}},  Phys. Rev.
  {\bf D79} (2009) 033001 [\href{http://xxx.lanl.gov/abs/0808.2782}{{\tt
  arXiv:0808.2782}}].

\bibitem{Antusch:2007vw}
S.~Antusch, S.~F. King and M.~Malinsky, {\it {Third Family Corrections to Quark
  and Lepton Mixing in SUSY Models with non-Abelian Family Symmetry}},  JHEP
  {\bf 05} (2008) 066 [\href{http://xxx.lanl.gov/abs/0712.3759}{{\tt
  arXiv:0712.3759}}].

\bibitem{Antusch:2007ib}
S.~Antusch, S.~F. King and M.~Malinsky, {\it {Third Family Corrections to
  Tri-bimaximal Lepton Mixing and a New Sum Rule}},  Phys. Lett. {\bf B671}
  (2009) 263--266 [\href{http://xxx.lanl.gov/abs/0711.4727}{{\tt
  arXiv:0711.4727}}].

\bibitem{King:2007pr}
S.~F. King, {\it {Parametrizing the lepton mixing matrix in terms of deviations
  from tri-bimaximal mixing}},  Phys. Lett. {\bf B659} (2008) 244--251
  [\href{http://xxx.lanl.gov/abs/0710.0530}{{\tt arXiv:0710.0530}}].

\bibitem{Antusch:2003kp}
S.~Antusch, J.~Kersten, M.~Lindner and M.~Ratz, {\it {Running neutrino masses,
  mixings and CP phases: Analytical results and phenomenological
  consequences}},  Nucl. Phys. {\bf B674} (2003) 401--433
  [\href{http://xxx.lanl.gov/abs/hep-ph/0305273}{{\tt hep-ph/0305273}}].

\bibitem{GonzalezGarcia:2010er}
M.~C. Gonzalez-Garcia, M.~Maltoni and J.~Salvado, {\it {Updated global fit to
  three neutrino mixing: status of the hints of theta13 > 0}},
  [\href{http://xxx.lanl.gov/abs/1001.4524}{{\tt arXiv:1001.4524}}].

\bibitem{Masina:2005hf}
I.~Masina, {\it {A maximal atmospheric mixing from a maximal CP violating
  phase}},  Phys. Lett. {\bf B633} (2006) 134--140
  [\href{http://xxx.lanl.gov/abs/hep-ph/0508031}{{\tt hep-ph/0508031}}].

\bibitem{Antusch:2007rk}
S.~Antusch, P.~Huber, S.~F. King and T.~Schwetz, {\it {Neutrino mixing sum
  rules and oscillation experiments}},  JHEP {\bf 04} (2007) 060
  [\href{http://xxx.lanl.gov/abs/hep-ph/0702286}{{\tt hep-ph/0702286}}].

\bibitem{Bandyopadhyay:2007kx}
{\bf ISS Physics Working Group} Collaboration, A.~Bandyopadhyay {\em et.~al.},
  {\it {Physics at a future Neutrino Factory and super-beam facility}},  Rept.
  Prog. Phys. {\bf 72} (2009) 106201
  [\href{http://xxx.lanl.gov/abs/0710.4947}{{\tt arXiv:0710.4947}}].

\bibitem{Charles:2004jd}
{\bf CKMfitter Group} Collaboration, J.~Charles {\em et.~al.}, {\it {CP
  violation and the CKM matrix: Assessing the impact of the asymmetric $B$
  factories}},  Eur. Phys. J. {\bf C41} (2005) 1--131
  [\href{http://xxx.lanl.gov/abs/hep-ph/0406184}{{\tt hep-ph/0406184}}].

\bibitem{Bona:2007qta}
{\bf UTfit} Collaboration, M.~Bona {\em et.~al.}, {\it {Improved determination
  of the CKM angle alpha from $B \to \pi \pi$ decays}},  Phys. Rev. {\bf D76}
  (2007) 014015 [\href{http://xxx.lanl.gov/abs/hep-ph/0701204}{{\tt
  hep-ph/0701204}}].

\bibitem{Sordini:2009gu}
{\bf UTfit} Collaboration, V.~Sordini, {\it {Status of the Unitarity Triangle
  analysis in UTfit}},  [\href{http://xxx.lanl.gov/abs/0905.3747}{{\tt
  arXiv:0905.3747}}].

\bibitem{Fritzsch:1997fw}
H.~Fritzsch and Z.-Z. Xing, {\it {Flavor symmetries and the description of
  flavor mixing}},  Phys. Lett. {\bf B413} (1997) 396--404
  [\href{http://xxx.lanl.gov/abs/hep-ph/9707215}{{\tt hep-ph/9707215}}].

\bibitem{Fritzsch:1997st}
H.~Fritzsch and Z.-z. Xing, {\it {On the parametrization of flavor mixing in
  the standard model}},  Phys. Rev. {\bf D57} (1998) 594--597
  [\href{http://xxx.lanl.gov/abs/hep-ph/9708366}{{\tt hep-ph/9708366}}].

\bibitem{Harrison:2007yn}
P.~F. Harrison, D.~R.~J. Roythorne and W.~G. Scott, {\it {Plaquette Invariants
  and the Flavour Symmetric Description of Quark and Neutrino Mixings}},  Phys.
  Lett. {\bf B657} (2007) 210--216
  [\href{http://xxx.lanl.gov/abs/0709.1439}{{\tt arXiv:0709.1439}}].

\bibitem{Harrison:2008ff}
P.~F. Harrison, D.~R.~J. Roythorne and W.~G. Scott, {\it {Flavour Permutation
  Symmetry and Fermion Mixing}},
  [\href{http://xxx.lanl.gov/abs/0805.3440}{{\tt arXiv:0805.3440}}].

\bibitem{Harrison:2009bb}
P.~F. Harrison, D.~R.~J. Roythorne and W.~G. Scott, {\it {Is the Unitarity
  Triangle Right?}},  [\href{http://xxx.lanl.gov/abs/0904.3014}{{\tt
  arXiv:0904.3014}}].

\bibitem{Harrison:2009bz}
P.~F. Harrison, S.~Dallison and W.~G. Scott, {\it {The Matrix of Unitarity
  Triangle Angles for Quarks}},  Phys. Lett. {\bf B680} (2009) 328--333
  [\href{http://xxx.lanl.gov/abs/0904.3077}{{\tt arXiv:0904.3077}}].

\bibitem{Harrison:2009pw}
P.~F. Harrison and W.~G. Scott, {\it {A Flavour-Symmetric Perspective on
  Neutrino Mixing}},  [\href{http://xxx.lanl.gov/abs/0906.2732}{{\tt
  arXiv:0906.2732}}].

\bibitem{Couture:2009it}
G.~Couture, C.~Hamzaoui, S.~S.~Y. Lu and M.~Toharia, {\it {Patterns in the
  Fermion Mixing Matrix, a bottom-up approach}},
  [\href{http://xxx.lanl.gov/abs/0910.3132}{{\tt arXiv:0910.3132}}].

\bibitem{Barr:1990td}
S.~M. Barr, {\it {A Predictive hierarchical mode of quark and lepton masses}},
  Phys. Rev. {\bf D42} (1990) 3150--3159.

\bibitem{Sogami:1992av}
I.~S. Sogami and T.~Shinohara, {\it {Universal seesaw mechanisms for quark
  lepton mass spectrum}},  Phys. Rev. {\bf D47} (1993) 2905--2917.

\bibitem{Ibanez:1994ig}
L.~E. Ibanez and G.~G. Ross, {\it {Fermion masses and mixing angles from gauge
  symmetries}},  Phys. Lett. {\bf B332} (1994) 100--110
  [\href{http://xxx.lanl.gov/abs/hep-ph/9403338}{{\tt hep-ph/9403338}}].

\bibitem{Jain:1994hd}
V.~Jain and R.~Shrock, {\it {Models of fermion mass matrices based on a flavor
  dependent and generation dependent U(1) gauge symmetry}},  Phys. Lett. {\bf
  B352} (1995) 83--91 [\href{http://xxx.lanl.gov/abs/hep-ph/9412367}{{\tt
  hep-ph/9412367}}].

\bibitem{Pomarol:1995xc}
A.~Pomarol and D.~Tommasini, {\it {Horizontal symmetries for the supersymmetric
  flavor problem}},  Nucl. Phys. {\bf B466} (1996) 3--24
  [\href{http://xxx.lanl.gov/abs/hep-ph/9507462}{{\tt hep-ph/9507462}}].

\bibitem{Mondragon:1998gy}
A.~Mondragon and E.~Rodriguez-Jauregui, {\it {The breaking of the flavour
  permutational symmetry: Mass textures and the CKM matrix}},  Phys. Rev. {\bf
  D59} (1999) 093009 [\href{http://xxx.lanl.gov/abs/hep-ph/9807214}{{\tt
  hep-ph/9807214}}].

\bibitem{Barbieri:1999pe}
R.~Barbieri, P.~Creminelli and A.~Romanino, {\it {Neutrino mixings from a U(2)
  flavour symmetry}},  Nucl. Phys. {\bf B559} (1999) 17--26
  [\href{http://xxx.lanl.gov/abs/hep-ph/9903460}{{\tt hep-ph/9903460}}].

\bibitem{Barbieri:1998em}
R.~Barbieri, L.~Giusti, L.~J. Hall and A.~Romanino, {\it {Fermion masses and
  symmetry breaking of a U(2) flavour symmetry}},  Nucl. Phys. {\bf B550}
  (1999) 32--40 [\href{http://xxx.lanl.gov/abs/hep-ph/9812239}{{\tt
  hep-ph/9812239}}].

\bibitem{Babu:2004tn}
K.~S. Babu and J.~Kubo, {\it {Dihedral families of quarks, leptons and
  Higgses}},  Phys. Rev. {\bf D71} (2005) 056006
  [\href{http://xxx.lanl.gov/abs/hep-ph/0411226}{{\tt hep-ph/0411226}}].

\bibitem{Masina:2006pe}
I.~Masina and C.~A. Savoy, {\it {Up quark masses from down quark masses}},
  Phys. Lett. {\bf B642} (2006) 472--477
  [\href{http://xxx.lanl.gov/abs/hep-ph/0606097}{{\tt hep-ph/0606097}}].

\bibitem{Bazzocchi:2007na}
F.~Bazzocchi, S.~Kaneko and S.~Morisi, {\it {A SUSY $A_4$ model for fermion
  masses and mixings}},  JHEP {\bf 03} (2008) 063
  [\href{http://xxx.lanl.gov/abs/0707.3032}{{\tt arXiv:0707.3032}}].

\bibitem{Babu:2009nn}
K.~S. Babu and Y.~Meng, {\it {Flavor Violation in Supersymmetric $Q_6$ Model}},
   Phys. Rev. {\bf D80} (2009) 075003
  [\href{http://xxx.lanl.gov/abs/0907.4231}{{\tt arXiv:0907.4231}}].

\bibitem{Ishimori:2009ns}
H.~Ishimori, Y.~Shimizu and M.~Tanimoto, {\it {$S_4$ Flavor Model of Quarks and
  Leptons}},  Prog. Theor. Phys. Suppl. {\bf 180} (2010) 61--71
  [\href{http://xxx.lanl.gov/abs/0904.2450}{{\tt arXiv:0904.2450}}].

\bibitem{Morisi:2010rk}
S.~Morisi and E.~Peinado, {\it {An $S_4$ model for quarks and leptons with
  maximal atmospheric angle}},  [\href{http://xxx.lanl.gov/abs/1001.2265}{{\tt
  arXiv:1001.2265}}].

\bibitem{Schwetz:2008er}
T.~Schwetz, M.~A. Tortola and J.~W.~F. Valle, {\it {Three-flavour neutrino
  oscillation update}},  New J. Phys. {\bf 10} (2008) 113011
  [\href{http://xxx.lanl.gov/abs/0808.2016}{{\tt arXiv:0808.2016}}].

\bibitem{Varzielas:2008jm}
I.~d.~M. Varzielas, G.~G. Ross and M.~Serna, {\it {Quasi-degenerate neutrinos
  and tri-bi-maximal mixing}},  Phys. Rev. {\bf D80} (2009) 073002
  [\href{http://xxx.lanl.gov/abs/0811.2226}{{\tt arXiv:0811.2226}}].

\bibitem{Nelson:1983zb}
A.~E. Nelson, {\it {Naturally Weak CP Violation}},  Phys. Lett. {\bf B136}
  (1984) 387.

\bibitem{Barr:1984fh}
S.~M. Barr, {\it {A Natural Class of Non-Peccei--Quin Models}},  Phys. Rev.
  {\bf D30} (1984) 1805.

\bibitem{KlapdorKleingrothaus:2000sn}
H.~V. Klapdor-Kleingrothaus {\em et.~al.}, {\it {Latest Results from the
  Heidelberg-Moscow Double Beta Decay Experiment}},  Eur. Phys. J. {\bf A12}
  (2001) 147--154 [\href{http://xxx.lanl.gov/abs/hep-ph/0103062}{{\tt
  hep-ph/0103062}}].

\bibitem{Smolnikov:2008fu}
{\bf GERDA} Collaboration, A.~A. Smolnikov, {\it {Status of the GERDA
  experiment aimed to search for neutrinoless double beta decay of 76Ge}},
  [\href{http://xxx.lanl.gov/abs/0812.4194}{{\tt arXiv:0812.4194}}].

\bibitem{Xing:2009eg}
Z.-z. Xing, {\it {Right Unitarity Triangles, Stable CP-violating Phases and
  Approximate Quark-Lepton Complementarity}},  Phys. Lett. {\bf B679} (2009)
  111--117 [\href{http://xxx.lanl.gov/abs/0904.3172}{{\tt arXiv:0904.3172}}].

\bibitem{Hamermesh:1962gt}
M.~Hamermesh, {Group theory}.
\newblock Addison-Wesley, Reading, Massachussets (1962).

\end{thebibliography}\endgroup
